\documentclass[11pt,a4paper, notitlepage, twoside]{book}
\usepackage[utf8]{inputenc}
\usepackage[english]{babel} 
\usepackage{cite}

\usepackage{authblk}
\usepackage{verbatim}
\usepackage{emptypage}		
\usepackage[normal,swapnames,norules]{frontespizio}
\usepackage{amsmath}
\usepackage{tensor}
\usepackage{bm}
\usepackage{slashed}
\usepackage[makeroom]{cancel}
\usepackage{color}
\usepackage{float}
\usepackage{graphicx}
\usepackage[margin=1.3in]{geometry}
\usepackage[margin=10pt,font=small,labelfont=bf]{caption}
\usepackage{chngcntr}
\counterwithout{figure}{chapter}
\usepackage{mathtools}
\DeclareMathAlphabet{\mathbbold}{U}{bbold}{m}{n}
\usepackage{amssymb}
\usepackage{accents}
\usepackage{makeidx}
\usepackage{enumitem}
\usepackage{braket}
\usepackage{mathrsfs}
\usepackage{tikz}
\usepackage[compat=1.1.0]{tikz-feynman}

\usepackage{fancyhdr}
\usepackage{bookmark, hyperref}

\DeclarePairedDelimiter{\abs}{\lvert}{\rvert}
\DeclarePairedDelimiter{\Tprod}{T\Big(}{\Big)}
\DeclarePairedDelimiter{\bangle}{\langle}{\rangle}

\newcommand{\Rs}{R_{\text{s}}}
\newcommand{\Rc}{\bm{\mathcal{R}}}
\newcommand{\ocirc}{\omega_{\text{c}}}
\newcommand{\oGW}{\omega_{\text{GW}}}
\newcommand{\GN}{G_{\text{N}}}
\newcommand{\Mpl}{M_{\text{pl}}}
\newcommand{\mpl}{m_{\text{pl}}}
\newcommand{\Diff}{\mathcal{D}}
\newcommand{\Path}{\int\!\!\Diff}
\newcommand{\Lagr}{\mathcal{L}}
\newcommand{\DLagr}{\mathscr{L}}
\newcommand{\grad}{\vec{\nabla}}
\newcommand{\xc}{\mathtt{x}}
\newcommand{\Ord}[1]{\mathit{O}\left(#1\right)}
\newcommand{\ord}[1]{^{(#1)}}
\newcommand{\delf}{\delta^{(4)}}
\newcommand{\Seff}[1]{S_{\text{eff}}^{\text{#1}}}
\newcommand{\Tr}[1]{\text{Tr}\left(#1\right)}
\newcommand{\Imm}[1]{\mathrm{Im}\left(#1\right)}
\newcommand{\Ree}[1]{\mathrm{Re}\left(#1\right)}
\newcommand{\frtr}[2]{\frac{d^{#1}#2}{(2\pi)^{#1}}}
\newcommand{\intvec}[1]{\int_{\bm{#1}}\!}
\newcommand{\potH}[2]{H_{{\bf{#1}}#2}}
\newcommand{\potHin}[2]{H_{\bf{#1}}^{#2}}
\newcommand{\potHmi}[3]{H{^#2}_{\!\!\!\!\!{\bf{#1}}\,\,\,#3}}
\newcommand{\radh}[1]{\bar{h}_{#1}}
\newcommand{\radhin}[1]{\bar{h}^{#1}}
\newcommand{\radhmi}[2]{\bar{h}{^#1}_{#2}}
\newcommand{\modul}[1]{\left|\bm{#1}\right|}

\DeclareRobustCommand{\rchi}{{\mathpalette\irchi\relax}}
\newcommand{\irchi}[2]{\raisebox{\depth}{$#1\chi$}}

\newcommand{\uwidehat}[1]{\mathpalette\douwidehat{#1}}

\makeatletter
\newcommand{\douwidehat}[2]{%
  \sbox0{$\m@th#1\widehat{\hphantom{#2}}$}%
  \sbox2{$\m@th#1x$}
  \sbox4{$\m@th#1#2$}
  \dimen0=\ht0
  \advance\dimen0 -.8\ht2
  \dimen2=\dp4
  \rlap{%
    \raisebox{\dimexpr\dimen0-\dimen2}{%
      \scalebox{1}[-1]{\box0}%
    }%
  }%
  {#2}%
}
\makeatother

\Preambolo{\renewcommand{\frontsmallfont}[1]{\small N. }}

\title{M2 thesis: Effective Field Theory for Gravitational Radiation in General Relativity and beyond\footnote{This work has been completed during the first part of my PhD at Institut de Physique Théorique, Université Paris Saclay, CEA, CNRS.}}
\author[1,2]{\textbf{Candidate:} Massimiliano Maria Riva}
\author[1]{\ \\ \textbf{Supervisors:} Claudio Destri}
\author[2]{Filippo Vernizzi}
\affil[1]{Dipartimento di Fisica, Università di Milano - Bicocca, Piazza della Scienza 3, I-20126 Milano, Italy}
\affil[2]{Institut de Physique Théorique, Université Paris Saclay, CEA, CNRS
91191 Gif-sur-Yvette, France}
\date{\textbf{Academic Year:} 2018--2019}  

\begin{document}

\maketitle
\begin{center}
\section*{Abstract}
\end{center}
\begin{small}

The topic of this thesis is the so-called Non-Relativistic General Relativity, an effective field theory approach proposed by Goldberger and Rothstein to study the conservative and dissipative dynamics of binary systems of compact objects in the post-Newtonian expansion. 

In the first part of the thesis we review this approach in the simplest possible case: a binary of non-spinning black holes at leading post-Newtonian order, both for the conservative and dissipative sector in general relativity. 

In the second part of the thesis we present the so-called Kaluza-Klein parametrization of the metric. Introduced by Kol and Smolkin, it allows to simplify the computations in the conservative sector at higher order in the post-Newtonian expansion. We then extend this parametrization
to (conformally-coupled) scalar-tensor gravity. In particular, we derive in details the gravitational action in a generic metric frame, including for the first time a conformally coupled scalar field. Using this formalism, we study the conservative and dissipative dynamics of a binary system of bodies conformally coupled to the metric, at leading post-Newtonian order, and compare this with the standard general relativity results. 

This thesis is aimed at a reader that approaches the field for the first time. Hence, we attempted to be as pedagogical, explicit and self-contained as possible.

\end{small}
\vspace{\fill}

\thispagestyle{empty}

\tableofcontents


\chapter*{Introduction}
\addcontentsline{toc}{chapter}{Introduction}

Gravitational Waves (GWs) open a completely new way of observing and studying signals from the universe and, since their first detection in 2015 \cite{firstGW}, they had already proved to be even more telling than expected. The main sources of GWs are binary systems of compact objects, Black Holes (BHs) or Neutron Stars, that eventually merge into one single body. The motion of this system can be divided in three phases: the inspiral phase, the merger and the ring-down of the final object. The inspiral phase actually contains most of the signal of gravitational waves, thus, a comprehensive knowledge of this stage is crucial to fully understand the GWs signal detected on Earth.

During the inspiral phase, the motion of the two compact objects is non relativistic, i.e. their relative velocity is $v/c \ll 1$. Hence, the binary inspiral problem can be solved in the so-called Post-Newtonian approximation to General Relativity (PN). In this approximation one finds corrections to the Newtonian theory as a perturbative series in the parameter $v/c$. When we say that a term is of ``$n$PN order'', we mean that it is of order $(v/c)^{2n}$ beyond the Newtonian approximation. The PN expansion up to 4PN order has been largely studied in the literature, see for instance Refs. \cite{Blachet-rev,megaBlanchet} and references therein for a comprehensive description of the PN study state of the art.

In 2006 W. D. Goldberger and I. Z. Rothstein proposed a Lagrangian approach to the PN expansion, based on an Effective Field Theory (EFT) reasoning, that they called \textit{Non Relativistic General Relativity} (NRGR). This approach has been proved to be more familiar to a reader with a quantum field theory background and insightful to understand several conceptual aspects such as the renormalization of infinities discussed in the previous approaches. Starting from the description of two non spinning objects \cite{EFT1,Diss.Eff},  this EFT has also been extended to include spins \cite{Levi-Review,EFT3,Porto-spin}. Recently, known results up to 4PN order have been reproduced in this framework \cite{4PN-1,4PN-2,4PN-3}. 

This work is divided in two parts. In the first one, we review the NRGR formalism in the simplest case possible: a binary system of two non spinning black holes. Despite being simplified, this model allows us to understand the main fundamental features of this EFT approach to the binary inspiral problem. In particular, first we re-derive the well known Einstein-Infeld-Hoffman (EIH) Lagrangian, which is nothing but the 1PN correction to the conservative Newtonian motion of two point particles. 

Then, we study the radiative part of the system, computing  not only the leading-order gravitational waveform, but also the gravitational radiation reaction, i.e. the dissipative corrections to the conservative dynamics, due to the fact that the system is loosing energy through emission of GWs. Finally, we re-derive the famous quadrupole emission of GWs, and the first Next-to-Leading Order correction to the power lost by the system. 

In the last chapter of this part, we justify the point-particle approximation for the two compact objects, briefly presenting the effacement theorem, and showing that finite size effects first enter in the series at $5$PN order.

The purpose of the second part of this work is devoted to the extension of NRGR to conformally coupled scalar-tensor theories. 

We first re-derive the Einstein-Hilbert action using the so-called Non-Relativistic-Gravity fields (NRG fields), a useful parametrisation of the metric in the EFT approach to the binary inspiral problem, first introduced in 2008 \cite{KK1,KK2}. We explain the physical interpretation of these fields, and we show the advantages of such a parametrisation by computing again the EIH Lagrangian, and the lowest order dissipative action. 

Then, we extend the NRGR to a generic conformally coupled scalar-tensor theory, with a single massless scalar field. This generalization has been recently worked out in Ref. \cite{NRGR+scalar} in the Einstein frame, i.e. the frame where the kinetic term for the metric is the usual Einstein-Hilbert action, and in Ref. \cite{Mass.scalar} where NRGR has been used to study the effects of light axions. 

In this work we shall obtain equivalent results. However, to simplify the calculations we work with the NRG fields parametrization and we perform our study in the so-called Jordan frame, where test particles are minimally coupled to the gravitational metric. This allows us not only to re-check the same results of \cite{NRGR+scalar}, but it is also a fundamental starting point to eventually extend this EFT formalism to higher-order scalar-tensor theories, where the Einstein frame does not always exist, see for instance Refs. \cite{EFT_HOST,Galileon,BeyHord}. 

We then derive for the first time the action for a conformally coupled scalar-tensor theory written in terms of NRG fields, with the corresponding gauge-fixing and matter actions. From these, we derive the Feynman rules needed to perform 1PN order computations in the conservative sector, and 2.5PN order in the dissipative sector. Defining mixed state propagators between the scalar and gravitational fields, we finally show how the advantages of the NRG fields parametrisation apply also in this case, allowing us to reduce the number of diagrams to compute at 1PN order, compared to Ref. \cite{NRGR+scalar}.

\chapter*{Conventions}\label{Ch:convent}
\addcontentsline{toc}{chapter}{Conventions}  
\begin{itemize}
\item We work in natural units $\hbar = c = 1$, unless otherwise noted.
\item  We define the Planck mass as follows: the upper case  $\Mpl^{-2} \equiv 32\pi\GN$ and the lower case one $\mpl^{-2} \equiv 8\pi\GN$, where $\GN$ is the Newton constant.
\item We use Einstein's summation over repeated indices. Greek and Latin indices ranges are $\mu, \nu, \dots =0,1,2,3$ and $i,j,\dots=1,2,3$.
\item We used round and square brackets to respectively symmetrise and anti-symmetrise indices, e.g. 
\begin{align*}
A^{(\mu\nu)} \equiv \frac{1}{2}\left(A^{\mu\nu}+A^{\nu\mu}\right) \; , & & A^{[\mu\nu]} \equiv \frac{1}{2}\left(A^{\mu\nu}-A^{\nu\mu}\right) \; .
\end{align*}
If symmetrised indices are not contiguous, we put straight lines to highlight them, e.g. 
\begin{align*}
A^{(\mu|\sigma\rho|\nu)} = \frac{1}{2}\left(A^{\mu\sigma\rho\nu}+A^{\nu\sigma\rho\mu}\right) \; , & & A^{[\mu|\sigma\rho|\nu]} = \frac{1}{2}\left(A^{\mu\sigma\rho\nu}-A^{\nu\sigma\rho\mu}\right) \; .
\end{align*} 
\item We use mostly-minus convention for the metric, i.e. the four-dimensional Minkowski metric is given by $\eta_{\mu\nu} \nolinebreak =\nolinebreak \text{diag}(1,-1,-1,-1)$. A generic metric $g_{\mu\nu}$ then keeps the same signature. We define $g\equiv \det g_{\mu\nu}$.
\item Concerning forms and vectors:
\begin{itemize}
\item We use $\underline{w}$ to denote a form and $\overline{v}$ to denote a vector, to distinguish them from their components; for instance, in terms of the usual  orthonormal basis of forms $dx^\mu$ and vectors $\partial_\mu$, we can write $\underline{w} = w_\mu dx^\mu$ and $\overline{v} = v^\mu\partial_\mu$. 
\item We denote with $\wedge$ the antisymmetric product of forms, e.g. $dx^\mu\wedge dx^\nu = - dx^\nu\wedge dx^\mu$. A $p$-form is then expanded as
\begin{equation*}
{}\ord{p}\underline{w} = \frac{1}{p!}w_{\mu_1\mu_2\dots\mu_p}dx^{\mu_1}\wedge dx^{\mu_2}\wedge dx^{\mu_p} \; .
\end{equation*}
\item We denote the external derivative of a $p$-form as $d{}\ord{p}\underline{w}$. This is given by
\begin{equation*}
d{}\ord{p}\underline{w} = \frac{1}{p!}\partial_\mu w_{\mu_1\mu_2\dots\mu_p}dx^\mu dx^{\mu_1}\wedge dx^{\mu_2}\wedge dx^{\mu_p} \; .
\end{equation*}
\item We denote 3-vectors in boldface, e.g. $\bm{x}, \bm{y}, \dots$, while we use non boldface plus a Latin index to denote a component of the 3-vector, e.g. $x^i, y^i, \dots$
\item Sometimes we use the compact notation introduced by L. Blanchet and T. Damour in which an upper case Latin index $N$ denotes a set of spatial indices $i_1\dots i_n$, e.g. $x^N \equiv x^{i_1}\dots x^{i_n}$, $x^{ijN-2} = x^ix^jx^{k_1}\dots x^{k_{n-2}}$. 
\end{itemize}
\item If $\omega{^\rho}_{\nu\mu}$ is an affine connection, we define the covariant derivative as follows $\nabla_{\mu}v^{\rho} = \partial_\mu v^{\rho}+\omega{^\rho}_{\nu\mu}v^\nu$ and of course $\nabla_{\mu}v_{\nu} = \partial_\mu v_{\nu}-\omega{^\rho}_{\nu\mu}v_\rho$. We use $\Gamma{^\rho}_{\mu\nu}$ to denote the torsion-free Levi-Civita connection.
\item We define the Riemann tensor as $\Rc{^\rho}_{\mu\sigma\nu} \equiv 2\partial_{[\sigma|}\Gamma{^\rho}_{\mu|\nu]}+2\Gamma{^\rho}_{\lambda[\sigma|}\Gamma{^\lambda}_{\mu|\nu]}$. The Ricci tensor and scalar are then respectively defined as follows $\Rc_{\mu\nu} \equiv \Rc{^\sigma}_{\mu\sigma\nu}$, $\Rc \equiv \Rc_{\mu\nu}g^{\mu\nu}$. Finally, pulling down all indices for convenience, we defined the Weyl tensor in four dimensions as 
\begin{equation*}
\bm{\mathcal{C}}_{\rho\sigma\mu\nu} \equiv \Rc_{\rho\sigma\mu\nu} - \left(g_{\rho[\mu}\Rc_{\nu]\sigma}-g_{\sigma[\mu}\Rc_{\nu]\rho}\right) +\frac{1}{3}g_{\rho[\mu}g_{\nu]\sigma}\Rc \; .
\end{equation*}
\item We use the compact notation \[\int\!\!\frac{d^3\!\bm{k}}{(2\pi)^3}\frac{d^3\bm{q}}{(2\pi)^3}\dots\equiv\int_{\bm{k,q,}\dots}\]
\end{itemize}

\part{NRGR for non-spinning objects}

\chapter{NRGR for a binary of non-spinning black holes}\label{ch:first}

\section{A tower of EFT}

EFT approach is very useful whenever the problem we are facing has a clear separation of scales. This is indeed the case of the binary inspiral problem, where we can distinguish three different length scales:
\begin{itemize}
\item the size of the compact object $\Rs \equiv 2\GN m$
\item the orbital radius of the binary\footnote{Throughout this work $r$ and $v$ are the only cases (unless otherwise noted) where a non bold lower case letter without indices denotes the modulus of a three vector, namely $r\equiv|\bm{r}|$, $v\equiv|\bm{v}|$.} $r$
\item the wavelength of the emitted radiation $\lambda$.
\end{itemize}
These three parameters are not independent, but they are in fact interconnected by means of the relative velocity of the system $v$, which is small during the inspiral phase, i.e. $v\ll 1$. Indeed, for gravitationally bound systems, one can use the virial theorem to roughly estimate
\begin{equation}
\frac{2\GN m}{r} =\frac{\Rs}{r} \sim v^2 \; .
\label{eq:rough.1}
\end{equation}
Moreover, due also to the sensitivity of the detector, the orbit of the binary is considered circular with frequency $\ocirc$. One can again roughly estimate\footnote{We are using the fact that the gravitational wave frequency $\oGW$ is proportional to $\ocirc$.}
\begin{equation}
\frac{1}{\lambda}=\oGW\sim \frac{v}{r} = \ocirc \; .
\label{eq:rough.2}
\end{equation}
Because of this link, we understand that the various orders of the PN expansion contain informations about the physics at different scales. From eqs. (\ref{eq:rough.1}) and (\ref{eq:rough.2}) it is also evident that the three scales respect the hierarchy $\Rs \ll r \ll \lambda$. In an EFT language we can say that they are \textit{decoupled}, and we can then treat them separately. This is realised by \textit{integrating out} one after the other the scales we can no longer distinguish.  

More precisely, we built a tower of three different effective theories in the following way:
\begin{enumerate}
\item First, we remove from the theory the size of the object $\Rs$. After this operation, we find an effective action $\Seff{}$ describing the motion of two point particles interacting via gravity only, plus terms containing finite size effects.
\item Then, we integrate out the orbit scale $r$, i.e. we remove from the theory the gravitational interaction between the two point particles. We then have a new effective action $\Seff{NR}$ describing a single point particle with suitable gravitational interactions.
\item Finally, we remove from the theory also the long scale $\lambda$, finding in this way our last effective action $\Seff{NRGR}$, which is in general a complex quantity. Looking at its real part we can find the Equation Of Motion (EOM) of the binary during the inspiral phase, while, by means of the optical theorem, the imaginary part of $\Seff{NRGR}$ gives us informations about the energy flux of the system\footnote{Here we will not treat other very important features of the theory like tail and dissipative effects, and the study of the renormalization procedure. We direct the reader to \cite{Diss.Eff,Tail.eff.,Rad-Corr,EFT3} and references therein.}. 
\end{enumerate}

\subsection{Brief review on EFT}\label{sec:EFT_rev}

Let's briefly review the main features of an effective approach to a theory. We will consider for simplicity a scalar case, because these steps are completely general and can be applied to any theory.

\subsubsection{Top-down VS Bottom-up}

Consider a scalar quantum theory in 4 dimensions for a field $\phi$, described by the action $S[\phi]$. Suppose we are interested in investigating only an energy regime $E\simeq\vartheta$, then, it is not necessary to keep track of the whole theory. In fact, it is convenient to split the field as $\phi =\Phi+\varphi$, such that: 
\begin{itemize}
\item $\varphi$ are light (i.e. long-distances) Degrees of Freedom (DOF) with a typical energy scale $\vartheta$
\item $\Phi$ are heavy (i.e. short-distances) DOF with a typical energy scale $\Lambda \gg \vartheta$.
\end{itemize}
$\Phi$ are too heavy to propagate as physical DOF, hence they will always be off-shell. Therefore, it is much more convenient to use an EFT approach, i.e. one can integrate out $\Phi$ and find an effective description of the theory as follows
\begin{equation}
e^{i\Seff{}[\varphi]}\equiv\Path\Phi\,e^{iS[\varphi, \Phi]} \; .
\end{equation}
If we are able to perform (usually perturbatively) this path integral, then the result can be in general written as
\begin{equation}
\Seff{}[\varphi] =\int\!\!d^4\!x\left\{\frac{1}{2}\partial_\mu\varphi\partial^\mu\varphi+\sum_n C_n\mathcal{O}_n(\varphi)\right\} \; .
\label{eq:first.EA}
\end{equation}
This is the so-called \textit{Top-down} approach first developed by Wilson. $\mathcal{O}_n$ are local operators; we denote their mass dimension as $\Delta_n$. The coefficients $C_n$ are call \textit{Wilson coefficients}, and they contain all the informations about the UV (i.e. high energies) behaviour of the theory. For dimensional reasons, we can say that they scale as
\begin{equation}
C_n \sim \frac{1}{\Lambda^{\Delta_n-4}} \; .
\end{equation}

There is another possible way of facing computations in an EFT: the so-called \textit{Bottom-up} approach. In this approach, one starts by more general considerations on the theory, namely symmetries and other properties of the long-distance physics, and writes an effective action as in (\ref{eq:first.EA}), where $\mathcal{O}$ now are the most general operators that respect all the properties mentioned earlier. In this case, however, the Wilson coefficients are completely general and unfixed. One needs to implement a \textit{matching procedure} in order to find an explicit expression. This can be done in two ways:
either comparing results with experimental data, or performing top-down computations and adjusting each $C_n$ in order to have coherent results. Therefore, whenever possible, it is convenient to apply both approaches in parallel. As we will see, this is actually the case of NRGR.

As we said, one can compute $\Seff{}$ perturbatively in the small parameter $\vartheta/\Lambda$, using for instance Feynman diagram technique. How many operators $\mathcal{O}_n$ do we have to include in our theory? In any EFT we always have to establish a \textit{power counting} procedure; this allows us to understand immediately which operators contribute at a certain perturbative order. First, we need to understand how quantities scale w.r.t. the perturbative parameter. In this case it is straightforward because
\begin{subequations}
\label{eqn:scheme1}
\begin{itemize}
\item First, $\forall \mu=0,\dots,3$
\begin{equation}
x^\mu\sim\vartheta^{-1} \; .
\end{equation}
\item Once we Fourier transform, $\partial_\mu\varphi\rightarrow k_\mu\varphi$, we can say $\forall \mu=0,\dots,3$
\begin{align}
k_\mu\sim \vartheta \longrightarrow \partial_\mu\varphi\sim \vartheta\varphi \; .
\end{align}
\item Finally, from the kinetic term, it is easy to see that, denoting with $\lfloor\bullet\rfloor$ the mass dimension of an object, $\lfloor\varphi\rfloor = 1$. Therefore we can say
\begin{equation}
\varphi\sim\vartheta \; .
\end{equation}
\end{itemize}
\end{subequations}
Given this set of simple rules, we can understand the scaling of any operator, for instance, we can see that
\begin{equation}
C_8\int\!\!d^4\!x\left(\partial_\mu\varphi\partial^\mu\varphi\right)^2\sim\left(\frac{\vartheta}{\Lambda}\right)^4 \; .
\end{equation}
The above equation means that this operator starts to be relevant at fourth perturbative order in the expansion.

\subsubsection{Classical and Quantum contributions}

For a generic scalar quantum theory described by an action $S[\phi]$, one can define the generating functional $Z[J]$, the generating connected functional $W[J]$, and the quantum effective action $\Gamma[\phi_{\text{cl}}]$ as
\begin{subequations}
\label{eqn:scheme2}
\begin{align}
Z[J] & = \Path\phi\,e^{iS[\phi]+i\int\!\!d^4\!x\,J(x)\phi(x)} \; , \\
e^{iW[J]} & = Z[J] ,\\
\Gamma[\phi_{\text{cl}}] & = W[J]-\int\!\!d^4\!x\,J\phi_{\text{cl}} \; ,
\end{align}
\end{subequations}
where $\phi_{\text{cl}}$ is defined as $\phi_{\text{cl}} =\delta W[J]/\delta J$. The quantum effective action is actually the most fundamental object of a QFT because:
\begin{itemize}
\item from its real part we can compute the EOM of the system
\item from its imaginary part, through the optical theorem, we can compute all the cross sections we want.
\end{itemize}
Again, $\Gamma$ can be computed perturbatively using Feynman diagrams. Schematically we can say
\begin{equation}
i\Gamma =\sum\begin{smallmatrix}\text{Tree-level}\\\text{diagrams}\end{smallmatrix}+\sum\begin{smallmatrix}\text{Loop}\\\text{diagrams}\end{smallmatrix} = i\Seff{cl}+i\Seff{Q} \; ,
\end{equation}
where:
\begin{itemize}
\item $\Seff{cl}$ comes from the sum of all tree level diagrams and is responsible for the description of the theory at classical (i.e. non quantum) level.
\item $\Seff{Q}$ comes from the sum of all loop diagrams and is responsible for the quantum corrections to the classical theory\footnote{Actually this distinction is no longer so clear when using on-shell scattering amplitudes for this kind of computations. See for instance Refs. \cite{Bern-Ampl,Cheung-Ampl,Kosower-Ampl}.}.
\end{itemize}

In NRGR we do a non relativistic expansion (PN expansion) of what above we called classical effective action $\Seff{cl}$.  It is known that quantum corrections come as a series in power of the (even smaller) parameter $\hbar$, therefore, in what follows \textit{we will not consider quantum corrections}, because they are for sure subleading w.r.t. PN corrections. 

\section{The NRGR machinery}

Let's now enter in the description of NRGR. As in every Lagrangian theory, the starting point is the construction of the action. In what follows we stop our computations at 1PN order. As we will see in Ch. \ref{ch:pp}, finite size effects start being relevant at order 5PN. Therefore, at this level, we can consider the two compact objects as point particles, thus
\begin{align}
\Seff{}[g_{\mu\nu},x^\mu_a] & = S_{\text{EH}}[g_{\mu\nu}]+S_{\text{pp}}[g_{\mu\nu},x_a^\mu] \notag \\
& = -2\Mpl^2\int\!\!d^4\!x\sqrt{-g}\Rc \,-\sum_{a=1}^2m_a\int\!\!\sqrt{g_{\mu\nu}(x_a)dx_a^\mu dx_a^\nu} \; .
\label{eq:Sstart}
\end{align}
This is nothing but the action describing two point particles ($a=1,2$) minimally coupled to a gravitational field $g_{\mu\nu}$. As we said in the introduction, we would like to do a non-relativistic study of this system, because, during the inspiral phase, the two objects have small relative velocity.

\subsection{The problems of covariant decomposition}

As usual one can think to treat this system considering small perturbations $h_{\mu\nu}$ around a flat background, i.e.
\begin{equation}
g_{\mu\nu} = \eta_{\mu\nu}+\frac{h_{\mu\nu}}{\Mpl} \; ,
\label{eq:cov.dec}
\end{equation}
where we normalize $h_{\mu\nu}$ in order to have $\lfloor h \rfloor=1$. Plugging this definition in (\ref{eq:Sstart}) one schematically obtains
\begin{subequations}
\label{eqn:scheme12}
\begin{align}
S_{\text{EH}}[h_{\mu\nu}]& =\int\!\!d^4\!x\bigg\{\left(\partial h\right)^2+\frac{h(\partial h)^2}{\Mpl}+\frac{h^2(\partial h)^2}{\Mpl^2}+\dots\Big\} \notag \\
& = \left(\begin{tikzpicture}[baseline]
\begin{feynman}
\vertex (z);
\vertex [below=0.1cm of z] (f);
\vertex [above=0.1cm of z] (f');
\vertex [right=1cm of f'] (c);
\diagram* {
(f') -- [photon] (c),
(f') -- [draw=none] (z) -- [draw=none] (f)
}; 
\end{feynman} 
\end{tikzpicture}\right)^{-1} +
\begin{tikzpicture}[baseline]
\begin{feynman}
\vertex (z);
\vertex [below=0.1cm of z] (f);
\vertex [above=0.1cm of z] (f');
\vertex [right=0.8cm of f'] (c);
\vertex [above left=0.8cm of f'] (a);
\vertex [below left=0.8cm of f'] (b);
\diagram* {
(a) -- [photon] (f') -- [photon] (c),
(b) -- [photon] (f'),
(f') -- [draw=none] (z) -- [draw=none] (f)
}; 
\end{feynman} 
\end{tikzpicture}+
\begin{tikzpicture}[baseline]
\begin{feynman}
\vertex (z);
\vertex [below=0.1cm of z] (f);
\vertex [above=0.1cm of z] (f');
\vertex [above right=0.8cm of f'] (c);
\vertex [below right=0.8cm of f'] (d);
\vertex [above left=0.8cm of f'] (a);
\vertex [below left=0.8cm of f'] (b);
\diagram* {
(a) -- [photon] (f') -- [photon] (c),
(b) -- [photon] (f') -- [photon] (d),
(f') -- [draw=none] (z) -- [draw=none] (f)
}; 
\end{feynman} 
\end{tikzpicture} + \dots 
\end{align}
\begin{align}
S_{\text{pp}}[h_{\mu\nu},x_a^\mu]& =-m_a\int\!\!d \bar{\tau}_a-\frac{m_a}{\Mpl}\int\!\!d \bar{\tau}_a \left(u_a\right)^2 \!\!h+\frac{m_a}{\Mpl^2}\int\!\!d \bar{\tau}_a\left[\left(u_a\right)^2 \!\!h\right]^2+\dots \notag \\
& = \begin{tikzpicture}[baseline]
\begin{feynman}
\vertex (z);
\vertex [below=0.4cm of z] (f);
\vertex [above=0.6cm of z] (f');
\diagram* {
(f') -- [plain] (f)
}; 
\end{feynman} 
\end{tikzpicture} \ +\ 
\begin{tikzpicture}[baseline]
\begin{feynman}
\vertex (z);
\vertex [above=0.1cm of z] (f);
\vertex [right=0.6cm of f] (c);
\vertex [above=0.6cm of z] (a);
\vertex [below=0.4cm of z] (b);
\diagram* {
(a) -- [plain] (f) -- [photon] (c),
(b) -- [plain] (f)
}; 
\end{feynman} 
\end{tikzpicture}\ +\ 
\begin{tikzpicture}[baseline]
\begin{feynman}
\vertex (z);
\vertex [above=0.1cm of z] (f);
\vertex [above right=0.6cm of f] (c);
\vertex [below right=0.6cm of f] (d);
\vertex [above=0.6cm of z] (a);
\vertex [below=0.4cm of z] (b);
\diagram* {
(a) -- [plain] (f) -- [photon] (c),
(b) -- [plain] (f) -- [photon] (d)
}; 
\end{feynman} 
\end{tikzpicture}\  + \  \dots \;\; ,
\end{align}
\end{subequations}
where $d\bar{\tau}^2_a\equiv\eta_{\mu\nu}dx_a^\mu dx_a^\nu$, $u^\mu_a=dx^\mu_a/d\bar{\tau}_a$, and we used wavy and straight lines to represent respectively the graviton and one of the point particles. In App. \ref{App:action}, in particular in Secs. \ref{sec:EHflatb} and \ref{sec:PPflatb}, we perform explicitly the expansion of the Einstein-Hilbert action and the point-particle one. 

As we said the massive point particles are moving non relativistically, hence they are characterised by a momentum $\modul{p}\sim mv$. On the other hand, reinserting $\hbar$ for convenience, we can say that the graviton has a momentum $\modul{k}\sim\hbar/r$. Hence, when one of the compact objects emits a graviton, the non relativistic particle recoils of roughly
\begin{equation}
\frac{|\Delta\bm{p}|}{\modul{p}}\simeq\frac{\modul{k}}{\modul{p}}\sim\frac{\hbar}{L} \ll 1\; ,
\end{equation}
where $L=mvr$ is the modulus of the orbital angular momentum. Thus, as far as the graviton dynamics is concerned, \textit{we can consider the two compact objects as static background sources of GWs}.

The main object we need to compute is the effective action of the theory; this can be computing using Feynman diagrams, namely 
\begin{equation}
i\Seff{cov} = \begin{tikzpicture}[baseline]
\begin{feynman}
\vertex (z);
\vertex [above=0.6cm of z] (c');
\vertex [below=0.4cm of z] (c);
\vertex [right=0.8cm of c] (a);
\vertex [left=0.8cm of c] (b);
\vertex [right=0.8cm of c'] (a');
\vertex [left=0.8cm of c'] (b');
\diagram* {
(a) -- [plain] (c) -- [plain] (b),
(a') -- [plain] (c') -- [plain] (b'),
(c) -- [photon] (c')
}; 
\end{feynman} 
\end{tikzpicture}+
\begin{tikzpicture}[baseline]
\begin{feynman}
\vertex (z);
\vertex [above=0.6cm of z] (c');
\vertex [below=0.4cm of z] (c);
\vertex [right=0.8cm of c] (a);
\vertex [left=0.8cm of c] (b);
\vertex [right=0.8cm of c'] (a');
\vertex [left=0.8cm of c'] (b');
\vertex [right=0.4cm of c] (a'');
\vertex [left=0.4cm of c] (b'');
\diagram* {
(a) -- [plain] (c) -- [plain] (b),
(a') -- [plain] (c') -- [plain] (b'),
(c') -- [photon] (a''),
(c') -- [photon] (b'')
}; 
\end{feynman} 
\end{tikzpicture}+
\begin{tikzpicture}[baseline]
\begin{feynman}
\vertex (z);
\vertex [above=0.1cm of z] (f);
\vertex [above=0.6cm of z] (c');
\vertex [below=0.4cm of z] (c);
\vertex [right=0.8cm of c] (a);
\vertex [left=0.8cm of c] (b);
\vertex [right=0.8cm of c'] (a');
\vertex [left=0.8cm of c'] (b');
\vertex [right=0.4cm of c] (a'');
\vertex [left=0.4cm of c] (b'');
\diagram* {
(a) -- [plain] (c) -- [plain] (b),
(a') -- [plain] (c') -- [plain] (b'),
(c') -- [photon] (f) -- [photon] (a''),
(f) -- [photon] (b''),
}; 
\end{feynman} 
\end{tikzpicture}\ +\ \dots \;\; .
\label{eq:first_act_diag}
\end{equation}
As we discussed in general in the introduction\footnote{See the paragraph "\textbf{Classical an Quantum contributions}" at p.~\pageref{eqn:scheme2}. See also standard QFT reference as \cite{Srednicki,Itsy}.}, we do not consider diagrams containing closed graviton loops, because these would scale as
\begin{equation}
\begin{tikzpicture}[baseline]
\begin{feynman}
\vertex (z');
\vertex [above = 0.1cm of z'] (z);
\vertex [above=0.4cm of z] (y);
\vertex [below=0.4cm of z] (x);
\vertex [above=0.5cm of y] (c);
\vertex [right=0.8cm of c] (a);
\vertex [left=0.8cm of c] (b);
\vertex [below=0.5cm of x] (c');
\vertex [right=0.8cm of c'] (a');
\vertex [left=0.8cm of c'] (b');
\diagram* {
(a) -- [plain] (c) -- [plain] (b),
(c) -- [photon] (y) -- [photon, half right] (x) -- [photon] (c'),
(y) -- [photon, half left] (x),
(a') -- [plain] (c') -- [plain] (b'),
}; 
\end{feynman} 
\end{tikzpicture} \sim \frac{\hbar}{L} \; .
\end{equation}
Clearly $\hbar/L \ll v$, hence these series of diagrams are for sure subleading w.r.t. to the PN expansion.

Let's try to compute the first diagram of (\ref{eq:first_act_diag}). We need the following Feynman rules
\begin{subequations}
\label{eqn:scheme13}
\begin{align}
\begin{tikzpicture}[baseline]
\begin{feynman}
\vertex (z);
\vertex [above=0.1cm of z,label=180:$\bar{\tau}_a$, label=45:$\mu\nu$] (f);
\vertex [right=1cm of f] (c);
\vertex [above=0.8cm of z] (a);
\vertex [below=0.6cm of z] (b);
\diagram* {
(a) -- [plain] (f) -- [photon] (c),
(b) -- [plain] (f)
}; 
\end{feynman} 
\end{tikzpicture} & = -i\frac{m_a}{2\Mpl}\int\!\!d \bar{\tau}_a\,u^\mu_au^\nu_a \; ,\label{eq:feyrulecov1}\\
\begin{tikzpicture}[baseline]
\begin{feynman}
\vertex (z);
\vertex [above=0.05cm, dot, label=90:$y$, label=270:$\rho\sigma$] (f) {};
\vertex [dot, label=90:$x$, right=1.5cm of f, label=270:$\mu\nu$] (w) {}; 
\diagram* {
(f) -- [draw=none] (z),
(f) -- [photon, edge label=$k$] (w)
}; 
\end{feynman} 
\end{tikzpicture} & = D(x-y)P_{\mu\nu\rho\sigma} \; ,\label{eq:feyrulecov2}
\end{align}
where
\begin{align}
D(x-y)\equiv \int\!\!\frtr{4}{k}\frac{i e^{-ik(x-y)}}{k^2+i\varepsilon} \; , & & P_{\mu\nu\rho\sigma} \equiv \frac{1}{2}(\eta_{\mu\rho}\eta_{\nu\sigma}+\eta_{\mu\sigma}\eta_{\nu\rho}-\eta_{\mu\nu}\eta_{\rho\sigma}) \; .
\label{eq:DandP}
\end{align}
\end{subequations}
Eq. (\ref{eq:feyrulecov1}) can be found by simply looking at the second term of (\ref{eq:Spp1}). Eq. (\ref{eq:feyrulecov2}) instead can be found directly from eq. (\ref{eq:quad.hbar}) replacing every $\bar{h}$ with $h$. Thus we obtain\footnote{Here we consider only the case $a\neq b$ in order to avoid self interactions which would lead to divergences that can be renormalized as usual, see Sec. \ref{sec:diverg}.} 
\begin{align}
\begin{tikzpicture}[baseline]
\begin{feynman}
\vertex [label=270:$\bar{\tau}_b$, label=20:$\rho\sigma$] (c);
\vertex [above=0.6cm of c] (z);
\vertex [above=0.8cm of z, label=90:$\bar{\tau}_a$, label=290:$\mu\nu$] (c');
\vertex [right=0.8cm of c] (a);
\vertex [left=0.8cm of c] (b);
\vertex [right=0.8cm of c'] (a');
\vertex [left=0.8cm of c'] (b');
\diagram* {
(a) -- [plain] (c) -- [plain] (b),
(a') -- [plain] (c') -- [plain] (b'),
(c) -- [photon] (c')
}; 
\end{feynman} 
\end{tikzpicture} & = \frac{(-i)^2}{2}\sum_{a\neq b}\frac{m_am_b}{4\Mpl^2}\int\!\!d \bar{\tau}_ad\bar{\tau}_bu^\mu_au^\nu_aP_{\mu\nu\rho\sigma}u^\rho_b u^\sigma_b D(x_a-x_b) = \notag \\
& = \frac{1}{2}\sum_{a\neq b}\frac{m_am_b}{8\Mpl^2}\int\!\!d \bar{\tau}_ad\bar{\tau}_b\Big(1-2(u_a\cdot u_{b})^2\Big) D(x_a-x_b) \; ,
\label{eq:covcomp}
\end{align}
where $u_a\cdot u_{b}=\eta_{\mu\nu}u_a^\mu u_b^\nu$, and we used the fact that, since we are using the proper time $d\bar{\tau}$ to parametrise everything, $\eta_{\mu\nu}u_a^\mu u_a^\nu=1$. The $1/2$ in front of everything is the symmetry factor of the diagrams

At this point we can ask ourself: at which perturbative order does this term contribute? We recall that we want to do a PN study of GR, hence we would like to have a perturbative series in power of the relative velocity $v$. However, it is obviously not possible to assign a unique power of $v$ to eq. (\ref{eq:covcomp}), mainly because the rules (\ref{eq:feyrulecov1}) and (\ref{eq:feyrulecov2}) are manifestly Lorentz covariant, while $v$ is obviously a non covariant parameter. Therefore, we need to implement another mechanism in order to better study the PN expansion using Feynman diagrams. As we shall see in the next few sections, the way to do this is to explicitly separate the scales $r$ and $\lambda$ already at the level of the graviton field. 

\subsection{Modes decomposition}\label{sec:mod.dec}

As we mentioned in the introduction, the binary problem can be separated in three regions, characterised by three length scales, $\Rs$, $r$ and $\lambda$, such that
\begin{equation}
\Rs \ll r \ll \lambda \; .
\end{equation}
The quantity $\Rs$ has already been integrated out, since we are considering point-like sources. In order to decouple the orbital and the radiation scales, we consider again eq. (\ref{eq:cov.dec}) and split the field $h_{\mu\nu}$ as 
\begin{equation}
h_{\mu\nu} = \potH{}{\mu\nu}+\radh{\mu\nu} \; ,
\label{eq:mode.dec}
\end{equation}
where:
\begin{itemize}
\item $\potH{}{\mu\nu}$ is responsible for the short-distance physics, namely the potential that keeps the binary  in a bound state. Therefore, it acts at the typical orbital scale $r$.
\item $\radh{\mu\nu}$ is responsible for the long-distance physics, namely the GWs radiated by the system. Hence, its typical scale is of course $\lambda$.
\end{itemize} 
The potential modes can never be detected, which implies that they are off-shell modes characterised by a momentum
\begin{align}
k_H^0\sim\frac{v}{r}\; , & & |\bm{k}_H|\sim\frac{1}{r} \quad \longrightarrow \quad \partial_0 \potH{}{\mu\nu}\sim\frac{v}{r}\potH{}{\mu\nu} \; , & &   \partial_i \potH{}{\mu\nu}\sim \frac{1}{r}\potH{}{\mu\nu} \; .
\end{align}
On the other hand, $\radh{\mu\nu}$ modes represent the GW signal we can detect from Earth, therefore they can be on-shell modes, i.e. such that
\begin{align}
k_{\radh{}}^0\sim\frac{v}{r} \; , \qquad |\bm{k}_{\radh{}}|\sim\frac{v}{r} \quad \longrightarrow \quad \partial_\rho \radh{\mu\nu}\sim\frac{v}{r}\radh{\mu\nu} \; .
\label{eq:kradh.scal}
\end{align}
Actually, it is convenient to work with the partially Fourier transformed field $\potH{k}{\mu\nu}$ which is such that
\begin{equation}
\potH{k}{\mu\nu} \longrightarrow \potH{}{\mu\nu}(x^0,\bm{x})=\intvec{k}e^{i\bm{k}\cdot\bm{x}}\potH{k}{\mu\nu}(x^0) \; .
\label{eq:Pot.Fourier}
\end{equation}
In this way we remove the large fluctuations $1/r$ coming from spatial derivatives of $\potH{}{}$, so that now all derivatives acting on any field of our theory scale in the same way.

Eventually we need to compute diagrams in this decomposition, thus, we expect the graviton propagator to be split as follows:
\begin{equation}
\begin{tikzpicture}[baseline]
\begin{feynman}
\vertex (z);
\vertex [below=0.1cm of z] (f);
\vertex [above=0.1cm of z] (f');
\vertex [right=1cm of f'] (c);
\diagram* {
(f') -- [photon] (c),
(f') -- [draw=none] (z) -- [draw=none] (f)
}; 
\end{feynman} 
\end{tikzpicture} \longrightarrow \begin{tikzpicture}[baseline]
\begin{feynman}
\vertex (z);
\vertex [below=0.1cm of z] (f);
\vertex [above=0.1cm of z] (f');
\vertex [right=1cm of f'] (c);
\diagram* {
(f') -- [scalar] (c),
(f') -- [draw=none] (z) -- [draw=none] (f)
}; 
\end{feynman} 
\end{tikzpicture} +
\begin{tikzpicture}[baseline]
\begin{feynman}
\vertex (z);
\vertex [below=0.1cm of z] (f);
\vertex [above=0.1cm of z] (f');
\vertex [right=1cm of f'] (c);
\diagram* {
(f') -- [gluon] (c),
(f') -- [draw=none] (z) -- [draw=none] (f)
}; 
\end{feynman} 
\end{tikzpicture} \;\; ,
\end{equation}
where we used a dashed line to represent potential modes $\potH{k}{\mu\nu}$, and a wiggle one for the radiation modes $\radh{\mu\nu}$. Let's look again at eq. (\ref{eq:feyrulecov2}); using the well known Plemelj-Sokhotski formula, one can see that
\begin{equation}
\frac{1}{(k^0)^2-\modul{k}^2+i\varepsilon} = P\left(\frac{1}{(k^0)^2-\modul{k}^2}\right)-i\pi\delta\Big((k^0)^2-\modul{k}^2\Big) \; .
\label{eq:Pl.Sok}
\end{equation}
Therefore we understand that, being off-shell i.e. $(k^0)^2\neq\modul{k}^2$, potential modes can give only real contributions. On the other hand, radiation modes give us in general complex quantities.

We also expect that the action (\ref{eq:Sstart}) will be decomposed in:
\begin{itemize}
\item a part depending only on the potential modes $S_{H}$, and one depending only on the radiation modes $S_{\radh{}}$
\item a part parametrising the non linear interactions between potential and radiation modes $S_{H\radh{}}$
\item a part representing the free propagation of the point particles $S^{0}_{\text{pp}}$
\item a part parametrising the interactions of the point particles with: the potential modes $S^{H}_{\text{pp}}$, the radiation modes $S^{\radh{}}_{\text{pp}}$ and the non linear interactions with both the potential and radiation modes $S^{H\radh{}}_{\text{pp}}$ .
\end{itemize}
See App. \ref{App:action} for an explicit derivation of all these actions. 

Now we can start the EFT procedure. We are going to integrate out the potential modes $\potH{k}{\mu\nu}$ first, hence
\begin{equation}
e^{i\Seff{NR}} = e^{i(S_{\radh{}}+S^{\radh{}}_{\text{pp}}+S^{0}_{\text{pp}})}\Path\potH{k}{\mu\nu}e^{i(S_{H}+S_{H\radh{}}+S^{\text{GF}}_H+S^{H}_{\text{pp}}+S^{H\radh{}}_{\text{pp}})} \; ,
\end{equation}
where we added to the path integral a gauge-fixing term, but no ghost fields because, as we said already, we consider a classical theory, hence no quantum loop diagrams. The gauge-fixing term is chosen in such a way that the final action is invariant under general diffeomorphisms of the background metric $\bar{g}_{\mu\nu}=\eta_{\mu\nu}+\radh{\mu\nu}/\Mpl$. Choosing to work in the so-called harmonic gauge, one has
\begin{align}
S^{\text{GF}}_{H}[\potH{}{\mu\nu},\radh{\mu\nu}] = \Mpl^2\int\!\!d^4\!x\sqrt{-\bar{g}}\bar{g}_{\mu\nu}\Gamma^\mu\Gamma^\nu \; , 
\label{eq:SGFpot}
\end{align}
where, calling $\nabla_\mu$ the derivative compatible with the background metric,
\begin{equation}
\Gamma_{\mu} \equiv\frac{1}{\Mpl}\bar{g}^{\rho\sigma}\left(\nabla_{\sigma}\potH{}{\rho\mu}-\frac{1}{2}\nabla_\mu\potH{}{\rho\sigma}\right) \; .
\label{eq:SGFpot2}
\end{equation}
The action $\Seff{NR}$ can be computed as the sum of all (tree-level) Feynman diagrams having $x_a^\mu$ and $\radh{\mu\nu}$ as external legs, and $\potH{k}{\mu\nu}$ as internal legs only. The resulting action will have schematically the following form
\begin{equation}
\Seff{NR}[\radh{\mu\nu},x^\mu_a] = S_\text{cons}[x_a^\mu]+S_{\radh{}}[\radh{\mu\nu}]+S_\text{diss}[\radh{\mu\nu},x^\mu_a] \; ,
\label{eq:SeffNR}
\end{equation}
where, as we said earlier, $S_{\radh{}}$ is the action depending only on $\radh{\mu\nu}$. In particular from this action we will extract the propagator of the radiation modes. For the other two terms, if we define our diagrammatic conventions as follows
\begin{align*}
& \begin{tikzpicture}[baseline]
\begin{feynman}
\vertex (f);
\vertex [above=0.1cm of f] (z);
\vertex [right=1.5cm of z] (w);  
\diagram* {
(z) -- [scalar] (w)
}; 
\end{feynman} 
\end{tikzpicture} \rightarrow \text {Potential mode} & & 
\begin{tikzpicture}[baseline]
\begin{feynman}
\vertex (f);
\vertex [above=0.1cm of f] (z);
\vertex [right=1.5cm of z] (w);  
\diagram* {
(z) -- [gluon] (w)
}; 
\end{feynman} 
\end{tikzpicture} \rightarrow \text{Radiation mode} \\
& \begin{tikzpicture}[baseline]
\begin{feynman}
\vertex (f);
\vertex [above=0.1cm of f] (z);
\vertex [right=1.5cm of z] (w);  
\diagram* {
(z) -- [plain] (w)
}; 
\end{feynman} 
\end{tikzpicture} \rightarrow \text{Extenal particle} & & 
\begin{tikzpicture}[baseline]
\begin{feynman}
\vertex (f);
\vertex [above=0.1cm of f] (z);
\vertex [empty dot, minimum size=0.4cm, right=0.65cm of z] (p) {$n$}; 
\vertex [right=0.75cm of p] (w); 
\diagram* {
(z) -- [plain] (p) -- [plain] (w)
}; 
\end{feynman} 
\end{tikzpicture} \rightarrow \text{Insertion of a } v^n \text{ factor} \; ,
\end{align*}
then we can write 
\begin{subequations}
\label{eqn:scheme14}
\begin{align}
iS_\text{cons}& = \left(\sum_{a=1}^2\sum_{n=0}^{\infty}\begin{tikzpicture}[baseline]
\begin{feynman}
\vertex (z);
\vertex [below=0.1cm of z] (f);
\vertex [above=0.1cm of z] (f');
\vertex [right=0.5cm of f', empty dot, minimum size=0.4cm] (c) {n};
\vertex [right=0.7cm of c] (d);
\diagram* {
(f') -- [plain] (c) -- [plain] (d),
(f') -- [draw=none] (z) -- [draw=none] (f)
}; 
\end{feynman} 
\end{tikzpicture}\right) +\begin{tikzpicture}[baseline]
\begin{feynman}
\vertex [blob, minimum size=0.7cm] (z) {};
\vertex [above=0.9cm of z, blob, minimum size=0.3cm] (c') {};
\vertex [below=0.9cm of z, blob, minimum size=0.3cm] (c) {};
\vertex [right=1.2cm of c] (a);
\vertex [left=1.2cm of c] (b);
\vertex [right=1.2cm of c'] (a');
\vertex [left=1.2cm of c'] (b');
\diagram* {
(a) -- [plain] (c) -- [plain] (b),
(a') -- [plain] (c') -- [plain] (b'),
(c) -- [scalar] (z) -- [scalar] (c')
}; 
\end{feynman} 
\end{tikzpicture} \label{eq:Sconsdiag} \\
iS_\text{diss} & = \left(\sum_{a=1}^2\sum_{n=0}^{\infty}\begin{tikzpicture}[baseline]
\begin{feynman}
\vertex (z);
\vertex [below=0.1cm of z] (f);
\vertex [above=0.1cm of z] (f');
\vertex [right=0.5cm of f', empty dot, minimum size=0.4cm] (c) {n};
\vertex [right=0.7cm of c] (d);
\vertex [above=0.5cm of d] (d');
\diagram* {
(f') -- [plain] (c) -- [plain] (d),
(f') -- [draw=none] (z) -- [draw=none] (f),
(d') -- [gluon] (c)
}; 
\end{feynman} 
\end{tikzpicture}\;\right) \!+\!\!\begin{tikzpicture}[baseline]
\begin{feynman}
\vertex [blob, minimum size=0.7cm] (z) {};
\vertex [above=0.9cm of z, blob, minimum size=0.3cm] (c') {};
\vertex [above=0.5cm of a'] (d);
\vertex [below=0.9cm of z, blob, minimum size=0.3cm] (c) {};
\vertex [right=1cm of c] (a);
\vertex [left=1cm of c] (b);
\vertex [right=1cm of c'] (a');
\vertex [left=1cm of c'] (b');
\diagram* {
(a) -- [plain] (c) -- [plain] (b),
(a') -- [plain] (c') -- [plain] (b'),
(c) -- [scalar] (z) -- [scalar] (c'),
(d) -- [gluon] (c')
}; 
\end{feynman} 
\end{tikzpicture}\!+\!
\begin{tikzpicture}[baseline]
\begin{feynman}
\vertex [blob, minimum size=0.7cm] (z) {};
\vertex [right=1cm of z] (d);
\vertex [above=0.9cm of z, blob, minimum size=0.3cm] (c') {};
\vertex [below=0.9cm of z, blob, minimum size=0.3cm] (c) {};
\vertex [right=1cm of c] (a);
\vertex [left=1cm of c] (b);
\vertex [right=1cm of c'] (a');
\vertex [left=1cm of c'] (b');
\diagram* {
(a) -- [plain] (c) -- [plain] (b),
(a') -- [plain] (c') -- [plain] (b'),
(c) -- [scalar] (z) -- [scalar] (c'),
(d) -- [gluon] (z)
}; 
\end{feynman} 
\end{tikzpicture}+\ {\begin{smallmatrix}\text{Similar diagrams} \\ \text{with arbitrary}\\\text{number of} \\ \text{external } \radh{}.\end{smallmatrix}}\label{eq:Sdissdiag}
\end{align}
\end{subequations}
A blob represents all possible interaction vertices with the potential gravitons. The first term in brackets of (\ref{eq:Sconsdiag}) represents $iS^0_{\text{pp}}$, while the term in brackets of (\ref{eq:Sdissdiag}) is a part of\footnote{The complete action $S^{\radh{}}_{\text{pp}}$ contains diagram with any number of external radiation fields.} $iS^{\radh{}}_{\text{pp}}$.  It is then clear that:
\begin{itemize}
\item $S_\text{cons}[x_a^\mu]$ is real, hence it takes into account the conservative dynamics of the binary. It is of the form kinetic plus potential term, $S_\text{cons}[x_a^\mu] = \int\!\!d t\left(K(x_a^\mu)-V(x_a^\mu)\right)$, and from it we can find the conservative EOM of the binary
\item $S_\text{diss}[\radh{\mu\nu},x_a^\mu]$ is in general complex, and it takes into account dissipative effects due to the emission of radiation gravitons.
\end{itemize}

Once we have $\Seff{NR}[\radh{\mu\nu},x^\mu_a]$, we then integrate out from the theory also the radiation field. In doing so, we find what we called $\Seff{NRGR}$ defined as
\begin{equation}
e^{i\Seff{NRGR}} =\Path\radh{\mu\nu}e^{i\Seff{NR}+S^{\text{GF}}_{\radh{}}} \; .
\label{eq:SNRGR.path}
\end{equation}
Here, we have added the needed gauge-fixing term for the radiation field $\radh{\mu\nu}$, which is
\begin{align}
S^{\text{GF}}_{\radh{}}[\radh{\mu\nu}] = \Mpl^2\int\!\!d^4\!x\bar{\Gamma}_\mu\bar{\Gamma}^\mu & & \text{where,} & & \bar{\Gamma}_{\mu} =\frac{1}{\Mpl}\left(\partial^{\rho}\radh{\rho\mu}-\frac{1}{2}\partial_\mu\radh{}\right) \; .
\label{eq:SGFrad}
\end{align}

In the rest of this chapter we are going to deal only with $S_\text{cons}$, postponing the discussion on dissipative effects in Ch. \ref{ch:rad}.

\section{Power counting}\label{sec:power.count}

Once we have decoupled UV and IR physics, we need to implement the other important tool of EFT: the power counting. As we said, in the top-down approach, power counting allows us to understand immediately which diagrams contribute to a certain PN order. We know already that
\begin{align}
\partial_0\potH{k}{\mu\nu}\sim\frac{v}{r}\potH{k}{\mu\nu} \; , & &  |\bm{k}_H|\sim\frac{1}{r} \; , & & \partial_\rho\radh{\mu\nu}\sim\frac{v}{r}\radh{\mu\nu} \; .
\end{align}
It is also clear that $t\sim (r/v)$ and $\modul{x} \sim r$. Moreover from the virial theorem we can roughly estimate
\begin{equation}
\frac{\GN m}{r}\simeq\frac{m}{r\Mpl^2}\sim v^2 \qquad \longrightarrow\qquad \frac{m}{\Mpl}\sim L^{1/2}v^{1/2} \; .
\end{equation}

Now, we just have to understand the scaling of the fields $\potH{}{\mu\nu}$ and $\radh{\mu\nu}$. In order to do so, we need the propagators of the two fields, hence the quadratic part of their actions. In App. \ref{App:action}, and in particular in Sec. \ref{sec:EHe}, we perform explicitly the expansion of the gauge fixed\footnote{See eqs. (\ref{eq:SGFpot}) and (\ref{eq:SGFrad}) for the explicit form of the gauge-fixing actions.} Einstein-Hilbert action in NRGR. The quadratic action for both $h_{\mu\nu} = \{\potH{}{\mu\nu}, \radh{\mu\nu}\}$ has exactly the same form which is 
\begin{align}
S^{(2)} &=\frac{1}{2}\int\!\!d^4\!x\partial_\alpha h_{\rho\sigma}\left(\eta^{\mu\rho}\eta^{\nu\sigma}-\frac{1}{2}\eta^{\mu\nu}\eta^{\rho\sigma}\right)\partial^\alpha h_{\mu\nu} \; .
\label{eq:quad.prop}
\end{align}
From this action one eventually finds a propagator equal to the one written in eq. (\ref{eq:feyrulecov2}). However, we know that potential modes are off-shell, and that $k^0\sim v/r$, therefore we can make the following expansion
\begin{align}
\bangle*{\Tprod{H_{\rho\sigma}(x) H_{\mu\nu}(y)}} & = \int\!\!\frtr{4}{k}\frac{i e^{-ik(x-y)}}{-|\bm{k}|^2}\left(1+\frac{(k^0)^2}{|\bm{k}|^2}+\dots\right)P_{\mu\nu\rho\sigma} \; ,
\label{eq:instantaneous_exp}
\end{align} 
where we dropped the $i\varepsilon$ prescription because there no possibility for the propagator to go on-shell in this case. We then take the first term as the potential propagator, describing instantaneous interactions. The terms proportional to $k^0$ are suppressed by powers of $v$, and will be considered as corrections to this propagator, which take into account the fact that the interactions are not really instantaneous. Eventually, performing the needed partial Fourier transformation for the potential sector, one finds the following form for the propagators
\begin{align}
\begin{tikzpicture}[baseline]
\begin{feynman}
\vertex (f);
\vertex [dot, above=0.05cm of f, label=75:$\bm{q}$, label=285:$\rho\sigma$, label=180:$t'$] (z) {};
\vertex [dot, right=1.6cm of z, label=105:$\bm{k}$, label=255:$\mu\nu$, label=0:$t$] (w) {}; 
\diagram* {
(z) -- [scalar] (w)
}; 
\end{feynman} 
\end{tikzpicture} & = -\frac{i}{|\bm{k}|^2}(2\pi)^3\delta^{(3)}\left(\bm{k}\!+\!\bm{q}\right)\delta(t\!-\!t')P_{\mu\nu\rho\sigma} \; ,\label{eq:Hprop}\\
\begin{tikzpicture}[baseline]
\begin{feynman}
\vertex (f);
\vertex [dot, above=0.05cm of f, label=90:$y$, label=285:$\rho\sigma$] (z) {};
\vertex [dot, label=90:$x$, right=2cm of z, label=255:$\mu\nu$] (w) {}; 
\diagram* {
(z) -- [gluon, edge label=$k$] (w)
}; 
\end{feynman} 
\end{tikzpicture} & = D(x\!-\!y) P_{\mu\nu\rho\sigma} \; , \label{eq:hbarprop}
\end{align}
where again $D(x-y)$ and $P_{\mu\nu\rho\sigma}$ are defined in eq. (\ref{eq:DandP}). To these, we have to add the corrections mentioned before. Denoting one of this with the insertion of a cross in the dashed line propagator, for the first correction one finds explicitly
\begin{align}
\begin{tikzpicture}[baseline]
\begin{feynman}
\vertex (f);
\vertex [dot, above=0.05cm of f, label=75:$\bm{q}$, label=285:$\rho\sigma$, label=180:$t'$] (z) {};
\vertex [crossed dot, right=1cm of z] (z') {};
\vertex [dot, right=1cm of z', label=105:$\bm{k}$, label=255:$\mu\nu$, label=0:$t$] (w) {}; 
\diagram* {
(z) -- [scalar] (z') -- [scalar] (w)
}; 
\end{feynman} 
\end{tikzpicture} & = -\frac{i}{|\bm{k}|^4}(2\pi)^3\delta^{(3)}\left(\bm{k}+\bm{q}\right)\frac{\partial^2}{\partial t' \partial t}\delta(t-t')P_{\mu\nu\rho\sigma} \; ,
\label{eq:corr_prop}
\end{align}
and so on for all the $k^0$ terms inside the round brackets of eq. (\ref{eq:instantaneous_exp}).
\begin{table}[t]
\begin{center}
\begin{tabular}{c|c|c|c|c|c|c|c}
\  & $t$ & $\delta \bm{x}$ & $m/\Mpl$ & $\potH{k}{\mu\nu}$ & $\radh{\mu\nu}$ & $\partial_0\potH{k}{\mu\nu}$ & $\partial_\sigma\radh{\mu\nu}$\\ \hline
\ & \ & \ & \ & \ & \ & \ & \\
Scaling & $r/v$ & $r$ & $L^{1/2}v^{1/2}$ & $r^2 v^{1/2}$ & $v/r$ & $(v/r)\potH{k}{\mu\nu}$ & $(v/r)\radh{\mu\nu}$
\end{tabular}
\end{center}
\caption{Scaling rules}\label{table:scaling}
\end{table}

Let's now go back to the scaling problem:
\begin{itemize}
\item from (\ref{eq:Hprop}) we see that
\begin{equation}
\potH{k}{\mu\nu}^2\sim\left(\frac{1}{r}\right)^{-2}\left(\frac{1}{r}\right)^{-3}\left(\frac{r}{v}\right)^{-1} =r^4v\ \longrightarrow\ \potH{k}{\mu\nu}\sim r^2v^{1/2} \; .
\end{equation}
\item from (\ref{eq:hbarprop}) we see that
\begin{equation}
\radh{\mu\nu}^2\sim\left(\frac{v}{r}\right)^{4}\left(\frac{v}{r}\right)^{-2} = \frac{v^2}{r^2}\ \longrightarrow\ \radh{\mu\nu}\sim \frac{v}{r} \; .
\end{equation}
\end{itemize}
All scaling rules are then collected in tab. \ref{table:scaling}

Now we can compute the scaling of any kind of vertex in NRGR. To make an example, let's consider the simplest interaction term between the point particle and the potential graviton; this is given by the first term of eq. (\ref{eq:SppH}) which is
\begin{equation}
-\frac{m}{2\Mpl}\int\!\!d t\intvec{k}e^{i\bm{k}\cdot\bm{x}}\potH{k}{00}\ \longrightarrow\ 
\begin{tikzpicture}[baseline]
\begin{feynman}
\vertex (f);
\vertex [above=0.1cm of f] (z);
\vertex [left=0.2cm of z] (c');
\vertex [above=0.6cm of c'] (a);
\vertex [above=0.2cm of a] (a');
\vertex [below=0.6cm of c'] (b);
\vertex [below=0.2cm of b] (b');
\vertex [right=0.7cm of z] (c); 
\diagram* {
(z) --[draw=none] (f),
(a') -- [draw=none] (a) -- [plain] (c') -- [plain] (b) -- [draw=none] (b'),
(c') -- [scalar] (c)
}; 
\end{feynman} 
\end{tikzpicture} \sim L^{1/2}v^{1/2}\left(\frac{r}{v}\right)\left(\frac{1}{r}\right)^3r^2v^{1/2} = L^{1/2} \; .
\end{equation}
One can then redo the same kind of computations for all vertices of NRGR. In Sec. \ref{App:FRscale} of App. \ref{App:action} we list the scaling and the Feynman rules of all vertices we need in the computations made in this work. At this point, we have all the ingredients we need to start performing diagrammatic computations in NRGR.  

\subsection{Scaling of the Newtonian order}\label{subsec:Newt}

As we said in the introduction, a term of $n$PN order means a term of order $v^{2n}$ beyond Newtonian physics. It is natural to ask then: what is the order of Newtonian physics?

We perfectly know that the classical dynamics of the binary system is described by the action
\begin{equation}
S_{\text{N}} = \int\!\!d t\left\{\frac{1}{2}\sum_{a=1}^2m_av_a^2+\frac{\GN m_1m_2}{r}\right\} \; ,
\end{equation}
where $r \equiv |\bm{x}_1-\bm{x}_2|$. It is trivial to see that this action is of order
\begin{equation}
S_{\text{N}} \sim \frac{r}{v} mv^2=L \; .
\end{equation}
This implies that \textit{in NRGR a contribution of order }$n$\textit{PN is given by the sum of all diagrams that are of order }$Lv^{2n}$. 

Of course NRGR has to contain also the Newtonian physics and, being this conservative, it has to be in $S_\text{cons}$. In order to verify this, let's construct all diagrams that are of order $Lv^0$. From the tables we wrote in App. \ref{App:FRscale}, and in particular form tabs. \ref{table:freepp} and \ref{table:pppot} (p.~\pageref{table:freepp}) we can see that the Newtonian level is given by
\begin{equation}
iS_\text{cons} = \left(\sum_a\begin{tikzpicture}[baseline]
\begin{feynman}
\vertex (z);
\vertex [below=0.1cm of z] (f);
\vertex [above=0.1cm of z] (f');
\vertex [right=1cm of f'] (c);
\diagram* {
(f') -- [plain, edge label=$x_a$] (c),
(f') -- [draw=none] (z) -- [draw=none] (f)
}; 
\end{feynman} 
\end{tikzpicture}\right)+
\begin{tikzpicture}[baseline]
\begin{feynman}
\vertex (z);
\vertex [above=0.6cm of z] (c');
\vertex [below=0.4cm of z] (c);
\vertex [right=0.8cm of c] (a);
\vertex [left=0.8cm of c] (b);
\vertex [right=0.8cm of c'] (a');
\vertex [left=0.8cm of c'] (b');
\diagram* {
(a) -- [plain] (c) -- [plain] (b),
(a') -- [plain] (c') -- [plain] (b'),
(c) -- [scalar] (c')
}; 
\end{feynman} 
\end{tikzpicture}+\Ord{Lv^2} \; .
\label{eq:Newton_diagonly}
\end{equation}
The first diagram, coming from $S_{\text{pp}}^0$, trivially gives the usual point particle kinetic term, while the second one is given by
\begin{align}
\begin{tikzpicture}[baseline]
\begin{feynman}
\vertex [label=270:$t_2$, label=160:$\rho\sigma$, label=20:$\bm{q}$] (c);
\vertex [above=0.6cm of c] (f);
\vertex [above=0.1cm of f] (z);
\vertex [above=0.9cm of z, label=90:$t_1$, label=200:$\mu\nu$, label=340:$\bm{k}$] (c');
\vertex [right=1.2cm of c] (a);
\vertex [left=1.2cm of c] (b) {$x_b$};
\vertex [right=1.2cm of c'] (a');
\vertex [left=1.2cm of c'] (b') {$x_a$};
\diagram* {
(a) -- [plain] (c) -- [plain] (b),
(a') -- [plain] (c') -- [plain] (b'),
(c) -- [scalar] (c')
}; 
\end{feynman} 
\end{tikzpicture} & = \frac{i}{2}\sum_{a\neq b}\frac{m_a m_b}{4\Mpl^2}\int\!\!d t\intvec{k}\frac{e^{-i\bm{k}\cdot(\bm{x}_a-\bm{x}_b)}}{\modul{k}^2}P_{0000}  \notag \\
& = \frac{i}{2}\sum_{a\neq b}\int\!\!d t\frac{\GN m_a m_b}{|\bm{x}_a-\bm{x}_b|} = i\int\!\!d t\frac{\GN m_1 m_2}{r} \; .
\label{eq:newton:NRGR}
\end{align}
The factor $1/2$ in front is the symmetry factor of the diagram, which comes from the fact that we have two equivalent diagrams. Moreover we stress the fact that, in keeping $a$ and $b$ completely generic, we mean that diagrams underline a sum over $a$ and $b$, with\footnote{We postpone the case $a = b$ to Sec. \ref{sec:diverg}.} $a\neq b$, i.e.
\begin{align*}
\begin{tikzpicture}[baseline]
\begin{feynman}
\vertex (f);
\vertex [above=0.1cm of f] (z);
\vertex [below=0.6cm of z](c);
\vertex [above=0.7cm of z] (c');
\vertex [right=1cm of c] (a);
\vertex [left=1cm of c] (b) {$x_b$};
\vertex [right=1cm of c'] (a');
\vertex [left=1cm of c'] (b') {$x_a$};
\diagram* {
(a) -- [plain] (c) -- [plain] (b),
(a') -- [plain] (c') -- [plain] (b'),
(c) -- [scalar] (c')
}; 
\end{feynman} 
\end{tikzpicture} =
\begin{tikzpicture}[baseline]
\begin{feynman}
\vertex (f);
\vertex [above=0.1cm of f] (z);
\vertex [below=0.6cm of z](c);
\vertex [above=0.7cm of z] (c');
\vertex [right=1cm of c] (a);
\vertex [left=1cm of c] (b) {$x_2$};
\vertex [right=1cm of c'] (a');
\vertex [left=1cm of c'] (b') {$x_1$};
\diagram* {
(a) -- [plain] (c) -- [plain] (b),
(a') -- [plain] (c') -- [plain] (b'),
(c) -- [scalar] (c')
}; 
\end{feynman} 
\end{tikzpicture} +
\begin{tikzpicture}[baseline]
\begin{feynman}
\vertex (f);
\vertex [above=0.1cm of f] (z);
\vertex [below=0.6cm of z](c);
\vertex [above=0.7cm of z] (c');
\vertex [right=1cm of c] (a);
\vertex [left=1cm of c] (b) {$x_1$};
\vertex [right=1cm of c'] (a');
\vertex [left=1cm of c'] (b') {$x_2$};
\diagram* {
(a) -- [plain] (c) -- [plain] (b),
(a') -- [plain] (c') -- [plain] (b'),
(c) -- [scalar] (c')
}; 
\end{feynman} 
\end{tikzpicture} \; .
\end{align*} 
Inserting eq. (\ref{eq:newton:NRGR}) in (\ref{eq:Newton_diagonly}) we immediately obtain
\begin{equation}
iS_\text{cons} = iS_{\text{N}}+\Ord{Lv^2} \; .
\end{equation}

Thus, we correctly find the Newtonian physics as the zeroth order term in the PN expansion in the framework of NRGR. This is a first important check that we did everything in a consistent way.

\section{1PN order contribution to $S_\text{cons}$: the EIH action}\label{sec: 1PN contr}

Now we are ready to compute the first PN contribution to $S_\text{cons}$. As we said the perturbative series of PN correction is of the form
\begin{equation}
L(1+v+v^2+v^3+v^4+\dots) \; .
\end{equation}
However, since at this stage time reversal is still a symmetry of the system, we can have only even power of $v$. Therefore the first correction to the Newtonian theory is of order 1PN, hence it comes from diagrams of order $Lv^2$. According to our scaling rules (see tab. \ref{table:scaling}, page~\pageref{table:scaling}), the graphic objects we can use to construct $Lv^2$ diagrams are the following (see also tabs. \ref{table:gravv}, \ref{table:freepp} and \ref{table:pppot} in App. \ref{App:action}, pp.~\pageref{table:gravv} -~\pageref{table:pppot}) 
\begin{align*}
& \begin{tikzpicture}[baseline]
\begin{feynman}
\vertex (f);
\vertex [above=0.1cm of f] (z);
\vertex [left=0.2cm of z] (c');
\vertex [above=0.6cm of c'] (a);
\vertex [above=0.2cm of a] (a');
\vertex [below=0.6cm of c'] (b);
\vertex [below=0.2cm of b] (b');
\vertex [right=0.7cm of z] (c); 
\diagram* {
(z) --[draw=none] (f),
(a') -- [draw=none] (a) -- [plain] (c') -- [plain] (b) -- [draw=none] (b'),
(c') -- [scalar] (c)
}; 
\end{feynman} 
\end{tikzpicture}\;\sim  L^{1/2} \;  , & & \begin{tikzpicture}[baseline]
\begin{feynman}
\vertex (f);
\vertex [above=0.1cm of f] (z);
\vertex [empty dot, minimum size=0.4cm, left=0.2cm of z] (c') {1};
\vertex [above=0.7cm of c'] (a);
\vertex [above=0.2cm of a] (a');
\vertex [below=0.7cm of c'] (b);
\vertex [below=0.2cm of b] (b');
\vertex [right=0.7cm of z] (c); 
\diagram* {
(a') -- [draw=none] (a) -- [plain] (c') -- [plain] (b) -- [draw=none] (b'),
(c') -- [scalar] (c),
(z) -- [draw=none] (f)
}; 
\end{feynman} 
\end{tikzpicture}\;\sim L^{1/2}v \; ,& & \begin{tikzpicture}[baseline]
\begin{feynman}
\vertex (f);
\vertex [above=0.1cm of f] (z);
\vertex [empty dot, minimum size=0.4cm, left=0.2cm of z] (c') {2};
\vertex [above=0.7cm of c'] (a);
\vertex [above=0.2cm of a] (a');
\vertex [below=0.7cm of c'] (b);
\vertex [below=0.2cm of b] (b');
\vertex [right=0.7cm of z] (c); 
\diagram* {
(a') -- [draw=none] (a) -- [plain] (c') -- [plain] (b) -- [draw=none] (b'),
(c') -- [scalar] (c),
(z) -- [draw=none] (f)
}; 
\end{feynman} 
\end{tikzpicture}\;\sim L^{1/2}v^2 \; ,\\
&\begin{tikzpicture}[baseline]
\begin{feynman}
\vertex (f);
\vertex [above=0.1cm of f] (z);
\vertex [left=0.2cm of z] (c');
\vertex [above=0.7cm of c'] (a);
\vertex [above=0.2cm of a] (a');
\vertex [below=0.7cm of c'] (b);
\vertex [below=0.2cm of b] (b');
\vertex [above right=0.7cm of z] (c); 
\vertex [below right=0.7cm of z] (d); 
\diagram* {
(a') -- [draw=none] (a) -- [plain] (c') -- [plain] (b) -- [draw=none] (b'),
(c') -- [scalar] (c),
(c') -- [scalar] (d),
(z) -- [draw=none] (f)
}; 
\end{feynman} 
\end{tikzpicture}\ \sim v^2 \; ,& & \begin{tikzpicture}[baseline]
\begin{feynman}
\vertex (f);
\vertex [above=0.1cm of f] (z);
\vertex [crossed dot, right=1.cm of z] (z') {};
\vertex [right=1.20cm of z'] (w); 
\diagram* {
(z) -- [scalar] (z') -- [scalar] (w),
(z) -- [draw=none] (f)
}; 
\end{feynman} 
\end{tikzpicture}\;\sim v^2 \; ,& & \begin{tikzpicture}[baseline]
\begin{feynman}
\vertex (f);
\vertex [above=0.1cm of f] (z);
\vertex [right=1.2cm of z] (c);
\vertex [above left=1.2cm of z] (a);
\vertex [below left=1.2cm of z] (b);
\diagram* {
(a) -- [scalar] (z) -- [scalar] (c),
(b) -- [scalar] (z),
(z) -- [draw=none] (f)
}; 
\end{feynman} 
\end{tikzpicture} \ \sim \frac{v^2}{L^{1/2}} \; .
\end{align*}
Therefore, including also the kinetic terms, we immediately write $S_\text{cons}$ at 1PN order 
\begin{align}
iS_\text{cons} = & iS_{\text{N}} + \left(\sum_a\begin{tikzpicture}[baseline]
\begin{feynman}
\vertex (z);
\vertex [below=0.1cm of z] (f);
\vertex [above=0.1cm of z] (f');
\vertex [right=0.5cm of f', empty dot, minimum size=0.4cm, label=90:$x_a$] (c) {2};
\vertex [right=0.7cm of c] (d);
\diagram* {
(f') -- [plain] (c) -- [plain] (d),
(f') -- [draw=none] (z) -- [draw=none] (f)
}; 
\end{feynman} 
\end{tikzpicture}\right) + 
\begin{tikzpicture}[baseline]
\begin{feynman}
\vertex [crossed dot, minimum size=0.25cm] (z) {};
\vertex [above=0.8cm of z] (c');
\vertex [below=0.8cm of z] (c);
\vertex [right=1cm of c] (a);
\vertex [left=1cm of c] (b) {$x_b$};
\vertex [right=1cm of c'] (a');
\vertex [left=1cm of c'] (b') {$x_a$};
\diagram* {
(a) -- [plain] (c) -- [plain] (b),
(a') -- [plain] (c') -- [plain] (b'),
(c) -- [scalar] (z),
(c') -- [scalar] (z)
}; 
\end{feynman} 
\end{tikzpicture}+
\begin{tikzpicture}[baseline]
\begin{feynman}
\vertex (z);
\vertex [above=0.8cm of z] (c');
\vertex [below=0.8cm of z] (c);
\vertex [right=1cm of c] (a);
\vertex [left=1cm of c] (b) {$x_b$};
\vertex [right=0.5cm of c] (a'');
\vertex [left=0.5cm of c] (b'');
\vertex [right=1cm of c'] (a');
\vertex [left=1cm of c'] (b') {$x_a$};
\diagram* {
(a) -- [plain] (c) -- [plain] (b),
(a') -- [plain] (c') -- [plain] (b'),
(c') -- [scalar] (z),
(a'') -- [scalar] (z),
(b'') -- [scalar] (z)
}; 
\end{feynman} 
\end{tikzpicture}+\notag \\
&+ \begin{tikzpicture}[baseline]
\begin{feynman}
\vertex (z);
\vertex [above=0.85cm of z] (c');
\vertex [below=0.8cm of z] (c);
\vertex [right=1cm of c] (a);
\vertex [left=1cm of c] (b) {$x_b$};
\vertex [right=0.6cm of c] (a'');
\vertex [left=0.6cm of c] (b'');
\vertex [right=1cm of c'] (a');
\vertex [left=1cm of c'] (b') {$x_a$};
\diagram* {
(a) -- [plain] (c) -- [plain] (b),
(a') -- [plain] (c') -- [plain] (b'),
(c') -- [scalar] (a''),
(c') -- [scalar] (b'')
}; 
\end{feynman} 
\end{tikzpicture}
+ \begin{tikzpicture}[baseline]
\begin{feynman}
\vertex (z);
\vertex [above=0.7cm of z, empty dot, minimum size=0.4cm] (c') {1};
\vertex [below=0.65cm of z, empty dot, minimum size=0.4cm] (c) {1};
\vertex [right=0.8cm of c] (a);
\vertex [left=1cm of c] (b) {$x_b$};
\vertex [right=0.8cm of c'] (a');
\vertex [left=1cm of c'] (b') {$x_a$};
\diagram* {
(a) -- [plain] (c) -- [plain] (b),
(a') -- [plain] (c') -- [plain] (b'),
(c) -- [scalar] (c')
}; 
\end{feynman} 
\end{tikzpicture}+
\begin{tikzpicture}[baseline]
\begin{feynman}
\vertex (z);
\vertex [above=0.7cm of z, empty dot, minimum size=0.4cm] (c') {2};
\vertex [below=0.8cm of z] (c);
\vertex [right=0.8cm of c] (a);
\vertex [left=0.8cm of c] (b) {$x_b$};
\vertex [right=0.8cm of c'] (a');
\vertex [left=1cm of c'] (b') {$x_a$};
\diagram* {
(a) -- [plain] (c) -- [plain] (b),
(a') -- [plain] (c') -- [plain] (b'),
(c) -- [scalar] (c')
}; 
\end{feynman} 
\end{tikzpicture} + \Ord{Lv^4} \; .
\label{eq:Scon.1PN}
\end{align}

Now we just have to compute the different diagrams using the Feynman rules listed in App. \ref{App:FRscale}. We start with the diagram coming from the correction to the potential propagator
\begin{align}
\begin{tikzpicture}[baseline]
\begin{feynman}
\vertex [label=270:$t_2$, label=20:$\bm{q}$] (c);
\vertex [above=0.6cm of c] (f);
\vertex [above=0.1cm of f, crossed dot] (z) {};
\vertex [above=0.9cm of z, label=90:$t_1$, label=340:$\bm{k}$] (c');
\vertex [right=1.2cm of c] (a);
\vertex [left=1.2cm of c] (b) {$x_b$};
\vertex [right=1.2cm of c'] (a');
\vertex [left=1.2cm of c'] (b') {$x_a$};
\diagram* {
(a) -- [plain] (c) -- [plain] (b),
(a') -- [plain] (c') -- [plain] (b'),
(c) -- [scalar] (z),
(c') -- [scalar] (z)
}; 
\end{feynman} 
\end{tikzpicture} & = \frac{i}{2}\sum_{a\neq b}\frac{m_a m_b}{4\Mpl^2}\int\!\!d t_1 d t_2\delta(t_1\!-\!t_2)P_{0000} \notag \\
&\qquad\qquad\qquad\qquad\times\intvec{k}\frac{1}{\modul{k}^4}\frac{\partial^2}{\partial t_1\partial t_2}e^{-i\bm{k}\cdot(\bm{x}_a(t_1)-\bm{x}_b(t_2))} \notag \\
& = \frac{i}{2}\sum_{a\neq b}\frac{m_a m_b}{8\Mpl^2}\int\!\!d t\,v_a^iv_b^j\intvec{k}\frac{k_ik_j}{\modul{k}^4}e^{-i\bm{k}\cdot\bm{x}_{ab}} \; ,
\label{eq:first.diag.step}
\end{align}
where we defined $\bm{x}_{ab}\equiv\bm{x}_a(t)-\bm{x}_b(t)$. Using now the well known result
\begin{equation}
\intvec{k}\frac{k_ik_j}{\modul{k}^4}e^{-i\bm{k}\cdot\bm{x}_{ab}} = \frac{1}{8\pi|\bm{x}_{ab}|}\left(\delta_{ij}-\frac{(x_{ab})_i(x_{ab})_j}{|\bm{x}_{ab}|^2}\right) \; ,
\label{eq:wkresult}
\end{equation}
where $\delta_{ij}$ is the Kronecker symbol, eq. (\ref{eq:first.diag.step}) becomes
\begin{align}
\begin{tikzpicture}[baseline]
\begin{feynman}
\vertex (c);
\vertex [above=0.6cm of c] (f);
\vertex [above=0.1cm of f, crossed dot] (z) {};
\vertex [above=0.9cm of z] (c');
\vertex [right=1cm of c] (a);
\vertex [left=1cm of c] (b) {$x_b$};
\vertex [right=1cm of c'] (a');
\vertex [left=1cm of c'] (b') {$x_a$};
\diagram* {
(a) -- [plain] (c) -- [plain] (b),
(a') -- [plain] (c') -- [plain] (b'),
(c) -- [scalar] (z),
(c') -- [scalar] (z)
}; 
\end{feynman} 
\end{tikzpicture} & = \frac{i}{2}\sum_{a\neq b}\frac{m_a m_b}{64\pi\Mpl^2}\!\int\!\!d t\frac{1}{|\bm{x}_{ab}|}\left[\bm{v}_a\cdot\bm{v}_b\!-\!\frac{(\bm{v}_a\cdot\bm{x}_{ab})(\bm{v}_b\cdot\bm{x}_{ab})}{|\bm{x}_{ab}|^2}\right] \notag \\
& = \frac{i}{2}\int\!\!d t\frac{\GN m_1m_2}{r}\left[\bm{v}_1\cdot\bm{v}_2-\frac{(\bm{v}_1\cdot\bm{r})(\bm{v}_2\cdot\bm{r})}{r^2}\right] \; .
\label{eq:diag1}
\end{align}
Here we recall that $\bm{r} =\bm{x}_1-\bm{x}_2$, and we denote with $r$ its modulus. As a consistency check, one can see that the result is really of order $Lv^2$, as expected.

The next diagram is the one with an insertion of the cubic interaction vertex for $\potH{}{}$. At 1PN order we just need the interaction between three potential gravitons all with polarization $00$. See the Feynman rule in tab. \ref{table:gravv} (p.~\pageref{table:gravv}).
\begin{align}
\begin{tikzpicture}[baseline]
\begin{feynman}
\vertex (c);
\vertex [above=0.8cm of c] (z);
\vertex [above=0.8cm of z, label=90:$t_1$] (c');
\vertex [right=1.2cm of c] (a);
\vertex [left=1.2cm of c] (b) {$x_b$};
\vertex [right=0.7cm of c, label=270:$t_3$] (a'');
\vertex [left=0.7cm of c, label=270:$t_2$] (b'');
\vertex [right=1.2cm of c'] (a');
\vertex [left=1.2cm of c'] (b') {$x_a$};
\diagram* {
(a) -- [plain] (c) -- [plain] (b),
(a') -- [plain] (c') -- [plain] (b'),
(c') -- [scalar, edge label=$\bm{k}_1$] (z),
(a'') -- [scalar, edge label'=$\bm{k}_3$] (z),
(b'') -- [scalar, edge label=$\bm{k}_2$] (z)
}; 
\end{feynman} 
\end{tikzpicture} & = \frac{i}{2}\sum_{a\neq b}\frac{m_a m^2_b}{8\Mpl^3}\int\!\!d t\!\int_{\bm{k}_1,\bm{k}_2,\bm{k}_3}\!\!\!\!(2\pi)^3\delta^{(3)}\left(\bm{k}_1\!+\!\bm{k}_2\!+\!\bm{k}_3\right) \notag \\
&\qquad\qquad\quad\times\!\bigg(\!-\!\frac{1}{4\Mpl}\frac{|\bm{k}_1|^2\!+\!|\bm{k}_2|^2\!+\!|\bm{k}_3|^2}{|\bm{k}_1|^2|\bm{k}_2|^2|\bm{k}_3|^2}\bigg)e^{i\bm{k}_1\cdot\bm{x}_a+i(\bm{k}_2+\bm{k}_3)\cdot\bm{x}_b}  \notag \\ 
& =-\frac{i}{2}\sum_{a\neq b}\frac{m_a m^2_b}{32\Mpl^4}\int\!\!d t\intvec{k,q}\frac{|\bm{k}+\bm{q}|2+|\bm{k}|^2+|\bm{q}|^2}{|\bm{k}+\bm{q}|^2|\bm{k}|^2|\bm{q}|^2}\,e^{-i(\bm{k}+\bm{q})\cdot\bm{x}_{ab}} \; .
\label{eq:diag2.step}
\end{align}
In the last step we implemented the $\delta^{(3)}$, hence $\bm{k}_1=-\bm{k}_2-\bm{k}_3$, and we redefined $\bm{k}\equiv\bm{k}_2$, $\bm{q}\equiv\bm{k}_3$. Now the two momentum integrals can be written as 
\begin{align}
\intvec{k,q}\frac{|\bm{k}+\bm{q}|^2+|\bm{k}|^2+|\bm{q}|^2}{|\bm{k}+\bm{q}|^2|\bm{k}|^2|\bm{q}|^2}\,e^{-i(\bm{k}+\bm{q})\cdot\bm{x}_{ab}} & = \intvec{k}\frac{e^{-i\bm{k}\cdot\bm{x}_{ab}}}{\modul{k}^2}\intvec{q}\frac{e^{-i\bm{q}\cdot\bm{x}_{ab}}}{\modul{q}^2}+\notag \\
& \qquad\qquad+2\intvec{k,q}\frac{e^{-i(\bm{k}+\bm{q})\cdot\bm{x}_{ab}}}{|\bm{k}+\bm{q}|^2|\bm{q}|^2} \; .
\end{align}
The term on the second line is actually a divergent quantity. We will briefly see in the next section that, as usual, divergences can be absorbed by suitable \textit{counterterms}. In particular in Sec. \ref{sec:diverg} we will see  that, using dimensional regularization, the divergent part of this integral is actually equal to zero. We can here neglect it, so that
\begin{align}
\begin{tikzpicture}[baseline]
\begin{feynman}
\vertex (c);
\vertex [above=0.8cm of c] (z);
\vertex [above=0.8cm of z] (c');
\vertex [right=1.2cm of c] (a);
\vertex [left=1.2cm of c] (b) {$x_b$};
\vertex [right=0.7cm of c] (a'');
\vertex [left=0.7cm of c] (b'');
\vertex [right=1.2cm of c'] (a');
\vertex [left=1.2cm of c'] (b') {$x_a$};
\vertex [above right=0.2cm of a'] (h');
\vertex [below right=0.2cm of a] (i);
\vertex [below=0.1cm of i] (h);
\diagram* {
(a) -- [plain] (c) -- [plain] (b),
(a') -- [plain] (c') -- [plain] (b'),
(c') -- [scalar] (z),
(a'') -- [scalar] (z),
(b'') -- [scalar] (z),
(h') -- [plain] (h)
}; 
\end{feynman} 
\end{tikzpicture}_{\text{finite}} & = -\frac{i}{2}\sum_{a\neq b}\frac{m_a m^2_b}{32\Mpl^4}\int\!\!d t\left[\intvec{k}\frac{e^{-i\bm{k}\cdot\bm{x}_{ab}}}{\modul{k}^2}\right]^2  \notag \\
& = -i\sum_{a\neq b}\frac{m_a m^2_b}{(32\pi)^2\Mpl^4}\int\!\!d t\,\frac{1}{|\bm{x}_{ab}|^2}  \notag \\
& = -i\int\!\!d t\,\frac{\GN^2 m_1m_2(m_1+m_2)}{r^2} \; .
\label{eq:diag2}
\end{align}

The last three diagrams are then
\begin{align}
\begin{tikzpicture}[baseline]
\begin{feynman}
\vertex (c);
\vertex [above=0.8cm of c] (z);
\vertex [above=0.8cm of z, label=90:$t_1$, label=185:$\bm{k}\;$, label=340:$\;\bm{q}$] (c');
\vertex [right=1.2cm of c] (a);
\vertex [left=1.2cm of c] (b) {$x_b$};
\vertex [right=0.7cm of c, label=270:$t_3$, label=90:$\quad\bm{q}'$] (a'');
\vertex [left=0.7cm of c, label=270:$t_2$, label=90:$\bm{k}'\quad$] (b'');
\vertex [right=1.2cm of c'] (a');
\vertex [left=1.2cm of c'] (b') {$x_a$};
\diagram* {
(a) -- [plain] (c) -- [plain] (b),
(a') -- [plain] (c') -- [plain] (b'),
(a'') -- [scalar] (c'),
(b'') -- [scalar] (c')
}; 
\end{feynman} 
\end{tikzpicture} & = \frac{i}{2}\sum_{a\neq b}\frac{m_a m_b^2}{16\Mpl^4}\int\!\!\!d t\!\!\intvec{k}\frac{e^{-i\bm{k}\cdot\bm{x}_{ab}}}{\modul{k}^2}\intvec{q}\frac{e^{-i\bm{q}\cdot\bm{x}_{ab}}}{\modul{q}^2} (P_{0000})^2 \notag \\
& = \frac{i}{2}\sum_{a\neq b}\frac{m_a m_b^2}{(32\pi)^2\Mpl^4}\int\!\!\!d t\frac{1}{|\bm{x}_{ab}|^2} \notag \\
& = \frac{i}{2}\int\!\!d t\,\frac{\GN^2 m_1m_2(m_1+m_2)}{r^2} \; .
\label{eq:diag3} \\
\begin{tikzpicture}[baseline]
\begin{feynman}
\vertex [label=270:$t_2$, label=20:$\bm{q}$,empty dot, minimum size=0.4cm,] (c) {1};
\vertex [above=0.6cm of c] (f);
\vertex [above=0.1cm of f] (z);
\vertex [above=0.9cm of z, empty dot, minimum size=0.4cm, label=90:$t_1$, label=340:$\bm{k}$] (c') {1};
\vertex [right=1cm of c] (a);
\vertex [left=1.2cm of c] (b) {$x_b$};
\vertex [right=1.2cm of c'] (a');
\vertex [left=1cm of c'] (b') {$x_a$};
\diagram* {
(a) -- [plain] (c) -- [plain] (b),
(a') -- [plain] (c') -- [plain] (b'),
(c) -- [scalar] (z),
(c') -- [scalar] (z)
}; 
\end{feynman} 
\end{tikzpicture} & = \frac{i}{2}\sum_{a\neq b}\frac{m_a m_b}{\Mpl^2}\int\!\!d t\frac{v^i_a v^j_b}{2}\left(P_{(0i)0j}\!+\!P_{(0i)j0}\right)\intvec{k}\frac{e^{-i\bm{k}\cdot\bm{x}_{ab}}}{\modul{k}^2} \notag \\
& = -\frac{i}{2}\sum_{a\neq b}\frac{m_a m_b}{8\pi\Mpl^2}\int\!\!d t\,\frac{\bm{v}_a\cdot\bm{v}_b}{|\bm{x}_{ab}|}\notag \\
& =-4i\int\!\!d t\,\frac{\GN m_1 m_2}{r}(\bm{v}_1\cdot\bm{v}_2) \; .
\label{eq:diag4}
\end{align}
\begin{align}
\begin{tikzpicture}[baseline]
\begin{feynman}
\vertex [label=270:$t_2$, label=20:$\bm{q}$] (c);
\vertex [above=0.6cm of c] (f);
\vertex [above=0.1cm of f] (z);
\vertex [above=0.9cm of z, empty dot, minimum size=0.4cm, label=90:$t_1$, label=340:$\bm{k}$] (c') {2};
\vertex [right=1.2cm of c] (a);
\vertex [left=1.2cm of c] (b) {$x_b$};
\vertex [right=1.2cm of c'] (a');
\vertex [left=1.2cm of c'] (b') {$x_a$};
\diagram* {
(a) -- [plain] (c) -- [plain] (b),
(a') -- [plain] (c') -- [plain] (b'),
(c) -- [scalar] (z),
(c') -- [scalar] (z)
}; 
\end{feynman} 
\end{tikzpicture} & = i\sum_{a\neq b}\frac{m_a m_b}{4\Mpl^2}\int\!\!d t\left(\frac{P_{0000}}{2}v_a^2+P_{ij00}v_a^iv_a^j\right)\intvec{k}\frac{e^{-i\bm{k}\cdot\bm{x}_{ab}}}{\modul{k}^2} \notag \\
& = i\frac{3}{2}\sum_{a\neq b}\frac{m_a m_b}{32\pi\Mpl^2}\int\!\!d t\,\frac{v_a^2}{|\bm{x}_{ab}|} \notag \\
& = i\frac{3}{2}\int\!\!d t\,\frac{\GN m_1 m_2}{r}\left(v_1^2+v_2^2\right) \; .
\label{eq:diag5}
\end{align}
Summing eqs. (\ref{eq:diag1}), (\ref{eq:diag2}), (\ref{eq:diag3}), (\ref{eq:diag4}) and (\ref{eq:diag5}), and adding the kinetic term coming from the diagram in round brackets of (\ref{eq:Scon.1PN}), we eventually obtain
\begin{align}
S_{\text{cons}}&=S_{\text{N}} + \int\!\!d t\bigg\{\frac{1}{8}\sum_{a=1}^2 m_av_a^4 + \notag \\
&\qquad\qquad\qquad\quad + \frac{1}{2}\frac{\GN m_1m_2}{r}\left[3(v_1^2+v_2^2)\!-\!7\bm{v}_1\cdot\bm{v}_2-\frac{(\bm{v}_1\cdot\bm{r})(\bm{v}_2\cdot\bm{r})}{r^2}\right]\notag \\
&\qquad\qquad\qquad\quad-\frac{\GN^2}{2}\frac{m_1m_2(m_1+m_2)}{r^2}\bigg\}\!+\!\Ord{Lv^4} \equiv S_{\text{EIH}}\!+\!\Ord{Lv^4} \; .
\label{eq:SEIH}
\end{align}
This is nothing but the well known Einstein-Infeld-Hoffmann action (EIH), known since 1938 \cite{EIH}, that precisely describes the 1PN correction to the Newtonian dynamics. Also at 1PN order then, NRGR techniques give us consistent results.

\subsection{What about divergences?}\label{sec:diverg}

In all computations performed so far, we postponed the discussion about divergent diagrams, which appear already at the Newtonian level. It is easy to prove that the following diagram is of order $Lv^0$
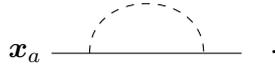
\begin{center}
\begin{tikzpicture}[baseline]
\begin{feynman}
\vertex (z);
\vertex [left=0.75cm of z] (w);
\vertex [left=0.5cm of w] (a) {$\bm{x}_a$};  
\vertex [right=0.75cm of z] (w');
\vertex [right=0.5cm of w'] (a');  
\diagram* {
(a) -- [plain] (a'),
(w) -- [scalar, half left] (w')
}; 
\end{feynman} 
\end{tikzpicture} \; .
\end{center}
Computing it in a straightforward way eventually leads us to 
\begin{align}
\begin{tikzpicture}[baseline]
\begin{feynman}
\vertex (z);
\vertex [left=0.75cm of z, label=100:$\bm{k}$, label=270:$t_1$] (w);
\vertex [left=0.5cm of w] (a) {$\bm{x}_a$};  
\vertex [right=0.75cm of z, label=80:$\bm{q}$, label=270:$t_2$] (w');
\vertex [right=0.5cm of w'] (a');  
\diagram* {
(a) -- [plain] (a'),
(w) -- [scalar, half left] (w')
}; 
\end{feynman} 
\end{tikzpicture}
& = i\frac{m_a^2}{16\Mpl^2}\int\!\!d t\!\intvec{k}\frac{1}{|\bm{k}|^2} \to\infty \; .
\end{align}
This is a recurrent situation in QFT whenever we have loop diagrams. Even if the above diagram is not a quantum loop diagram, it behaves exactly in the same way. Therefore we can solve the problem of divergences using the usual \textit{renormalization procedure}. 

We choose to work in dimensional regularization, hence we go to $D=4-2\varepsilon$ dimensions. We can then go back to $D=4$ by simply sending $\varepsilon\to 0$. The only thing we should be careful with is the mass dimension of fields and couplings. In fact in $D$ dimensions, the mass dimension of the graviton (either potential or radiation) is 
\begin{equation}
\lfloor h\rfloor=\frac{D-2}{2} \; .
\end{equation}
However, we want to keep the metric $g_{\mu\nu}=\eta_{\mu\nu}+h_{\mu\nu}/\Mpl$ dimensionless, thus we need to impose
\begin{equation}
\lfloor\Mpl\rfloor =\lfloor h\rfloor=\frac{D-2}{2} = 1-\varepsilon \; .
\label{eq:mass.dim}
\end{equation}
Rather than change the dimension of the Planck mass, we redefine
\begin{equation}
\Mpl\longrightarrow\frac{\Mpl}{\mu^{\varepsilon}} \; ,
\end{equation}
where $\mu$ is an arbitrary parameter with mass dimension 1. In this way the combination $\Mpl/\mu^\varepsilon$ has the right mass dimension given by (\ref{eq:mass.dim}), and, at the same time, $\Mpl$ has still mass dimension 1, hence it deserves the name Planck ``mass''. Moreover we stress that when we work in $D$ dimensions we decompose a $D$ vector $k^\mu$ as 
\begin{equation}
k^\mu=(k^0,\bm{k}) \; ,
\end{equation} 
where now $\bm{k}$ represent a $D-1$ dimensional vector.

Let's try to compute the previous diagram in $D$ dimensions. We get
\begin{align}
\begin{tikzpicture}[baseline]
\begin{feynman}
\vertex (z);
\vertex [left=0.75cm of z, label=100:$\bm{k}$, label=270:$t_1$] (w);
\vertex [left=0.5cm of w] (a) {$\bm{x}_a$};  
\vertex [right=0.75cm of z, label=80:$\bm{q}$, label=270:$t_2$] (w');
\vertex [right=0.5cm of w'] (a');  
\diagram* {
(a) -- [plain] (a'),
(w) -- [scalar, half left] (w')
}; 
\end{feynman} 
\end{tikzpicture}
 & = \frac{i}{2}\frac{m^2_a\mu^{2\varepsilon}}{4\Mpl^2}\int\!\!d t\int\!\!\frtr{D-1}{\bm{k}}\frtr{D-1}{\bm{q}}\frac{e^{i(\bm{k}+\bm{q})\cdot\bm{x}_a}}{|\bm{k}|^2}\notag\\
 &\qquad\qquad\qquad\times(2\pi)^{D-1}\delta^{D-1}P_{0000} \notag \\
 & = i\frac{m_a^2\mu^{2\varepsilon}}{16\Mpl^2}\int\!\!d t\!\int\!\!\frtr{D-1}{\bm{k}}\frac{1}{|\bm{k}|^2} \; .
\label{eq:diverg.dim}
\end{align}
The above integral can be seen as the limit $\bm{x}\to 0$ of
\begin{equation}
\int\!\!\frtr{D-1}{\bm{k}}\frac{e^{-i\bm{k}\cdot\bm{x}}}{(\modul{k}^2)^\alpha}=\frac{1}{(4\pi)^{(D-1)/2}}\frac{\Gamma\left(\frac{D-1}{2}-\alpha\right)}{\Gamma(\alpha)}\left(\frac{\modul{x}^2}{4}\right)^{\alpha -\frac{D-1}{2}} \; ,
\end{equation}
where $\Gamma(z)$ is the Euler gamma function, and $\modul{\bm{k}}$ represents the modulus of the $k$-vector in $D-1$ dimensions. Therefore one can easily conclude that
\begin{equation}
\!\int\!\!\frtr{D-1}{\bm{k}}\frac{1}{|\bm{k}|^2} = 0 \; .
\end{equation}
Thus, eq. (\ref{eq:diverg.dim}) is really equal to zero, and this diagram does not add any contribution to the Newtonian physics.

This situation occurs in general at every PN order. In fact, in the computations of the previous section, we have already met a divergent quantity. This came from the diagram (\ref{eq:diag2.step}), which divergent part is
\begin{align}
\begin{tikzpicture}[baseline]
\begin{feynman}
\vertex (c);
\vertex [above=0.8cm of c] (z);
\vertex [above=0.8cm of z] (c');
\vertex [right=1.2cm of c] (a);
\vertex [left=1.2cm of c] (b) {$x_b$};
\vertex [right=0.7cm of c] (a'');
\vertex [left=0.7cm of c] (b'');
\vertex [right=1.2cm of c'] (a');
\vertex [left=1.2cm of c'] (b') {$x_a$};
\vertex [above right=0.2cm of a'] (h');
\vertex [below right=0.2cm of a] (i);
\vertex [below=0.1cm of i] (h);
\diagram* {
(a) -- [plain] (c) -- [plain] (b),
(a') -- [plain] (c') -- [plain] (b'),
(c') -- [scalar] (z),
(a'') -- [scalar] (z),
(b'') -- [scalar] (z),
(h') -- [plain] (h)
}; 
\end{feynman} 
\end{tikzpicture}_{\,\infty} & = -i\sum_{a\neq b}\frac{m_a m^2_b}{32\Mpl^4}\int\!\!d t\intvec{q}\frac{e^{-i\bm{q}\cdot\bm{x}_{ab}}}{|\bm{q}|^2}\intvec{k}\frac{1}{|\bm{k}|^2} \; .
\end{align}
However we had just seen that in dimensional regularization
\begin{equation}
\intvec{k}\frac{1}{|\bm{k}|^2}\longrightarrow\int\!\!\frtr{D-1}{\bm{k}}\frac{1}{|\bm{k}|^2} = 0 \; .
\end{equation}
Again we do not get any contribution from the divergent part of this diagram.

Of course, the fact that divergence contributions vanish in dimensional regularization is not a rule. In general, as in any QFT, we need to add suitable counterterms to remove the divergences form the theory\footnote{See \cite{EFT1} and \cite{EFT3} for deepen discussion on renormalization in NRGR.}.

\chapter{The radiative sector of NRGR}\label{ch:rad}

In the previous chapter we introduced the formalism of NRGR. After the mode decomposition (\ref{eq:mode.dec}), we integrated out the potential modes $\potH{k}{\mu\nu}$, and we wrote the resulting action schematically in eq. (\ref{eq:SeffNR}). In particular, we explicitly computed $S_\text{cons}$ at 1PN order.

In this chapter we are going to add also the contribution coming from the radiation field $\radh{\mu\nu}$, hence compute the complete $\Seff{NR}$. Once we have this, we can integrate out also the radiation field $\radh{\mu\nu}$, obtaining in this way what we called $\Seff{NRGR}$, which will depend only on the two world-lines $x_a^\mu$. Having this action, we will be able to compute the cross section for unit of energy $\omega$ and solid angle $\Omega$ through the optical theorem
\begin{equation}
\frac{1}{T}\Imm{\Seff{NRGR}[x_a^\mu]} = \frac{1}{2}\int\!\!d \omega d \Omega\frac{d^2\Gamma}{d \omega d \Omega} \; ,
\label{eq:Opt.th}
\end{equation}
where $T$ is an arbitrary time interval. From eq. (\ref{eq:Opt.th}) we derive the flux of energy lost by the system due to the emission of GWs, 
\begin{equation}
\mathcal{P}=\int\!\!d \omega d \Omega\left(\omega \frac{d^2\Gamma}{d \omega d \Omega}\right) \; .
\label{eq:Ploss}
\end{equation}

\section{Computing $\Seff{NR}$}\label{sec:SNR}

We want to add dissipative contributions to the conservative sector. Looking at eq. (\ref{eq:SeffNR}), we understand that we need to compute $S_\text{diss}$, which is diagrammatically defined in eq. (\ref{eqn:scheme14}b). At this stage we still do not need the propagator of $\radh{\mu\nu}$, nor its interaction vertices, hence we will not care about $S_{\radh{}}$ in this section. An explicit derivation of the quantity inside the round brackets of eq. (\ref{eqn:scheme14}b), which is nothing but $S^{\radh{}}_\text{pp}$, is provided in App. \ref{sec:PPe}. For the diagrammatic computations we will consider all possible diagrams with only one external radiation graviton\footnote{Graviton multi emission would contribute at higher PN orders.}. 

At this point, however, we have a significant issue. For sure we have diagrams with the following kind of vertices.
\begin{center}
\begin{tikzpicture}[baseline]
\begin{feynman}
\vertex (z);
\vertex [above=1.4cm of z] (c);
\vertex [right=2cm of z] (b) {$(p^0\!-\!k^0,\bm{p}\!-\!\bm{k})$};
\vertex [right=1.7cm of z] (b');
\vertex [left=2cm of z] (a) {$(p^0,\bm{p})$};
\vertex [left=1.7cm of z] (a');
\diagram* {
(a') -- [scalar, edge label=$\longrightarrow$] (z) -- [gluon, edge label'=$\uparrow (k^0\text{,}\,\bm{k})$] (c),
(z) -- [scalar, edge label=$\longrightarrow$] (b')
}; 
\end{feynman} 
\end{tikzpicture}
\end{center}
The final potential propagator, then, contains a factor of
\begin{equation}
\frac{1}{|\bm{p}-\bm{k}|^2}\simeq\frac{1}{\modul{p}^2}\left(1+2\frac{\bm{k}\cdot\bm{p}}{\modul{p}^2}+\dots\right) \; .
\label{eq:inf.series}
\end{equation}
As we wrote in eq. (\ref{eq:kradh.scal}), the radiation momentum $\bm{k}$ scales as $\bm{k}\sim v/r$, therefore the above term is actually an infinite series in powers of $v$, hence \textit{diagrams with this type of vertices do not have an homogeneous power of} $v$. As a consequence, we are not able to understand at which PN order they contribute, and we can not use the procedure set up in the previous chapter.

\subsection{Multipole expansion and computation of the diagrams}\label{sec:multipole}

Luckily, there is a tool that allows us to restore the power counting established in the previous chapter. We perform a \textit{multipole expansion} of the radiation field, i.e.
\begin{equation}
\radh{\mu\nu}(x^0,\bm{x}) =\radh{\mu\nu}(x^0,\bm{X})+\delta x^i\partial_i\radh{\mu\nu}(x^0,\bm{X})+\frac{1}{2}\delta x^i\delta x^j\partial_i\partial_j\radh{\mu\nu}(x^0,\bm{X})+\dots \; ,
\label{eq:multipole_generic}
\end{equation}
where $\bm{X}$ is some reference point, for instance the position of the center of mass, and $\delta\bm{x}\equiv \bm{x}-\bm{X}$. In this way $\radh{\mu\nu}$ is now only a function of time, and it will not transfer any 3-momentum in case of a vertex as the one sketched at the end of the previous page. The infinite series of eq. (\ref{eq:inf.series}) is replaced by a sequence of different vertices proportional to 
\begin{equation}
\delta x^i\partial_i\radh{\mu\nu}\sim v\radh{\mu\nu} \; .
\end{equation}
Thus, each of these vertices has a definite power of $v$, and the power counting established in the previous chapter can be used again. Of course then a similar expansion is induced also in vertices of the following form.
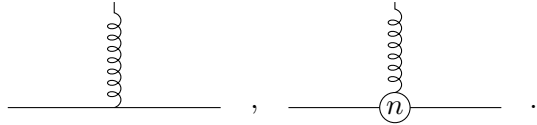
\begin{center}
\begin{tikzpicture}[baseline]
\begin{feynman}
\vertex (z);
\vertex [above=1.4cm of z] (c);
\vertex [right=1.4cm of z] (b);
\vertex [left=1.4cm of z] (a);
\diagram* {
(a) -- [plain] (z) -- [gluon] (c),
(z) -- [plain] (b)
}; 
\end{feynman} 
\end{tikzpicture} $\ $ , $\ $  
\begin{tikzpicture}[baseline]
\begin{feynman}
\vertex [empty dot, minimum size=0.4cm] (z) {$n$};
\vertex [above=1.4cm of z] (c);
\vertex [right=1.4cm of z] (b);
\vertex [left=1.4cm of z] (a);
\diagram* {
(a) -- [plain] (z) -- [gluon] (c),
(z) -- [plain] (b)
}; 
\end{feynman} 
\end{tikzpicture} \; .
\end{center}

After this multipole expansion, we are able to compute $S_\text{diss}$ order by order in the PN formalism. From eq. (\ref{eqn:scheme14}b) we can write
\begin{align}
iS_\text{diss} & = \underset{\sim L^{1/2}v^{1/2}}{\underbrace{\sum_a\begin{tikzpicture}[baseline]
\begin{feynman}
\vertex (z);
\vertex [below=0.1cm of z] (f);
\vertex [above=0.1cm of z] (f');
\vertex [right=0.5cm of f'] (c);
\vertex [right=0.7cm of c] (d);
\vertex [above=0.5cm of d] (d');
\diagram* {
(f') -- [plain] (c) -- [plain] (d),
(f') -- [draw=none] (z) -- [draw=none] (f),
(d') -- [gluon] (c)
}; 
\end{feynman} 
\end{tikzpicture}}} +
\underset{\sim L^{1/2}v^{3/2}}{\underbrace{\sum_a\begin{tikzpicture}[baseline]
\begin{feynman}
\vertex (z);
\vertex [below=0.1cm of z] (f);
\vertex [above=0.1cm of z] (f');
\vertex [right=0.5cm of f', empty dot, minimum size=0.4cm] (c) {1};
\vertex [right=0.7cm of c] (d);
\vertex [above=0.5cm of d] (d');
\diagram* {
(f') -- [plain] (c) -- [plain] (d),
(f') -- [draw=none] (z) -- [draw=none] (f),
(d') -- [gluon] (c)
}; 
\end{feynman} 
\end{tikzpicture}}} +
\underset{\sim L^{1/2}v^{5/2}}{\underbrace{\sum_a\begin{tikzpicture}[baseline]
\begin{feynman}
\vertex (z);
\vertex [below=0.1cm of z] (f);
\vertex [above=0.1cm of z] (f');
\vertex [right=0.5cm of f', empty dot, minimum size=0.4cm] (c) {2};
\vertex [right=0.7cm of c] (d);
\vertex [above=0.5cm of d] (d');
\diagram* {
(f') -- [plain] (c) -- [plain] (d),
(f') -- [draw=none] (z) -- [draw=none] (f),
(d') -- [gluon] (c)
}; 
\end{feynman} 
\end{tikzpicture}}} \notag \\
& \qquad\qquad +\underset{\sim L^{1/2}v^{5/2}}{\underbrace{\begin{tikzpicture}[baseline]
\begin{feynman}
\vertex (z);
\vertex [above=0.7cm of z] (c');
\vertex [below=0.6cm of z] (c);
\vertex [right=1cm of c] (a);
\vertex [left=1cm of c] (b);
\vertex [right=1cm of c'] (a');
\vertex [below=0.52cm of a'] (d);
\vertex [left=1cm of c'] (b');
\diagram* {
(a) -- [plain] (c) -- [plain] (b),
(a') -- [plain] (c') -- [plain] (b'),
(c) -- [scalar] (z) -- [scalar] (c'),
(d) -- [gluon] (c')
}; 
\end{feynman} 
\end{tikzpicture}\!+\!
\begin{tikzpicture}[baseline]
\begin{feynman}
\vertex (z);
\vertex [right=1cm of z] (d);
\vertex [above=0.7cm of z] (c');
\vertex [below=0.6cm of z] (c);
\vertex [right=1cm of c] (a);
\vertex [left=1cm of c] (b);
\vertex [right=1cm of c'] (a');
\vertex [left=1cm of c'] (b');
\diagram* {
(a) -- [plain] (c) -- [plain] (b),
(a') -- [plain] (c') -- [plain] (b'),
(c) -- [scalar] (z) -- [scalar] (c'),
(d) -- [gluon] (z)
}; 
\end{feynman} 
\end{tikzpicture}}} \!+\Ord{L^{1/2}v^{7/2}} \; .
\label{eq:scal.1/2.rad}
\end{align}
Under each diagram we wrote its scaling computed following tabs. \ref{table:gravv}, \ref{table:ppradmulti} and \ref{table:pppotrad} (pp.~\pageref{table:gravv},~\pageref{table:ppradmulti}). We decide to stop our computations at order $v^{5/2}$. In what follows, we also fix the origin of our frame in the center of mass (CoM) and do the multipole expansion around it, hence
\begin{equation}
\bm{X} =\bm{x}_{\text{cm}} = 0 \; .
\end{equation}
Therefore for the rest of this section, all the $\radh{\mu\nu}$ fields are evaluated at $(x^0,0)$.

The first three diagrams are very easy to compute; in fact, each of them is equal to its Feynman rule as written in tab. \ref{table:ppradmulti}, hence
\begin{subequations}
\label{eqn:scheme18}
\begin{align}
\sum_a\begin{tikzpicture}[baseline]
\begin{feynman}
\vertex (z);
\vertex [below=0.1cm of z] (f);
\vertex [above=0.1cm of z] (f');
\vertex [right=0.5cm of f'] (c);
\vertex [right=0.7cm of c] (d);
\vertex [above=0.5cm of d] (d');
\diagram* {
(f') -- [plain] (c) -- [plain] (d),
(f') -- [draw=none] (z) -- [draw=none] (f),
(d') -- [gluon] (c)
}; 
\end{feynman} 
\end{tikzpicture} & = -i\sum_a\frac{m_a}{2\Mpl}\int\!\!d t\,\bar{h}_{00} \; , \label{eq:diss1}\ \\
\sum_a\begin{tikzpicture}[baseline]
\begin{feynman}
\vertex (z);
\vertex [below=0.1cm of z] (f);
\vertex [above=0.1cm of z] (f');
\vertex [right=0.5cm of f', empty dot, minimum size=0.4cm] (c) {1};
\vertex [right=0.7cm of c] (d);
\vertex [above=0.5cm of d] (d');
\diagram* {
(f') -- [plain] (c) -- [plain] (d),
(f') -- [draw=none] (z) -- [draw=none] (f),
(d') -- [gluon] (c)
}; 
\end{feynman} 
\end{tikzpicture} & = -i\sum_a\frac{m_a}{2\Mpl}\int\!\!d t\,\left(x_a^i\partial_i\bar{h}_{00}+2\bar{h}_{0i}v_a^i\right) \; , \label{eq:nocontr} \\
\sum_a\begin{tikzpicture}[baseline]
\begin{feynman}
\vertex (z);
\vertex [below=0.1cm of z] (f);
\vertex [above=0.1cm of z] (f');
\vertex [right=0.5cm of f', empty dot, minimum size=0.4cm] (c) {2};
\vertex [right=0.7cm of c] (d);
\vertex [above=0.5cm of d] (d');
\diagram* {
(f') -- [plain] (c) -- [plain] (d),
(f') -- [draw=none] (z) -- [draw=none] (f),
(d') -- [gluon] (c)
}; 
\end{feynman} 
\end{tikzpicture} & = -i\sum_a\frac{m_a}{2\Mpl}\int\!\!d t \bigg(\frac{1}{2}\bar{h}_{00}v_a^2+\bar{h}_{ij}v_a^i v_a^j\notag \\
&\qquad\qquad\qquad\qquad\qquad +2 x_a^\ell\partial_\ell \bar{h}_{0i}v^i_a+\frac{1}{2}x_a^i x_a^j\partial_i\partial_j\bar{h}_{00}\bigg) \; .\label{eq:diss2}
\end{align}
\end{subequations}
Let us look at eq. (\ref{eq:nocontr}), we can easily recognise
\begin{align}
\sum_{a}m_ax_a^i = mx_{\text{cm}}^i \; ,& &  \sum_am_av_a^i=p_{\text{cm}}^i \; ,
\end{align}
where we defined $m\equiv\sum_a m_a$. By definition $\bm{p}_{\text{cm}} = 0$, and, since we have chosen to put the origin of our frame in the CoM, we have also $\bm{x}_{\text{cm}}=0$. From this considerations we see that (\ref{eq:nocontr}) gives a vanishing contribution to the action.

Then, we have to compute the last two diagrams
\begin{align}
\begin{tikzpicture}[baseline]
\begin{feynman}
\vertex [label=270:$t_2$, label=170:$\bm{q}$](c);
\vertex [above=0.6cm of c] (z);
\vertex [above=0.8cm of z,label=90:$t_1\ $, label=190:$\bm{k}$] (c');
\vertex [right=1cm of c] (a);
\vertex [left=1cm of c] (b);
\vertex [right=1cm of c'] (a');
\vertex [below=0.65cm of a'] (d);
\vertex [left=1cm of c'] (b');
\diagram* {
(a) -- [plain] (c) -- [plain] (b),
(a') -- [plain] (c') -- [plain] (b'),
(c) -- [scalar] (z) -- [scalar] (c'),
(d) -- [gluon] (c')
}; 
\end{feynman} 
\end{tikzpicture} & = -i\sum_{a\neq b}\frac{m_am_b}{8\Mpl^3}\int\!\!d tP_{0000}\intvec{k}\frac{e^{-i\bm{k}\cdot\bm{x}_{ab}}}{\modul{k}^2}\radh{00}  \notag \\
& =-\frac{i}{2\Mpl}\sum_{a\neq b}\frac{m_am_b}{32\pi\Mpl^2}\int\!\!d t\frac{1}{|\bm{x}_{ab}|}\radh{00}  \notag \\
& = -\frac{i}{\Mpl}\int\!\!d t\frac{\GN m_1m_2}{r}\radh{00} \; .
\label{eq:disdiag} \\
\begin{tikzpicture}[baseline]
\begin{feynman}
\vertex [label=270:$t_2$, label=170:$\bm{q}$] (c);
\vertex [above=0.7cm of c] (z);
\vertex [right=1cm of z] (d);
\vertex [above=0.7cm of z, label=90:$t_1\ $, label=190:$\bm{k}$] (c');
\vertex [right=1cm of c] (a);
\vertex [left=1cm of c] (b);
\vertex [right=1cm of c'] (a');
\vertex [left=1cm of c'] (b');
\diagram* {
(a) -- [plain] (c) -- [plain] (b),
(a') -- [plain] (c') -- [plain] (b'),
(c) -- [scalar] (z) -- [scalar] (c'),
(d) -- [gluon] (z)
}; 
\end{feynman} 
\end{tikzpicture} & = \frac{i}{2}\sum_{a\neq b}\frac{m_am_b}{8\Mpl^3}\int\!\!d t\intvec{k}\frac{e^{-i\bm{k}\cdot\bm{x}_{ab}}}{\modul{k}^4}\left[\left(\frac{3}{2}\radh{00}+\frac{\radhmi{i}{i}}{2}\right)\modul{k}^2-k^ik^j\radh{ij}\right]  \notag \\
& = \frac{i}{2} \sum_{a\neq b}\frac{m_am_b}{(32\pi)\Mpl^3}\int\!\!d t\left\{\frac{3}{2}\frac{\radh{00}}{|\bm{x}_{ab}|}\!+\!\cancel{\frac{1}{2}\frac{\radhmi{i}{i}}{|\bm{x}_{ab}|}}\!-\!\cancel{\frac{1}{2}\frac{\radhmi{i}{i}}{|\bm{x}_{ab}|}}\!+\!\frac{1}{2}\frac{x_{ab}^ix_{ab}^j}{|\bm{x}_{ab}|^3}\radh{ij}\right\} \notag \\
& = \frac{i}{2\Mpl}\int\!\!d t\left\{3\frac{\GN m_1m_2}{r}\radh{00}+\frac{\GN m_1m_2}{|\bm{x}_{12}|^3}x_{12}^ix_{12}^j\radh{ij}\right\} \; .
\label{eq:disdiag.int}
\end{align}
In order to simplify the last term, we recall that at leading order the world-lines satisfy the following EOM
\begin{align}
\ddot{x}_1^i=\frac{\GN m_2}{|\bm{x}_{12}|^3}x_{21}^i +\Ord{v} \; ,& & \ddot{x}_2^i=\frac{\GN m_1}{|\bm{x}_{12}|^3}x_{12}^i+\Ord{v} \; .
\label{eq:EOM_1PN}
\end{align}
Therefore, eq. (\ref{eq:disdiag.int}) can be written eventually as
\begin{equation}
\begin{tikzpicture}[baseline]
\begin{feynman}
\vertex (z);
\vertex [below=0.7cm of z, label=270:$t_2$, label=170:$\bm{q}$] (c);
\vertex [right=1cm of z] (d);
\vertex [above=0.7cm of z, label=90:$t_1\ $, label=190:$\bm{k}$] (c');
\vertex [right=1cm of c] (a);
\vertex [left=1cm of c] (b);
\vertex [right=1cm of c'] (a');
\vertex [left=1cm of c'] (b');
\diagram* {
(a) -- [plain] (c) -- [plain] (b),
(a') -- [plain] (c') -- [plain] (b'),
(c) -- [scalar] (z) -- [scalar] (c'),
(d) -- [gluon] (z)
}; 
\end{feynman} 
\end{tikzpicture} = \frac{i}{2\Mpl}\int\!\!d t\left\{3\frac{\GN m_1m_2}{r}\radh{00}-\sum_{a}m_ax_a^i\ddot{x}_a^j\radh{ij}\right\} \; .
\label{eq:disdiag2}
\end{equation}
Summing up eqs. (\ref{eq:diss1}), (\ref{eq:diss2}), (\ref{eq:disdiag}) and (\ref{eq:disdiag2}), we finally obtain
\begin{align}
iS_{\text{diss}} = & \!-i\frac{m}{2\Mpl}\int\!\!d t\,\bar{h}_{00} \notag \\
&-\frac{i}{2\Mpl}\int\!\!d t\Bigg\{\left[\frac{1}{2}\sum_a m_av_a^2\!-\frac{\GN m_1m_2}{r}\right]\radh{00}+2\sum_a m_ax_a^iv_a^j\partial_i\radh{0j}\notag \\
& \quad+\sum_a m_a\left(x_a^i\ddot{x}_a^j\radh{ij}\!+\!v_a^iv_a^j\radh{ij}\!+\!\frac{1}{2}x_a^ix_a^j\partial_i\partial_j\radh{00}\right)\Bigg\} \!+\! \Ord{L^{1/2}v^{7/2}} \; .
\label{eq:Sdissfirst}
\end{align}

Now, let's try to rewrite the above equation in a more convenient way. The last term of the second line can be recast as
\begin{align}
2\sum_a m_ax_a^iv_a^j\partial_i\radh{0j} & = \sum_a m_a(x_a^iv_a^j\partial_i\radh{0j} + x_a^iv_a^j\partial_i\radh{0j}) \notag \\
& \simeq \varepsilon^{ijk}L_k\partial_j\radh{0i}-\sum_a m_ax_a^ix_a^j\partial_0\partial_i\radh{0j} \; ,
\label{eq:last.term}
\end{align}
Where $\simeq$ means equal up to surface term; indeed to go from the first to the second line we integrated by parts one of the term in round brackets, and dropped terms proportional to a total derivatives. $L_k = \sum_a m_a\varepsilon_{kij}x^i v^j$ is the $k$ component of the orbital angular momentum.
If we add the last term of eq. (\ref{eq:last.term}) to the third line of (\ref{eq:Sdissfirst}), and we integrate by parts we obtain
\begin{equation}
\sum_am_a\left(-\cancel{v_a^iv_a^j\radh{ij}}-x_a^iv_a^j\partial_0\radh{ij}\!+\!\cancel{v_a^iv_a^j\radh{ij}}\!+\!\frac{1}{2}x_a^ix_a^j\partial_i\partial_j\radh{00}-x_a^ix_a^j\partial_0\partial_i\radh{0j}\right) \; .
\end{equation}
The quantity inside brackets of the above equation is actually equivalent to $-x_a^ix_a^j\tilde{\Rc}_{0i0j}$, where $\tilde{\Rc}_{\mu\nu\rho\sigma}$ is define as
\begin{equation}
\tilde{\Rc}_{\rho\mu\sigma\nu}\equiv \Mpl \Rc_{\rho\mu\sigma\nu}[\bar{g}] =\frac{1}{2}(\partial_\sigma\partial_\nu\radh{\mu\rho}+\partial_\rho\partial_\mu\radh{\nu\sigma}-\partial_\sigma\partial_\rho\radh{\mu\nu}-\partial_\mu\partial_\nu\radh{\rho\sigma}) \;.
\label{eq:R.Mpl.fact}
\end{equation} 
This is nothing but the linearised Riemann tensor for the metric $\bar{g}_{\mu\nu} = \eta_{\mu\nu}+\radh{\mu\nu}/\Mpl$, in which we isolated a factor of $\Mpl$ for convenience. Then, we eventually rewrite the action (\ref{eq:Sdissfirst}) as
\begin{align}
iS_{\text{diss}} = & \!-i\frac{m}{2\Mpl}\int\!\!d t\,\bar{h}_{00}-\frac{i}{2\Mpl}\int\!\!d t\!\left\{E_{\text{N}}\radh{00}\!+\!\varepsilon^{ijk}L_{k}\partial_j\radh{0i} \right\}\notag \\
& +i\sum_a\frac{m_a}{2\Mpl}\int\!\!d t\,x_a^ix_a^j\tilde{\Rc}_{0i0j} \!+\!\Ord{L^{1/2}v^{7/2}} \; ,
\end{align}
where we recognised the total Newtonian energy $E_{\text{N}}$ as the coefficient in front of $\radh{00}$. Inserting this expression into (\ref{eq:SeffNR}), and recalling the conservative sector given in eq. (\ref{eq:SEIH}), we get our final result for $\Seff{NR}$
\begin{align}
\Seff{NR}[x_a^\mu, \radh{\mu\nu}] &= S_{\radh{}}[\radh{}]\!+\!S_{\text{EIH}}[x_a^\mu]\!-\!\frac{m}{2\Mpl}\int\!\!d t\,\bar{h}_{00}\!-\!\frac{1}{2\Mpl}\int\!\!d t\!\left\{E_{\text{N}}\radh{00}\!+\!\varepsilon^{ijk}L_{k}\partial_j\radh{0i}\right\}\notag \\
& \qquad\qquad+\sum_a\frac{m_a}{2\Mpl}\int\!\!d t\,x_a^ix_a^j\tilde{\Rc}_{0i0j} + \Ord{L^{1/2}v^{7/2}} \; .
\label{eq:SNRfin}
\end{align}

If everything has been carried on consistently, we expect eq. (\ref{eq:SNRfin}) to be invariant under gauge transformation of the background metric, i.e.
\begin{equation}
\radh{\mu\nu}\rightarrow\radh{\mu\nu}+2\partial_{(\mu}\xi_{\nu)} \; .
\end{equation}
The conservative part $S_{\text{EIH}}$ does not depend on $\radh{\mu\nu}$, and $S_{\radh{}}[\radh{}]$ is obviously gauge invariant, along with the the term proportional to the Riemann tensor. The term proportional to the mass monopole $m$ transforms into a total derivative. We are left with 
\begin{equation}
E_{\text{N}}\radh{00}\rightarrow E_{\text{N}}\radh{00} + 2E_{\text{N}}\partial_0\xi_0= E_{\text{N}}\radh{00} + 2\partial_0\left(E_{\text{N}}\xi_0\right) \; .
\end{equation}
In the last step we were able to pull $\partial_0$ in front of everything because, at this order, $E_{\text{N}}$ is a conserved quantity. Thus, this term transforms into a total derivative as well. Moreover, we can see that
\begin{align}
\varepsilon^{ijk}L_{k}\partial_j\radh{0i} & \rightarrow \varepsilon^{ijk}L_{k}\partial_j\radh{0i} + \cancelto{0}{\varepsilon^{ijk}L_{k}\partial_j\partial_i\xi_0}+\varepsilon^{ijk}L_{k}\partial_j\partial_0\xi_i  \notag \\
& = \varepsilon^{ijk}L_{k}\partial_j\radh{0i} +\varepsilon^{ijk}\partial_0\left(L_{k}\partial_j\xi_i\right) \; .
\end{align}
The second term on the right hand side goes to zero because we are contracting a symmetric tensor $\partial_i\partial_j$ with an antisymmetric one $\varepsilon^{ijk}$. The third one instead becomes a total derivative, again, because at this order also the total orbital angular momentum is conserved.

We immediately see that, in order to preserve gauge invariance, it is crucial to add to the computation the two non linear diagrams (\ref{eq:disdiag}) and (\ref{eq:disdiag2}). In a more classical approach to GR, we would have had to add to the matter stress-energy tensor a \textit{pseudo-stress-energy tensor}, which takes into account the contributions to the energy of the system coming from the gravitational field itself. The non linear diagrams mentioned above, which entered naturally in our computations because of the power counting rules of NRGR, play physically the same role of this pseudo-stress-energy tensor\footnote{We will see this explicitly in Sec. \ref{sec:matching} when we will implement the matching procedure in the bottom-up approach to the radiative sector of NRGR.}.  

\section{Leading order power loss}\label{sec:power_leading}

We are now ready to compute the power loss of the system. In order to do so we have to integrate out also the radiation field, and find $\Seff{NRGR}$ defined in eqs.  (\ref{eq:SNRGR.path}). At this stage, we have already completely removed the orbit scale $r$, hence we can not really distinguish the two particles any more. The system is well described by a single particle (the CoM of the previous system) in a gravitational field $\eta_{\mu\nu}+\radh{\mu\nu}/\Mpl$. Therefore we slightly change our diagrammatic conventions and we represent the two-particle system as a double plain line, i.e.
\begin{equation*}
\begin{tikzpicture}[baseline]
\begin{feynman}
\vertex (f);
\vertex [above=0.1cm of f] (z);
\vertex [right=1.5cm of z] (w);  
\diagram* {
(z) -- [double] (w)
}; 
\end{feynman} 
\end{tikzpicture} \rightarrow \text{Two particle system.}
\end{equation*}
We can then rewrite 
\begin{equation}
i\Seff{NR}[x_a^\mu, \radh{\mu\nu}] = iS_{\radh{}}[\radh{}]+\sum_{n} \begin{tikzpicture}[baseline]
\begin{feynman}
\vertex (f);
\vertex [above=0.1cm of f] (z);
\vertex [right=0.6cm of z, empty dot, minimum size=0.4cm] (c) {n};
\vertex [right=1.6cm of z] (w);  
\diagram* {
(z) -- [double] (c) -- [double] (w)
}; 
\end{feynman} 
\end{tikzpicture} + \sum_{n} \begin{tikzpicture}[baseline]
\begin{feynman}
\vertex (f);
\vertex [above=0.1cm of f] (z);
\vertex [right=0.6cm of z, empty dot, minimum size=0.4cm] (c) {n};
\vertex [above=0.8cm of c] (c');
\vertex [right=1.6cm of z] (w);  
\diagram* {
(z) -- [double] (c) -- [double] (w),
(c') -- [gluon] (c)
}; 
\end{feynman} 
\end{tikzpicture} + \ {\begin{smallmatrix}\text{Similar diagrams with}\\ \text{arbitrary number of} \\ \text{external } \radh{} .\end{smallmatrix}}
\end{equation}
Comparing this expression with eq. (\ref{eq:SeffNR}), we immediately see that
\begin{align}
iS_{\text{cons}}[x_a^\mu]  = \sum_{n} \begin{tikzpicture}[baseline]
\begin{feynman}
\vertex (f);
\vertex [above=0.1cm of f] (z);
\vertex [right=0.6cm of z, empty dot, minimum size=0.4cm] (c) {n};
\vertex [right=1.6cm of z] (w);  
\diagram* {
(z) -- [double] (c) -- [double] (w)
}; 
\end{feynman} 
\end{tikzpicture} \; , & & 
iS_{\text{diss}}[x_a^\mu,\radh{\mu\nu}] = \sum_{n} \begin{tikzpicture}[baseline]
\begin{feynman}
\vertex (f);
\vertex [above=0.1cm of f] (z);
\vertex [right=0.6cm of z, empty dot, minimum size=0.4cm] (c) {n};
\vertex [above=0.8cm of c] (c');
\vertex [right=1.6cm of z] (w);  
\diagram* {
(z) -- [double] (c) -- [double] (w),
(c') -- [gluon] (c)
}; 
\end{feynman} 
\end{tikzpicture} + \dots  
\end{align}
In particular from eq. (\ref{eq:SNRfin}) we can say 
\begin{subequations}
\label{eqn:scheme21}
\begin{align}
\begin{tikzpicture}[baseline]
\begin{feynman}
\vertex (f);
\vertex [above=0.1cm of f] (z);
\vertex [right=0.8cm of z] (c);
\vertex [above=0.8cm of c] (c');
\vertex [right=1.6cm of z] (w);  
\diagram* {
(z) -- [double] (c) -- [double] (w),
(c') -- [gluon] (c)
}; 
\end{feynman} 
\end{tikzpicture} & = \frac{im}{2\Mpl}\int\!\!d t\,\bar{h}_{00} \label{eq:vert.rad1}\\
\begin{tikzpicture}[baseline]
\begin{feynman}
\vertex (f);
\vertex [above=0.1cm of f] (z);
\vertex [right=0.6cm of z, empty dot, minimum size=0.4cm] (c) {1};
\vertex [above=0.8cm of c] (c');
\vertex [right=1.6cm of z] (w);  
\diagram* {
(z) -- [double] (c) -- [double] (w),
(c') -- [gluon] (c)
}; 
\end{feynman} 
\end{tikzpicture} & = 0\ \text{ in the CoM} \\
\begin{tikzpicture}[baseline]
\begin{feynman}
\vertex (f);
\vertex [above=0.1cm of f] (z);
\vertex [right=0.6cm of z, empty dot, minimum size=0.4cm] (c) {2};
\vertex [above=0.8cm of c] (c');
\vertex [right=1.6cm of z] (w);  
\diagram* {
(z) -- [double] (c) -- [double] (w),
(c') -- [gluon] (c)
}; 
\end{feynman} 
\end{tikzpicture} & = -\frac{i}{2\Mpl}\int\!\!d t\left\{E_{\text{N}}\radh{00}+\varepsilon^{ijk}L_{k}\partial_j\radh{0i}-Q^{ij}\tilde{\Rc}_{0i0j}\right\}
\label{eq:vert.rad2}
\end{align}
\end{subequations}
In the last equation, we introduced the quadrupole moment
\begin{equation}
Q^{ij}(t)\equiv\sum_a m_ax_a^i(t)x_a^j(t) \; .
\end{equation} 
Before going on, let's underline that now our scaling rules are
\begin{align}
\begin{tikzpicture}[baseline]
\begin{feynman}
\vertex (f);
\vertex [above=0.1cm of f] (z);
\vertex [right=0.8cm of z] (c);
\vertex [above=0.8cm of c] (c');
\vertex [right=1.6cm of z] (w);  
\diagram* {
(z) -- [double] (c) -- [double] (w),
(c') -- [gluon] (c)
}; 
\end{feynman} 
\end{tikzpicture}\sim L^{1/2}v^{1/2} & & 
\begin{tikzpicture}[baseline]
\begin{feynman}
\vertex (f);
\vertex [above=0.1cm of f] (z);
\vertex [right=0.6cm of z, empty dot, minimum size=0.4cm] (c) {n};
\vertex [above=0.8cm of c] (c');
\vertex [right=1.6cm of z] (w);  
\diagram* {
(z) -- [double] (c) -- [double] (w),
(c') -- [gluon] (c)
}; 
\end{feynman} 
\end{tikzpicture}\sim L^{1/2}v^{n+1/2}
\end{align}

Having this ``new'' diagrammatic representation, we can compute $\Seff{NRGR}$ as the sum of all (tree-level) diagrams with only the two point particles as external lines, i.e.
\begin{align}
i\Seff{NRGR}[x_a^\mu] & = \sum_n \begin{tikzpicture}[baseline]
\begin{feynman}
\vertex (f);
\vertex [above=0.1cm of f] (z);
\vertex [right=0.6cm of z, empty dot, minimum size=0.4cm] (c) {n};
\vertex [right=1.6cm of z] (w);  
\diagram* {
(z) -- [double] (c) -- [double] (w)
}; 
\end{feynman} 
\end{tikzpicture} + \underset{\sim Lv}{\underbrace{\begin{tikzpicture}[baseline]
\begin{feynman}
\vertex (f);
\vertex [above=0.1cm of f] (z);
\vertex [right=0.6cm of z] (c);
\vertex [right=1.2cm of c] (c');
\vertex [right=0.6cm of c'] (w);  
\diagram* {
(z) -- [double] (c) -- [double] (c') -- [double] (w),
(c') -- [gluon, half right] (c)
}; 
\end{feynman} 
\end{tikzpicture}}} + \underset{\sim Lv^3}{\underbrace{\begin{tikzpicture}[baseline]
\begin{feynman}
\vertex (f);
\vertex [above=0.1cm of f] (z);
\vertex [right=0.6cm of z] (c);
\vertex [right=1.2cm of c, empty dot, minimum size=0.4cm] (c') {2};
\vertex [right=0.6cm of c'] (w);  
\diagram* {
(z) -- [double] (c) -- [double] (c') -- [double] (w),
(c') -- [gluon, half right] (c)
}; 
\end{feynman} 
\end{tikzpicture}}} \notag \\
&\qquad + \underset{\sim Lv^5}{\underbrace{\begin{tikzpicture}[baseline]
\begin{feynman}
\vertex (f);
\vertex [above=0.1cm of f] (z);
\vertex [right=0.6cm of z, empty dot, minimum size=0.4cm] (c) {2};
\vertex [right=1.2cm of c, empty dot, minimum size=0.4cm] (c') {2};
\vertex [right=0.6cm of c'] (w);  
\diagram* {
(z) -- [double] (c) -- [double] (c') -- [double] (w),
(c') -- [gluon, half right] (c)
}; 
\end{feynman} 
\end{tikzpicture} + \begin{tikzpicture}[baseline]
\begin{feynman}
\vertex (f);
\vertex [above=0.1cm of f] (z);
\vertex [right=0.6cm of z] (c);
\vertex [right=1.2cm of c, empty dot, minimum size=0.4cm] (c') {4};
\vertex [right=0.6cm of c'] (w);  
\diagram* {
(z) -- [double] (c) -- [double] (c') -- [double] (w),
(c') -- [gluon, half right] (c)
}; 
\end{feynman} 
\end{tikzpicture}}} \notag \\ 
&\qquad + \underset{\sim Lv^5}{\underbrace{\begin{tikzpicture}[baseline]
\begin{feynman}
\vertex (f);
\vertex [above=0.1cm of f] (z);
\vertex [right=0.4cm of z] (c);
\vertex [right=0.8cm of c] (x);
\vertex [right=0.8cm of x] (c');
\vertex [above left=0.1cm of c'] (y1);
\vertex [above right=0.06cm of c'] (y2);
\vertex [right=0.8cm of c'] (x');
\vertex [right=0.8cm of x'] (c'');
\vertex [right=0.4cm of c''] (w);  
\diagram* {
(z) -- [double] (c) -- [double] (c') -- [double] (w),
(c'') -- [gluon, half right] (y2),
(y1) -- [gluon, half right] (c),
(y2) -- [gluon] (c'),
(c') -- [gluon] (y1),
}; 
\end{feynman} 
\end{tikzpicture} +\begin{tikzpicture}[baseline]
\begin{feynman}
\vertex (f);
\vertex [above=0.1cm of f] (z);
\vertex [right=0.6cm of z] (c);
\vertex [right=1.1cm of c] (c');
\vertex [above=0.7cm of c', empty dot, minimum size=0.4cm] (x) {1};
\vertex [right=1.1cm of c'] (c'');
\vertex [right=0.6cm of c''] (w);  
\diagram* {
(z) -- [double] (c) -- [double] (c') -- [double] (w),
(c'') -- [gluon, out=90, in=0] (x),
(x) -- [gluon, out=180, in=90] (c),
(x) -- [gluon] (c')
}; 
\end{feynman} 
\end{tikzpicture}}} + \dots \; 
\label{eq:SNRGRdiag}
\end{align}
In (\ref{eq:SNRGRdiag}) we decide to stop at order $Lv^5$, hence at $2.5$PN order. Luckily, we do not have to compute most of the diagrams we listed  above. It has been shown in Ref. \cite{rad.rec} that the coupling of a generic vertex to a conserved quantity gives a vanishing contribution. More concretely this means that if $V^{\rho\sigma}(t)$ is a completely generic vertex and $\mathcal{C}^{\alpha\beta}$ is a conserved quantity, at a certain PN order, one can prove 
\begin{equation}
\begin{tikzpicture}[baseline]
\begin{feynman}
\vertex (f);
\vertex [above=0.1cm of f] (z);
\vertex [right=0.8cm of z, label=270:$\mathcal{C}^{\alpha\beta}$] (c);
\vertex [right=1.6cm of c, blob, minimum size=0.3cm, label=270:$V^{\rho\sigma}(t)$] (c') {};
\vertex [right=0.8cm of c'] (w);  
\diagram* {
(z) -- [double] (c) -- [double] (c') -- [double] (w),
(c') -- [gluon, half right] (c)
}; 
\end{feynman} 
\end{tikzpicture} = 0 \; .
\label{eq:cons.quant}
\end{equation}
We reproduce the computations of Ref. \cite{rad.rec} in App. \ref{App:cons.quant}. Since at our PN order the mass monopole $m$ is a conserved quantity, we can discard from eq. (\ref{eq:SNRGRdiag}) the second, the third, the fifth, the sixth and the seventh diagram. Then, we get
\begin{equation}
i\Seff{NRGR}[x_a^\mu] = iS_{\text{cons}}[x_a^\mu] +\begin{tikzpicture}[baseline]
\begin{feynman}
\vertex (f);
\vertex [above=0.1cm of f] (z);
\vertex [right=0.6cm of z, empty dot, minimum size=0.4cm] (c) {2};
\vertex [right=1.2cm of c, empty dot, minimum size=0.4cm] (c') {2};
\vertex [right=0.6cm of c'] (w);  
\diagram* {
(z) -- [double] (c) -- [double] (c') -- [double] (w),
(c') -- [gluon, half right] (c)
}; 
\end{feynman} 
\end{tikzpicture}+\Ord{Lv^6} \; .
\label{eq:SNRGR.2.5PN}
\end{equation}
In particular, this implies that we do not have any correction coming from the $0.5$ nor the $1.5$ PN order. Moreover, since the Newtonian energy $E_{\text{N}}$ and the orbital angular momentum $\bm{L}$ are also conserved at this order, in the above diagram only the term proportional to the Riemann tensor gives a non-vanishing contribution.

Let's then compute this diagram; it is easy to see that
\begin{align}
\!\!\begin{tikzpicture}[baseline]
\begin{feynman}
\vertex (f);
\vertex [above=0.1cm of f] (z);
\vertex [right=0.6cm of z, empty dot, minimum size=0.4cm, label=105:$t$, label=270:$ij$] (c) {2};
\vertex [right=1.5cm of c, empty dot, minimum size=0.4cm, label=75:$t'$, label=270:$k\ell$] (c') {2};
\vertex [right=0.6cm of c'] (w);  
\diagram* {
(z) -- [double] (c) -- [double] (c') -- [double] (w),
(c') -- [gluon, half right, edge label'=$k$] (c)
}; 
\end{feynman} 
\end{tikzpicture} \!=\! \frac{1}{2}\frac{i^2}{4\Mpl^2}\int\!\!d td t'Q^{ij}(t)\bangle*{\Tprod{\tilde{\Rc}_{0i0j}(t,\bm{x})\tilde{\Rc}_{0k0\ell}(t',\bm{x}')}} Q^{k\ell}(t') \; ,
\label{eq:diag.2.5PN}
\end{align}
where at the end we have to impose $\bm{x}=\bm{x}'=0$ because of the multipole expansion. The longest thing to compute is the correlator of the two Riemann tensors. Recalling the propagator of the radiation graviton given in eq. (\ref{eq:hbarprop}), one can find
\begin{align}
&\bangle*{\Tprod{\tilde{\Rc}_{0i0j}(t,0)\tilde{\Rc}_{0k0\ell}(t',0)}} = \frac{i}{4}\int\!\!\frtr{4}{k}\frac{e^{-ik^0(t-t')}}{k^2+i\varepsilon}\notag \\
&\qquad\qquad\qquad\qquad\times\bigg\{\frac{(k^0)^4}{2}\left(\eta_{ik}\eta_{j\ell}+\eta_{i\ell}\eta_{jk}-\eta_{ij}\eta_{k\ell}\right) \notag \\ 
&\qquad\qquad\qquad\qquad+\frac{1}{2}k_i k_j k_k k_\ell+(k^0)^2\left(-\frac{k_kk_\ell}{2}\eta_{ij}-\frac{k_ik_j}{2}\eta_{k\ell}\right)\notag \\
&\qquad\qquad\qquad\qquad+(k^0)^2\left(\frac{k_ik_k}{2}\eta_{j\ell}\!+\!\frac{k_ik_\ell}{2}\eta_{jk}\!+\!\frac{k_jk_k}{2}\eta_{i\ell}\!+\!\frac{k_jk_\ell}{2}\eta_{ik}\right)\bigg\}
\end{align}
Using rotational invariance of the integral, we can simplify the above integral by considering
\begin{align}
k_ik_j \rightarrow\frac{\modul{k}^2}{3}\left(-\eta_{ij}\right) \; ,& & k_ik_jk_kk_\ell\rightarrow\frac{\modul{k}^4}{15}\left(\eta_{ij}\eta_{k\ell}+\eta_{ik}\eta_{j\ell}+\eta_{i\ell}\eta_{jk}\right)\; .
\end{align}
We stress that this replacement can be done only under the integral sign. Recalling also that the radiation gravitons are on-shell, i.e. $k^0 =\modul{k}$, we obtain
\begin{align}
\bangle*{\Tprod{\tilde{\Rc}_{0i0j}(t,0)\tilde{\Rc}_{0k0\ell}(t',0)}} & = \frac{i}{20}\int\!\!\frtr{4}{k}\frac{(k^0)^4e^{-ik^0(t-t')}}{k^2+i\varepsilon} \notag \\
& \qquad\qquad\qquad\times\bigg\{\eta_{ik}\eta_{j\ell}+\eta_{i\ell}\eta_{jk}-\frac{2}{3}\eta_{ij}\eta_{k\ell}\bigg\} \; .
\end{align}
Therefore eventually eq. (\ref{eq:diag.2.5PN}) becomes
\begin{align}
\begin{tikzpicture}[baseline]
\begin{feynman}
\vertex (f);
\vertex [above=0.1cm of f] (z);
\vertex [right=0.6cm of z, empty dot, minimum size=0.4cm] (c) {2};
\vertex [right=1.2cm of c, empty dot, minimum size=0.4cm] (c') {2};
\vertex [right=0.6cm of c'] (w);  
\diagram* {
(z) -- [double] (c) -- [double] (c') -- [double] (w),
(c') -- [gluon, half right] (c)
}; 
\end{feynman} 
\end{tikzpicture} & = -\frac{i}{80\Mpl^2}\int\!\!d td t'\left(Q_{ij}(t)Q^{ij}(t')-\frac{1}{3}Q(t)Q(t')\right)\notag \\
&\qquad\qquad\qquad\times\intvec{k}\int\!\!\frac{d k^0}{2\pi}\frac{(k^0)^4}{k^2+i\varepsilon}e^{-ik^0(t-t')}  \notag \\
& = -\frac{i}{80\Mpl^2}\int\!\!d td t'I_{ij}(t)I^{ij}(t')\intvec{k}\left\{P\int\!\!\frac{d k^0}{2\pi}\frac{(k^0)^4e^{-ik^0(t-t')}}{(k^0)^2-\modul{k}^2}\right.\notag \\
&\qquad\qquad - \left.i\pi\int\!\!\frac{d k^0}{2\pi}\delta\Big((k^0)^2\!-\!\modul{k}^2\Big)(k^0)^4e^{-ik^0(t-t')}\right\} \notag \\
& \equiv -\frac{i}{80\Mpl^2}\left(\mathcal{I}_1+\mathcal{I}_2\right) \; ,
\label{eq:ste.ref.it}
\end{align}
where $Q$ is the trace of the quadrupole moment. In the second line we also introduced the traceless quadrupole moment defined as
\begin{equation}
I_{ij}(t)\equiv Q_{ij}(t)-\frac{\delta_{ij}}{3}Q(t) \; ,
\label{eq:TF.quad.mom}
\end{equation}
and we used eq. (\ref{eq:Pl.Sok}) to split the last integral in two. In fact, we could have used $I_{ij}$ rather than $Q_{ij}$ from the beginning, because the trace of the linearised Riemann tensor vanishes for on-shell gravitons. From now on we also make explicit the fact that  $\eta_{ij} = -\delta_{ij}$, hence we can use the Kronecker symbol to raise/lowered indices.

Now we solve the two integrals separately:
\begin{itemize}
\item replacing $t'=t+s$ in the first line of eq. (\ref{eq:ste.ref.it}) we get
\begin{align}
\mathcal{I}_1 & = \int\!\!d td s'I_{ij}(t)I^{ij}(t\!+\!s)\intvec{k}P\int\!\!\frac{d k^0}{2\pi}\frac{(k^0)^4e^{-ik^0s}}{(k^0)^2-\modul{k}^2}  \notag \\
& = \int\!\!d tI_{ij}(t)\sum_{n=0}^{\infty}\frac{1}{n!}\frac{d^n}{d t^n}I^{ij}(t)\int\!\!d s\intvec{k}P\int\!\!\frac{d k^0}{2\pi}\frac{s^n(k^0)^4e^{ik^0s}}{(k^0)^2-\modul{k}^2} \; ,
\label{eq:ste.ref.2}
\end{align}
where in the second step we Taylor expanded $I^{ij}(t\!+\!s)$ and sent $k^0\to -k^0$. In the second line of the above equation we recognise the integral $I^{(4)}_{\text{M}}(n,4,0)$ defined in App. \ref{sec:mast.int}, thus according to eq. (\ref{eq:mast.int}) (p.~\pageref{eq:mast.int}) we obtain, eventually,
\begin{equation}
\mathcal{I}_1 = \frac{1}{4\pi}\int\!\!d tI_{ij}\frac{d^5I^{ij}}{d t^5} \; .
\label{eq:I.1}
\end{equation}
\item The integral $\mathcal{I}_2$ of (\ref{eq:ste.ref.it}) is more straightforward
\begin{align}
\mathcal{I}_2 & = -\frac{i}{2}\int d k^0\!\intvec{k}\frac{(k^0)^4}{\modul{k}}\tilde{I}_{ij}(k^0)\tilde{I}^{ij}(-k^0)\left(\delta\Big(k^0\!+\!\modul{k}\Big)+\delta\Big(k^0\!-\!\modul{k}\Big)\right) \notag \\
& = -\frac{i}{2}\intvec{k}\,\modul{k}^3\Big|\tilde{I}_{ij}(\modul{k})\Big|^2 \; ,
\label{eq:I.2}
\end{align}
where we denote with $\tilde{I}_{ij}$ the time Fourier transform of the traceless quadrupole moment.
\end{itemize}
Inserting eqs. (\ref{eq:I.1}) and (\ref{eq:I.2}) in (\ref{eq:ste.ref.it}), one obtains
\begin{equation}
\begin{tikzpicture}[baseline]
\begin{feynman}
\vertex (f);
\vertex [above=0.1cm of f] (z);
\vertex [right=0.6cm of z, empty dot, minimum size=0.4cm] (c) {2};
\vertex [right=1.2cm of c, empty dot, minimum size=0.4cm] (c') {2};
\vertex [right=0.6cm of c'] (w);  
\diagram* {
(z) -- [double] (c) -- [double] (c') -- [double] (w),
(c') -- [gluon, half right] (c)
}; 
\end{feynman} 
\end{tikzpicture} = -\frac{i\GN}{10}\int\!\!d tI_{ij}\frac{d^5I^{ij}}{d t^5}-\frac{\pi\GN}{5}\intvec{k}\,\modul{k}^3\Big|\tilde{I}_{ij}(\modul{k})\Big|^2 \; .
\label{eq:final_diag_rad}
\end{equation}
This result allows us to write explicitly the action (\ref{eq:SNRGR.2.5PN}) as
\begin{align}
\Seff{NRGR}[x_a^\mu] & = S_{\text{cons}}[x_a^\mu] \notag \\
& -\frac{\GN}{10}\int\!\!d tI_{ij}\frac{d^5I^{ij}}{d t^5}+\frac{i\pi\GN}{5}\intvec{k}\,\modul{k}^3\Big|\tilde{I}_{ij}(\modul{k})\Big|^2+\Ord{Lv^6} \; .
\label{eq:SNRGR2.5.comp}
\end{align}

This is our final result for $\Seff{NRGR}$. Having this, we can compute the leading order power loss of the system through eq. (\ref{eq:Ploss}). To this end, we need to compute the imaginary part of the above action. As we said already, $S_{\text{cons}}$ is a real quantity, therefore, with a little effort, we obtain
\begin{equation}
2\,\Imm{\Seff{NRGR}} = \frac{2\pi\GN}{5}\intvec{k}\,\modul{k}^3\Big|\tilde{I}_{ij}(\modul{k})\Big|^2 \; .
\end{equation}
The $\bm{k}$ integral can be easily computed in spherical coordinates. Replacing $\omega =\modul{k}$ and using $d \Omega$ to denote the infinitesimal solid angle, we get
\begin{equation}
2\,\Imm{\Seff{NRGR}} = \int_0^\infty\!\!\!d \omega\int\!\!d \Omega\left[\frac{\GN}{20\pi^2}\omega^5\Big|\tilde{I}_{ij}(\omega)\Big|^2\right] = T\int_0^\infty\!\!\!d \omega\int\!\!d \Omega\left[\frac{d^2\Gamma}{d \omega d \Omega}\right] \; .
\end{equation}
In the last step we used the optical theorem (\ref{eq:Opt.th}). Now we can also understand that $T$ can be interpreted as the typical period of the binary, hence what we will compute is really the average over a period of the power loss\footnote{In fact, one expects the total power loss to be divergent, see Chaps. 3 and 4 of Ref. \cite{Maggiore} for more details.}. Using then eq. (\ref{eq:Ploss}), we arrive to our final result
\begin{align}
\mathcal{P} & = \frac{1}{T}\frac{\GN}{20\pi^2}\int\!\!d \Omega\int_{-\infty}^\infty\!\!\!d \omega\omega^6\Big|\tilde{I}_{ij}(\omega)\Big|^2 \notag \\
& = \frac{1}{T}\frac{\GN}{5\pi}\int\!\!d \omega\int\!\!d td t'e^{i\omega(t-t')}\frac{d^3I_{ij}(t)}{d t^3}\frac{d^3I^{ij}(t')}{d t'^3} \notag \\
& = \frac{\GN}{5}\frac{1}{T}\int\!\!d t\dddot{I}_{ij}(t)\dddot{I}^{\,ij}(t) \notag \\
& = \frac{\GN}{5}\left\langle\dddot{I}_{ij}(t)\dddot{I}^{\,ij}(t)\right\rangle_T \; ,
\label{eq:quad.formula.P}
\end{align}
where a dot means a derivative w.r.t. time and $\langle\bullet\rangle_T$ stands for a time average. This is the famous quadrupolar emission of gravitational waves already found by Einstein in 1916. This formula agrees also with the results of Refs. \cite{Ddim-Rules,EFT1}. This result confirms once again that NRGR is an alternative, consistent and elegant way to solve the binary inspiral problem using well established EFT techniques. 

\section{Radiation reaction and waveform}\label{sec:GWandGR}

This is not the end of the story. There are in fact other very interesting objects that we can compute in the radiative sector. In the following section we focus our attention on two quantities:
\begin{enumerate}
\item From eq. (\ref{eq:SNRGR2.5.comp}) we see that $S_{\text{diss}}$ contributes also to the real part of $\Seff{NRGR}$, thus it affects also the EOM of the system. These extra terms represent the so-called \textit{gravitational radiation reaction} effects on the dynamics of the system.
\item Consider the one point diagrams which gives the amplitude of emitting one radiation graviton
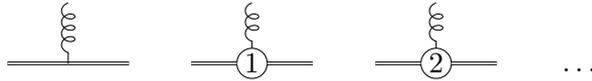
\begin{align*}
\begin{tikzpicture}[baseline]
\begin{feynman}
\vertex (f);
\vertex [above=0.1cm of f] (z);
\vertex [right=0.8cm of z] (c);
\vertex [above=0.8cm of c] (c');
\vertex [right=1.6cm of z] (w);  
\diagram* {
(z) -- [double] (c) -- [double] (w),
(c') -- [gluon] (c)
}; 
\end{feynman} 
\end{tikzpicture} \qquad
\begin{tikzpicture}[baseline]
\begin{feynman}
\vertex (f);
\vertex [above=0.1cm of f] (z);
\vertex [right=0.6cm of z, empty dot, minimum size=0.4cm] (c) {1};
\vertex [above=0.8cm of c] (c');
\vertex [right=1.6cm of z] (w);  
\diagram* {
(z) -- [double] (c) -- [double] (w),
(c') -- [gluon] (c)
}; 
\end{feynman} 
\end{tikzpicture} \qquad 
\begin{tikzpicture}[baseline]
\begin{feynman}
\vertex (f);
\vertex [above=0.1cm of f] (z);
\vertex [right=0.6cm of z, empty dot, minimum size=0.4cm] (c) {2};
\vertex [above=0.8cm of c] (c');
\vertex [right=1.6cm of z] (w);  
\diagram* {
(z) -- [double] (c) -- [double] (w),
(c') -- [gluon] (c)
}; 
\end{feynman} 
\end{tikzpicture} \qquad \dots
\end{align*}
Summing all these diagrams, one can compute the \textit{gravitational waveform} $\bangle*{\radh{\mu\nu}(x)}$ at every PN order.
\end{enumerate} 

Before going on with our discussion, we recall that, up until now, we have always worked in the harmonic gauge (\ref{eq:SGFrad}). However, this is not enough to completely fix the gauge of our problem. To do so, one needs to work in the so-called \textit{transverse-traceless gauge}, that, in our case, means
\begin{align}
\radh{0\mu} = 0 \; ,& & \radhmi{j}{j} = 0 \; ,& & \partial_j\radhmi{j}{i} = 0 \; .
\label{eq:TTgauge}
\end{align}
Let's introduce the following projector    
\begin{align}
\Lambda_{ijk\ell}(\hat{\bm{n}}) = P_{ik}P_{j\ell}-\frac{1}{2}P_{ij}P_{k\ell} \; ,
\end{align}
where $\hat{\bm{n}}$ is the unit vector pointing in the direction of propagation, and $P_{ij} = \delta_{ij}-n_i n_j$. Then it is known\footnote{For a deepen discussion on this subject see for instance Sec. 1.2 of Ref. \cite{Maggiore}.} that, if one has $\radh{\mu\nu}$ in the harmonic gauge, the GW in the transverse-traceless gauge can be found by applying the above projector to the spatial components of $\radh{\mu\nu}$, i.e.
\begin{equation}
\bar{h}^{\text{TT}}_{ij}(x) = \Lambda_{ijk\ell}(\hat{\bm{n}})\radhin{k\ell} (x) \; .
\label{eq:Grav_Wave_proj}
\end{equation}
Once we have the gravitational waveform in transverse-traceless gauge, we can compute again the power loss\footnote{For an explicit derivation of (\ref{eq:Ploss.GWf}) see again Ref. \cite{Maggiore}, Sec. 1.4.} of the system as
\begin{equation}
\frac{d\mathcal{P}}{d \Omega}= \modul{x}^2\bangle{\dot{\bar{h}}_{ij}^{\text{TT}}(t,\bm{x})\dot{\bar{h}}^{ij}_{\text{TT}}(t,\bm{x})}_T \; .
\label{eq:Ploss.GWf}
\end{equation}

In the following discussion, however, we shall see that the formalism used in the previous sections gives non physical results for the two quantities described above. We will then solve this problem in Sec. \ref{sec:inin_form_solved}.

\subsection{The need for the in-in formalism}\label{subsec:neddinin}

\subsubsection{No radiation reaction}

Let's then try to compute the radiation reaction. Since in the action (\ref{eq:SNRGR2.5.comp}) $S_{\text{cons}}$  take into account the conservative part of the motion of the two compact objects, any dissipative effect can only come from the real part of the radiative contribution we computed in the previous section, i.e. from the real part of the diagram (\ref{eq:final_diag_rad}). Let us then define the \textit{dissipative potential} $V_{\text{diss}}(t)$ as
\begin{equation}
\int\!\!d t V_{\text{diss}}(t) \equiv -\frac{1}{i}\Ree{\begin{tikzpicture}[baseline]
\begin{feynman}
\vertex (f);
\vertex [above=0.1cm of f] (z);
\vertex [right=0.6cm of z, empty dot, minimum size=0.4cm] (c) {2};
\vertex [right=1.2cm of c, empty dot, minimum size=0.4cm] (c') {2};
\vertex [right=0.6cm of c'] (w);  
\diagram* {
(z) -- [double] (c) -- [double] (c') -- [double] (w),
(c') -- [gluon, half right] (c)
}; 
\end{feynman} 
\end{tikzpicture}} \; ,
\label{eq:diss_pot_def}
\end{equation}
which takes into account any dissipative effect on the conservative motion of the binary. From eq. (\ref{eq:final_diag_rad}) one immediately reads
\begin{equation}
\int\!\!d t V_{\text{diss}}(t) = \frac{\GN}{10}\int\!\!d tI_{ij}\frac{d^5I^{ij}}{d t^5} \; .
\end{equation} 
Integrating by parts two times we obtain
\begin{equation}
\int\!\!d t V_{\text{diss}}(t) \simeq \frac{\GN}{10}\int\!\!d t\ddot{I}_{ij}(t)\dddot{I}^{\,ij}(t) = \frac{\GN}{20}\int\!\!d t\frac{d}{d t}\left(\ddot{I}^2_{ij}(t)\right) = 0 \; ,
\end{equation}
where again $\simeq$ means equal up to a surface term. Given this, one eventually concludes that
\begin{equation}
\Ree{\Seff{NRGR}} = S_{\text{cons}} \; .
\end{equation}

If the above equation were true, then the system would be described by conservative EOM and, subsequently, the two compact objects would stay on their orbits indefinitely. However this is in contrast with the fact that the system is loosing energy, and then that eventually has to decay.

\subsubsection{Non causal gravitational waveform}

Another way of seeing that there is a ``problem'' is looking at the gravitational waveform. As we said the relevant DOF are transverse-traceless. Therefore looking at the vertices (\ref{eq:vert.rad1}) and (\ref{eq:vert.rad2}), we immediately understand that the only contribution to the gravitational waveform at this PN order comes from 
\begin{equation}
\!\!\!\bangle*{\bar{h}^{\text{TT}}_{ij}(x)}\!=\!\Lambda_{ij\ell k}\ \begin{tikzpicture}[baseline]
\begin{feynman}
\vertex [empty dot, minimum size=0.4cm, label=45:$t'$] (z) {2};
\vertex [above=0.6cm of z] (a);
\vertex [right=1.2cm of z, label=270:$x$, label=90:$k\ell$] (c);
\vertex [below=0.6cm of z] (b);  
\diagram* {
(a) -- [double] (z) -- [double] (b),
(c) -- [gluon] (z)
}; 
\end{feynman} 
\end{tikzpicture} \!=\! \frac{i}{2\Mpl}\Lambda_{ij\ell k}\int\!\!d t'\bangle*{\Tprod{\radhin{k\ell}(x)\tilde{\Rc}_{0m0n}(t',0)}}I^{mn}(t') \, .
\label{eq:htt.step}
\end{equation}
From eq. (\ref{eq:TTgauge}), we understand that the only relevant part of the correlator inside the integral is
\begin{equation}
\frac{1}{2}\partial^2_{t'}\bangle*{\Tprod{\radhin{k\ell}(x)\radh{mn}(t',0)}} \!=\! \frac{1}{2}\partial^2_{t'}D(t-t',\bm{x})P{^{k\ell}}_{mn} \; .
\end{equation}
Inserting this in (\ref{eq:htt.step}) and integrating by parts two times we get
\begin{equation}
\bangle*{\bar{h}^{\text{TT}}_{ij}(x)}=\frac{1}{4\Mpl}\Lambda_{ij\ell k}\int\!\!d t'\left(iD(t\!-\!t',\bm{x})\right)\ddot{I}^{k\ell}(t') \; .
\end{equation}
The right hand side of the above equation is in general a complex quantity. Let's consider only the real part, which is the physically relevant one. We can then use the following distributional relation
\begin{equation}
\Ree{iD(x-x')}\!=\!\frac{1}{2}\left[D_{\text{Rt}}(x,x')\!+\!D_{\text{Ad}}(x,x')\right] \!=\! \frac{1}{4\pi}\delta\left((t-t')^2-|\bm{x}-\bm{x}'|^2\right) \; ,
\end{equation}
where $D_{\text{Rt}}$ and $D_{\text{Ad}}$ are respectively the retarded and the advanced propagator, that in direct space have the following form
\begin{subequations}
\label{eqn:scheme22}
\begin{align}
D_{\text{Rt}}(x,x') & = \frac{1}{2\pi}\vartheta(t-t')\delta\left((t-t')^2-|\bm{x}-\bm{x}'|^2\right) \; ,\label{eq:Dret.mass0}\\
 D_{\text{Ad}}(x,x') & = \frac{1}{2\pi}\vartheta\big(-(t-t')\big)\delta\left((t-t')^2-|\bm{x}-\bm{x}'|^2\right) \; .
\end{align}
\end{subequations}
Multiplying both sides by $\Mpl^{-1}$ for convenience, it is not hard to see that the real part of (\ref{eq:htt.step}) is given by
\begin{align}
\frac{1}{\Mpl}\Ree{\bangle*{\bar{h}^{\text{TT}}_{ij}(x)}} & = 2\GN\Lambda_{ij\ell k}\int\!\!d t'\frac{1}{2\modul{x}}\left[\delta\left(t'\!-\!t_\text{Rt}\right)+\delta\left(t'\!-\!t_\text{Ad}\right)\right]\ddot{I}^{k\ell }(t') \notag \\
& = \frac{\GN}{\modul{x}}\Lambda_{ijk\ell }\left[I^{k\ell }(t_\text{Rt})+I^{k\ell }(t_\text{Ad})\right] \; .
\end{align}
In the last two steps we defined for simplicity $t_\text{Rt} \equiv t-\modul{x}$ and $t_\text{Ad} \equiv t+\modul{x}$. There must be again something wrong because, since on the right hand side we have both $t_\text{Rt} $ and $t_\text{Ad}$, we find as a result a graviton which is propagating non causally.

These kind of problems rely in the approach we used to compute diagrams in our EFT formalism rather than in NRGR itself. In fact up until now, to compute our effective actions we used the so-called \textit{in-out} path integral formalism. The problem is basically that this approach does not take into account the fact that the system is no longer time symmetric. We shall now see that implementing the \textit{in-in} formalism in NRGR indeed solves the two problems we have just found. In order not to deviate too much from the main theme of this work, we will (briefly) talk about the comparison between in-in and in-out formalism in App. \ref{App:inininout}. A nice discussion and a good set of references on this subject are presented in Ref. \cite{rad.rec}.

\subsection{Implementing the in-in formalism in NRGR}\label{sec:inin_form_solved}

Let's now try to implement the in-in formalism in NRGR. First, we have to derive the expression of the generator of connected diagrams in order to understand how Feynman rules change in this formalism. Following the brief description we have done in App. \ref{App:inininout}, we can implement the in-in formalism by adding a forward-evolution current $J_1^{\mu\nu}$ and a backward-evolution current $J_2^{\mu\nu}$ so that
\begin{align}
e^{iW[x^\mu_{a\,1},x^\mu_{a\,2}, J_1^{\mu\nu},J_2^{\mu\nu}]} = \int\!\!\Diff\radh{\mu\nu\,1}\Diff\radh{\mu\nu\,2}\exp & \Bigg\{i\Seff{NR}[x^\mu_{a\,1},\radh{\mu\nu\,1}]- i\Seff{NR}[x^\mu_{a\,2},\radh{\mu\nu\,2}] \notag \\
&\!\!+ iS^{\text{GF}}[\radh{\mu\nu\,1}]-iS^{\text{GF}}[\radh{\mu\nu\,2}] \notag \\
&\!\!+i\int\!\!d^4\!x\left(J_1^{\mu\nu}\radh{\mu\nu\,1}\!-\!J_2^{\mu\nu}\radh{\mu\nu\,2}\right)\Bigg\} \; ,
\label{eq:inin.path}
\end{align}
with the condition that all forward and backward fields coincide at infinity. Recalling eq. (\ref{eq:SeffNR}) for $\Seff{NR}$, for both forward and backward evolution field, we will now split it as follows
\begin{equation}
\Seff{NR}[x^\mu_a,\radh{\mu\nu}]+S^{\text{GF}}[\radh{\mu\nu}] = S_\text{cons}[x_a^\mu]+S^{(2)}[\radh{\mu\nu}]+S_\text{int}[x^\mu_a,\radh{\mu\nu}] \; ,
\end{equation}
where $S^{(2)}$ is the gauge fixed quadratic part of the action for the graviton, and $S_\text{int}$ contains $S_{\text{diss}}$ and all pure graviton interaction terms. In this way we can write (\ref{eq:inin.path}) as follows
\begin{align}
e^{iW[\dots]} & =  e^{i\left(S_\text{cons}[x_{a\,1}^\mu]-S_\text{cons}[x_{a\,2}^\mu]\right)} \notag \\
&\qquad\qquad\qquad\times \exp\left\{iS_{\text{int}}\left[x^\mu_{a\,1},x^\mu_{a\,2},\frac{\delta}{i\delta J_1^{\mu\nu}},\frac{\delta}{i\delta J_2^{\mu\nu}}\right]\right\}Z_{0}[J_1^{\mu\nu},J_2^{\mu\nu}]
\label{eq:inin_feyn_NRGR}
\end{align}
where $Z_{0}$ is the functional generator for the free graviton theory. 

We will now work in the Keldysh representation (see App. \ref{subsec:Keldysh}), hence we define\footnote{We define the Keldysh variables also for the zero component of $x^\mu_a$, even though it is obvious that $x_{a\,-}^0=0$ and $x_{a\,+}^0=x_a^0$ which is nothing but the time, which is absolute in this non relativistic system.}
\begin{subequations}
\label{eqn:scheme26}
\begin{align}
& x_{a+}^\mu = \frac{1}{2}\left(x_{a\,1}^\mu+x_{a\,2}^\mu\right) \; ,& &  x_{a-}^\mu = x_{a\,1}^\mu-x_{a\,2}^\mu \; ,\\
& J_{+}^{\mu\nu} = \frac{1}{2}\left(J_1^{\mu\nu}+J_2^{\mu\nu}\right) \; , & &  J_{-}^{\mu\nu} = J_{1}^{\mu\nu}-J_2^{\mu\nu} \; ,
\end{align}
\end{subequations}  
with the condition that at infinity
\begin{align}
x_{a+}^\mu = x_a^\mu & & x_{a-}^\mu =0 \; .
\label{eq:cond.+.-}
\end{align}
We can then write 
\begin{equation}
Z_{0}[J_+^{\mu\nu},J_-^{\mu\nu}] = \exp\left\{\frac{1}{2}\int\!\!d^4\!xd^4\!y\,iJ_A^{\mu\nu}D^{AB}(x-y)\,iJ_B^{\rho\sigma}P_{\mu\nu\rho\sigma}\right\} \; ,
\end{equation}
where $A,B=+,-$, and $D^{AB}$ is given by eq. (\ref{eq:matrix.prop})
\begin{equation}
D^{AB} = \left(\begin{matrix}
D^{++} & D^{+-} \\
D^{-+} & D^{--}
\end{matrix}\right) = \left(\begin{matrix}
0 & -iD_{\text{Ad}} \\
-iD_{\text{Rt}} & \frac{1}{2}D_{\text{H}}
\end{matrix}\right) \; .
\end{equation}
See Secs. \ref{sec:inin.app} and \ref{subsec:Keldysh} of App. \ref{App:inininout} for an explicit derivation of this equation. The quantity $D_{\text{H}}$ is the Hadamard two point function, defined in eq. (\ref{eq:had.2point}) of the same appendix. Therefore from (\ref{eq:inin_feyn_NRGR}) we understand that Feynman rules are slightly different from the ones depicted in the previous chapter. For instance the graviton propagator becomes
\begin{equation} 
\begin{tikzpicture}[baseline]
\begin{feynman}
\vertex [dot, label=180:$y$, label=270:$\rho\sigma$, label=90:$A$] (z) {};
\vertex [dot, label=0:$x$, right=2.5cm of z, label=270:$\mu\nu$, label=90:$B$] (w) {}; 
\diagram* {
(z) -- [gluon, edge label=$k$] (w)
}; 
\end{feynman} 
\end{tikzpicture} = D^{AB}(x-y) P_{\mu\nu\rho\sigma} \, .
\end{equation}
Moreover, any vertex coupled to a graviton has an additional index $A=+,-$ which contracts with the one coming from the propagator written above.

\subsection{Radiation reaction}

Once we have this construction, we are now ready to compute again the second diagram of eq. (\ref{eq:SNRGR.2.5PN}). As we said in the previous section, this is the only one contributing to the dissipative part of $\Seff{NRGR}$. With our new Feynman rules we get
\begin{align}
\!\!\begin{tikzpicture}[baseline]
\begin{feynman}
\vertex (f);
\vertex [above=0.1cm of f] (z);
\vertex [right=0.6cm of z, empty dot, minimum size=0.4cm, label=120:$t$, label=270:$\{A\text{, }ij\}$] (c) {2};
\vertex [right=1.5cm of c, empty dot, minimum size=0.4cm, label=60:$t'$, label=270:$\{B\text{, }k\ell \}$] (c') {2};
\vertex [right=0.6cm of c'] (w);  
\diagram* {
(z) -- [double] (c) -- [double] (c') -- [double] (w),
(c') -- [gluon, half right] (c)
}; 
\end{feynman} 
\end{tikzpicture} \!=\! \frac{1}{2}\frac{i^2}{4\Mpl^2}\int\!\!d td t'\,I_A^{ij}(t)\langle\Tprod{\tilde{\Rc}^A_{0i0j}(t,\bm{x})\tilde{\Rc}^B_{0k0\ell }(t',\bm{x}')}\rangle I_B^{k\ell }(t') \; ,
\end{align}
where we have already written everything in terms of the trace-free quadrupole moment\footnote{See the brief discussion around eq. (\ref{eq:TF.quad.mom}), p.~\pageref{eq:TF.quad.mom}.} (\ref{eq:TF.quad.mom}). Recall also that eventually we have to evaluate everything at $\bm{x}=\bm{x}'=0$, because of the multipole expansion. Doing computations similar to those that we carried out in the previous section, we find
\begin{align}
\begin{tikzpicture}[baseline]
\begin{feynman}
\vertex (f);
\vertex [above=0.1cm of f] (z);
\vertex [right=0.6cm of z, empty dot, minimum size=0.4cm] (c) {2};
\vertex [right=1.5cm of c, empty dot, minimum size=0.4cm] (c') {2};
\vertex [right=0.6cm of c'] (w);  
\diagram* {
(z) -- [double] (c) -- [double] (c') -- [double] (w),
(c') -- [gluon, half right] (c)
}; 
\end{feynman} 
\end{tikzpicture} & = -\frac{1}{80\Mpl^2}\int\!\!d td t'\,I_A^{ij}(t)\frac{d^4D^{AB}(t\!-\!t',0)}{d^2td^2t'}I_{B\,ij}(t')  \notag \\
& = \frac{1}{80\Mpl^2}\int\!\!d td t'\Bigg\{2iI_-^{ij}(t)\frac{d^4D_{\text{Rt}}(t\!-\!t',0)}{d^2td^2t'}I_{+\,ij}(t') \notag \\
&\qquad\qquad\qquad\qquad\quad-\frac{1}{2}I^{ij}_-(t)\frac{d^4D_{\text{H}}(t\!-\!t',0)}{d^2td^2t'}I_{-\,ij}(t')\Bigg\} \; .
\label{eq:diag.inin.sec}
\end{align}
To go from the first to the second line in the above equation, we used the fact that $D_{\text{Ad}}(t\!-\!t',0) = D_{\text{Rt}}(t'\!-\!t,0)$. Moreover, since we are working in the Keldysh representation, we introduced
\begin{align}
I_+^{ij}(t) = \frac{1}{2}\big(I_1^{ij}(t)+I_2^{ij}(t)\big) \; ,& & I_-^{ij}(t) = I_1^{ij}(t)-I_2^{ij}(t) \; .
\end{align}
These two quantities can be written explicitly in terms of $\bm{x}_{a\,+}$ and $\bm{x}_{a\,-}$ as
\begin{subequations}
\label{eqn:scheme27}
\begin{align}
I_+^{ij} & = \sum_a m_a\left(x_{a\,+}^i x_{a\,+}^j-\frac{1}{3}\delta^{ij}|\bm{x}_{a\,+}|^2\right) +\Ord{\bm{x}_{a\,-}^3} \; , \\
I_-^{ij} & = \sum_a m_a\left(x_{a\,-}^i x_{a\,+}^j+x_{a\,+}^i x_{a\,-}^j-\frac{2}{3}\delta^{ij}\bm{x}_{a\,-}\cdot\bm{x}_{a\,+}\right) \; . 
\end{align}
\end{subequations}
From now on we will not consider any $\Ord{\bm{x}_{a\,-}^2}$ contribution; it can be showed that these are effects of the quantum fluctuations of the radiation graviton field on the trajectory of the binary\footnote{This discussion is beyond the scope of this work. We redirect the reader to Ref. \cite{rad.rec} and references therein  for more informations on this subject.}. Given this consideration, one could have considered from the beginning 
\begin{equation}
D^{AB}  = \left(\begin{matrix}
0 & -iD_{\text{Ad}} \\
-iD_{\text{Rt}} & 0
\end{matrix}\right) \; .
\end{equation}
Being $I_+^{ij}$ symmetric and traceless, eq. (\ref{eq:diag.inin.sec}) becomes
\begin{equation}
\!\!\begin{tikzpicture}[baseline]
\begin{feynman}
\vertex (f);
\vertex [above=0.1cm of f] (z);
\vertex [right=0.5cm of z, empty dot, minimum size=0.4cm] (c) {2};
\vertex [right=1.3cm of c, empty dot, minimum size=0.4cm] (c') {2};
\vertex [right=0.5cm of c'] (w);  
\diagram* {
(z) -- [double] (c) -- [double] (c') -- [double] (w),
(c') -- [gluon, half right] (c)
}; 
\end{feynman} 
\end{tikzpicture} = \frac{i}{20\Mpl^2}\int\!\!d td t'\frac{d^4D_{\text{Rt}}(t\!-\!t',0)}{d^2td^2t'}\left(\sum_a m_ax_{a\,-}^i(t) x_{a\,+}^j(t)\!\right)I^+_{ij}(t') \; .
\end{equation}
Finally, using the fact that
\begin{equation}
\int\!\!d t'\frac{d^4D_{\text{Rt}}(t\!-\!t',0)}{d^2td^2t'}I_{+\,ij}(t') = -\frac{1}{4\pi}\frac{d^5}{dt^5}I_{+\,ij}(t) \, ,
\end{equation}
we eventually obtain
\begin{equation}
\begin{tikzpicture}[baseline]
\begin{feynman}
\vertex (f);
\vertex [above=0.1cm of f] (z);
\vertex [right=0.6cm of z, empty dot, minimum size=0.4cm] (c) {2};
\vertex [right=1.5cm of c, empty dot, minimum size=0.4cm] (c') {2};
\vertex [right=0.6cm of c'] (w);  
\diagram* {
(z) -- [double] (c) -- [double] (c') -- [double] (w),
(c') -- [gluon, half right] (c)
}; 
\end{feynman} 
\end{tikzpicture} = -\frac{i}{80\pi\Mpl^2}\int\!\!d t\left(\sum_a m_ax_{a\,-}^i(t) x_{a\,+}^j(t)\right)\frac{d^5}{dt^5}I_{+\,ij}(t) \; .
\end{equation}

At this point, we can use eq. (\ref{eq:diss_pot_def}) to compute the dissipative potential $V_{\text{diss}}$. This is explicitly given by
\begin{equation}
\int\!\!dt\,V_{\text{diss}}(t) = \frac{2\GN}{5}\int\!\!d t\left(\sum_a m_ax_{a\,-}^i(t) x_{a\,+}^j(t)\right)\frac{d^5}{dt^5}I_{+\,ij}(t) \; .
\label{eq:V_diss_explic_inin}
\end{equation}
Once we have this, we can find the total EOM of the system as usual, i.e.
\begin{align}
\frac{\delta}{\delta x_{a\,-}^i}\Ree{\Seff{NRGR}}\Bigg|_{\begin{smallmatrix}x_{a\,-}^i =0\\ \ \\ \ x_{a\,+}^i =x_a^i \end{smallmatrix}} = 0 \; ,& & \frac{\delta}{\delta x_{a\,+}^i}\Ree{\Seff{NRGR}}\Bigg|_{\begin{smallmatrix}x_{a\,-}^i =0\\ \ \\ \ x_{a\,+}^i =x_a^i \end{smallmatrix}} = 0 \; .
\end{align}
Since eventually one has to set $x_{a\,-}^i =0$, the second of the above equations is always trivial, hence we do not consider it any more. We are interesting in computing the dissipative part of the EOM, therefore here we do not care about the variation of $S_\text{cons}$. We just need to compute the extrema of $V_{\text{diss}}$ obtained in eq. (\ref{eq:V_diss_explic_inin}). It is not hard to see that
\begin{align}
\frac{\delta}{\delta x_{a\,-}^i}\left(\int\!\!dt\,V_{\text{diss}}(t)\right)\Bigg|_{\begin{smallmatrix}x_{a\,-}^i =0\\ \ \\ \ x_{a\,+}^i =x_a^i \end{smallmatrix}} & = \frac{2m_a\GN}{5}x_a^j(t)\frac{d^5}{dt^5}I_{ij}(t) \; .
\end{align}

Implementing the in-in path integral formalism in the context of NRGR, we were able to compute the correct form of the radiation reaction force experienced by the system due to the emission of GWs, solving in this way one of the two inconsistency found in Sec. \ref{subsec:neddinin}. This result coincides with the one that had already been obtained by Burke and Thorne in 1970 \cite{Thorne.Burke}.

\subsection{Gravitational waveform}\label{subsec:GWf}

The other quantity we can now correctly compute using this formalism is the gravitational waveform. As we wrote in eq. (\ref{eq:Grav_Wave_proj}), this can be found by looking at the one point Green function projected onto the transverse-traceless gauge, i.e.
\begin{equation}
\bangle{\radh{ij}^{\text{TT}}(t,\bm{x})} = \Lambda_{ij}{^{k\ell }}\bangle*{\radh{k\ell} (t, \bm{x})}=\Lambda_{ij}{^{k\ell }}\frac{\delta iW[\dots]}{i \delta J^{k\ell }_-(t,\bm{x})}\bigg|_{\text{sources}=0} \; .
\end{equation}
We used eq. (\ref{eq:one_point_inin_K}) of App. \ref{App:inininout} to compute the one point expectation value in the Keldysh representation. The quantity $W[\dots]$ is given by (\ref{eq:inin_feyn_NRGR}), rewritten in terms of Keldysh variables. At this order, the only non-vanishing contribution comes from the following term of $S_{\text{int}} $
\begin{align}
\begin{tikzpicture}[baseline]
\begin{feynman}
\vertex (f);
\vertex [above=0.1cm of f] (z);
\vertex [right=0.6cm of z, empty dot, minimum size=0.4cm] (c) {2};
\vertex [above=0.8cm of c] (c');
\vertex [right=1.6cm of z] (w);  
\diagram* {
(z) -- [double] (c) -- [double] (w),
(c') -- [gluon] (c)
}; 
\end{feynman} 
\end{tikzpicture} & \longrightarrow \frac{i}{2\Mpl}\int\!\!d t'\,I_A^{mn}(t')\frac{1}{2}\frac{d^2}{d t'^2}\radh{mn}^A(t',0)\notag \\
& \longrightarrow  \frac{i}{4\Mpl}\int\!\!d t'\,\ddot{I}_A^{mn}(t')\left(\frac{\delta}{i\delta J^{mn}_A(t',0)}\right) \; ,
\end{align}
where in the second step we integrated by parts two times. Therefore, we can compute
\begin{align}
\bangle*{\radh{k\ell} (t, \bm{x})} & = \frac{\delta}{i\delta J^{k\ell }_-(t,\bm{x})}\left.\left(\frac{i}{4\Mpl}\int\!\!d t'\,\ddot{I}_A^{mn}(t')\frac{\delta}{i\delta J^{mn}_A(t',0)}\right)Z_{0}\right|_{\text{sources = 0}} \notag \\
& = \frac{i}{4\Mpl^2}\int\!\!d t'\,D^{-A}(t-t',\bm{x})P_{k\ell mn}\ddot{I}^{mn}_A(t')\Bigg|_{x_{a\,-}^i =0 \, , \, x_{a\,+}^i =x_a^i} \notag \\
& = 8\pi\GN\int\!\!d t'\,D_{\text{Rt}}(t-t',\bm{x})\ddot{I}_{k\ell } \; .
\end{align}
Projecting this result onto the transverse-traceless gauge and dividing both sides by $\Mpl$ for convenience, we eventually find
\begin{equation}
\frac{1}{\Mpl}\bangle{\radh{ij}^{\text{TT}}(t,\bm{x})} = \frac{2\GN}{\modul{x}}\Lambda_{ijk\ell}\frac{d^2}{dt^2}I^{k\ell}(t\!-\!\modul{x}) \; ,
\end{equation}
where in the last step we used the explicit expression of the retarded propagator given in eq. (\ref{eq:Dret.mass0}). We immediately see that, thanks to the fact that we used the in-in formalism, we find a pure real and causal gravitational waveform. In this way we solved also the other inconsistency appeared in Sec. \ref{subsec:neddinin}.

We can verify this result by computing again the power loss of the system using eq. (\ref{eq:Ploss.GWf}), hence
\begin{align}
\frac{d \mathcal{P}}{d\Omega} & =\frac{\GN}{8\pi}\frac{1}{T}\int_{-T}^T\!\!d t\Lambda_{ijk\ell }\Lambda^{ijmn}\dddot{I}^{\,k\ell }(t\!-\!\modul{x})\dddot{I}_{mn}(t\!-\!\modul{x}) \; . 
\end{align}
Being $\Lambda_{ijk\ell}$ a projector, it is not hard to verify that $\Lambda_{ijk\ell }\Lambda^{ijmn} = \Lambda{_{k\ell}}^{mn}$. Then, we should compute the integral on the solid angle to get the total power loss; using rotational invariance of the integrals, one can writes
\begin{align}
\int\!\!d\Omega\; n_i n_j = \frac{4\pi}{3}\delta_{ij} \; , && \int\!\!d\Omega\; n_i n_j n_k n_\ell = \frac{4\pi}{15}\left(\delta_{ij}\delta_{k\ell}+\delta_{ik}\delta_{j\ell}+\delta_{i\ell}\delta_{jk}\right) \; .
\end{align}
Recalling that $I_{ij}$ is traceless, one eventually finds
\begin{align}
\mathcal{P} = \frac{\GN}{2}\frac{2}{5}\bangle{\dddot{I}^{\,ij}\dddot{I}_{ij}}_T = \frac{\GN}{5}\bangle{\dddot{I}^{\,ij}\dddot{I}_{ij}}_T \; .
\end{align}

This result agrees with the one we found in the previous section. This represents a strong internal check of the correct implementation of the in-in formalism for NRGR.

\section{The bottom-up approach to the radiative sector}\label{sec:bott.up}

Up until now we used a top-down approach in our EFT for the binary system. In the radiative part of NRGR, a bottom-up approach has been extensively used in literature, see for instance Refs. \cite{Levi-Review,EFT3,Rad-Corr}. In this section we show how to implement it, and how to derive again physical results.

\subsection{The action and the power loss}\label{subsec:expE.B}

\subsubsection{Linear action for a generic gravitation theory}

Let's start by a generic example in GR and consider fluctuations of the metric $g_{\mu\nu} = \eta_{\mu\nu}+h_{\mu\nu}/\Mpl$ in presence of a single body. It is known that, at linear level in $h_{\mu\nu}$, one can describe interactions between gravity and matter as
\begin{equation}
S_{\text{int}} = -\frac{1}{2\Mpl}\int\!\!d^4\!x\,T^{\mu\nu}h_{\mu\nu} \; ,
\label{eq:Sint.nomult}
\end{equation}
where $T^{\mu\nu}$ is the usual stress-energy tensor of the system. We can do a multipole expansion of this action, i.e. 
\begin{equation}
h_{\mu\nu}(t,\bm{x}) = \sum_{n=0}^{\infty}\frac{1}{n!}x^{i_1}\dots x^{i_n}\partial_{i_1}\dots\partial_{i_n}h_{\mu\nu}(t,0) = \sum_{n=0}^{\infty}\frac{1}{n!}x^N\partial_N h_{\mu\nu}(t,0) \; .
\end{equation}
Here we perform the multipole expansion at the origin $\bm{x}=0$, which is where we can put our single body. Moreover, we consider a frame in which the body is at rest, i.e. its four-velocity is $u^{\mu}=(1,0,0,0)$. 

In Ref. \cite{Ross-multi} it is showed explicitly that, after extensive use of symmetries and Young tableaux, one can write (\ref{eq:Sint.nomult}) in terms of symmetric trace-free (STF) multipole moments, i.e.
\begin{align}
S_{\text{int}} & =-\frac{1}{2\Mpl}\int\!\!d t\left[M h_{00}+2P^i h_{0i}+MX^i\partial_i h_{00}+\varepsilon_{ijk}L^i\partial^j h^{0k}\right]  \notag \\
&\quad +\int\!\!d t\sum_{\ell=2}^{\infty}\frac{1}{\ell!}I^L\partial_{L-2}E_{k_{\ell-1}k_\ell}-\int\!\!d t\sum_{\ell=2}^{\infty}\frac{2\ell}{(\ell+1)!}J^L\partial_{L-2}B_{k_{\ell-1}k_\ell} \; .
\label{eq:Sint-mult-lin}
\end{align}
The above action is written in a locally-flat co-moving frame $e^{\mu}_{\alpha}$ such that, being $u^\mu$ the four-velocity of our body,
\begin{subequations}
\label{eqn:scheme3}
\begin{align}
e^{\mu}_{0} & = u^\mu  \; ,\\
g^{\mu\nu} & = e^\mu_{0}e^\nu_{0}-\delta^{ij}e^\mu_{i}e^\nu_{j} \; .
\label{eq:loc.flat}
\end{align}
\end{subequations}
The explicit expressions of the multipole moments as functions of $T^{\mu\nu}$ are the following 
\begin{subequations}
\label{eqn:scheme28}
\begin{align}
M & = \int\!\!d^3\!\bm{x}\,T^{00} \; ,\label{eq:multiM}\\
P^{i} & = \int\!\!d^3\!\bm{x}\,T^{0i} \; ,\label{eq:multiP}\\
MX^{i} & = \int\!\!d^3\!\bm{x}\,T^{00}x^i \; ,\label{eq:multiMX} \\
L^{i} & = -\int\!\!d^3\!\bm{x}\,\varepsilon^{ijk}T_{0j}x_k \; ,\label{eq:multiL} 
\end{align}
\begin{align}
I^L & =  \sum_{p=0}^\infty\frac{(2\ell\!+\!1)!!}{(2p)!!(2\ell\!+\!2p\!+\!1)!!}\Bigg\{\left(1\!+\!\frac{8p(\ell\!+\!p\!+\!1)}{(\ell\!+\!1)(\ell\!+\!2)}\right)\left[\int\!\!d^3\!\bm{x}\partial_0^{2p}T^{00}\modul{x}^{2p}x^L\right]_{\text{STF}} \notag \\
&\qquad\qquad\quad +\left(1+\frac{4p}{(\ell+1)(\ell+2)}\right)\left[\int\!\!d^3\!\bm{x}\partial_0^{2p}T{^k}_{k}\modul{x}^{2p}x^L\right]_{\text{STF}} \notag \\
&\qquad\qquad\quad -\frac{4}{\ell+1}\left(1+\frac{2p}{\ell+2}\right)\left[\int\!\!d^3\!\bm{x}\partial_0^{2p+1}T_{0k}\modul{x}^{2p}x^{kL}\right]_{\text{STF}} \notag \\
&\qquad\qquad\quad +\frac{2}{(\ell+1)(\ell+2)}\left[\int\!\!d^3\!\bm{x}\partial_0^{2p+2}T_{k_1k_2}\modul{x}^{2p}x^{k_1k_2L}\right]_{\text{STF}}\Bigg\} \label{eq:multI} \; ,\\
J^L & 	\!=\!  \sum_{p=0}^\infty\frac{(2\ell\!+\!1)!!}{(2p)!!(2\ell\!+\!2p+1)!!}\Bigg\{\left(\!1\!+\!\frac{2p}{\ell\!+\!2}\right)\left[\int\!\!d^3\!\bm{x}\,\varepsilon^{k_\ell mn}\partial_0^{2p}T_{0n}\modul{x}^{2p}\!x_mx^{L-1}\!\right]_{\text{STF}} \notag \\
&\qquad\qquad\quad -\frac{1}{\ell+2}\left[\int\!\!d^3\!\bm{x}\,\varepsilon^{k_\ell mn}\partial_0^{2p+1}T_{nq}\modul{x}^{2p}x_mx^{qL-1}\right]_{\text{STF}}\Bigg\} \; .
\label{eq:multJ}
\end{align}
\end{subequations}
The label STF means that we have to take the symmetric trace-free part of the tensor inside the square brackets. Finally, we recall that in eq. (\ref{eq:Sint-mult-lin}) $E_{\mu\nu}$ and $B_{\mu\nu}$ are respectively the electric and magnetic part of the Weyl tensor $\bm{\mathcal{C}}_{\mu\nu\rho\sigma}$, defined as
\begin{align}
E_{\mu\nu} = \bm{\mathcal{C}}_{\mu\alpha\nu\beta}u^{\alpha}u^{\beta} \; , & &  B_{\mu\nu} = \frac{1}{2}\varepsilon_{\mu\rho\sigma\alpha}\bm{\mathcal{C}}{^{\rho\sigma}}_{\nu\beta}u^{\alpha}u^{\beta} \; .
\label{eq:El-Mag}
\end{align}
By definition, the two tensors $E_{\mu\nu}$ and $B_{\mu\nu}$ satisfy
\begin{align}
E_{\mu\nu}g^{\mu\nu} = 0 = B_{\mu\nu}g^{\mu\nu} \; , & &  E_{\mu\nu}u^\nu = 0 = B_{\mu\nu}u^\nu \; .
\end{align}
Then, let us remember that, on-shell in the vacuum (that is $\Rc_{\mu\nu} =\Rc = 0$), the Weyl and the Riemann tensors coincide, and, since we chose a system in which $u^\mu=(1,0,0,0)$, the only non-zero components of $E_{\mu\nu}$ and $B_{\mu\nu}$ are
\begin{subequations}
\label{eqn:scheme29}
\begin{align}
E_{ij}& =\Rc_{0i0j} =\frac{1}{2\Mpl}\left(\partial_0\partial_j h_{0i}+\partial_0\partial_i h_{0j}-\partial_i\partial_j h_{00}-\partial^2_0 h_{ij}\right) \; ,\label{eq:linear_E}\\
B_{ij} & = \frac{1}{2}\varepsilon_{imn}\Rc_{0j}{^{mn}}=\frac{1}{2\Mpl}\varepsilon_{imn}\left(\partial_0\partial^n h_{j}{^m}+\partial_j\partial^m h_{0}{^n}\right) \; . \label{eq:linear_B}
\end{align}
\end{subequations}

In the next paragraph we shall extend the above action and momenta to the radiative sector of NRGR.

\subsubsection{Going back to the two the body problem of NRGR}

As we said already, once we have integrated out the potential modes $\potH{k}{\mu\nu}$, our two-body problem reduces to the study of a single body (described by the CoM of the old binary system) in the external gravitational field 
\begin{equation}
\bar{g}_{\mu\nu} = \eta_{\mu\nu}+\frac{\radh{\mu\nu}}{\Mpl} \; .
\end{equation}
The symmetries of this theory are basically world-line reparametrisation and invariance under general diffeomorphisms of the metric $\bar{g}_{\mu\nu}$. If we put the origin of our system of reference in the CoM, then the bottom-up constructed action $\Seff{NR}$ has to have the following form
\begin{equation}
\Seff{NR}[x_a^\mu,\radh{\mu\nu}] = S_{\text{EH}}^{(\text{GF})}[\radh{\mu\nu}]+S_{\text{cons}}[x^\mu_a]+S_{\text{mult}}[x_a^\mu,\radh{\mu\nu}] \; .
\end{equation}
Here $S_{\text{EH}}^{(\text{GF})}$ is the gauge-fixed Einstein-Hilbert action for $\radh{\mu\nu}$, $S_{\text{cons}}$ is the usual action describing the conservative dynamics, while $S_{\text{mult}}$ is given by
\begin{align}
\!S_{\text{mult}}[x_a^\mu,\radh{\mu\nu}] \!=\! \int\!\!d t\sqrt{\bar{g}_{00}}\bigg\{-M(t)\!+\!\frac{1}{2\Mpl}\sum_{\ell=2}^{\infty}&\bigg(\frac{1}{\ell!}I^L(t)\nabla_{L-2}\tilde{E}_{k_{\ell-1}k_\ell}\notag \\
&\!\!\!\!-\frac{2\ell}{(\ell+1)!}J^L(t)\nabla_{L-2}\tilde{B}_{k_{\ell-1}k_\ell}\bigg)\bigg\} \; .
\label{eq:S-int-mult-NR}
\end{align}
This last action has been constructed as follows:
\begin{itemize}
\item Starting from the linear action (\ref{eq:Sint-mult-lin}) still written in locally-flat co-moving frame similar to (\ref{eq:loc.flat}), we ``turn on'' non linearities by covariantising the theory w.r.t. the metric $\bar{g}_{\mu\nu}$, hence we trade ordinary derivatives with covariant derivatives $\nabla_\mu$, compatible with $\bar{g}_{\mu\nu}$. This ensure invariance under diffeomorphism. The expressions of the symmetric trace free multipole moments $I^L$ and $J^L$ are the same given in (\ref{eq:multI}) and (\ref{eq:multJ}), where now $T^{\mu\nu}$ is the pseudo stress-energy tensor containing contributions from the point particles and the gravitational energy coming from the integration of $\potH{}{\mu\nu}$.  
\item $\tilde{E}_{\mu\nu}$ and $\tilde{B}_{\mu\nu}$ are the electric and magnetic gravity tensors, constructed with a procedure similar to the one we used for $\tilde{\Rc}_{\rho\mu\sigma\nu}$ in eq. (\ref{eq:R.Mpl.fact}), i.e.
\begin{subequations}
\label{eqn:scheme31} 
\begin{align}
\tilde{E}_{\mu\nu} \equiv 2\Mpl E_{\mu\nu}[\bar{g}] \; ,& &   \tilde{B}_{\mu\nu} \equiv 2\Mpl B_{\mu\nu}[\bar{g}] \; .
\end{align}
\end{subequations}
In this way we do not have any implicit $\Mpl$ in the definition of the electric and magnetic gravity tensor.
\item Then we integrate w.r.t. the proper time $\sqrt{\bar{g}_{\mu\nu}dx^\mu dx^\nu}$ to ensure world-line reparametrisation invariance. We decide to put the CoM of the original binary at the origin of our system of reference. Moreover, as we said, we put ourself in a system such that the single body we are watching is at rest, hence described by a velocity $u^{\mu}=(1,0,0,0,)$. Therefore, we immediately write
\begin{equation}
\sqrt{\bar{g}_{\mu\nu}dx^\mu dx^\nu} = d t\sqrt{\bar{g}_{00}}
\end{equation}
This is the reason of the integration measure in (\ref{eq:S-int-mult-NR}).
\item Finally, since we are working in the CoM system of the binary, and since we put this CoM at the origin of our frame, we can easily see that in this case multipole moments (\ref{eq:multiP}) (\ref{eq:multiMX}) and (\ref{eq:multiL}) are actually equal to zero.
\end{itemize}
In fact, we could insert in eq. (\ref{eq:S-int-mult-NR}) an entire set of operators that contains higher powers of the graviton $\radh{\mu\nu}$. These terms, however, contribute to a multi-graviton-emission amplitude which we will not be considered in this work.

Once we have the action (\ref{eq:S-int-mult-NR}), we are ready to compute observables in terms of the STF multipole moments. Actually, we are going to write the expression of observables only at linear level, which means that in (\ref{eq:S-int-mult-NR}) we expand
\begin{subequations}
\label{eqn:scheme30}
\begin{align}
dt\sqrt{\bar{g}_{00}} & = dt + \text{ non linearities} \, \\
\nabla_{\mu} & = \partial_\mu + \text{ non linearities} \; ,
\end{align}
\end{subequations}
and we ignore all non-linearities terms\footnote{For a more complete study of non linearities we redirect the reader to Sec. III of Ref. \cite{Rad-Corr}, and Sec. 7.4 of Ref. \cite{EFT3}.}. The relevant part of the action (\ref{eq:S-int-mult-NR}) is then
\begin{equation}
S_{\text{lin}} \!=\! \frac{1}{2\Mpl}\int\!\!d t\left\{\sum_{\ell=2}^{\infty}\left(\frac{1}{\ell!}I^L(t)\partial_{L-2}\tilde{E}_{k_{\ell-1}k_\ell}\!-\!\frac{2\ell}{(\ell+1)!}J^L(t)\partial_{L-2}\tilde{B}_{k_{\ell-1}k_\ell}\right)\right\}
\label{eq:S-int-mult-NR2}
\end{equation}

From this, we would like to derive an expression for the power loss of the system at every PN order in terms of the multipole moments. To do so, we shall now see how to compute the single-graviton emission amplitude $\mathcal{A}_h$, where $h = \pm 2$ denotes the elicity of the graviton. This quantity is clearly related to the probability of emitting one graviton\footnote{See for instance Ref. \cite{Ross-multi}.} 
\begin{equation}
d\Gamma = \sum_{h}\frac{1}{T}\frac{d^3\!\bm{k}}{(2\pi)^3 2\modul{k}}\Big|\mathcal{A}_h(\modul{k},\bm{k})\Big|^2 \; .
\label{eq:Prob_em_1GW}
\end{equation}
From this, we can compute the power loss of the system as
\begin{equation}
\mathcal{P}=\int\!\modul{k}d\Gamma \; .
\label{eq:Plost_last_expr}
\end{equation} 

\subsubsection{The power loss $\mathcal{P}$}

We can compute diagrammatically  $\mathcal{A}_h$ as follows
\begin{equation}
i\mathcal{A}_h = \sum_{\ell=2}^{\infty}\left.\left(\begin{tikzpicture}[baseline]
\begin{feynman}
\vertex (f);
\vertex [above=0.1cm of f] (z);
\vertex [right=0.6cm of z, dot, label=90:$I^L$] (c) {};
\vertex [below=1cm of c] (c');
\vertex [right=1.6cm of z] (w);  
\diagram* {
(z) -- [double] (c) -- [double] (w),
(c') -- [gluon] (c)
}; 
\end{feynman} 
\end{tikzpicture} + \begin{tikzpicture}[baseline]
\begin{feynman}
\vertex (f);
\vertex [above=0.1cm of f] (z);
\vertex [right=0.6cm of z, dot, label=90:$J^L$] (c) {};
\vertex [below=1cm of c] (c');
\vertex [right=1.6cm of z] (w);  
\diagram* {
(z) -- [double] (c) -- [double] (w),
(c') -- [gluon] (c)
}; 
\end{feynman} 
\end{tikzpicture}\right)\right|_{\text{on-shell}} \; .
\label{eq:On-shell_ampl}
\end{equation}
The explicit expression of the above diagrams can be found by deriving the Feynman rules starting from the action (\ref{eq:S-int-mult-NR2}). From what we said in the previous section, we work in the transverse traceless gauge (\ref{eq:TTgauge}), so that actually only the spatial components $\radh{ij}$ can propagate.

Let's start from the electric term. First, we Fourier transform it w.r.t. time\footnote{Again we explicit $\eta_{ij}=-\delta_{ij}$, hence we use raised and lowered indices just to stress that they are contracted, but one can actually move them up or down without inserting any additional sign.}, i.e. for each $\ell$
\begin{align}
\int\!\!d t\frac{1}{\ell!}I^L(t)\partial_{L-2}\tilde{E}_{k_{\ell-1}k_\ell}&=\int\!\!d t\frac{d \omega}{2\pi}\frac{d \omega'}{2\pi}\frtr{3}{\bm{k}}\frac{1}{\ell!}e^{-i(\omega+\omega')t}e^{i\bm{k}\cdot\bm{x}}I^L(\omega) \notag \\
&\qquad\qquad \times (-i)^{\ell-2}k_{L-2}(-i\omega')^2\radh{i_{\ell-1}i_\ell}(\omega',\bm{k})\bigg|_{\bm{x}=0} \notag \\
& =\int\!\!\frtr{4}{k}\frac{(-i)^{\ell-2}}{\ell!}\omega^2I^L(\omega)k_{L-2}\radh{i_{\ell-1}i_\ell}(-\omega,\bm{k}) \; ,
\end{align}
where, up to a factor $(2\Mpl)^{-1}$, we use an expression analogous to (\ref{eq:linear_E}) for the electric tensor. From here, we immediately find for each $\ell$
\begin{equation}
\begin{tikzpicture}[baseline]
\begin{feynman}
\vertex (f);
\vertex [above=0.1cm of f] (z);
\vertex [right=0.6cm of z, dot, label=90:$I^L$] (c) {};
\vertex [below=1cm of c] (c');
\vertex [right=1.6cm of z] (w);  
\diagram* {
(z) -- [double] (c) -- [double] (w),
(c') -- [gluon] (c)
}; 
\end{feynman} 
\end{tikzpicture} = \frac{i}{2\Mpl}\frac{(-i)^{\ell-2}}{\ell!}\omega^2I^L(\omega)k_{L-2}\epsilon^*_{i_{\ell-1}i_\ell}(\bm{k}, h) \; ,
\label{eq:Feyn.IL}
\end{equation}
where the star denote the complex conjugate operation. Here we inserted the polarization tensor\footnote{In order for the graviton to be real we have $\epsilon^*_{\mu\nu}(\omega,\bm{k},h) = \epsilon_{\mu\nu}(-\omega,\bm{k},h)$. From here on, we will not write any more the dependence on $\omega$ of the polarization tensor.} $\epsilon_{\mu\nu}(\bm{k},h)$, that in the transverse traceless gauge (\ref{eq:TTgauge}) satisfies
\begin{align}
\epsilon_{0\mu}(\bm{k},h)=0 \; ,& &  \epsilon{^i}_{i}(\bm{k},h) = 0 \; , & &  k^i\epsilon_{ij}(\bm{k},h) = 0 \; .
\end{align}
Then, we do the same thing for the magnetic term, thus
\begin{align}
\int\!\!d t\frac{2\ell}{(\ell+1)!}&J^L(t)\partial_{L-2}\tilde{B}_{k_{\ell-1}k_\ell}=\int\!\!d t\frac{d \omega}{2\pi}\frac{d \omega'}{2\pi}\frtr{3}{\bm{k}}\frac{2\ell}{(\ell+1)!}e^{-i(\omega+\omega')t}e^{i\bm{k}\cdot\bm{x}}J^L(\omega) \notag \\
&\qquad  \times \varepsilon_{i_{\ell-1}mn}(-i)^{\ell-2}k_{L-2}(-i\omega')(-ik^n)\radhmi{m}{i_\ell}(\omega',\bm{k})\bigg|_{\bm{x}=0} \notag \\
& =\int\!\!\frtr{4}{k}\frac{(-i)^{\ell-2}2\ell}{(\ell+1)!}\omega J^L(\omega)\varepsilon_{i_{\ell-1}mn}k_{L-2}k^n\radhmi{m}{i_\ell}(-\omega,\bm{k}) \; ,
\end{align}
where, again, we used an expression analogous to (\ref{eq:linear_B}) for the magnetic tensor. From here we get the following Feynman rules for each $\ell$
\begin{equation}
\begin{tikzpicture}[baseline]
\begin{feynman}
\vertex (f);
\vertex [above=0.1cm of f] (z);
\vertex [right=0.6cm of z, dot, label=90:$J^L$] (c) {};
\vertex [below=1cm of c] (c');
\vertex [right=1.6cm of z] (w);  
\diagram* {
(z) -- [double] (c) -- [double] (w),
(c') -- [gluon] (c)
}; 
\end{feynman} 
\end{tikzpicture} = -\frac{i}{2\Mpl}(-i)^{\ell-2}\frac{2\ell}{(\ell+1)!}\omega J^L(\omega)\varepsilon_{i_{\ell-1}mn}k_{L-2}k^n\epsilon{^{*m}}_{i_\ell}(\bm{k},h) \; .
\label{eq:Feyn.JL}
\end{equation}
Inserting now eqs. (\ref{eq:Feyn.IL}) and (\ref{eq:Feyn.JL}) into (\ref{eq:On-shell_ampl}), and imposing the on-shell condition for the external graviton $\omega =\modul{k}$, we eventually obtain
\begin{align}
i\mathcal{A}_h(\modul{k},\bm{k}) = \frac{i}{2\Mpl}\sum_{\ell=2}^{\infty}&\frac{(-i)^{\ell-2}}{\ell!}\bigg(\modul{k}^2I^L(\modul{k})k_{L-2}\epsilon^*_{i_{\ell-1}i_\ell}(\bm{k}, h)\notag \\
&\ -\frac{2\ell}{(\ell+1)}\modul{k} J^L(\modul{k})\varepsilon_{i_{\ell-1}mn}k_{L-2}k^n\epsilon{^{*m}}_{i_\ell}(\bm{k}, h)\bigg) \; .
\end{align}

Now we are ready to compute (\ref{eq:Prob_em_1GW}). To do so, we need to compute the modulo square of the above quantity, and sum over the elicity $h$. Therefore, let us first remember the following identity
\begin{align}
\sum_h\epsilon_{ij}(\bm{k},h)\epsilon^*_{k\ell}(\bm{k},h) & =\frac{1}{2}\Bigg\{\delta_{ik}\delta_{j\ell}+\delta_{i\ell}\delta_{jk}-\delta_{ij}\delta_{k\ell}+\frac{1}{\modul{k}^2}\left(\delta_{ij}k_{k}k_{\ell}+\delta_{k\ell}k_ik_j\right)\notag \\
&\qquad\quad -\frac{1}{\modul{k}^2}\left(\delta_{ik}k_j k_\ell+\delta_{i\ell}k_jk_k+\delta_{jk}k_i k_\ell+\delta_{j\ell}k_ik_k\right)\notag \\
&\qquad\quad+\frac{1}{\modul{k}^4}k_ik_jk_kk_\ell\Bigg\}
\end{align}
One can eventually find the general formula for the power loss of the system written in term of the multipole moments
\begin{align}
\mathcal{P} & = \GN\sum_{\ell=2}^{\infty}\Bigg\{\frac{(\ell\!+\!1)(\ell\!+\!2)}{\ell(\ell\!-\!1)\ell!(2\ell\!+\!1)!!}\bangle*{\left(\frac{d^{\ell+1}\!I^L}{d t^{\ell+1}}\right)^2}_T\notag \\
&\qquad\qquad\qquad\qquad\qquad+\frac{4\ell(\ell\!+\!2)}{(\ell\!-\!1)(\ell\!+\!1)!(2\ell\!+\!1)!!}\bangle*{\left(\frac{d^{\ell+1}\!J^L}{d t^{\ell+1}}\right)^2}_T\Bigg\} = \notag \\
& = \frac{\GN}{T}\int_0^\infty\!\!\frac{d \omega}{\pi}\left[\frac{\omega^6}{5}\abs*{I^{ij}(\omega)}^2+\frac{16}{45}\omega^6\abs{J^{ij}(\omega)}^2+\frac{\omega^8}{189}\abs*{I^{ijk}(\omega)}^2+\dots\right] \; .
\label{eq:power.loss.multi}
\end{align}
In the second line, we wrote the first few terms of the infinite $\ell$ sum. We immediately see that the Leading-Order (LO) part is giving us the quadrupole emission in Fourier space, result that we had already found in the previous sections.

The aim of the next few sections is to find a Next-to-Leading Order (NLO) contribution to the power loss of the system. However, from the above expression, it is not trivial to understand which terms contribute to a specific PN order. In order to restore a series in the parameter $v$, we first need to understand the scaling of the various multipole moments.

\section{Scaling rules for the multipole moments}

We would like to implement a power counting similar to the one we constructed in the previous chapter, in order to understand immediately which are the terms of the series (\ref{eq:power.loss.multi}) that contributes to a specific PN order. Clearly, we need to understand the scaling of the multipole moments $I^L$ and $J^L$, hence the scaling of the pseudo stress-energy tensor $T^{\mu\nu}$. Ignoring non linearities and denoting with a generic $M$ the scale of the total mass-energy of the system, it is clear that
\begin{equation}
\int\!\!d^3\!\bm{x}T^{\mu\nu} \sim
\begin{cases}
M & \quad(\mu=0,\nu=0) \\ 
Mv &\quad (\mu=0,\nu=i) \\
Mv^2 &\quad (\mu=i,\nu=j) 
\end{cases} \; .
\end{equation}
Therefore, we can easily see that the momenta scale as
\begin{equation}
\frac{1}{\Mpl}\int\!\!d^3\!\bm{x}T^{\mu\nu}x^L\sim
\begin{cases}
L^{1/2}r^\ell v^{1/2} &\quad (\mu=0,\nu=0) \\ 
L^{1/2}r^\ell v^{3/2} &\quad (\mu=0,\nu=i) \\
L^{1/2}r^\ell v^{5/2} &\quad (\mu=i,\nu=j)
\end{cases} \; .
\label{eq:scaling_mom_T}
\end{equation}

As a consistency check, we can compare this result with the scaling found in the previous section using the top-down approach. This is realised by looking at how each single term of the action (\ref{eq:S-int-mult-NR2}) scales. Schematically, we can write both the electric and magnetic term as 
\begin{equation}
\frac{1}{\Mpl}\int\!\!d t\int\!\!d^3\!\bm{x}T^{\mu\nu}x^L\partial_L\radh{\mu\nu}\sim
\begin{cases}
L^{1/2}v^\ell v^{1/2} &\quad (\mu=0,\nu=0) \\ 
L^{1/2}v^{\ell+1} v^{1/2} &\quad (\mu=0,\nu=i) \\
L^{1/2}v^{\ell+2} v^{1/2} &\quad (\mu=i,\nu=j)
\end{cases} \; .
\end{equation}
We recover the result found in eq. (\ref{eq:scal.1/2.rad}), i.e. the radiative contributions to $\Seff{NR}$ scale as $L^{1/2}v^{n+(1/2)}$. 

Now in order to find which terms contribute to the NLO power loss, we first need to understand how the LO scales. Looking at (\ref{eq:power.loss.multi}), we understand that the LO power loss is given by
\begin{equation}
\mathcal{P}_{\text{LO}} = \frac{\GN}{T}\int_0^\infty\!\!\frac{d \omega}{\pi}\frac{\omega^6}{5}\abs*{I_{\text{LO}}^{ij}(\omega)}^2
\end{equation}
where $I_{\text{LO}}^{ij}$ can be found by looking at eq. (\ref{eq:multI}) for $\ell = 2$, and taking all the least scaling terms according to the scaling rules (\ref{eq:scaling_mom_T}), i.e.
\begin{equation}
I_{\text{LO}}^{ij}(t) =\int\!\!d^3\!\bm{x}T^{00}(t,\bm{x})\left[x^ix^j\right]_{\text{STF}}\sim Mr^2 \; .
\end{equation}
Performing a time Fourier transform, we find
\begin{equation}
I_{\text{LO}}^{ij}(t) \xrightarrow[\text{time Fourier}]{} I_{\text{LO}}^{ij}(\omega)\sim\frac{Mr^3}{v} \; ,
\end{equation}
so that, eventually, 
\begin{equation}
\mathcal{P}_{\text{LO}} \sim \frac{M^2}{\Mpl^2}\left(\dfrac{v}{r}\right)^2\left(\dfrac{v}{r}\right)^6\frac{r^6}{v^2}\sim\frac{L}{r^2}v^{7} \; .
\label{eq:scaling_LO_P}
\end{equation}
This result is again consistent with the one we found in the previous section using the top-down approach. Indeed, we saw that the diagram contributing to the LO power loss was (\ref{eq:SNRGR.2.5PN}), which scales as $Lv^5$. Using then the optical theorem given in eq. (\ref{eq:Opt.th}), it is not hard to find the same result of eq. (\ref{eq:scaling_LO_P}). 

Having this, we expect that the NLO scales with an additional $v^2$, thus we have to find all terms in (\ref{eq:power.loss.multi}) that scales as $v^{9}$. These terms are the following:
\begin{itemize}
\item the first order correction to $I^{ij} = I_{\text{LO}}^{ij}+I_{\text{NLO}}^{ij}+\dots$, that has an additional $v^2$ in the scaling. Explicitly this can be found again from eq. (\ref{eq:multI}) taking $\ell = 2$
\begin{subequations}
\label{eqn:scheme67}
\begin{equation}
I_{\text{NLO}}^{ij}(t)\! = \!\int\!\!d^3\!\bm{x}\left(T{^k}_{k}\!-\!\frac{4}{3}\dot{T}_{0k}x^k\!+\!\frac{11}{42}\ddot{T}^{00}\modul{x}^2\right)\left[x^ix^j\right]_{\text{STF}}\sim Mv^2r^2 \; .
\end{equation}
Fourier transforming w.r.t. time, we obtain as expected
\begin{equation}
I_{\text{NLO}}^{ij}(\omega) \sim Mvr^3 
\end{equation}
\end{subequations}
\item the leading order of $J^{ij} = J_{\text{LO}}^{ij}+\dots$, which is given by the least scaling terms of eq. (\ref{eq:multJ}) for $\ell = 2$
\begin{subequations}
\label{eqn:scheme68}
\begin{equation}
J_{\text{LO}}^{ij}(t) = -\int\!\!d^3\!\bm{x}\,\varepsilon{^i}_{k\ell}\left[T^{0k}x^jx^\ell\right]_{\text{STF}}\sim Mvr^2 \; . 
\end{equation}
Going again in Fourier space, we find 
\begin{equation}
J_{\text{LO}}^{ij}(\omega)\sim Mr^3 \; .
\end{equation} 
\end{subequations}
\item the leading order of $I^{ijk} = I_{\text{LO}}^{ijk}+\dots$, which can be find by looking at the least scaling terms of eq. (\ref{eq:multI}) for $\ell = 3$
\begin{subequations}
\label{eqn:scheme69}
\begin{equation}
I_{\text{LO}}^{ijk}(t) = \int\!\!d^3\!\bm{x}T^{00}\left[x^ix^jx^k\right]_{\text{STF}}\sim Mr^3\; . 
\label{eq:need.pag}
\end{equation}
Performing as usual a time Fourier transform we obtain 
\begin{equation}
I_{\text{LO}}^{ijk}(\omega)\sim \frac{Mr^4}{v} \; .
\end{equation}
\end{subequations}
\end{itemize}
Given this, defining the power loss as $\mathcal{P}=\mathcal{P}_{\text{LO}} +\mathcal{P}_{\text{NLO}}+\dots $, then we have
\begin{equation}
\mathcal{P}_{\text{NLO}} \!=\!\frac{\GN}{T}\!\int\!\!\frac{d \omega}{\pi}\left\{\frac{2}{5}\omega^6\,I_{\text{LO}}^{ij}I_{\text{NLO}\,ij}\!+\!\frac{16}{45}\omega^6\abs{J_{\text{LO}}^{ij}(\omega)}^2\!+\!\frac{\omega^8}{189}\abs*{I_{\text{LO}}^{ijk}(\omega)}^2\right\}\!\sim\!\frac{L}{r^2}v^{9}
\label{eq:NLOPLO}
\end{equation}

Of course, at this level, the multipole moments are written in term of the pseudo stress-energy tensor, which exact expression is still not known. As we briefly discuss is Sec. \ref{sec:EFT_rev}, we need to implement also a matching procedure to obtain an explicit result for the multipole moments. For this reason, we shall now implement in parallel a top-down procedure.

\section{Computing $T^{\mu\nu}$}\label{sec:matching}

The computation of the pseudo stress-energy tensor is not a trivial problem. $T^{\mu\nu}$ is in fact the sum of two contributions
\begin{equation}
T^{\mu\nu} = T_{\text{pp}}^{\mu\nu}+T_H^{\mu\nu} \; .
\end{equation}
The first term comes from the point-particle part of the system, while $T_H^{\mu\nu}$ takes into account the energy of the gravitational field itself, coming from the fact that we have already integrated out the potential modes $\potH{}{\mu\nu}$. The point particle part is well known, and it is given by
\begin{equation}
T_{\text{pp}}^{\mu\nu}(t,\bm{x}) = \sum_a m_a\frac{u^\mu(t)u^\nu(t)}{\sqrt{1-v^2}}\delta^{(3)}\left(\bm{x}-\bm{x}_a(t)\right) \, , 
\label{eq:Tpp.expl}
\end{equation}
where $u^\mu(t) = (1, \bm{v}(t))$. This expression can be easily expanded in power of $v$. We can find diagrammatically the contribution to each order of $v$ imposing that
\begin{equation}
\sum_{a=1}^2\sum_{n=0}\begin{tikzpicture}[baseline]
\begin{feynman}
\vertex (z);
\vertex [below=0.1cm of z] (f);
\vertex [above=0.1cm of z] (f');
\vertex [right=0.5cm of f', empty dot, minimum size=0.4cm] (c) {n};
\vertex [right=0.7cm of c] (d);
\vertex [above=0.5cm of d] (d');
\diagram* {
(f') -- [plain] (c) -- [plain] (d),
(f') -- [draw=none] (z) -- [draw=none] (f),
(d') -- [gluon] (c)
}; 
\end{feynman} 
\end{tikzpicture} = -\frac{i}{2\Mpl}\int\!\!d^4\!xT^{\mu\nu}_{\text{pp}}\radh{\mu\nu}
\end{equation}
For the $\potH{}{}$ part, instead, we can look at diagrams of the following form
\begin{equation}
\begin{tikzpicture}[baseline]
\begin{feynman}
\vertex [blob, minimum size=0.7cm] (z) {};
\vertex [above=0.9cm of z, blob, minimum size=0.3cm] (c') {};
\vertex [right=1.2cm of c'] (a');
\vertex [above=0.5cm of a'] (d);
\vertex [below=0.9cm of z, blob, minimum size=0.3cm] (c) {};
\vertex [right=1.2cm of c] (a);
\vertex [left=1.2cm of c] (b);
\vertex [left=1.2cm of c'] (b');
\diagram* {
(a) -- [plain] (c) -- [plain] (b),
(a') -- [plain] (c') -- [plain] (b'),
(c) -- [scalar] (z) -- [scalar] (c'),
(d) -- [gluon] (c')
}; 
\end{feynman} 
\end{tikzpicture}+ 
\begin{tikzpicture}[baseline]
\begin{feynman}
\vertex [blob, minimum size=0.7cm] (z) {};
\vertex [right=1.2cm of z] (d);
\vertex [above=0.9cm of z, blob, minimum size=0.3cm] (c') {};
\vertex [below=0.9cm of z, blob, minimum size=0.3cm] (c) {};
\vertex [right=1.2cm of c] (a);
\vertex [left=1.2cm of c] (b);
\vertex [right=1.2cm of c'] (a');
\vertex [left=1.2cm of c'] (b');
\diagram* {
(a) -- [plain] (c) -- [plain] (b),
(a') -- [plain] (c') -- [plain] (b'),
(c) -- [scalar] (z) -- [scalar] (c'),
(d) -- [gluon] (z)
}; 
\end{feynman} 
\end{tikzpicture}= -\frac{i}{2\Mpl}\int\!\!d^4\!xT^{\mu\nu}_{H}\radh{\mu\nu} \; .
\end{equation}
This equation is the mathematical expression of what we said about the non linear diagrams at the end of Sec. \ref{sec:multipole}, p.~\pageref{eq:SNRfin}. Here, we see even more clearly that the non linear diagrams are taking into account the contributions to the stress-energy tensor coming from the gravitational energy that keep the system bounded.

The above diagrams are computed using the still non multipole expanded Feynman rules, that are listed in tab. \ref{table:pprad} of App. \ref{App:action} (p.~\pageref{table:pprad}). Once we computed the diagrams, we need to perform a partial Fourier transform and then multipole expand in $\bm{k}$, i.e.
\begin{align}
T^{\mu\nu}(t,\bm{k}) &=\int\!\!d^3\!\bm{x}T^{\mu\nu}(t,\bm{x})e^{-i\bm{k}\cdot\bm{x}} = \sum_{\ell=0}^{\infty}\frac{(-i)^\ell}{\ell!}\left(\int\!\!d^3\!\bm{x}T^{\mu\nu}(t,\bm{x})x^L\right)k_L \; .
\label{eq:part_fourier_transf}
\end{align}
In this way, order by order, we can easily see the contribution to the $\ell^{\text{th}}$ multipole moment from each diagram.

\subsection{Computation of $J^{ij}$ and $I^{ijk}$}

Let's now be more concrete and compute the various elements needed to find $\mathcal{P}_{\text{NLO}}$.  We start from the magnetic quadrupole $J^{ij}$ and the electric octupole $I^{ijk}$. Computing their expression is indeed easier because, as we said, we only need their leading order expressions. This implies that we do not need to compute any diagram, but only the expansion of the point-particle part of the pseudo stress-energy tensor. So let's isolate the leading order part of  eq. (\ref{eq:Tpp.expl}), i.e.
\begin{subequations}
\label{eqn:scheme32}
\begin{align}
T^{00}(t,\bm{x}) & \simeq \sum_a m_a\delta^{(3)}\!\left(\bm{x}-\bm{x}_a(t)\right) \; , \\
T^{0i}(t,\bm{x}) & \simeq \sum_a m_av^i\delta^{(3)}\!\left(\bm{x}-\bm{x}_a(t)\right) \, .
\end{align}
\end{subequations}
Inserting these expression in eqs. (\ref{eqn:scheme68}a) and (\ref{eqn:scheme69}a), we eventually obtain
\begin{align}
J^{ij} & = -\frac{1}{2}\int\!\!d^3\!\bm{x}\sum_a m_a\left[\varepsilon^{ik\ell}v_kx_\ell x^j+\varepsilon^{jkl}v_kx_\ell x^i\right]_{\text{TF}}\delta^{3}\!\left(\bm{x}-\bm{x}_a\right) \notag \\
& = \sum_a m_a\left[(\bm{x}_a\times\bm{v}_a)^ix_a^j\right]_{\text{STF}} \; ,\label{eq:Jij.fin}\\
I^{ijk} & = \int\!\!d^3\!\bm{x}\sum_a m_a\delta^{(3)}\!\left(\bm{x}-\bm{x}_a\right)\left[x^ix^jx^k\right]_{\text{STF}} = \sum_a m_a\left[x_a^ix_a^jx_a^k\right]_{\text{STF}}  \; .\label{eq:Iijk.fin}
\end{align} 

\subsection{Computation of $I^{ij}$}

The computation of the NLO correction to the electric quadrupole moment is a bit more involved. We recall that at this order the complete expression of this multipole moment is
\begin{equation}
I^{ij} = \int\!\!d^3\!\bm{x}\left(T^{00}+T{^k}_{k}-\frac{4}{3}\dot{T}_{0k}x^k+\frac{11}{42}\ddot{T}^{00}\modul{x}^2\right)\left[x^ix^j\right]_{\text{STF}} \; .
\label{eq:I.NLO.comp}
\end{equation}
For the last two terms inside the square brackets we just need the LO expression of $T^{\mu\nu}$, because the time derivatives provide already the needed extra powers of $v$. Thus, we can use again eqs. (\ref{eqn:scheme32}) to find
\begin{subequations}
\label{eqn:scheme33}
\begin{align}
\int\!\!d^3\!\bm{x}\,\frac{4}{3}\dot{T}_{0k}x^k & = -\frac{4}{3}\sum_a m_a\int\!\!d^3\!\bm{x}\,\frac{d}{d t}\Big(v_k\delta^{(3)}\!\left(\bm{x}-\bm{x}_a\right)\Big)x^k \label{eq:Tdot.contr} \\
\int\!\!d^3\!\bm{x}\,\frac{11}{42}\ddot{T}^{00}\modul{x}^2 & = \frac{11}{42}\sum_a m_a\int\!\!d^3\!\bm{x}\,\frac{d^2}{d t^2}\Big(\delta^{(3)}\!\left(\bm{x}-\bm{x}_a\right)\Big) \modul{x}^2 \label{eq:Tddot.contr}
\end{align}
\end{subequations}

On the other hand, for the combination $T^{00}+T{^k}_{k}$ we have to compute corrections coming from the following diagrams:
\begin{align*}
\text{from the expansion of } T^{\mu\nu}_{\text{pp}} & \longrightarrow \ \ 
\begin{tikzpicture}[baseline]
\begin{feynman}
\vertex (z);
\vertex [below=0.1cm of z] (f);
\vertex [above=0.1cm of z] (f');
\vertex [right=0.9cm of f', empty dot, minimum size=0.4cm] (c) {2};
\vertex [right=1cm of c] (d);
\vertex [above=0.5cm of d] (d');
\diagram* {
(f') -- [plain] (c) -- [plain] (d),
(f') -- [draw=none] (z) -- [draw=none] (f),
(d') -- [gluon] (c)
}; 
\end{feynman} 
\end{tikzpicture} \equiv -\frac{i}{2\Mpl}\int\!\!d^4\!x\,{\ord{A}}T^{\mu\nu}\radh{\mu\nu}  \\
\text{from the expansion of } T^{\mu\nu}_{H} & \longrightarrow \begin{cases}
\begin{tikzpicture}[baseline]
\begin{feynman}
\vertex (z);
\vertex [below=0.6cm of z] (c);
\vertex [above=0.8cm of z] (c');
\vertex [right=1cm of c] (a);
\vertex [left=1cm of c] (b);
\vertex [right=1cm of c'] (a');
\vertex [right=1cm of z] (d);
\vertex [left=1cm of c'] (b');
\diagram* {
(a) -- [plain] (c) -- [plain] (b),
(a') -- [plain] (c') -- [plain] (b'),
(c) -- [scalar] (z) -- [scalar] (c'),
(d) -- [gluon] (c')
}; 
\end{feynman} 
\end{tikzpicture} \equiv \displaystyle -\frac{i}{2\Mpl}\int\!\!d^4\!x\,{\ord{B}}T^{\mu\nu}\radh{\mu\nu}& \ \\
\ & \ \\
\begin{tikzpicture}[baseline]
\begin{feynman}
\vertex (z);
\vertex [below=0.7cm of z] (c);
\vertex [right=1cm of z] (d);
\vertex [above=0.7cm of z] (c');
\vertex [right=1cm of c] (a);
\vertex [left=1cm of c] (b);
\vertex [right=1cm of c'] (a');
\vertex [left=1cm of c'] (b');
\diagram* {
(a) -- [plain] (c) -- [plain] (b),
(a') -- [plain] (c') -- [plain] (b'),
(c) -- [scalar] (z) -- [scalar] (c'),
(d) -- [gluon] (z)
}; 
\end{feynman} 
\end{tikzpicture} \equiv \displaystyle -\frac{i}{2\Mpl}\int\!\!d^4\!x\,{\ord{C}}T^{\mu\nu}\radh{\mu\nu} & \ \\
\end{cases}
\end{align*}
We stress again that these diagrams are slightly different from the one we computed in (\ref{eq:disdiag}) and (\ref{eq:disdiag.int}), because we are using Feynman rules coming from the non-multipole-expanded action, listed in tab. \ref{table:pprad} of App. \ref{App:action}, p.~\pageref{table:pprad}. 

The first diagram follows immediately from the Feynman rule
\begin{align}
\begin{tikzpicture}[baseline]
\begin{feynman}
\vertex (z);
\vertex [below=0.1cm of z] (f);
\vertex [above=0.1cm of z] (f');
\vertex [right=0.9cm of f', empty dot, minimum size=0.4cm] (c) {2};
\vertex [right=1cm of c] (d);
\vertex [above=0.5cm of d] (d');
\diagram* {
(f') -- [plain] (c) -- [plain] (d),
(f') -- [draw=none] (z) -- [draw=none] (f),
(d') -- [gluon] (c)
}; 
\end{feynman} 
\end{tikzpicture} & = -\frac{i}{2\Mpl}\sum_a m_a\!\int\!\!d t\left[\frac{v_a^2}{2}\radh{00}+v_a^iv_a^j\radh{ij}\right] \notag \\
&\equiv-\frac{i}{2\Mpl}\int\!\!d^4\!x\,\ord{A}T^{\mu\nu}\radh{\mu\nu} \; .
\end{align}
Therefore, we immediately understand that
\begin{subequations}
\label{eqn:scheme70}
\begin{align}
\ord{A}T^{00}(t,\bm{x}) & = \sum_a \frac{m_a}{2}v^2\,\delta^{(3)}\!\left(\bm{x}-\bm{x}_{a}(t)\right) \; , \\
\ord{A}T^{ij}(t,\bm{x}) & =\sum_a m_a v^iv^j\,\delta^{(3)}\!\left(\bm{x}-\bm{x}_{a}(t)\right) \; .
\end{align}
\end{subequations}
We just need the non-vanishing contribution to the combination $T^{00}+T{^k}_{k}$, thus, summing and performing a partial Fourier transform as in eq. (\ref{eq:part_fourier_transf}), we obtain
\begin{equation}
\ord{A}T^{00}(t,\bm{k})+\ord{A}T{^k}_{k}(t,\bm{k}) = \frac{3}{2}\sum_a m_av_a^2\,e^{-i\bm{k}\cdot\bm{x}_a(t)} \; .
\end{equation} 
Then, we have to multipole expand this quantity and take the contribution to the quadrupole moment, hence 
\begin{equation}
\int\!\!d^3\!x\left(\ord{A}T^{00}(t,\bm{x})+\ord{A}T{^k}_{k}(t,\bm{x})\right)x^ix^j = \left(\frac{3}{2}\sum_a m_a v_a^2\right)x_a^ix_a^j \; .
\label{eq:TA.contr}
\end{equation}

The first non linear diagram is given by
\begin{align}
\begin{tikzpicture}[baseline]
\begin{feynman}
\vertex [label=270:$t_2$, label=170:$\bm{q}$](c);
\vertex [above=0.6 of c] (z);
\vertex [above=0.8cm of z,label=90:$t_1\ $, label=190:$\bm{k}$] (c');
\vertex [right=1cm of c] (a);
\vertex [left=1cm of c] (b);
\vertex [right=1cm of c'] (a');
\vertex [right=1cm of z] (d);
\vertex [left=1cm of c'] (b');
\diagram* {
(a) -- [plain] (c) -- [plain] (b),
(a') -- [plain] (c') -- [plain] (b'),
(c) -- [scalar] (z) -- [scalar] (c'),
(d) -- [gluon] (c')
}; 
\end{feynman} 
\end{tikzpicture} & = -\frac{i}{2\Mpl}\left(\sum_{a\neq b}\frac{\GN m_am_b}{\abs*{\bm{x}_{ab}}}\right)\radh{00}\equiv-\frac{i}{2\Mpl}\int\!\!d^4\!x\,\ord{B}T^{\mu\nu}\radh{\mu\nu} \; .
\end{align}
We omitted the explicit steps of the computation because they are equivalent to the one done in eq. (\ref{eq:disdiag}). The only non-vanishing component of $T^{\mu\nu}_B$ already partially Fourier transform is then
\begin{equation}
T^{00}_B(t,\bm{k}) = \sum_{a\neq b}\frac{\GN m_am_b}{\abs*{\bm{x}_{ab}}}e^{-i\bm{k}\cdot\bm{x}_{a}} \; .
\end{equation}
Then we multipole expand to find the contribution to $I^{ij}$ 
\begin{equation}
\int\!\!d^3\!\bm{x}\left(\ord{B}T^{00}(t,\bm{x})\right)x^ix^j=\sum_{a\neq b}\frac{\GN m_am_b}{\abs*{\bm{x}_{ab}}}x_a^ix_a^j
\label{eq:TB.contr}
\end{equation}
Finally, we compute the contribution coming from $\ord{C}T^{\mu\nu}$
\begin{align}
\begin{tikzpicture}[baseline]
\begin{feynman}
\vertex [label=270:$t_2$, label=170:$\bm{q}$] (c);
\vertex [above=0.7cm of c] (z);
\vertex [right=1cm of z] (d);
\vertex [above=0.7cm of z, label=90:$t_1\ $, label=190:$\bm{k}$] (c');
\vertex [right=1cm of c] (a);
\vertex [left=1cm of c] (b);
\vertex [right=1cm of c'] (a');
\vertex [left=1cm of c'] (b');
\diagram* {
(a) -- [plain] (c) -- [plain] (b),
(a') -- [plain] (c') -- [plain] (b'),
(c) -- [scalar] (z) -- [scalar] (c'),
(d) -- [gluon] (z)
}; 
\end{feynman} 
\end{tikzpicture} & = \frac{i}{2\Mpl}\sum_{a\neq b}\int\!\!d t\frac{\GN m_am_b}{\abs*{\bm{x}_{ab}}}\left(\frac{3}{2}\radh{00}+\frac{x_{ab}^ix_{ab}^j}{\abs*{\bm{x}_{ab}}^2}\radh{ij}\right) \notag \\
&\equiv-\frac{i}{2\Mpl}\int\!\!d^4\!x\,\ord{C}T^{\mu\nu}\radh{\mu\nu} \; .
\end{align}
Again computations are equivalent to the one of eq. (\ref{eq:disdiag.int}). Performing the partial Fourier transform, we see that this diagram contributes to
\begin{equation}
\left(\ord{C}T^{00}(t,\bm{k})+\ord{C}T{^k}_{k}(t,\bm{k})\right) =-2\sum_{a\neq b}\frac{\GN m_am_b}{\abs*{\bm{x}_{ab}}}e^{-i\bm{k}\cdot\bm{x}_a} \; .
\end{equation}
Then, we multipole expand as usual, so that
\begin{equation}
\int\!\!d^3\!\bm{x}\left(\ord{C}T^{00}(t,\bm{x})+\ord{C}T{^k}_{k}(t,\bm{x})\right)x^ix^j =\left(-2\sum_{a\neq b}\frac{\GN m_am_b}{\abs*{\bm{x}_{ab}}}\right)x_a^ix_a^j \; .
\label{eq:TC.contr}
\end{equation}

Now we have all the contributions we need, therefore, inserting (\ref{eq:Tdot.contr}), (\ref{eq:Tddot.contr}), (\ref{eq:TA.contr}), (\ref{eq:TB.contr}) and (\ref{eq:TC.contr}) in (\ref{eq:I.NLO.comp}) we eventually get
\begin{align}
I^{ij}(t) & =\sum_a m_a\left[x_a^ix_a^j\right]_{\text{STF}} \notag \\
&\quad + \sum_a m_a\left(\frac{3}{2}v_a^2-\sum_{a\neq b}\frac{\GN m_b}{\abs*{\bm{x}_{ab}}}\right)\left[x_a^ix_a^j\right]_{\text{STF}} \notag \\
&\quad -\frac{4}{3}\sum_a m_a\frac{d}{d t}\Big(\bm{v}_a\cdot\bm{x}_a\left[x_a^ix_a^j\right]_{\text{STF}}\Big)+  \frac{11}{42}\sum_a m_a\frac{d^2}{d t^2}\Big(\abs*{\bm{x}_a}^2\left[x_a^ix_a^j\right]_{\text{STF}}\Big) \; .
\label{eq:Iij.NLO.final}
\end{align}
This is the result we were looking for. Now that we have the explicit expression of the various multipole moments, we are ready to compute the NLO power loss given in eq. (\ref{eq:NLOPLO}).

\section{Next-to-Leading Order power loss}

As we said many times, the final system behaves as a single body placed in the CoM of the original binary system emitting GWs. Then, we have to rewrite $J^{ij}$, $I^{ijk}$ and $I^{ij}$ given respectively by (\ref{eq:Jij.fin}), (\ref{eq:Iijk.fin}) and (\ref{eq:Iij.NLO.final}) in term of CoM variables. Let's then introduce
\begin{align}
m \equiv\sum_a m_a \; ,&& \mu \equiv \frac{m_1 m_2}{m} \; ,&& \nu \equiv \frac{\mu}{m} \;  , && \bm{r} \equiv \bm{x}_1-\bm{x}_2 \; , && \bm{v} \equiv \dot{\bm{r}} \; .
\end{align}
We have just to be careful because, even if we did not compute it explicitly, the previous diagrams modified also the zero and the linear multipole moment given by (\ref{eq:multiM}) and (\ref{eq:multiMX}) as follows\footnote{See Ref. \cite{Rad-Corr} for an explicit computation of these terms.}
\begin{subequations}
\label{eqn:scheme34}
\begin{align}
M & = \sum_a m_a\left(1+\frac{1}{2}v_a^2-\frac{1}{2}\sum_{b\neq a}\frac{\GN m_b}{\abs*{\bm{x}_a-\bm{x}_b}}\right) +\Ord{v^4} \; , \\
X^i & = \frac{1}{m}\sum_a m_a\left(1+\frac{1}{2}v_a^2-\frac{1}{2}\sum_{b\neq a}\frac{\GN m_b}{\abs*{\bm{x}_a-\bm{x}_b}}\right)x_a^i +\Ord{v^4} \; . 
\end{align}
\end{subequations}
In particular then, we have to use the above $\bm{X}$ as the real position of the CoM. Choosing the origin of the system in the CoM implies $\bm{X} = 0$, which means that eventually
\begin{subequations}
\label{eqn:scheme34a}
\begin{align}
\bm{x}_1 &= \left[\frac{m_2}{m}+\frac{\nu}{m}(m_1 - m_2)\left(\frac{v^2}{2}-\frac{\GN m}{r}\right)\right]\bm{r} +\Ord{v^4} \; ,\\
\bm{x}_2 & = \left[-\frac{m_1}{m}+\frac{\nu}{m}(m_1 - m_2)\left(\frac{v^2}{2}-\frac{\GN m}{r}\right)\right]\bm{r} +\Ord{v^4} \; .
\end{align}
\end{subequations}
To write these expressions as function of CoM variables only, we used that at LO
\begin{align}
\bm{v}_1 & = \frac{m_2}{m}\bm{v}+\Ord{v^3} \; , & \bm{v}_2 & = -\frac{m_1}{m}\bm{v}+\Ord{v^3} \; .
\end{align}
Inserting eqs. (\ref{eqn:scheme34a}) into (\ref{eq:Jij.fin}), (\ref{eq:Iijk.fin}) and (\ref{eq:Iij.NLO.final}) and considering a circular orbit\footnote{Note that in this case $\bm{v}\cdot\bm{r} = 0$.} of frequency $\ocirc$, one eventually gets
\begin{align}
I^{ij}(t) & = \mu\left\{1-\left(\frac{1}{42}+\frac{39}{42}\nu\right)\xc\,\right\}\left[r^ir^j\right]_{\text{STF}}+\frac{11}{42}\mu r^2(1-3\nu)\left[v^iv^j\right]_{\text{STF}} \; ,\\
J^{ij}(t) & = \mu\sqrt{1-4\nu}\left[(\bm{r}\times\bm{v})^ir^j\right]_{\text{STF}} \; ,\\
I^{ijk}(t) & =  \mu\sqrt{1-4\nu}\left[r^ir^jr^k\right]_{\text{STF}} \; .
\end{align}
In the above equations we used the 1PN order EOM of the binary to replace any acceleration term in the final expression, and we defined the PN expansion parameter
\begin{equation}
\xc \equiv (\GN m \ocirc)^{2/3} \; .
\end{equation}
This comes from the Kepler law $\ocirc^2 r^3 = \GN m$, which implies that $\xc = (\ocirc r)^2 \sim v^2$, hence this is indeed a good PN expansion parameter.

To compute the power loss, we have to Fourier transform these quantities and to compute their modulo square. Since we are analysing non spinning objects, we can consider the motion to happen in a plane. Therefore, for a circular orbit\footnote{See Chap. 4 of \cite{Maggiore} to see how to generalise this to the elliptic case at 1PN.}, we can then parametrise $\bm{r}$ as follows
\begin{align}
r_x(t) = r \cos(\ocirc t) \; , && r_y(t) = r \sin(\ocirc t) \; ,  && r_z(t) = 0 \; ,
\end{align} 
Then, after some computations one gets
\begin{align}
\abs{I^{ij}(\omega)}^2 & =\frac{\pi T\nu^2\xc^5}{2\GN^2\ocirc^6}\left[1+\left(-\frac{107}{21}+\frac{55}{21}\nu\right)\xc\right]\delta(\omega-2\ocirc) \; ,\\
\abs*{J^{ij}(\omega)}^2 & = \frac{\pi T\nu^2\xc^6}{2\GN^2\ocirc^6}(1-4\nu)\delta(\omega-\ocirc)  \; ,\\
\abs*{I^{ijk}(\omega)}^2 & =  \frac{\pi T\nu^2\xc^6}{4\GN^2\ocirc^8}(1-4\nu)\bigg[\delta(\omega-3\ocirc) +\frac{3}{5}\delta(\omega-\ocirc)\bigg] \; .
\end{align}
In the above formula we called $T=2\pi\delta(0)$, and we discarded terms proportional to $\delta(\omega)$ or to $\delta(\omega+\alpha\ocirc)$ with $\alpha>0$, because we are considering $\omega \in (0, +\infty)$, hence they do not contribute to the power loss. From here we understand that the LO power loss is quadrupolar with $\omega\simeq 2\ocirc$ and is given explicitly by
\begin{equation}
\mathcal{P}_{\text{LO}} = \frac{32}{5}\frac{\nu^2}{\GN}\xc^5 \; .
\end{equation}
Moreover we also find the NLO power loss which is
\begin{equation}
\frac{\mathcal{P}_{\text{NLO}}}{\mathcal{P}_{\text{LO}}}=1-\left(\frac{1247}{336}+\frac{35}{12}\nu\right)\xc \; .
\end{equation}
After all this computations, we were able to reproduce the result for the NLO power loss already found by other technique, see Ref. \cite{megaBlanchet}. Therefore, we see again that NRGR gives results that are consistent with other computation methods. These steps continue to all orders, up to some non-linear effects in the radiation zone\footnote{In this work we did not discuss non linearities, we redirect the reader to Sec. 7.4 of Ref. \cite{EFT3}.}.

\chapter{Finite size effects in NRGR (brief discussion)}\label{ch:pp}

In the previous two chapters we saw how to applied NRGR to a binary of compact objects. We have always considered the two body as point-like sources of GWs, and we justified this assumption saying that the finite size effects coming from the integration of the size of the compact objects $R_S = 2\GN m$ enter at a higher PN order. In this chapter we shall prove that, indeed, these effects enter with a power of $v^{10}$, hence at the 5PN order.

\section{EFT for gravitationally interactive extended objects}

Let's consider an extended isolated compact object of mass $m_a$. We perfectly know that the space outside this mass is completely described by the pure Einstein-Hilbert action 
\begin{equation}
S_{\text{EH}}[\tilde{g}_{\mu\nu}] = -2\Mpl^2\int\!\!d^4\!x\sqrt{-\tilde{g}}\tilde{\Rc} \; ,
\end{equation}
where $\tilde{\Rc}$ is the Ricci scalar associated to the four dimensional metric $\tilde{g}_{\mu\nu}$. As we said, we would like to remove from the problem the small scale $\Rs=2\GN m_a$. We shall use again an EFT approach.

In principle, we could use a top-down approach similar to the one shown in Ch. \ref{ch:first}, hence we can split the metric $\tilde{g}_{\mu\nu}$ as
\begin{equation}
\tilde{g}_{\mu\nu} = g\ord{S}_{\mu\nu} + g_{\mu\nu} \; .
\end{equation}
The quantity $g\ord{S}_{\mu\nu}$ takes into account the short-distance physics at scale $R_S$, while $g_{\mu\nu}$ is the metric of space-time far from the object. However, since we know that the physics outside a non spinning BH has to be invariant under general diffeomorphism and under world-line reparametrisation, we can use a bottom-up approach, and write the most general action containing operators that respect these symmetries. These operators depend on the long-distances (w.r.t. the size of the object $\Rs$) DOF, which are the metric $g_{\mu\nu}$, and the position of the object $x^{\mu}_a$. Therefore it is easy to see that the more general effective action can be written as
\begin{align}
\Seff{}\left[g_{\mu\nu},x^\mu_a\right] = S_{\text{EH}}\left[g_{\mu\nu}\right]+S_{\text{pp}}\left[g_{\mu\nu},x^\mu_a\right] + S_{\text{fin}}\left[g_{\mu\nu},x^\mu_a\right] \; ,
\label{eq:Seff.pp}
\end{align}
where $S_{\text{EH}}\left[g_{\mu\nu}\right]$ is the usual Einstein-Hilbert action for the metric $g_{\mu\nu}$, $S_{\text{pp}}\left[g_{\mu\nu},x^\mu_a\right]$ is the point-particle action written in (\ref{eq:Sstart}), and $S_{\text{fin}}\left[g_{\mu\nu},x^\mu_a\right]$ is defined as
\begin{equation}
S_{\text{fin}}\left[g_{\mu\nu},x^\mu_a\right]  \equiv \sum_n C_n(\Rs)\int\!\!d \tau_a\mathcal{O}_n(\tau_a) \; .
\end{equation}
Reparametrisation invariance it is here made explicit by the use of the proper time $d\tau_a^2 =g_{\mu\nu}dx_a^\mu dx_a^\nu$. All informations about the UV (i.e. short-distance) behaviour of the theory is contained in the Wilson coefficients $C_n(\Rs)$ of the action $S_{\text{fin}}$. Since we are dealing with non spinning BH, our Wilson coefficients depend only on the size of the object\footnote{In the case of a neutron star, for instance, the Wilson coefficients will also depend on the equation of state of the neutron gas.} $\Rs$. What we have done in eq. (\ref{eq:Seff.pp}) was basically adding a series of world-line operators to the standard Einstein-Hilbert action. In particular we have:
\begin{itemize}
\item the minimal coupling $S_{\text{pp}} = -m\int\!\!d \tau_a$ describing the geodesic motion of the point particle
\item a set of non minimal couplings contained in $S_{\text{fin}}$ describing deviation from the geodesic motion, hence \textit{finite size effects}.
\end{itemize} 

\subsection{Redundant couplings}

As we said, the operators $\mathcal{O}_n$ have to be invariant under diffeomorphisms. Therefore, we can easily construct the first non minimal couplings using invariant quantities like the Ricci scalar and tensor, e.g.
\begin{align}
C_{R}\int\!\!d \tau_a\,\Rc\big(x_a(\tau_a)\big) \; ,& & C_V\int\!\!d \tau_a\,\Rc_{\mu\nu}\big(x_a(\tau_a)\big)u_a^\mu u_a^\nu  \; ,
\end{align}
where $u_a^\mu =dx_a^\mu/d\tau_a$ is the four-velocity of the point particle. 

These two are in fact redundant operators, i.e. they do not contribute to any measurable quantity. We can easily understand this if, in the action (\ref{eq:Seff.pp}), we perform the field shift
\begin{align}
g^{\mu\nu} \rightarrow g^{\mu\nu}+\delta g^{\mu\nu} & &\text{with} & & \delta g^{\mu\nu}=\frac{\epsilon}{2\Mpl^2}\int\!\!d \tau_a\,\frac{\delf\big(x-x_a(\tau_a)\big)}{\sqrt{-g}}g^{\mu\nu} \; ,
\label{eq:delta.redun}
\end{align}
where $\epsilon$ is a completely arbitrary parameter. Up to surface terms we get that 
\begin{equation}
S_{\text{EH}}[g_{\mu\nu}] \rightarrow S_{\text{EH}}[g_{\mu\nu}]+\delta S_{\text{EH}}[g_{\mu\nu}] \; ,
\end{equation}
where the second term is given by
\begin{align}
\delta S_{\text{EH}}[g_{\mu\nu}] & = -2\Mpl^2\int\!\!d^4\!x\sqrt{-g(x)}G_{\mu\nu}(x)\delta g^{\mu\nu}(x) \notag \\
& = -\epsilon\left\{\int\!\!d \tau_a\,\Rc_{\mu\nu}\big(x(\tau_a)\big)g^{\mu\nu}\big(x(\tau_a)\big)-2\int\!\!d \tau_a\,\Rc\big(x(\tau_a)\big)\right\} \notag \\
& =  \epsilon\int\!\!d \tau_a\,\Rc\big(x(\tau_a)\big) \; .
\end{align}
As a result we have that the coefficient $C_R$ it is actually not fixed, but can be shifted by an arbitrary quantity
\begin{equation}
C_R \rightarrow C_R+\epsilon \; .
\end{equation}
Therefore one can always set $C_R$ to zero by tuning $\epsilon = -C_R$. That's why we call this term a redundant operator. 

In order to remove also the coupling $C_V$, we just have to slightly modify the expression (\ref{eq:delta.redun}) for $\delta g^{\mu\nu}$ as follows
\begin{equation}
\delta g^{\mu\nu}\rightarrow \delta g^{\mu\nu} =\frac{1}{2\Mpl^2}\int\!\!d \tau_a\,\frac{\delf\big(x-x_a(\tau_a)\big)}{\sqrt{-g}}\left(\epsilon_1g^{\mu\nu}+\epsilon_2u^\mu u^\nu\right) \; .
\end{equation}
We recall that, since we are using the proper time $\tau_a$ to parametrise the world-line, $g_{\mu\nu}u^\mu u^\nu=1$. Therefore it is easy to see that in this case we obtain
\begin{equation}
\delta S_{\text{EH}}[g_{\mu\nu}] = \left(\epsilon_1+\frac{\epsilon_2}{2}\right)\int\!\!d \tau_a\,\Rc-\epsilon_2\int\!\!d \tau_a\,\Rc_{\mu\nu}u^\mu u^\nu \; ,
\end{equation}
in which everything is evaluated at the world-line $x^\mu_a(\tau_a)$. Choosing $\epsilon_1$ and $\epsilon_2$ properly, we are able to remove both the $C_R$ and the $C_V$ couplings. 

Actually, as pointed out in App. A of Ref. \cite{EFT3}, after this field shift there is a left-over piece in the effective action proportional to $\delta^{(3)}\big(\bm{x}_1(\tau)-\bm{x}_2(\tau)\big)$. This is nothing but the analogous of the ``Darwin term'' in the hydrogen atom, responsible for the fine structure of the energy levels. However, since we are considering the compact objects as classical objects, we can consistently remove and neglect this contribution.

\section{Finite size effects}

Once we discarded terms proportional to $\Rc_{\mu\nu}$ and $\Rc$, we have to take into account possible non minimal couplings that contain the Riemann tensor. However, rather than considering $\Rc_{\mu\nu\rho\sigma}$, it is easier to consider again the electric and magnetic part  of the Weyl tensor $\bm{\mathcal{C}}_{\mu\nu\rho\sigma}$ defined in eq. (\ref{eq:El-Mag}). Therefore, using the proper time to parametrise every quantities so that world-line reparametrisation invariance is ensured, we can immediately write new non minimal couplings as
\begin{equation}
S_{\text{fin}} = \int\!\!d \tau_a\!\int\!\!d^4\!x\,\delf\big(x-x_a(\tau_a)\big)\left(Q_E^{\mu\nu}(\tau_a)E_{\mu\nu}(x)+Q_B^{\mu\nu}(\tau_a)B_{\mu\nu}(x)\right) \; .
\label{eq:NM1}
\end{equation}
Going now in a locally-flat comoving system of reference, similarly to eq. (\ref{eq:loc.flat}), we can write
\begin{equation}
S_{\text{fin}} = -\int\!\!d \tau_a\!\int\!\!d^4\!x\,\delf\big(x-x_a(\tau_a)\big)\left(Q_E^{ij}(\tau_a)E_{ij}(x)+Q_B^{ij}(\tau_a)B_{ij}(x)\right) \; .
\label{eq:Snm}
\end{equation}

It is now easy to understand that $Q_{E(B)}^{\mu\nu}$ are the usual symmetric trace-free quadrupole moments of the compact object, that we have already use in \ref{sec:bott.up}. We call them $Q^L$ rather than $I^L$ to stress that they are different from the one defined in \ref{sec:bott.up}; in fact $Q^L$ are multipole moments of the single black hole, while the ones used in the study of the radiative sector were the multipole moments of the binary system. In general of course these multipole moments can depend on time, hence we can decompose\footnote{Of course we could do a similar decomposition also for the multipole moments $I^L$ defined in \ref{sec:bott.up}. In that case we used only what we called here background modes, ignoring the induced moments due to the presence of an external gravitational field. We redirect the reader to part III of Ref. \cite{EFT3} to see a field in which this induced multipole moments are relevant.} them as
\begin{equation}
Q_{E(B)}^{ij} = \left\langle Q_{E(B)}^{ij}\right\rangle_S+\left(Q_{E(B)}^{ij}\right)_R \; .
\end{equation}
The quantity $\left\langle Q_{E(B)}^{ij}\right\rangle_S$ is the expectation value computed in the background of the short mode. Actually for a non spinning body $\left\langle Q_{E(B)}^{ij}\right\rangle_S= 0$. The second term $\left(Q_{E(B)}^{ij}\right)_R$ is the induced quadrupole due to the presence of the long-distance gravitational field. Let's focus on $\left(Q_{E}^{ij}\right)_R$, the computations for the magnetic quadrupole momentum are analogous. The gravitational field $E_{ij}$ acts as a source for the quadrupole moment; therefore, using linear response theory we can compute
\begin{align}
\left(Q_{E}^{ij}\right)_R(\tau_a) = \frac{1}{2}\int\!\!d \tau'_a\!\int\!\!d^4\!x\delf\big(x-x_a(\tau'_a)\big)\left(iG_{Q,\text{ret}}^{ijk\ell }(\tau_a,\tau'_a)\right) E_{k\ell }(x) \; ,
\label{eq:fse.first}
\end{align}
where we defined
\begin{equation}
iG_{Q,\text{Rt}}^{ijk\ell }(\tau_a,\tau'_a)\equiv \vartheta(\tau_a-\tau'_a) \left\langle\Tprod{Q_{E}^{ij}(\tau_a)Q_{E}^{k\ell }(\tau'_a)}\right\rangle \; .
\end{equation}

In general the result will be a complex quantity; the real part $\Ree{\left(Q_{E}^{ij}\right)_R}$ contributes to \textit{finite size effects}, while the imaginary part $\Imm{\left(Q_{E}^{ij}\right)_R}$ contributes to \textit{absorption phenomena} at the horizon. Ignoring absorption phenomena\footnote{In Sec. 7.8 of Ref. \cite{EFT3} it is shown that this kind of terms become relevant at 6.5PN order.}, we focus only on the real part of (\ref{eq:fse.first}). As we said, $\left(Q_{E}^{ij}\right)_R$ is symmetric trace-free tensor, therefore, going in Fourier space for convenience, we can parametrise
\begin{equation}
\Ree{iG_{Q,\text{ret}}^{ijk\ell }(\omega)} = \mathcal{Q}^{ijk\ell }\Ree{f(\omega)} \; ,
\label{eq:Green.omega} 
\end{equation}
where $\mathcal{Q}^{ijk\ell }$ is the projector onto the symmetric trace-free tensor, i.e.
\begin{equation}
\mathcal{Q}^{ijk\ell } = \delta^{ik}\delta^{j\ell }+\delta^{i\ell }\delta^{jk}-\frac{2}{3}\delta^{ij}\delta^{k\ell } \; .
\end{equation}
Now, we are interest in the long-distance physics w.r.t. $\Rs$, namely we can consider $\omega\Rs\ll 1$ and expand
\begin{equation}
\Ree{f(\omega)} = C_E+\cancelto{0}{C_{\dot{E}}\Rs\omega}+\dots \; .
\end{equation}
All the odd powers in $\omega$ vanish because, at this stage, time reversal is still a symmetry of the theory. Inserting this expression in (\ref{eq:Green.omega}) and going back to the direct space we get
\begin{equation}
\Ree{iG_{Q,\text{ret}}^{ijk\ell }(\tau_a,\tau'_a)} \simeq \mathcal{Q}^{ijk\ell }C_E\delta(\tau_a-\tau'_a) \; .
\end{equation}
Then inserting this expression into (\ref{eq:fse.first}), and taking only the real part, we arrive at our final result
\begin{equation}
\Ree{\left(Q_{E}^{ij}\right)_R(\tau_a)} \!\simeq\! \int\!\!d^4\!x\delf\big(x-x_a(\tau_a)\big)\,C_E E^{ij}(x) \; .
\end{equation}

A similar computation leads to an analogous expression for $Q_{B}^{ij}$, with $E_{ij}$ replaced by $B_{ij}$. Substituting these expressions into (\ref{eq:Snm}), and considering only the first leading terms, we eventually obtain the following expression
\begin{equation}
S_{\text{fin}} \!=\! -\int\!\!d \tau_a\!\int\!\!d^4\!x\,\delf\big(x-x_a(\tau_a)\big)\left(C_E E^{ij}(x)E_{ij}(x)+C_B B^{ij}(x)B_{ij}(x)\right) \; .
\label{eq:Sfse}
\end{equation}
These are the first non-vanishing terms describing finite size effects to the point-particle theory. 

\section{Effacement Theorem}

We now want to understand at which order in the PN expansion these terms start to be relevant. Let's first consider the leading term of the action (\ref{eq:Sfse}). Once we have decomposed the graviton  field $h_{\mu\nu}$ as in eq. (\ref{eq:mode.dec}), and once we have established a power counting rule in the PN parameter $v$ (see Sec. \ref{sec:power.count}), we understand that the term of the action (\ref{eq:Sfse}) with the smallest scaling in $v$ is given by
\begin{equation}
S_{\text{fin}} \longrightarrow \frac{C_E}{\Mpl^2}\int\!\!d t\left(\partial_i \partial_j H_{00}\right)^2 \; .
\label{eq:scaling1_pp}
\end{equation}
We need to understand how the coefficient $C_{E}$ scales; due to dimensional analysis we expect it to have mass dimension equal to $-3$. A matching procedure for the EFT described in this chapter is implemented explicitly in Sec. 6.6 of Ref. \cite{EFT3}, where it is shown that the scaling of the coefficient $C_E$ is
\begin{equation}
C_E\sim \Mpl^2\Rs^5 \; .
\label{eq:scaling2_pp}
\end{equation}
Using this result and the usual scaling rules derived in Sec. \ref{sec:power.count}, we find that the vertex associated to this first finite size effects scales as follows
\begin{equation}
\begin{tikzpicture}[baseline]
\begin{feynman}
\vertex (f);
\vertex [above=0.1cm of f] (z);
\vertex [left=0.4cm of z, label=180:$C_E$] (c');
\vertex [above=0.9cm of c'] (a);
\vertex [above=0.4cm of a] (a');
\vertex [below=0.9cm of c'] (b);
\vertex [below=0.4cm of b] (b');
\vertex [above right=0.9cm of z] (c); 
\vertex [below right=0.9cm of z] (d); 
\diagram* {
(a') -- [draw=none] (a) -- [plain] (c') -- [plain] (b) -- [draw=none] (b'),
(c') -- [scalar] (c),
(c') -- [scalar] (d),
(z) -- [draw=none] (f)
}; 
\end{feynman} 
\end{tikzpicture} \sim \frac{\Mpl^2\Rs^5}{\Mpl^2}\frac{r}{v}\left(\frac{v^{1/2}}{r^3}\right)^2 \sim \left(\frac{\Rs}{r}\right)^5 \sim v^{10} \; .
\end{equation}

If we now consider a binary of compact objects, it is not hard to see that the lowest order diagram (in power of $v$) responsible for finite size effects is 
\begin{equation}
\begin{tikzpicture}[baseline]
\begin{feynman}
\vertex (z);
\vertex [below=0.8cm of z] (c);
\vertex [above=0.8cm of z, label=90:$C_E$] (c');
\vertex [right=1.2cm of c] (a);
\vertex [left=1.2cm of c] (b);
\vertex [right=0.7cm of c] (a'');
\vertex [left=0.7cm of c] (b'');
\vertex [right=1.2cm of c'] (a');
\vertex [left=1.2cm of c'] (b');
\diagram* {
(a) -- [plain] (c) -- [plain] (b),
(a') -- [plain] (c') -- [plain] (b'),
(a'') -- [scalar] (c'),
(b'') -- [scalar] (c')
}; 
\end{feynman} 
\end{tikzpicture} \sim L\left(\frac{\Rs}{r}\right)^5 \sim Lv^{10} \; ,
\label{eq:Eff_th}
\end{equation}
Recalling what we said in Sec. \ref{subsec:Newt}, we immediately see that \textit{finite size effects first enter at $5PN$ order}. Eq. (\ref{eq:Eff_th}) is basically the content of the \textit{Effacement Theory}, which allows to reduce the problem outside the two (or more) bodies to the study of point-like sources interacting via gravity. It turns out that the coefficients $C_{E(B)}$ actually vanish for a non-spinning BH in four dimension; in particular, this is shown explicitly in Ref. \cite{KK_BH} for the electric-type coefficient. Therefore, for a non-spinning BH-BH binary, the first non-vanishing finite size effect enter in the PN expansion at an even higher order. This is consistent with previously known results, see Refs. \cite{Damour-Nagar,LoveNumb}. On the other hand, for neutron stars, the coefficients $C_{E(B)}$ are finite and depend on the so-called \textit{Love numbers}, which phenomenology have been largely study in the literature. We redirect the reader to Refs. \cite{Damour-Nagar,LoveNumb,Love-2}.

\chapter*{Conclusions of the first part}

\addcontentsline{toc}{chapter}{Conclusions of the first part} 

In the first part of this work, we presented the NRGR, namely an EFT to study the two-body inspiral problem first introduced in 2006 \cite{EFT1,Diss.Eff}. Our intention was to make a pedagogical and, as much as possible, self-contained introduction to this approach, mainly because it has proved to be a successful and sometimes more clear procedure to the PN study of the binary inspiral problem.

In Ch. \ref{ch:first} we described the main features of NRGR. We showed how this EFT is implemented, and how to compute the effective action in the PN formalism. In particular, in Secs. \ref{sec:power.count} and \ref{sec: 1PN contr}, we computed explicitly the 1PN order action describing the conservative dynamic of the binary, finding the same conclusions of the original Ref. \cite{EFT1}, which are consistent with known results \cite{EIH, MTW}. 

Then, in Ch. \ref{ch:rad} we completed the description of the previous chapter, computing also the radiative part of the motion of the binary system in the context of NRGR. In particular, in Sec. \ref{sec:SNR} we derived the complete action at order $L^{1/2}v^{5/2}$, and we used it to find our final effective action $\Seff{NRGR}$ at 2.5PN order. Then, we computed physical results like the LO power lost by the system (Sec. \ref{sec:power_leading}), the gravitational radiation reaction and the gravitational waveform (Sec. \ref{sec:GWandGR}). In particular, in order to find physically consistent results for these last two quantities, we saw how to implement the in-in formalism in NRGR, following Ref. \cite{rad.rec}. All physical conclusions coincide with previously known results \cite{EFT1, EFT3, EIH, Thorne.Burke}. After this, we also presented the bottom-up approach to NRGR. Implementing a matching procedure through top-down computation of Feynman diagrams, we were finally able to compute also the NLO contribution to the power loss of the system. 

Finally in Ch. \ref{ch:pp}, we explained how to deal with extended objects in this EFT approach. In particular, we briefly presented the effacement theorem and legitimised the use of point-particle approximation to described the two BHs of the binary.

This is not the end of the story. In fact, a new parametrisation of the metric introduced in Refs. \cite{KK2,KK1}, that we shall present in the next part of this work, has significantly simplified the computations at higher PN orders. This led to find explicitly result for the 2PN \cite{2PN}, the 3PN \cite{3PN} and the 4PN \cite{4PN-1,4PN-2,4PN-3} corrections to the conservative dynamics of the non spinning BH-BH binary. These results coincide with the one found with other techniques, see Ref. \cite{Blachet-rev,megaBlanchet} and references therein.

Moreover, this EFT approach has been proved to be very useful in extending this kind of computations to spinning binary of compact objects. First studied in \cite{Porto-spin}, this system has been largely explored still in the context of PN expansion. We redirect the reader to the two reviews \cite{EFT3, Levi-Review} and references therein for a complete discussion on this subject. Finally, we point out that in Ref. \cite{EFTofPNG} is presented the Mathematica package \textit{EFTofPNG}, which is a package for high precision computations in this EFT approach to the PN gravity, including spins.

\part{NRGR in scalar-tensor theory}

\chapter{Parametrisation with NRG fields }\label{ch:NRGR_NRG}

In the second part of this work, we shall extend the NRGR formalism seen in the previous part to alternative theories of gravity. In particular, in Ch. \ref{ch:conf.coupl} we generalise NRGR to a conformally coupled scalar-tensor theory with a single scalar field. This extension has recently been done by in Ref. \cite{NRGR+scalar} for the case of a massless scalar field,  and in \cite{Mass.scalar} where NRGR is used to study the effects of light axion. In \cite{NRGR+scalar}, the authors considered the theory in the so-called \textit{Einstein frame}, while here we re-derive their results using the equivalent description of the theory in the \textit{Jordan frame}.

Before that, however, we want to introduce another tool of ordinary NRGR. Up until now we parametrised the graviton $h_{\mu\nu}$ following the original approach designed by Goldberger and Rothstein \cite{EFT1}. In 2008 Kol and Smolkin \cite{KK2,KK1} proposed a different parametrisation of the metric, based on a Kaluza-Klein reduction, that turned out to be very useful, especially in the conservative sector. In this first chapter we shall present their work, re-deriving their results and showing the advantages of this parametrisation.

We finally stress that, following the usual convention of scalar-tensor theory and the one of Ref. \cite{NRGR+scalar}, we use as Planck mass the following definition $\mpl^{-2} \equiv 8\pi\GN$.

\section{The NRG fields}

Let's introduce the so-called Kaluza-Klein parametrisation of the metric in 4 dimensions
\begin{equation}
ds^2=e^{2\phi}\left(dt-A_idx^i\right)^2-e^{-2\phi}\gamma_{ij}dx^idx^j \;.
\end{equation}
We are rewriting the 10 components metric $g_{\mu\nu}$ using three fields: a scalar $\phi$, a three vector $\bm{A}$ and a three dimensional metric $\gamma_{ij}$. These are called \textit{Non-Relativistic-Gravity fields} (NRG fields). 

First we note that this parametrisation resembles the usual  Arnowitt-Deser-Misner (ADM) formalism \cite{ADM}
\begin{equation}
ds^2 = N^2dt^2-\gamma^{\text{ADM}}_{ij}(dx^i+N^idt)(dx^j+N^jdt) \;.
\label{eq:ADM-dec}
\end{equation}
In fact, the two coincide at linear level for all fields, and at all levels whenever $A_i\nolinebreak=\nolinebreak 0\nolinebreak=\nolinebreak N_i$. However, there are some marked differences between these two parametrisations:
\begin{itemize}
\item ADM is designed for initial value problems and to study the evolution of the metric in  time. It is defined starting by a Kaluza-Klein spatial reduction, and the $(N,\bm{N}, \gamma^{\text{ADM}}_{ij})$ fields transform nicely under space-independent change of coordinates
\item NRG fields have been specifically used to study the two-body problem in the PN approach \cite{KK2,KK1}. Their definition relies on a Kaluza-Klein temporal reduction, and $(\phi,\bm{A}, \gamma_{ij})$ transform nicely under time-independent change of coordinates.
\end{itemize}

The next two sections are devoted to the computation of the action of NRGR in terms of NRG fields. In particular in Sec. \ref{sec:EH.NRG} we derive the expression of the Einstein-Hilbert action in terms of NRG fields, with the corresponding needed harmonic gauge-fixing term. While this may seems a purely mathematical exercise, some steps of this derivation will be very useful when we will derive the action of NRGR for the case of gravity conformally coupled to a scalar field in Ch. \ref{ch:conf.coupl}.

\section{Einstein-Hilbert action with NRG fields}\label{sec:EH.NRG}

As we said earlier our goal is to compute the Einstein-Hilbert action using NRG fields; this section is basically a more explicit version of appendix A of reference \cite{KK2}. In what follows we will use the so-called \textit{vielbein formalism} of GR\footnote{For a review on this topic, we redirect to Refs. \cite{arXiv-Cartan,VanProyen}.}, therefore we start our discussion by reviewing some basic properties of this technique, postponing the actual computation to Secs. \ref{sec:explic_comp} and \ref{sec:NRG-fields_expl}.

\subsection{Connection and Curvature in non-coordinate basis}\label{susec:non-coord-basis}

Up until now we have used the canonical coordinate basis for forms and vectors, given respectively by $dx^\mu$ and $\partial_\mu$. Another common way to perform computations in GR is to introduce the vielbein fields $e{_a}{^\mu}$ and $\vartheta{^a}{_\mu}$, which in general are space-time dependent quantities. With the help of these auxiliary fields, we can introduce a new basis for vectors $\overline{e}_a$ and forms $\underline{\vartheta}^a$ given by
\begin{align}
\overline{e}_a \equiv e{_a}{^\mu}\partial_\mu \; , && \underline{\vartheta}^a \equiv \vartheta{^a}{_\mu}dx^\mu \;.
\label{eq:new_basis}
\end{align}
Since we know that $dx^\mu \partial_\nu  = \delta{^\mu}_\nu$, we have that
\begin{align}
e{_a}{^\mu}\vartheta{^a}{_\nu} = \delta{^\mu}_\nu \; , && e{_a}{^\mu}\vartheta{^b}{_\mu} = \delta{_a}^b \;.
\label{eq:delta_vielbein}
\end{align}
We expect each vector $\overline{v}$ and form $\underline{w}$ to have a unique expansion in both the old coordinate system and the new frame, i.e. $\overline{v} = v^a\overline{e}_a = v^\mu\partial_\mu$, and  $\underline{w} = w_a\underline{\vartheta}^a = w_\mu dx^\mu$. Therefore, thanks to eq. (\ref{eq:new_basis}), we can use the vielbein fields to compute the components of the various quantities in the new frame fields, e.g. 
\begin{align}
v^a  = \vartheta{^a}{_\mu} v^\mu \; , &&  w_a  = e{_a}{^\mu} w_\mu \; .
\end{align}
In particular we can find the metric and its inverse in the new frame as
\begin{align}
\hat{g}_{ab} =  e{_a}{^\mu}g_{\mu\nu}e{_b}{^\nu} \; , && \hat{g}^{ab} =  \vartheta{^a}{_\mu}g^{\mu\nu}\vartheta{^b}{_\nu} \; .
\end{align}
In a sense we are not really using the vielbein methods, because usually one chooses $e{_a}{^\mu}$ so that the new frame has a flat metric, i.e. $e{_a}{^\mu}g_{\mu\nu}e{_b}{^\nu} = \eta_{ab}$. However considering a non-trivial frame metric is what we need to do when we will use the NRG fields in Sec. \ref{sec:explic_comp} (cf. eq. (\ref{eq:non.trivial.metric})).

In general this basis \textit{does not introduce a new coordinate system}. This can be easily noticed by looking at the Lie bracket of the frame derivative $\overline{e}_a = e{_a}{^\mu}\partial_\mu $, i.e.
\begin{align}
\left[\overline{e}_a,\overline{e}_b\right] = \left[e{_a}{^\mu}\partial_\mu,e{_b}{^\nu}\partial_\nu\right] = \left(e{_a}{^\mu}\partial_\mu e{_b}{^\nu}-e{_b}{^\mu}\partial_\mu e{_a}{^\nu}\right)\partial_\nu \; .
\label{eq:comm_der_frame}
\end{align}
If there were local coordinates $y_a$ such that $\overline{e}_a = \partial/\partial y_a = \partial_a$, these differential operators would commute. In the next sections, however, we will sometimes abuse a bit of our notation and use the symbol $\partial_a$ to denote a derivative in the non-coordinate basis, hence we basically define 
\begin{equation}
\partial_a \equiv \overline{e}_a = e{_a}{^\mu}\partial_\mu \; .
\label{eq:def_der_frame}
\end{equation} 

Let's go back to eq. (\ref{eq:comm_der_frame}); for the properties of the Lie bracket, we expect the result of this equation to be still a vector, hence it can be written as a linear combination of the basis. There exist then some coefficients $\Omega{_{ab}}^c$ such that
\begin{equation}
\left[\overline{e}_a,\overline{e}_b\right] = -\Omega{_{ab}}^c \overline{e}_c \; .
\label{eq:Olon_coeff_def}
\end{equation}
These are often called \textit{anholonomy coefficients}. Comparing this equation with (\ref{eq:comm_der_frame}) and using eq. (\ref{eq:delta_vielbein}), it is not hard to see that the explicit expression of these coefficients is
\begin{equation}
\Omega{_{ab}}^c = - 2 e{_{[a|}}{^\mu}\partial_\mu e{_{|b]}}{^\nu}\vartheta{^c}_\nu 
\label{eq:anol_coeff_viel}
\end{equation}

We want to find a way of computing the components of the affine connection and the curvature tensor in the new frame. A priori we do not know if, also in this frame, we can use the affine Christoffel connection. It is known that, given three vector fields $\overline{u}$, $\overline{v}$ and $\overline{z}$, we can define the torsion and the curvature tensor of a manifold as
\begin{subequations}
\label{eqn:schem58}
\begin{align}
T(\overline{u}, \overline{v}) & \equiv \nabla_{\overline{u}}\overline{v} - \nabla_{\overline{v}}\overline{u} - [\overline{u}, \overline{v}] \; , \\
\Rc(\overline{u}, \overline{v})\overline{z} & \equiv \nabla_{\overline{u}}\nabla_{\overline{v}}\overline{z} - \nabla_{\overline{v}}\nabla_{\overline{u}}\overline{z} - \nabla_{[\overline{u}, \overline{v}]}\overline{z} \; ,
\end{align}
\end{subequations}
where $\nabla$ denotes as usual the covariant derivative. Let's work in the non-coordinate system, the three vector fields $\overline{u}$, $\overline{v}$ and $\overline{z}$ are completely arbitrary, so we can take them to be exactly equal to the basis vectors, i.e. $\overline{u}= \overline{e}_a , \overline{v}= \overline{e}_b $ and  $\overline{z}=\overline{e}_c \,$. Using $\omega{^c}_{ab}$ to denote the affine connection, we know that by definition that $\nabla_a \overline{e}_b \equiv \omega{^c}_{ba} \overline{e}_c$. Therefore, using this and the anholonomy coefficients (\ref{eq:anol_coeff_viel}), we can rewrite the above equations in components\footnote{We follow the conventions of Refs. \cite{KK2,MTW} to define the components of the torsion and the curvature tensors. See App. \ref{sec:diff_tors_def} for a list of conventions for the torsion tensor used in the most common GR books.}, i.e.
\begin{subequations}
\label{eqn:scheme62}
\begin{align}
T{_{ab}}{^c} & = 2\omega{^c}_{[ba]} + \Omega{_{ab}}^c \; , \\
\Rc{^d}_{cab} & = 2\partial_{[a|}\omega{^d}{_{c|b]}} + 2\omega{^d}{_{e[a|}}\omega{^e}{_{c|b]}} +\Omega{_{ab}}{^e}\omega{^d}_{cf}\; .
\end{align}
\end{subequations}

Let's focus on (\ref{eqn:scheme62}a). We know that the torsion tensor is required to be zero in GR, hence we immediately see that
\begin{equation}
2\omega{^c}_{[ba]} = - \Omega{_{ab}}^c \; .
\label{eq:nedd_app_tor}
\end{equation}
This equation implies that, even if the theory is torsion-free, the affine connection in the \textit{non-coordinate basis is not symmetric in the exchange of the two down indices}. As a consequence, the computations of the affine connection in this frame can be carried out as if we have a non-vanishing torsion, which components are given by the anholonomy coefficients (\ref{eq:anol_coeff_viel}). We stress that this is not a physical torsion, but simply a consequence of the fact that the new basis, defined by the vielbein fields, does not introduce a new coordinate system. 

In App. \ref{ch.forms.torsion}, in particular in Sec. \ref{sec:torsion}, we show explicitly that, imposing the metricity condition $\nabla_a \hat{g}_{ab} = 0$, one eventually finds
\begin{equation}
\omega_{abc} = \Gamma_{abc}+\Sigma_{abc} \; ,
\label{eq:aff.conn.def}
\end{equation}
where $\Gamma_{abc}$ is the usual Christoffel connection, and $\Sigma_{abc}$ is the so-called contorsion tensor. These two are respectively given by 
\begin{subequations}
\label{eqn:scheme64}
\begin{align}
\Gamma_{abc} & = \frac{1}{2}\left(\partial_b \hat{g}_{ac}+\partial_c \hat{g}_{ab}-\partial_a \hat{g}_{bc}\right) \; , \\
\Sigma_{abc} & = -\frac{1}{2}\left(\Omega_{abc}-\Omega_{bca}-\Omega_{cab}\right) \; .
\end{align}
\end{subequations}
Having the explicit expression of the affine connection, we can immediately compute the curvature tensor in the new frame using (\ref{eqn:scheme62}b). Notice that the only difference with usual expression of the Riemann tensor is, again, the presence of the anoholonmy coefficients.

At this point, we have all the ingredients we need to do computations in the non-coordinate basis (\ref{eq:new_basis}). We end this section stressing that here we used a slightly different approach with respect to Ref. \cite{KK2}, where these kind of computations in the context of NRGR are done for the first time\footnote{In this reference there is also a different convention for the anholonomy coefficient. Indeed there it is used $C{_{ab}}{^c} = 2\omega{^c}_{[ba]} =  -\Omega{_{ab}}{^c}$.}. In this reference, to compute the affine connection and the curvature tensor, the authors used the so-called \textit{Cartan structure equations}. In App. \ref{sec:Cart-eq} we derive these two equations starting from eqs. (\ref{eqn:scheme62}), showing in this way that the two methods are completely equivalent.

\subsection{Rewriting the Einstein-Hilbert action}\label{sec:explic_comp}

Before making any explicit computation using NRG fields, let's rewrite the usual Einstein-Hilbert action using a non coordinate basis in a simpler way. From eq. (\ref{eqn:scheme62}b) one can easily compute the Ricci scalar; the other ingredient we need in order to write the action is the canonical volume form that is given by
\begin{align}
dV & = \sqrt{-\hat{g}}\underline{\vartheta}^{\hat{t}}\wedge\underline{\vartheta}^{\hat{1}}\wedge\underline{\vartheta}^{\hat{2}}\wedge\underline{\vartheta}^{\hat{3}} \notag \\
& = \frac{\sqrt{-\hat{g}}}{4!}\varepsilon_{abcd}\underline{\vartheta}^{a}\wedge\underline{\vartheta}^{b}\wedge\underline{\vartheta}^{c}\wedge\underline{\vartheta}^{d} = \sqrt{-\hat{g}} \vartheta d^4\!x \; ,
\label{eq:int_meas}
\end{align}
where $\hat{g}\equiv \det\hat{g}_{ab}$, $d^4\!x =  dx^0\wedge dx^1 \wedge dx^2\wedge dx^3$ and
\begin{equation}
\vartheta\equiv \det\left(\vartheta{^a}_{\mu}\right) = \varepsilon_{abcd}\varepsilon^{\mu\nu\rho\sigma}\vartheta{^a}_{\mu}\vartheta{^b}_{\nu}\vartheta{^c}_{\rho}\vartheta{^d}_{\sigma} \; .
\end{equation}
Therefore the Einstein-Hilbert action is now proportional to 
\begin{align}
\!\!\int\!\!d V\Rc \! = \!\!\int\!\!d^4\!x\!\sqrt{-\hat{g}}\vartheta \!\left\{\hat{g}^{ab}\partial_{c}\omega{^c}_{ab}\!-\!\hat{g}^{ab}\partial_{b}\hat{\omega}_{a}\!+\!\hat{\omega}_d\omega{^d}\!-\!\omega^{cab}\omega_{abc}\!+\!\omega^{cab}\Omega_{cab}\right\} \; ,
\label{eq:step.Ricci}
\end{align}
where, following what is done in Ref. \cite{KK2}, we defined the two quantities
\begin{subequations}
\label{eqn:scheme4b}
\begin{align}
\omega^{a} &\equiv \omega{^a}_{bc}\hat{g}^{bc} = \Gamma{^a}_{bc}\hat{g}^{bc}+\Omega{_c}{^{ac}} \; ,\label{eq:oma}\\
\hat{\omega}_{a}& \equiv \omega{^c}_{ac} = \Gamma{^c}_{ac}-\Omega{_{ca}}{^c} \; .\label{eq:omhata}
\end{align}
\end{subequations}

In order to simplify even more the above equation, we are now going to perform an integration by parts of the first two terms 
\begin{align}
A \equiv \int\!\!d^4\!x\sqrt{-\hat{g}}\vartheta\hat{g}^{ab}\partial_{c}\omega{^c}_{ab} \; , && B \equiv \int\!\!d^4\!x\sqrt{-\hat{g}}\vartheta\hat{g}^{ab}\partial_{b}\hat{\omega}_{a} \; .
\label{eq:def_A_B}
\end{align}
We recall that the derivative defined in (\ref{eq:def_der_frame}) contains the vielbein field $e{_a}^{\mu}$, hence, before doing any integration by parts, we need to make explicit every frame derivative. Therefore, writing $\partial_c = e{_c}^\mu\partial_\mu$, we obtain
\begin{align}
A & \simeq -\int\!\!d^4x\vartheta \omega{^c}_{ab}\hat{g}^{ab}\partial_{c}\Big(\sqrt{-\hat{g}}\Big)  -\int\!\!d^4x\sqrt{-\hat{g}}\vartheta \omega{^c}_{ab}\partial_{c}\Big(\hat{g}^{ab}\Big) \notag \\
&\qquad- \int\!\!d^4x\sqrt{-\hat{g}}\omega{^c}_{ab}\hat{g}^{ab}\partial_{\mu}\Big(\vartheta e{_c}^{\mu}\Big) \; ,
\label{eq:1.int.part}
\end{align}
where $\simeq$ means equal up to a surface term. The first term of the above equation can be simplified using the first of the two following identities
\begin{align}
\partial_{a}(\sqrt{-\hat{g}}) = \sqrt{-\hat{g}}\Gamma{^c}_{ac} \; ,& &  \partial_b(\sqrt{-\hat{g}}\hat{g}^{ab}) = -\sqrt{-\hat{g}}\Gamma{^a}_{bc}\hat{g}^{bc} \; .  \label{eq:der.Chr}
\end{align}
The third term of eq. (\ref{eq:1.int.part}) can instead be easily computed once we rewrite the determinant $\vartheta = \exp\left(\text{Tr}\log\left(\vartheta{^d}_{\nu}\right)\right)$. One eventually arrives to 
\begin{equation}
\partial_{\mu}\Big(\vartheta e{_c}^{\mu}\Big) = \vartheta\Big(\partial_{\mu}e{_c}^{\mu}-e{_c}^{\mu}\partial_\mu e{_d}^{\nu}\vartheta{^d}_{\nu}\Big) = -\vartheta\Omega{_{dc}}^d \; ,
\label{eq:JacobisFormula}
\end{equation}
where we use the fact that the inverse of $\vartheta{^d}_{\nu}$ is nothing but $e{_d}^{\nu}$. Therefore, $A$ can be rewritten as follows
\begin{align}
A = -\int\!\!d^4x\sqrt{-\hat{g}}\vartheta \omega^c\hat{\omega}_c+\int\!\!d^4x\sqrt{-\hat{g}}\vartheta\omega^{cab}\partial_{c}\hat{g}_{ab} \; ,
\label{eq:A}
\end{align}
where we also used the fact that $\omega{^c}_{ab}\partial_{c}\hat{g}^{ab}  = -\omega^{cab}\partial_c \hat{g}_{ab}$. Using the second identity of eq. (\ref{eq:der.Chr}) and eq. (\ref{eq:JacobisFormula}), one can integrate by parts also the term $B$ defined in eq. (\ref{eq:def_A_B}), obtaining eventually
\begin{equation}
B \simeq \int\!\!d^4x\sqrt{-\hat{g}}\vartheta\hat{\omega}_a\omega^a \; ,
\label{eq:B}
\end{equation}
where again $\simeq$ means equal up to a total derivative term. 

Inserting (\ref{eq:A}) and (\ref{eq:B}) into (\ref{eq:step.Ricci}), we obtain
\begin{align}
\int\!\!dV\Rc &  = \int\!\!d^4x\sqrt{-\hat{g}}\vartheta\left\{-\omega^a\hat{\omega}_a+\omega^{cab}\left(\partial_{c}\hat{g}_{ab}-\omega_{abc}+\Omega_{cab}\right)\right\} \; .
\label{eq:almost_final}
\end{align}
Recalling eqs. (4.13) and (4.15), we can see that 
\begin{equation}
\partial_{c}\hat{g}_{ab}-\omega_{abc}+\Omega_{cab} = \omega_{bca} \; .
\end{equation}
In this way we arrive to our final result\footnote{There is a typo in the first term of eq. (A.13) of Ref. \cite{KK2}. In this article one reads $\omega^{abc}\omega_{abc}$ for the first term of $\hat{\Rc}$, but doing computations using this term leads to wrong results.}
\begin{align}
S_{\text{EH}} \simeq -\frac{\mpl^2}{2}\int\!\!d^4x\sqrt{-\hat{g}}\vartheta\hat{\Rc}\; , & &  \hat{\Rc} \equiv \omega^{abc}\omega_{cab}-\omega^a\hat{\omega}_a \; ,
\label{eq:final.Ricci}
\end{align}
where we used $\simeq$ to stress that this equality is valid up to surface terms. 

Now we can see the advantages of such a procedure. In eq. (\ref{eq:final.Ricci}) we have written the Einstein-Hilbert action in term of the affine connection only; in this way, we do not have to compute any derivative of the latter, but simply the components $\omega_{abc}$ in order to get the explicit result in term of NRG fields.

\subsection{Making explicit the NRG fields}\label{sec:NRG-fields_expl}

We are now ready to write the Einstein-Hilbert action using NRG fields, hence parametrising the metric as follows
\begin{subequations}
\label{eqn:36}
\begin{align}
g_{\mu\nu} & =
\begin{pmatrix}
e^{2\phi} & -e^{2\phi}A_j \\
-e^{2\phi}A_i & -e^{-2\phi}\gamma_{ij}+e^{2\phi}A_iA_j
\end{pmatrix} \; ,\\
g^{\mu\nu} & =
\begin{pmatrix}
e^{-2\phi}-e^{2\phi}A_kA^k & -e^{2\phi}A^j \\
-e^{2\phi}A^i & -e^{2\phi}\gamma^{ij}
\end{pmatrix} \; .
\label{eq:gin.NRG}
\end{align}
\end{subequations}
Clearly $\gamma^{ik}\gamma_{kj} =\delta{^i}_j$, and we use $\gamma^{ij}$ and $\gamma_{ij}$ to raise and lower Latin indices. Trying to compute the Ricci scalar directly by its definition in this case is very complicated, so we need to follow an alternative procedure. As we shall see, using vielbein fields helps us simplifying considerably these computations.

In fact, the NRG parametrisation naturally defines a new form basis $\underline{\vartheta}^a$ and the corresponding dual vectors $\overline{e}_{a}$ given by
\begin{align}
\begin{cases}
\underline{\vartheta}^{\hat{t}} = e^\phi\left(dt-A_idx^i\right) & \\
\underline{\vartheta}^{\hat{i}} = \delta{^{\hat{i}}}_{i}dx^i &
\end{cases} \; , & & 
\begin{cases}
\overline{e}_{\hat{t}} = e^{-\phi}\partial_t & \\
\overline{e}_{\hat{i}} = \delta{_{\hat{i}}}{^i}\left(\partial_i+A_i\partial_t\right) = \delta{_{\hat{i}}}{^i}D_i & 
\end{cases} \; .
\label{eq:vielbein_NRG}
\end{align}
As usual, we use the first letters of the Latin alphabet to label the new frame, i.e. $a,b,\text{...} = (\hat{t},\hat{i})$; we also introduce a Kronecker delta needed to go from the physical indices $i$ to the frame ones $\hat{i}$. Finally, we define for simplicity $D_i \equiv \partial_i+A_i\partial_t$. The systems in eq. (\ref{eq:vielbein_NRG}) can be rewritten using vielbein fields, i.e. $\underline{\vartheta}^{a} =\vartheta{^a}_{\mu}dx^{\mu}$ and $\overline{e}_{a} =e{_a}{^\mu}\partial_{\mu}$, where, explicitly,
\begin{align}
\vartheta{^a}_{\mu} = \begin{pmatrix}
e^{\phi} & -e^{\phi}A_i \\
0 & \delta{^{\hat{i}}}_{i}
\end{pmatrix}\; , & & e{_a}{^\mu} = \begin{pmatrix} 
e^{-\phi} & 0 \\
\delta{_{\hat{i}}}{^i}A_i & \delta{_{\hat{i}}}{^i}
\end{pmatrix} \; .
\label{eq:theta}
\end{align}
As a consistency check, it is easy to see that $e{_a}{^\mu}$ and  $\vartheta{^a}{_\mu}$ satisfied eq. (\ref{eq:delta_vielbein}) as expected.
In this new frame, the metric and its inverse take the following form\footnote{Actually the easiest way to derived the inverse metric $g^{\mu\nu}$ written in (\ref{eq:gin.NRG}) is to use these vielbein fields.}
\begin{align}
\hat{g}_{ab} = 
\begin{pmatrix}
1 & 0 \\
0 & -\tilde{\gamma}_{\hat{i}\hat{j}}
\end{pmatrix} \; ,&& \hat{g}^{ab} = 
\begin{pmatrix}
1 & 0 \\
0 & -\tilde{\gamma}^{\hat{i}\hat{j}}
\end{pmatrix} \; ,
\label{eq:non.trivial.metric}
\end{align}
where we defined 
\begin{align}
\tilde{\gamma}_{\hat{i}\hat{j}}\equiv e^{-2\phi}\delta{_{\hat{i}}}{^i}\tilde{\gamma}_{ij}\delta{^j}_{\hat{j}} \; ,& & \tilde{\gamma}^{\hat{i}\hat{j}}\equiv e^{2\phi}\delta{^{\hat{i}}}{_i}\tilde{\gamma}^{ij}\delta{_j}^{\hat{j}} \; .
\end{align}
From now on we use this metric to raise/lower spatial indices of the new frame.

We immediately see that we are precisely in the situation described in Sec. \ref{susec:non-coord-basis}, with now an explicit expression of the vielbein fields. Therefore we can first compute the anholonomy coefficients (\ref{eq:anol_coeff_viel}), and then the affine connection using respectively eqs. (\ref{eq:aff.conn.def}) and (\ref{eqn:scheme64}). This is all we need to know to write explicitly the Einstein-Hilbert action (\ref{eq:final.Ricci}).

The only non-zero components of the anholonomy coefficients $\Omega{_{ab}}^{c}$ are 
\begin{align}
\Omega{_{\hat{t}\hat{i}}}{^{\hat{t}}} = - \left(D_i\phi+\dot{A}_i\right)\delta{^i}_{\hat{i}} \; , & &
\Omega{_{\hat{i}\hat{j}}}{^{\hat{t}}} = -e^{\phi}\bar{F}_{ij}\,\delta{^{i}}_{\hat{i}}\delta{^{j}}_{\hat{j}} \; ,
\label{eq:anol}
\end{align}
where we denote with a dot the derivative w.r.t. time, e.g. $\dot{\phi} \equiv \partial_t\phi$. In the last step we also defined
\begin{equation}
\bar{F}_{ij} \equiv 2D_{[i}A_{j]} = 2\partial_{[i}A_{j]}+2A_{[i}\dot{A}_{j]} \; .
\end{equation}
Computing the non-vanishing components of the contorsion tensor $\Sigma_{abc}$ through eq. (\ref{eqn:scheme64}b) is now straightforward. These are given by\footnote{Recall that, by definition, both the anholonomy coefficients and the contorsion tensor are antisymmetric in the first two indices.}
\begin{subequations}
\label{eqn:scheme41}
\begin{align}
\Sigma_{\hat{t}\hat{i}\hat{t}} & = \left(D_i\phi+\dot{A}_i\right)\delta{^i}_{\hat{i}} \; ,\\ 
\Sigma_{\hat{i}\hat{t}\hat{j}} &  =  \frac{1}{2}e^{\phi}\bar{F}_{ij}\,\delta{^{i}}_{\hat{i}}\delta{^{j}}_{\hat{j}} = \Sigma_{\hat{i}\hat{j}\hat{t}}\; .
\end{align}
\end{subequations}

Then we need to compute the usual Christoffel symbols using eq. (\ref{eqn:scheme64}a). Recalling that the frame derivatives are given by (\ref{eq:def_der_frame}), one immediately sees that the only non trivial components of $\Gamma_{abc}$ are
\begin{subequations}
\label{eqn:scheme42}
\begin{align}
\Gamma_{\hat{t}\hat{i}\hat{j}} & = \frac{1}{2}e^{-\phi}\dot{\tilde{\gamma}}_{ij}\,\delta{^i}_{\hat{i}}\delta{^j}_{\hat{j}} = - \Gamma_{\hat{i}\hat{t}\hat{j}}\; ,\\
\Gamma_{\hat{i}\hat{j}\hat{k}} & =  -\tilde{\Gamma}[\tilde{\gamma}]_{ijk}\,\delta{^i}_{\hat{i}}\delta{^j}_{\hat{j}}\delta{^k}_{\hat{k}}  \; ,
\end{align}
\end{subequations}
where we introduced
\begin{equation}
\tilde{\Gamma}[\tilde{\gamma}]_{ijk} \equiv \frac{1}{2}\left(D_{j}\tilde{\gamma}_{ik}+D_{k}\tilde{\gamma}_{ij}-D_{i}\tilde{\gamma}_{jk}\right) \; ,
\end{equation}
which is nothing but the three dimensional Christoffel symbol constructed using the metric $\tilde{\gamma}_{ij}$ and replacing every ordinary derivative with $D_i$.

Inserting eqs. (\ref{eqn:scheme41}) and  (\ref{eqn:scheme42}) in eq. (\ref{eq:aff.conn.def}) and recalling the symmetry properties of $\Gamma_{abc}$ and $\Sigma_{abc}$, one can eventually compute the non-zero components of the affine connection $\omega_{abc}$. From eq. (\ref{eq:theta}) we immediately compute $\vartheta= e^{\phi}$, so that eventually we can write eq. (\ref{eq:final.Ricci}) explicitly as
\begin{align}
\int\!\!d^4x & \sqrt{-\hat{g}} \vartheta\hat{\Rc}  = -\int\!\!d^4x\sqrt{-\hat{g}}e^{\phi}\bigg\{\frac{e^{2\phi}}{4}\bar{F}^{ij}\bar{F}_{ij} +\frac{e^{-2\phi}}{4}\left(\tilde{\gamma}^{ik}\tilde{\gamma}^{j\ell}\dot{\tilde{\gamma}}_{ij}\dot{\tilde{\gamma}}_{k\ell}-\left(\tilde{\gamma}^{ij}\dot{\tilde{\gamma}}_{ij}\right)^2\right)\notag\\
&\quad +\left[\tilde{\Gamma}^{ijk}\tilde{\Gamma}_{kij}\!-\! \tilde{\Gamma}{^\ell}_{i\ell}\tilde{\Gamma}{^i}_{jk}\tilde{\gamma}^{jk} \!+\!\left(\tilde{\Gamma}{^\ell}_{j\ell}\tilde{\gamma}^{ji}\!-\!\tilde{\Gamma}{^i}_{jk}\tilde{\gamma}^{jk}\right)\left(D_i\phi\!+\!\dot{A}_i\right)\right]\bigg\} \, .
\label{eq:final.omega.prod}
\end{align}

The last thing we need to do now is to rewrite in a better way the second line of (\ref{eq:final.omega.prod}). In fact, up to surface terms, this term is nothing but the three dimensional Ricci scalar of the metric $\tilde{\gamma}_{ij}$.

\subsubsection{Extracting the three dimensional Ricci}

First of all we define the following quantities
\begin{subequations}
\label{eqn:scheme65}
\begin{align}
C_1 \equiv \int\!\!d^4x\sqrt{-\hat{g}}e^{\phi}\left[\tilde{\Gamma}{^\ell}_{j\ell}\tilde{\gamma}^{ji}\!-\!\tilde{\Gamma}{^i}_{jk}\tilde{\gamma}^{jk}\right]\left(D_i\phi\!+\!\dot{A}_i\right) \; ,\\
C \equiv \int\!\!d^4x\sqrt{-\hat{g}}e^{\phi}\left(\tilde{\Gamma}^{ijk}\tilde{\Gamma}_{kij}\!-\! \tilde{\Gamma}{^\ell}_{i\ell}\tilde{\Gamma}{^i}_{jk}\tilde{\gamma}^{jk}\right) + C_1 \; .
\end{align}
\end{subequations}
Let's look closely to the last term in round bracket of (\ref{eqn:scheme65}a). Recalling eqs. (\ref{eq:JacobisFormula}) and (\ref{eq:anol}), one can see that $e^\phi\left(D_i\phi\!+\!\dot{A}_i\right) = \partial_{\mu}\left(e^{\phi}e{_{\hat{i}}}{^\mu}\right)\delta{^{\hat{i}}}_i$. Therefore, integrating by parts and using two identities similar to (\ref{eq:der.Chr}), namely
\begin{align}
D_{i}(\sqrt{-\hat{g}}) = \sqrt{-\hat{g}}\tilde{\Gamma}{^\ell}_{i\ell} \; ,& & D_i(\sqrt{-\hat{g}}\hat{g}^{ji}) = -\sqrt{-\hat{g}}\tilde{\Gamma}{^i}_{jk}\tilde{\gamma}^{jk} \; ,
\end{align}
one eventually arrives to
\begin{align}
C_1 \simeq \int\!\!d^4x\sqrt{-\hat{g}}e^{\phi}\left\{2\tilde{\gamma}^{ij}D_{[k|}\tilde{\Gamma}{^k}_{i|j]}+2\tilde{\Gamma}{^k}_{\ell k}\tilde{\Gamma}{^\ell}_{ij}\tilde{\gamma}^{ij}-\tilde{\Gamma}^{ijk}D_i\tilde{\gamma}_{jk}\right\} \; ,
\label{eq:final.C1}
\end{align}
where, again, $\simeq$ means equal up to a surface term. To write the above equation, we also used the relation $\tilde{\Gamma}{^i}_{jk}D_i\tilde{\gamma}^{jk}=-\tilde{\Gamma}^{ijk}D_i\tilde{\gamma}_{jk}$.
Finally, inserting eq. (\ref{eq:final.C1}) in (\ref{eqn:scheme65}b), and using the fact that $\tilde{\Gamma}_{kij}-D_i\tilde{\gamma}_{jk} = -\tilde{\Gamma}_{jik}$, we obtain
\begin{align}
C & = \int\!\!d^4x\sqrt{-\hat{g}}e^{\phi}\, \tilde{\Rc}[\tilde{\gamma}] \; ,
\end{align}
were we defined
\begin{equation}
\tilde{\Rc}[\tilde{\gamma}] \equiv 2\,\tilde{\gamma}^{ij}D_{[k|}\tilde{\Gamma}{^k}_{i|j]}+ 2\tilde{\Gamma}{^k}_{\ell [k|}\tilde{\Gamma}{^\ell }_{i|j]}\tilde{\gamma}^{ij} \; ,
\end{equation}
which is nothing but the three dimensional Ricci scalar for the metric $\tilde{\gamma}_{ij}$ computed replacing every ordinary derivative with a $D_i$.

Inserting this result into (\ref{eq:final.omega.prod}), we can eventually rewrite the Einstein-Hilbert action (\ref{eq:final.Ricci}) as
\begin{equation}
\!\!\!S_{\text{EH}} \!=\! \frac{\mpl^2}{2}\int\!\!d^4\!x\!\sqrt{\tilde{\gamma}}e^{\phi}\left\{\frac{e^{2\phi}}{4}\bar{F}^{ij}\!\bar{F}_{ij} \!+\!\frac{e^{-2\phi}}{4}\left(\tilde{\gamma}^{ik}\tilde{\gamma}^{j\ell}\dot{\tilde{\gamma}}_{ij}\dot{\tilde{\gamma}}_{k\ell}\!-\!\left(\tilde{\gamma}^{ij}\dot{\tilde{\gamma}}_{ij}\right)^2\right)\!+\!\tilde{\Rc}[\tilde{\gamma}]\right\} \; ,
\label{eq:almost_done_SEH_NRG}
\end{equation}
where we used $\sqrt{-\hat{g}} =\sqrt{\tilde{\gamma}}$. This is almost our final result. However we need to make explicit $\tilde{\gamma}_{ij}=e^{-2\phi}\gamma_{ij}$ and $\tilde{\gamma}^{ij}=e^{2\phi}\gamma^{ij}$, hence, we need to perform a final conformal transformation. 

\subsubsection{Final conformal transformation}

The trickiest quantities of eq. (\ref{eq:almost_done_SEH_NRG}) to transform  is the three dimensional Ricci scalar $\tilde{\Rc}[\tilde{\gamma}]$. With a little effort, we can show that
\begin{equation}
\tilde{\Gamma}{^k}{_{ij}}[\tilde{\gamma}] = \bar{\Gamma}{^k}{_{ij}}-\left(\delta{^k}{_j}D_i\phi\!+\!\delta{^k}{_i}D_j\phi\!-\!\gamma^{kl}\gamma_{ij}D_l\phi\right) \; ,
\end{equation}
where we defined
\begin{equation}
\bar{\Gamma}_{ijk} \equiv \frac{1}{2}\left(D_{j}\gamma_{ik}+D_{k}\gamma_{ij}-D_{i}\gamma_{jk}\right) \; ,
\label{eq:three_dim_chris_D}
\end{equation}
which is nothing but the three dimensional Christoffel symbol for $\gamma_{ij}$, where we replaced every ordinary derivative with $D_i$. One can then arrive to the following equation
\begin{equation}
\tilde{\Rc}[\tilde{\gamma}] = e^{2\phi}\left\{\bar{\Rc}_\gamma+4\gamma^{ij}D_iD_j\phi-4\gamma^{ij}\bar{\Gamma}{^k}_{ij}D_k\phi-2\gamma^{ij}D_i\phi D_j\phi\right\} \; .
\end{equation}
Again, we called $\bar{\Rc}_\gamma$ the three dimensional Ricci scalar for the metric $\gamma_{ij}$, where each ordinary derivative is replaced with a $D_i$. Observing that $\sqrt{\tilde{\gamma}}e^\phi =\sqrt{\gamma}e^{-2\phi}$, one arrives to rewrite eq. (\ref{eq:almost_done_SEH_NRG}) as
\begin{align}
S_{\text{EH}} = \frac{\mpl^2}{2}\!\int\!\!d^4\!x\sqrt{\tilde{\gamma}}& \left\{ -2\gamma^{ij}D_i\phi D_j\phi+\frac{e^{4\phi}}{4}\bar{F}^{ij}\bar{F}_{ij} + \bar{\Rc}_\gamma\right. \notag \\
& \ \left.+\frac{e^{-4\phi}}{4}\left(\tilde{\gamma}^{ik}\tilde{\gamma}^{j\ell}\dot{\tilde{\gamma}}_{ij}\dot{\tilde{\gamma}}_{k\ell}\!-\!\left(\tilde{\gamma}^{ij}\dot{\tilde{\gamma}}_{ij}\right)^2\right)\right. +\notag \\
&\ \left. +e^{-4\phi}\left(2\dot{\phi}\gamma^{ij}\dot{\gamma}_{ij}-6\dot{\phi}^2\right)\!+\!4\gamma^{ij}\left(D_iD_j\phi\!-\!\bar{\Gamma}{^k}_{ij}D_k\phi\right)\right\} \; .
\label{eq:almost.done}
\end{align}

This could be considered as our final result. However, in Ref. \cite{KK2} it is pointed out that the last term in round brackets can be rewritten in a more compact form. Defining $\bar{\Gamma}{^k}\equiv\gamma^{ij}\bar{\Gamma}{^k}_{ij}$ and performing a last integration by parts, we can indeed rewrite this term as
\begin{align}
\int\!\!d^4\!x & \sqrt{\gamma}\left(4\gamma^{ij}D_iD_j\phi-\bar{\Gamma}{^k}D_k\phi\right) \!\simeq\! -4\int\!\!d^4\!x\sqrt{\gamma}\gamma^{ij}\dot{A}_jD_i\phi \; ,
\label{eq:final.int.parts}
\end{align}
After all these steps we can finally write
\begin{align}
S_{\text{EH}} = \frac{\mpl^2}{2}\int\!\!d^4\!x\sqrt{\gamma}\bigg\{ & -2\gamma^{ij}D_i\phi D_j\phi+\frac{e^{4\phi}}{4}\bar{F}_{ij}\bar{F}^{ij}+\bar{\Rc}_{\gamma}+\notag \\
& +\frac{e^{-4\phi}}{4}\left[\dot{\gamma}_{ij}\dot{\gamma}_{k\ell}\gamma^{ik}\gamma^{j\ell}-\left(\gamma^{ij}\dot{\gamma}_{ij}\right)^2\right]-4\gamma^{ij}\dot{A}_iD_j\phi+ \notag \\
& +e^{-4\phi}\left(2\dot{\phi}\gamma^{ij}\dot{\gamma}_{ij}-6\dot{\phi}^2\right)\bigg\} \; .
\label{eq:SEH.kk}
\end{align}
This is precisely the \textit{Einstein-Hilbert action written in terms of NRG fields} that we were looking for.

\subsection{Physical interpretation of the NRG fields}

In order to better understand the physical interpretation of NRG fields, let's first look at the static limit of (\ref{eq:SEH.kk})
\begin{equation}
S^{\text{stat}}_{\text{EH}} =\frac{\mpl^2}{2}\int\!\!d^4\!x\sqrt{\gamma}\left(\Rc_{\gamma}-2\gamma^{ij}\partial_i\phi\partial_j\phi+\frac{e^{4\phi}}{4}F_{ij}F^{ij}\right) \; ,
\label{eq:KK.EH.stat}
\end{equation}
where $\Rc_{\gamma}$ is the three dimensional Ricci scalar for $\gamma_{ij}$ and $F_{ij}$ is the usual field strength $F_{ij} = 2\partial_{[i}A_{j]}$. First let's focus on the scalar sector of eq. (\ref{eq:KK.EH.stat}). When $\gamma_{ij}=\delta_{ij}+\sigma_{ij}$, this is
\begin{equation}
S^{\text{stat}}_{\text{EH}}\longrightarrow -\frac{1}{8\pi\GN}\int\!\!d td^3\!\bm{x}\left(\grad\phi^2\right)+\dots \; ,
\end{equation}
where $\grad$ is the usual flat gradient operator in three dimension. The above equation is nothing but the Newton action for pure gravity without matter. Therefore we can identify $\phi$ with the Newton potential.

Let's then look at the part of (\ref{eq:KK.EH.stat}) depending on $\bm{A}$. A part from an overall sign, this resembles the magnetic part of Maxwell's action for the electromagnetic field. One usually calls $\bm{A}$ the \textit{gravito-magnetic} field, which, as in the case of an electromagnetic system, is needed to add dynamics to the static Newtonian physics. 

Finally we immediately note that the field $\gamma_{ij}$ is described by the usual Einstein-Hilbert action for a three dimensional metric. This can be seen as the pure GR part  of (\ref{eq:KK.EH.stat}).

\subsection{The harmonic gauge fixing}

Now we shall find also the explicit expression for the gauge-fixing action in terms of NRG fields. We recall that in NRGR one usually work with the harmonic gauge\footnote{In Ref. \cite{KK2} there is also an interesting discussion on the reasons why the harmonic gauge seems to be the best for these kind of computations.}, hence with the following action
\begin{align}
S_{\text{GF}} = \frac{\mpl^2}{4}\int\!\!d^4\!x\sqrt{-g}g_{\mu\nu}\Gamma^{\mu}\Gamma^{\nu} \; ,
\label{eq:GF.for.NRG}
\end{align}
where as usual\footnote{Notice also that when working with NRG fields, one usually uses the full harmonic gauge, not the linearised version given in eqs. (\ref{eq:SGFpot2}) and (\ref{eq:SGFrad}).}
\begin{align}
\Gamma^{\mu}\equiv\Gamma{^\mu}_{\rho\sigma}g^{\rho\sigma}=-\frac{1}{\sqrt{-g}}\partial_\nu\left(\sqrt{-g}g^{\mu\nu}\right) \; .
\label{eq:def.gammamu}
\end{align}
To find the explicit expression of the gauge-fixing action in terms of NRG fields, we use again the non orthonormal frame defined by eq. (\ref{eq:theta}),  and (\ref{eq:non.trivial.metric}). it is easy to see that in this frame the action (\ref{eq:GF.for.NRG}) becomes
\begin{equation}
S_{\text{GF}} = \frac{\mpl^2}{4}\int\!\!d^4\!x\sqrt{-\hat{g}}e^{\phi}\hat{g}_{ab}\hat{\Gamma}^{a}\hat{\Gamma}^{b} \; ,
\label{eq:GF.gamma.hat.a}
\end{equation}
where we use the fact that $\sqrt{-g}=e^{\phi}\sqrt{-\hat{g}}$, and we defined $\hat{\Gamma}^a\equiv\vartheta{^a}{_\mu}\Gamma^{\mu}$. Our goal then is to find the explicit expression $\hat{\Gamma}^a$. 

Recalling the second equality of eq. (\ref{eq:der.Chr}), from (\ref{eq:def.gammamu}) it is not hard to obtain
\begin{align}
\Gamma^\mu & = -\frac{e^{-\phi}}{\sqrt{-\hat{g}}}\vartheta{^c}{_\nu}\partial_c\left(e^{\phi}\sqrt{-\hat{g}}\hat{g}^{ab}e{_a}{^\mu}e{_b}{^\nu}\right) \notag \\
& = e{_a}^{\mu}\left[\Gamma^a-\hat{g}^{ac}\partial_c\phi+\hat{g}^{ac}e{_c}^{\nu}\partial_b\vartheta{^b}_\nu\right] -\hat{g}^{ac}\partial_c e{_a}{^\mu} \; ,
\end{align}
where $\Gamma^a \equiv \Gamma{^a}_{bc}\hat{g}^{bc}$. Multiplying by $\vartheta{^a}_{\mu}$ both sides and recalling that the vielbein fields satisfy (\ref{eq:delta_vielbein}), one can eventually find the components of $\hat{\Gamma}^a$
\begin{subequations}
\label{eqn:scheme71}
\begin{align}
\hat{\Gamma}^{\hat{t}} & = e^{\phi}\left(e^{2\phi}\gamma^{ij}D_{i}A_{j}-\frac{e^{-2\phi}}{2}\left(\gamma^{ij}\dot{\gamma}_{ij}-8\dot{\phi}\right)\right) \; , \\
 \hat{\Gamma}^{\hat{i}} & = -e^{2\phi}\left(\bar{\Gamma}^k-\gamma^{kj}\dot{A}_j\right)\delta{_k}{^{\hat{i}}} \; .
\end{align}
\end{subequations}
We defined as usual $\bar{\Gamma}^k \equiv \bar{\Gamma}{^k}_{ij}\gamma^{ij}$. Inserting these in eq. (\ref{eq:GF.gamma.hat.a}), and writing explicitly $\sqrt{-\hat{g}}=\sqrt{\gamma}e^{-3\phi}$, we eventually get the harmonic gauge-fixing action in terms of NRG fields
\begin{align}
\!\!\!S_{\text{GF}} \!=\! \frac{\mpl^2}{4}\int\!\!d^4\!x\sqrt{\gamma}\left\{\left(e^{2\phi}\gamma^{ij}D_iA_j\!+\!4e^{-2\phi}\dot{\phi}\!-\frac{e^{-2\phi}}{2}\left(\gamma^{ij}\dot{\gamma}_{ij}\right)\right)^2\!\!\!-\left|\bar{\Gamma}_{i}\!-\!\dot{A}_i\right|^2\right\} \; .
\label{eq:SGF.NRG}
\end{align}
For convenience, we used the compact notation $\left|\bar{\Gamma}_{i}-\dot{A}_i\right|^2\equiv\gamma^{ij}\left(\bar{\Gamma}_{i}-\dot{A}_i\right)\left(\bar{\Gamma}_{j}-\dot{A}_j\right)$.

\section{Scaling and advantages of NRG fields in NRGR}\label{sec:adv}

Now that we have the gravity action, with its corresponding gauge-fixing term, one can start the NRGR machinery. Before doing so however, let's see the advantages that NRG fields bring in this theory. 

Splitting again $\gamma_{ij} = \delta_{ij}+\sigma_{ij}$, and dealing with $\sigma_{ij}$ rather than $\gamma_{ij}$ from now on, first, we rescale the gravity fields as follows
\begin{align}
\phi\longrightarrow\frac{\phi}{\mpl}\; , & & A_i\longrightarrow\frac{A_i}{\mpl} \; ,& & \sigma_{ij}\longrightarrow\frac{\sigma_{ij}}{\mpl} \; .
\label{eq:norm.field.NRG}
\end{align}
In this way,  they all have mass dimension one. Then, as in eq. (\ref{eq:mode.dec}), we should split each of these fields into potential plus radiation modes, i.e.
\begin{subequations}
\label{eqn:scheme42b}
\begin{align}
\phi & =\varphi+\bar{\phi}_{\,\,\,\,\,\,\,\,\,\,} = \begin{tikzpicture}[baseline]
\begin{feynman}
\vertex (f);
\vertex [above=0.1cm of f] (z);
\vertex [right=1cm of z] (c); 
\diagram* {
(z) -- [dotted] (c)
}; 
\end{feynman} 
\end{tikzpicture} + \begin{tikzpicture}[baseline]
\begin{feynman}
\vertex (f);
\vertex [above=0.1cm of f] (z);
\vertex [right=1cm of z] (c); 
\diagram* {
(z) -- [photon] (c)
}; 
\end{feynman} 
\end{tikzpicture} \; ,\\
A_i & = \mathcal{A}_i+\bar{A}_{i\,\,} = \begin{tikzpicture}[baseline]
\begin{feynman}
\vertex (f);
\vertex [above=0.1cm of f] (z);
\vertex [right=1cm of z] (c); 
\diagram* {
(z) -- [dashed] (c)
}; 
\end{feynman} 
\end{tikzpicture} +\begin{tikzpicture}[baseline]
\begin{feynman}
\vertex (f);
\vertex [above=0.1cm of f] (z);
\vertex [above=0.04cm of z] (z');
\vertex [right=1cm of z] (c); 
\vertex [above=0.04cm of c] (c');
\diagram* {
(z) -- [photon] (c),
(z') -- [photon] (c')
}; 
\end{feynman} 
\end{tikzpicture} \; ,\\
\sigma_{ij} & =\varsigma_{ij}+\bar{\sigma}_{ij} = \begin{tikzpicture}[baseline]
\begin{feynman}
\vertex (f);
\vertex [above=0.1cm of f] (z);
\vertex [right=1cm of z] (c); 
\diagram* {
(z) -- [double] (c)
}; 
\end{feynman} 
\end{tikzpicture} + \begin{tikzpicture}[baseline]
\begin{feynman}
\vertex (f);
\vertex [above=0.1cm of f] (z);
\vertex [right=1cm of z] (c); 
\diagram* {
(z) -- [gluon] (c)
}; 
\end{feynman} 
\end{tikzpicture} \; .
\end{align}
\end{subequations}
Here we introduced also the diagrammatic conventions we are going to use in doing computations with NRG fields. To make equations more compact, in the following we may use $\xi\equiv\left\{\varphi,\mathcal{A}_i,\varsigma_{ij}\right\}$ and $\bar{\Xi}\equiv\left\{\bar{\phi},\bar{A}_i,\bar{\sigma}_{ij}\right\}$. Recall that for the potential fields we should perform a partial Fourier transform, hence
\begin{align}
\xi(x^0,\bm{x})& =\intvec{k}e^{i\bm{k}\cdot\bm{x}}\xi_{\bm{k}}(x^0) \; ,
\label{eq:frvec_NRG}
\end{align}
where analogously $\xi_{\bm{k}} \equiv \left\{\varphi_{\bm{k}},\mathcal{A}_{\bm{k}\,i},\varsigma_{\bm{k}\,ij}\right\}$. In this way, as we said in  Sec. \ref{sec:mod.dec}, one removes from the theory the large fluctuations coming from spatial derivatives acting on potential gravitons. The next step is to find the propagators of these fields, from which one finds the scaling rules of the various gravity fields. The procedure is exactly equivalent to the one we picture in Sec. \ref{sec:power.count}, and leads to similar results that we collect in the table \ref{table:scaling.NRG}. Having this, we are able to associate a unique scaling in the PN parameter $v$ to all possible vertices (hence diagrams) in NRGR. 
\begin{table}[H]
\begin{center}
\begin{tabular}{c|c|c|c|c|c|c|c}
"Fields" & $t$ & $\delta \bm{x}$ & $m/\Mpl$ & $\xi_{\bm{k}}$ & $\bar{\Xi}$ & $\partial_0\xi_{\bm{k}}$ & $\partial_\mu\bar{\Xi}$\\ \hline
\ & \ & \ & \ & \ & \ & \ & \\
Scaling & $r/v$ & $r$ & $L^{1/2}v^{1/2}$ & $r^2 v^{1/2}$ & $v/r$ & $(v/r)\xi_{\bm{k}}$ & $(v/r)\bar{\Xi}$
\end{tabular}
\end{center}
\caption{Scaling rules with NRG fields}\label{table:scaling.NRG}
\end{table}

Let's now expand also the point-particle action; recalling that $d \tau_a^2 = g_{\mu\nu}dx_a^\mu dx_a^\nu$, we can immediately write
\begin{align}
S_{\text{pp}} & \!=\! -\sum_a m_a\int\!\!d \tau_a  \!=\!  -\sum_a m_a\int\!\!d te^{\phi}\left((1-\bm{A}\cdot\bm{v})^2-e^{-4\phi}\gamma_{ij}v^iv^j\right)^{\frac{1}{2}} \; ,
\label{eq:firt.pp_NRG}
\end{align}
where $a=1,2$ label again the the two BHs of the binary, considered as point-particles sources. Re-parametrising as usual $\gamma_{ij}=\delta_{ij}+\sigma_{ij}$, and normalizing the fields as in eq. (\ref{eq:norm.field.NRG}), we easily obtain
\begin{align}
S_{\text{pp}} \!=\! \sum_a m_a\int\!\!d t & \left(-1\!+\frac{1}{2}v_a^2\!+\frac{1}{8}v_a^4+\dots\right) -\sum_a m_a\!\!\int\!\!d t\left(\frac{\varphi}{\mpl}+\frac{\varphi^2}{2\mpl^2}\right. \notag \\
&\left.\qquad-\frac{1}{\mpl}\mathcal{A}_i\cdot v^i_a+\frac{3\varphi}{2\mpl}v_a^2 -\frac{1}{2\mpl}\varsigma_{ij}v_a^iv_a^j+\dots\right) \; .
\label{eq:PP.kk}
\end{align}
In the above equation we focused only on the potential sector of the point-particle action. Once we performed the partial Fourier transform (\ref{eq:frvec_NRG}), from eq. (\ref{eq:PP.kk}) we understand that the lowest order vertices for each field are
\begin{subequations}
\label{eqn:scheme46}
\begin{align}
\begin{tikzpicture}[baseline]
\begin{feynman}
\vertex (f);
\vertex [above=0.1cm of f] (z);
\vertex [left=0.2cm of z] (c');
\vertex [above=0.6cm of c'] (a);
\vertex [above=0.2cm of a] (a');
\vertex [below=0.6cm of c'] (b);
\vertex [below=0.2cm of b] (b');
\vertex [right=0.7cm of z] (c); 
\diagram* {
(z) --[draw=none] (f),
(a') -- [draw=none] (a) -- [plain] (c') -- [plain] (b) -- [draw=none] (b'),
(c') -- [dotted] (c)
}; 
\end{feynman} 
\end{tikzpicture} & \longrightarrow -\frac{m_a}{\mpl}\int\!\!d t\!\intvec{k}e^{i\bm{k}\cdot\bm{x}_a}\varphi_{\bm{k}} \sim  L^{1/2} 
\label{eq:vertex.phi} \; ,\\
\begin{tikzpicture}[baseline]
\begin{feynman}
\vertex (f);
\vertex [above=0.1cm of f] (z);
\vertex [left=0.2cm of z] (c');
\vertex [above=0.6cm of c'] (a);
\vertex [above=0.2cm of a] (a');
\vertex [below=0.6cm of c'] (b);
\vertex [below=0.2cm of b] (b');
\vertex [right=0.7cm of z] (c); 
\diagram* {
(z) --[draw=none] (f),
(a') -- [draw=none] (a) -- [plain] (c') -- [plain] (b) -- [draw=none] (b'),
(c') -- [scalar] (c)
}; 
\end{feynman} 
\end{tikzpicture} & \longrightarrow  \ \frac{m_a}{\mpl}\int\!\!d t\!\intvec{k}e^{i\bm{k}\cdot\bm{x}_a}\mathcal{A}_{\bm{k}\, i}\cdot v^i_a\sim L^{1/2}v 
\label{eq:vertex.A} \; ,\\
\begin{tikzpicture}[baseline]
\begin{feynman}
\vertex (f);
\vertex [above=0.1cm of f] (z);
\vertex [left=0.2cm of z] (c');
\vertex [above=0.6cm of c'] (a);
\vertex [above=0.2cm of a] (a');
\vertex [below=0.6cm of c'] (b);
\vertex [below=0.2cm of b] (b');
\vertex [right=0.7cm of z] (c); 
\diagram* {
(z) --[draw=none] (f),
(a') -- [draw=none] (a) -- [plain] (c') -- [plain] (b) -- [draw=none] (b'),
(c') -- [double] (c)
}; 
\end{feynman} 
\end{tikzpicture} & \longrightarrow  \ \frac{m_a}{2\mpl}\int\!\!d t\!\intvec{k}e^{i\bm{k}\cdot\bm{x}_a}\varsigma_{\bm{k}\, ij}v_a^iv_a^j\sim  L^{1/2}v^2 \; ,
\label{eq:vertex.sigma}
\end{align}
\end{subequations}
where we represent again the compact objects as straight lines. From the above relations we immediately see that:
\begin{itemize}
\item at 1PN order, a diagram involving a correction to the propagators can come only from the $\phi$ propagator. Indeed any insertion of a correction $\otimes$ implies an additional factor $v^2$ in the scaling. Corrected propagator for $\bm{A}$ or $\sigma_{ij}$ has to eventually end on a vertex like (\ref{eq:vertex.A}) or (\ref{eq:vertex.sigma}), which means at least another power of $v$. Hence, the resulting diagram would contribute to a higher PN order.
\item The cubic vertex in this case can have different forms
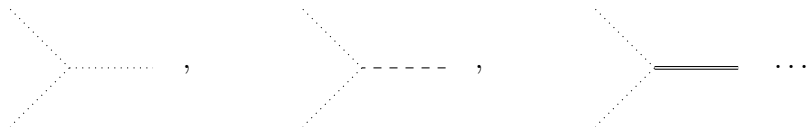
\begin{align*}
\begin{tikzpicture}[baseline]
\begin{feynman}
\vertex (z);
\vertex [above left=1.1cm of z] (a);
\vertex [below left=1.1cm of z] (b);
\vertex [right=1.1cm of z] (c);
\diagram* {
(a) --[dotted] (z) -- [dotted] (c),
(z) -- [dotted] (b)
}; 
\end{feynman} 
\end{tikzpicture} \quad , & &  \begin{tikzpicture}[baseline]
\begin{feynman}
\vertex (z);
\vertex [above left=1.1cm of z] (a);
\vertex [below left=1.1cm of z] (b);
\vertex [right=1.1cm of z] (c);
\diagram* {
(a) --[dotted] (z),
(c) -- [scalar] (z),
(z) -- [dotted] (b)
}; 
\end{feynman} 
\end{tikzpicture} \quad , & & \begin{tikzpicture}[baseline]
\begin{feynman}
\vertex (z);
\vertex [above left=1.1cm of z] (a);
\vertex [below left=1.1cm of z] (b);
\vertex [right=1.1cm of z] (c);
\diagram* {
(a) --[dotted] (z) -- [double] (c),
(z) -- [dotted] (b)
}; 
\end{feynman} 
\end{tikzpicture} \quad \dots
\end{align*}
In any case, a three vertex has schematically a structure $\xi(\partial\xi)(\partial\xi)$. The one that has the minimum scaling in $v$ can not contain any time derivative, hence it is proportional to a factor of the form
\begin{equation}
\frac{1}{\mpl}\int\!\!d^4\!x\,\xi\partial_i\xi\partial^i\xi=\frac{1}{\mpl}\int\!\!d t\!\!\intvec{k,q}\,\bm{k}\cdot\bm{q}\xi_{-\bm{k}-\bm{q}}\xi_{\bm{k}}\xi_{\bm{q}}\sim\frac{v^2}{L^{1/2}} \; .
\end{equation}
Hence we understand that the only cubic vertex that could give a contribution at 1PN order is an interaction between three $\varphi_{\bm{k}}$, because if we had even just one $\mathcal{A}_{\bm{k}\,i}$ or $\varsigma_{\bm{k}\,ij}$, then we would eventually end on a vertex like (\ref{eq:vertex.A}) or (\ref{eq:vertex.sigma}), which would add another power of $v$. Looking at  action (\ref{eq:SEH.kk}), we see that such a vertex can come only from the expansion of the term proportional to
\begin{equation}
\int\!\!d^4\!x\,e^{-4(\phi/\mpl)}\dot{\phi}^2\longrightarrow\frac{1}{\mpl}\int\!\!d^4\!x\,\phi\dot{\phi}^2 \; .
\label{eq:three_vertex_phi}
\end{equation}
This term contributes of course to an order higher than 1PN, because it contains two time derivatives. It is then clear that \textit{no cubic vertex diagram contributes to 1PN order when we use the NRG fields to parametrise the metric}. 
\end{itemize}

We saw that thanks to NRG fields we can reduce the number of diagrams entering at 1PN order computations by excluding all three-vertex-interaction topology diagrams. In fact this reduction has been proved to hold also at higher PN orders. For instance in Ref. \cite{2PN}\footnote{See Ch. I, Secs. C and D of this reference.} it has been shown that NRG fields parametrisation allows to remove 4 out of 9 possible topologies of diagrams at 2PN. Clearly, as we proceed to higher order in the PN expansion, the advantages of the NRG fields parametrisation become more and more important.

Finally, due to the similarities described at the beginning of this chapter, one can ask if the usual ADM decomposition written in eq. (\ref{eq:ADM-dec}) gives advantages similar to the ones we have just shown. A comparison between these two parametrisations in the context of NRGR has been explicitly done in Ref. \cite{ADM-NRG-comp}. Here, the authors showed that, with a suitable redefinition of the lapse $N$ and the three dimensional metric $\gamma^{\text{ADM}}_{ij}$, one can obtains basically the same simplification at order 1PN. However, already at 2PN order, the ADM decomposition requires to compute one extra diagram: at higher order, the computational cost of the ADM decomposition grows, making NRG fields the most efficient parametrisation for computations in NRGR so far.

\section{NRGR with NRG fields}

Let's now be more concrete and compute again explicitly the conservative dynamics of the binary at 1PN order, and the radiative contribution up to $2.5$PN order using these NRG fields .

Firs of all we need the explicit expression of the various propagators, hence we should expand actions (\ref{eq:SEH.kk}) and (\ref{eq:SGF.NRG}) up to quadratic order. It is not hard to see that, for the potential and the radiation sector one obtains respectively 
\begin{subequations}
\label{eqn:scheme44}
\begin{align}
S\ord{2}_{\text{pot.}} & = \int\!\!d t\!\!\intvec{k}\left(-\frac{|\bm{k}|^2}{2}\right)\left\{2\varphi_{\bm{k}}\varphi_{-\bm{k}}-\frac{1}{2}\mathcal{A}_{\bm{k}\,i}\mathcal{A}^i_{-\bm{k}}+\frac{1}{4}\varsigma_{\bm{k}\,ij}\left(\delta^{ik}\delta^{j\ell}-\frac{1}{2}\delta^{ij}\delta^{k\ell}\right)\varsigma_{-\bm{k}\,k\ell}\right\}\notag \\
& \;+\frac{1}{2}\int\!\!d t\!\!\intvec{k}\left\{2\dot{\varphi}_{\bm{k}}\dot{\varphi}_{-\bm{k}}-\frac{1}{2}\dot{\mathcal{A}}_{\bm{k}\,i}\dot{\mathcal{A}}^i_{-\bm{k}}+\frac{1}{4}\dot{\varsigma}_{\bm{k}\,ij}\left(\delta^{ik}\delta^{j\ell}-\frac{1}{2}\delta^{ij}\delta^{k\ell}\right)\dot{\varsigma}_{-\bm{k}\,k\ell}\right\} \; ,
\label{eq:SEH.kk.2}\\
S\ord{2}_{\text{rad.}} & \!=\! \frac{1}{2}\int\!\!d^4\!x\!\!\left\{2\partial_\mu\bar{\phi}\partial^\mu\bar{\phi}-\frac{1}{2}\partial_\mu \bar{A}_i\partial^\mu \bar{A}^i\!+\frac{1}{4}\partial_{\mu}\bar{\sigma}_{ij}\left(\delta^{ik}\delta^{j\ell}\!-\!\frac{1}{2}\delta^{ij}\delta^{k\ell}\right)\partial^{\mu}\bar{\sigma}_{k\ell}\right\} \, .
\end{align}
\end{subequations}
In the above actions we used the flat metric $\eta_{\mu\nu}$ to raise/lower indices. 

From eqs. (\ref{eqn:scheme44}) we can find the explicit expression for the propagators. In the conservative sector we have
\begin{subequations}
\label{eqn:scheme17}
\begin{align}
\bangle*{\Tprod{\varphi_{\bm{q}}(t')\varphi_{\bm{k}}(t)}} & = \begin{tikzpicture}[baseline]
\begin{feynman}
\vertex (f);
\vertex [above=0.1cm of f, label=75:$\bm{q}$, label=180:$t'$] (z);
\vertex [right=1.8cm of z, label=105:$\bm{k}$, label=0:$t$] (w);  
\diagram* {
(z) -- [dotted] (w)
}; 
\end{feynman} 
\end{tikzpicture} = -\frac{i}{2\modul{k}^2}(2\pi)^3\delta^{(3)}(\bm{k}\!+\!\bm{q})\delta(t\!-\!t') \; ,\\
\bangle*{\Tprod{\mathcal{A}_{\bm{q} i}(t')\mathcal{A}_{\bm{k} j}(t)}} & = \begin{tikzpicture}[baseline]
\begin{feynman}
\vertex (f);
\vertex [above=0.1cm of f, label=75:$\bm{q}$, label=285:$i$, label=180:$t'$] (z);
\vertex [right=1.8cm of z, label=105:$\bm{k}$, label=255:$j$, label=0:$t$] (w); 
\diagram* {
(z) -- [scalar] (w)
}; 
\end{feynman} 
\end{tikzpicture} = \frac{2i}{\modul{k}^2}(2\pi)^3\delta^{(3)}(\bm{k}\!+\!\bm{q})\delta(t\!-\!t')\delta_{ij} \; ,\\
\bangle*{\Tprod{\varsigma_{\bm{q} k\ell}(t')\varsigma_{\bm{k} ij}(t)}} & =\!\! \begin{tikzpicture}[baseline]
\begin{feynman}
\vertex (f);
\vertex [above=0.1cm of f, label=75:$\bm{q}$, label=285:$k\ell$, label=180:$t'$] (z);
\vertex [right=1.8cm of z, label=105:$\bm{k}$, label=255:$ij$, label=0:$t$] (w); 
\diagram* {
(z) -- [double] (w)
}; 
\end{feynman} 
\end{tikzpicture} \!\!= -\frac{4i}{\modul{k}^2}(2\pi)^3\delta^{(3)}(\bm{k}\!+\!\bm{q})\delta(t\!-\!t')P_{ijk\ell} \, ,
\end{align}
\end{subequations}
where $P_{ijk\ell} =\frac{1}{2}(\delta_{ik}\delta_{j\ell}+\delta_{i\ell}\delta_{jk}-2\delta_{ij}\delta_{k\ell})$. We also have corrections to these propagators of the form
\begin{align}
\bangle*{\Tprod{\varphi_{\bm{q}}(t')\varphi_{\bm{k}}(t)}}_\otimes & = \begin{tikzpicture}[baseline]
\begin{feynman}
\vertex (f);
\vertex [above=0.1cm of f, label=75:$\bm{q}$, label=180:$t'$] (z);
\vertex [right=0.75cm of z, crossed dot] (c) {};
\vertex [right=1.8cm of z, label=105:$\bm{k}$, label=0:$t$] (w);  
\diagram* {
(z) -- [dotted] (c) -- [dotted] (w)
}; 
\end{feynman} 
\end{tikzpicture} \!=\! -\frac{i}{2\modul{k}^4}(2\pi)^4\delta^{(3)}(\bm{k}\!+\!\bm{q})\frac{\partial^2}{\partial_t\partial_{t'}}\delta(t\!-\!t') \, ,
\end{align}
and similarly for the other two fields. On the other hand, for the radiative sector one finds
\begin{subequations}
\label{eqn:scheme45}
\begin{align}
\bangle*{\Tprod{\bar{\phi}(y)\bar{\phi}(x)}} & = \begin{tikzpicture}[baseline]
\begin{feynman}
\vertex (f);
\vertex [above=0.1cm of f, label=90:$y$] (z);
\vertex [right=1.8cm of z, label=90:$x$] (w);  
\diagram* {
(z) -- [photon] (w)
}; 
\end{feynman} 
\end{tikzpicture} = \frac{1}{2}D(x-y) \; ,\\
\bangle*{\Tprod{\bar{A}_{i}(y)\bar{A}_{j}(x)}} & = \begin{tikzpicture}[baseline]
\begin{feynman}
\vertex (f);
\vertex [above=0.1cm of f, label=90:$y$, label=270:$i$] (z);
\vertex [right=1.8cm of z, label=90:$x$, label=270:$j$] (w); 
\vertex [above=0.04cm of z] (z');
\vertex [above=0.04cm of w] (w');
\diagram* {
(z) -- [photon] (w),
(z') -- [photon] (w')
}; 
\end{feynman} 
\end{tikzpicture} = -2D(x-y)\delta_{ij} \; ,\\
\bangle*{\Tprod{\bar{\sigma}_{k\ell}(y)\bar{\sigma}_{ij}(x)}} & = \begin{tikzpicture}[baseline]
\begin{feynman}
\vertex (f);
\vertex [above=0.1cm of f, label=90:$y$, label=270:$k\ell$] (z);
\vertex [right=1.8cm of z, label=90:$x$, label=270:$ij$] (w); 
\diagram* {
(z) -- [gluon] (w)
}; 
\end{feynman} 
\end{tikzpicture} = 4D(x-y)P_{ijk\ell} \; .
\end{align}
\end{subequations} 
Here we used again $D(x-y)$ defined in  eq. (\ref{eq:DandP}).

\subsection{The conservative sector at 1PN order}

\subsubsection{Feynman rules} 

\begin{table}[t]
\begin{center}
\begin{tabular}{|c|c|c|}
\hline
Diagrammatic expression & Scaling & Explicit expression \\
\hline
\begin{tikzpicture}[baseline]
\begin{feynman}
\vertex (z);
\vertex [left=0.2cm of z, label=180:$t$, label=45:$\bm{k}$] (c');
\vertex [above=0.6cm of c'] (a);
\vertex [above=0.2cm of a] (a');
\vertex [below=0.6cm of c'] (b);
\vertex [below=0.2cm of b] (b');
\vertex [right=0.7cm of z] (c); 
\diagram* {
(a') -- [draw=none] (a) -- [plain] (c') -- [plain] (b) -- [draw=none] (b'),
(c') -- [dotted] (c)
}; 
\end{feynman} 
\end{tikzpicture} & $\sim L^{1/2}$ & \(\displaystyle
-i\frac{m_a}{\mpl}\int\!\!d t\!\!\intvec{k}e^{i\bm{k}\cdot\bm{x}_a} \) \\
\hline 
\begin{tikzpicture}[baseline]
\begin{feynman}
\vertex (z);
\vertex [empty dot, minimum size=0.4cm, left=0.2cm of z, label=180:$t$, label=45:$\bm{k}$] (c') {2};
\vertex [above=0.7cm of c'] (a);
\vertex [above=0.2cm of a] (a');
\vertex [below=0.7cm of c'] (b);
\vertex [below=0.2cm of b] (b');
\vertex [right=0.7cm of z] (c); 
\diagram* {
(a') -- [draw=none] (a) -- [plain] (c') -- [plain] (b) -- [draw=none] (b'),
(c') -- [dotted] (c)
}; 
\end{feynman} 
\end{tikzpicture} & $\sim L^{1/2}v^2$ & 
\(\displaystyle
-i\frac{3m_a}{2\mpl}\int\!\!d t\!\!\intvec{k}e^{i\bm{k}\cdot\bm{x}_a}v^2_a\) \\
\hline 
\begin{tikzpicture}[baseline]
\begin{feynman}
\vertex (z);
\vertex [left=0.2cm of z, label=180:$t$, label=75:$\bm{k}$, label=280:$\bm{q}$] (c');
\vertex [above=0.7cm of c'] (a);
\vertex [above=0.2cm of a] (a');
\vertex [below=0.7cm of c'] (b);
\vertex [below=0.2cm of b] (b');
\vertex [right=0.7cm of z] (z');
\vertex [above=0.5cm of z'] (c); 
\vertex [below=0.5cm of z'] (d); 
\diagram* {
(a') -- [draw=none] (a) -- [plain] (c') -- [plain] (b) -- [draw=none] (b'),
(c') -- [dotted] (c),
(c') -- [dotted] (d)
}; 
\end{feynman} 
\end{tikzpicture} & $\sim v^2$ & 
\(\displaystyle
-i\frac{m_a}{\mpl^2}\int\!\!d t\!\!\intvec{k,q}e^{i(\bm{k}+\bm{q})\cdot\bm{x}_a} \) \\
\hline
\begin{tikzpicture}[baseline]
\begin{feynman}
\vertex (z);
\vertex [left=0.2cm of z, label=75:$\bm{k}$, label=180:$t$, label=290:$\bm{i}$] (c');
\vertex [above=0.7cm of c'] (a);
\vertex [above=0.2cm of a] (a');
\vertex [below=0.7cm of c'] (b);
\vertex [below=0.2cm of b] (b');
\vertex [right=0.7cm of z] (c); 
\diagram* {
(a') -- [draw=none] (a) -- [plain] (c') -- [plain] (b) -- [draw=none] (b'),
(c') -- [scalar] (c)
}; 
\end{feynman} 
\end{tikzpicture} & $\sim L^{1/2}v$ & 
\(\displaystyle i\frac{m_a}{\mpl}\!\int\!\!d t\!\!\intvec{k}e^{i\bm{k}\cdot\bm{x}_a}v_a^i
\)\\
\hline 
\end{tabular}
\caption{Point-particle potential NRG fields rules}
\label{table:pot.NRG}
\end{center}
\end{table}

Let's first focus on the conservative sector of NRGR. It is not hard to see that the part of (\ref{eq:PP.kk}) which is relevant for the conservative sector at 1PN order is given by 
\begin{align}
S^{\text{pp}}_{\text{cons}} = & \sum_a m_a\int\!\!d t \left(-1+\frac{1}{2}v_a^2+\frac{1}{8}v_a^2\right)-\sum_a\frac{m_a}{\mpl}\int\!\!d t\!\!\intvec{k}e^{i\bm{k}\cdot\bm{x}_a}\left(\varphi_{\bm{k}}+\frac{3v_a^2}{2}\varphi_{\bm{k}}\right) \notag \\
&-\sum_a\frac{m_a}{2\mpl}\int\!\!d t\!\!\intvec{k,q}e^{i(\bm{k}+\bm{q})\cdot\bm{x}_a}\phi^2+\sum_a\frac{m_a}{\mpl}\int\!\!d t\!\!\intvec{k}e^{i\bm{k}\cdot\bm{x}_a}\mathcal{A}_{\bm{k}\, i} v^i_a \; .
\end{align}
From this we find the Feynman rules listed in tab. \ref{table:pot.NRG}. Therefore, if we split the conservative action as $S_{\text{cons}} = S_{\text{N}}+S_{\text{1PN}}+\Ord{Lv^4}$, it is not hard to see that diagrammatically we can write
\begin{align}
iS_{\text{N}} & = \left(\sum_a\begin{tikzpicture}[baseline]
\begin{feynman}
\vertex (z);
\vertex [below=0.1cm of z] (f);
\vertex [above=0.1cm of z] (f');
\vertex [right=1cm of f'] (c);
\diagram* {
(f') -- [plain, edge label=$x_a$] (c),
(f') -- [draw=none] (z) -- [draw=none] (f)
}; 
\end{feynman} 
\end{tikzpicture}\right)+
\begin{tikzpicture}[baseline]
\begin{feynman}
\vertex (z);
\vertex [above=0.6cm of z] (c');
\vertex [below=0.4cm of z] (c);
\vertex [right=0.8cm of c] (a);
\vertex [left=0.8cm of c] (b);
\vertex [right=0.8cm of c'] (a');
\vertex [left=0.8cm of c'] (b');
\diagram* {
(a) -- [plain] (c) -- [plain] (b),
(a') -- [plain] (c') -- [plain] (b'),
(c) -- [dotted] (c')
}; 
\end{feynman} 
\end{tikzpicture}  \; .
\end{align}
We realise that at the Newtonian level we do not have any differences w.r.t. the previous parametrisation given in eq. (\ref{eq:mode.dec}). 

\noindent The 1PN order action, on the other hand, is given by 
\begin{align}
iS_{\text{1PN}} & = \left(\sum_a\begin{tikzpicture}[baseline]
\begin{feynman}
\vertex (z);
\vertex [below=0.1cm of z] (f);
\vertex [above=0.1cm of z] (f');
\vertex [right=0.5cm of f', empty dot, minimum size=0.4cm, label=90:$x_a$] (c) {2};
\vertex [right=0.7cm of c] (d);
\diagram* {
(f') -- [plain] (c) -- [plain] (d),
(f') -- [draw=none] (z) -- [draw=none] (f)
}; 
\end{feynman} 
\end{tikzpicture}\right) +  \begin{tikzpicture}[baseline]
\begin{feynman}
\vertex [crossed dot, minimum size=0.25cm] (z) {};
\vertex [above=0.8cm of z] (c');
\vertex [below=0.8cm of z] (c);
\vertex [right=1cm of c] (a);
\vertex [left=1cm of c] (b);
\vertex [right=1cm of c'] (a');
\vertex [left=1cm of c'] (b');
\diagram* {
(a) -- [plain] (c) -- [plain] (b),
(a') -- [plain] (c') -- [plain] (b'),
(c) -- [dotted] (z),
(c') -- [dotted] (z)
}; 
\end{feynman} 
\end{tikzpicture} + 
\begin{tikzpicture}[baseline]
\begin{feynman}
\vertex (z);
\vertex [above=0.8cm of z] (c');
\vertex [below=0.8cm of z] (c);
\vertex [right=1cm of c] (a);
\vertex [left=1cm of c] (b);
\vertex [right=0.6cm of c] (a'');
\vertex [left=0.6cm of c] (b'');
\vertex [right=1cm of c'] (a');
\vertex [left=1cm of c'] (b');
\diagram* {
(a) -- [plain] (c) -- [plain] (b),
(a') -- [plain] (c') -- [plain] (b'),
(c') -- [dotted] (a''),
(c') -- [dotted] (b'')
}; 
\end{feynman} 
\end{tikzpicture} \notag \\
& \qquad\qquad\qquad\qquad\ +
\begin{tikzpicture}[baseline]
\begin{feynman}
\vertex (z);
\vertex [above=0.6cm of z, empty dot, minimum size=0.3cm] (c') {2};
\vertex [below=0.8cm of z] (c);
\vertex [right=1cm of c] (a);
\vertex [left=1cm of c] (b);
\vertex [right=1cm of c'] (a');
\vertex [left=1cm of c'] (b');
\diagram* {
(a) -- [plain] (c) -- [plain] (b),
(a') -- [plain] (c') -- [plain] (b'),
(c) -- [dotted] (z),
(c') -- [dotted] (z)
}; 
\end{feynman} 
\end{tikzpicture}  +
\begin{tikzpicture}[baseline]
\begin{feynman}
\vertex (z);
\vertex [above=0.8cm of z] (c');
\vertex [below=0.8cm of z] (c);
\vertex [right=1cm of c] (a);
\vertex [left=1cm of c] (b);
\vertex [right=1cm of c'] (a');
\vertex [left=1cm of c'] (b');
\diagram* {
(a) -- [plain] (c) -- [plain] (b),
(a') -- [plain] (c') -- [plain] (b'),
(c) -- [scalar] (c')
}; 
\end{feynman} 
\end{tikzpicture} \; .
\label{eq:scheme72}
\end{align}
As we discussed in the previous section, we have one less diagram to compute compared to eq. (\ref{eq:Scon.1PN}), due to the fact that NRG fields parametrisation ruled out the three-vertex-interaction topology at this order.

\subsubsection{Diagrams computation}

The Newtonian level is straightforward, so we do not compute it explicitly. The 1PN order diagrams sketched in eq. (\ref{eq:scheme72}) can be computed following steps similar to the ones performed in Sec. \ref{sec: 1PN contr}. One obtains, eventually, 
\begin{subequations}
\label{eqn:scheme73}
\begin{align}
\begin{tikzpicture}[baseline]
\begin{feynman}
\vertex [crossed dot, minimum size=0.25cm] (z) {};
\vertex [above=0.8cm of z] (c');
\vertex [below=0.8cm of z] (c);
\vertex [right=1cm of c] (a);
\vertex [left=1cm of c] (b) {$x_b$};
\vertex [right=1cm of c'] (a');
\vertex [left=1cm of c'] (b') {$x_a$};
\diagram* {
(a) -- [plain] (c) -- [plain] (b),
(a') -- [plain] (c') -- [plain] (b'),
(c) -- [dotted] (z),
(c') -- [dotted] (z)
}; 
\end{feynman} 
\end{tikzpicture}  & = \frac{i}{2}\int\!\!d t\,\frac{\GN m_1 m_2}{r}\left[(\bm{v}_1\cdot\bm{v}_2)-\frac{(\bm{v}_1\cdot\bm{r})(\bm{v}_2\cdot\bm{r})}{r^2}\right] \; , \\
\begin{tikzpicture}[baseline]
\begin{feynman}
\vertex (z);
\vertex [above=0.8cm of z] (c');
\vertex [below=0.8cm of z] (c);
\vertex [right=1cm of c] (a);
\vertex [left=1cm of c] (b) {$x_b$};
\vertex [right=0.6cm of c] (a'');
\vertex [left=0.6cm of c] (b'');
\vertex [right=1cm of c'] (a');
\vertex [left=1cm of c'] (b') {$x_a$};
\diagram* {
(a) -- [plain] (c) -- [plain] (b),
(a') -- [plain] (c') -- [plain] (b'),
(c') -- [dotted] (a''),
(c') -- [dotted] (b'')
}; 
\end{feynman} 
\end{tikzpicture} & = -\frac{i}{2}\int\!\!d t\,\frac{\GN^2 m_1m_2(m_1+m_2)}{r^2} \; , 
\end{align}
\begin{align}
\begin{tikzpicture}[baseline]
\begin{feynman}
\vertex (z);
\vertex [above=0.6cm of z, empty dot, minimum size=0.3cm] (c') {2};
\vertex [below=0.8cm of z] (c);
\vertex [right=1cm of c] (a);
\vertex [left=1cm of c] (b) {$x_b$};
\vertex [right=1cm of c'] (a');
\vertex [left=1.3cm of c'] (b') {$x_a$};
\diagram* {
(a) -- [plain] (c) -- [plain] (b),
(a') -- [plain] (c') -- [plain] (b'),
(c) -- [dotted] (z),
(c') -- [dotted] (z)
}; 
\end{feynman} 
\end{tikzpicture} &  = i\int\!\!d t\frac{3\GN m_1m_2}{2r}(v_1^2\!+\!v_2^2) \; ,\\
\begin{tikzpicture}[baseline]
\begin{feynman}
\vertex (z);
\vertex [above=0.8cm of z] (c');
\vertex [below=0.8cm of z] (c);
\vertex [right=1cm of c] (a);
\vertex [left=1cm of c] (b) {$x_b$};
\vertex [right=1cm of c'] (a');
\vertex [left=1cm of c'] (b') {$x_a$};
\diagram* {
(a) -- [plain] (c) -- [plain] (b),
(a') -- [plain] (c') -- [plain] (b'),
(c) -- [scalar] (c')
}; 
\end{feynman} 
\end{tikzpicture} & =-i\int\!\!d t4\frac{\GN m_1m_2}{r}\bm{v}_1\cdot\bm{v}_2 \; ,
\end{align}
\end{subequations}
where we called again $r \equiv \abs{\bm{x}_1-\bm{x}_2}$. Summing up all these diagrams, as expected, one obtains the Einstein-Infeld-Hoffmann action (\ref{eq:SEIH}). This result also check that all the manipulations we did with NRG fields are indeed correct.

\subsection{The radiative sector at 2.5PN order}\label{sec:Rad_sec_NRG}

Now let's briefly see how to deal with the radiative sector using the NRG fields parametrisation. Again we can isolate from (\ref{eq:PP.kk}) the part which is relevant for the dissipative sector up to 2.5PN order
\begin{align}
S^{\text{pp}}_{\text{diss}} = & -\sum_a\frac{m_a}{\mpl}\int\!\!d t\!\!\left(\bar{\phi}+\frac{3}{2}\bar{\phi}v_a^2\right)-\sum_a\frac{m_a}{\mpl^2}\int\!\!d t\!\!\intvec{k}e^{i\bm{k}\cdot\bm{x}_a}\varphi_{\bm{k}}\bar{\phi}  \notag \\
&+\sum_a\frac{m_a}{\mpl}\int\!\!d t\bar{A}_iv_a^i+\sum_a\frac{m_a}{2\mpl}\int\!\!d t\bar{\sigma}_{ij}v_a^iv_a^j \; .
\label{eq:Spp.diss.NRG1}
\end{align}
As explained in Sec. \ref{sec:multipole}, we also need to perform a multipole expansion of the radiative fields in order to have vertices with a definite power of $v$, hence
\begin{equation}
\bar{\Xi}(x^0,\bm{x}) =\bar{\Xi}(x^0,0)+x^i\partial_i\bar{\Xi}(x^0,0)+\frac{1}{2} x^i x^j\partial_i\partial_j\bar{\Xi}(x^0,0)+\dots \; ,
\end{equation}
for each $\bar{\Xi} =\left\{\bar{\phi},\bar{A}_i,\bar{\sigma}_{ij}\right\}$. We put the origin of our frame in the center of mass of the binary, and performed the expansion around this point. Eq. (\ref{eq:Spp.diss.NRG1}) then becomes
\begin{align}
S^{\text{pp}}_{\text{diss}} = & -\sum_a\frac{m_a}{\mpl}\int\!\!d t\!\!\left[\bar{\phi}\!+\!x_a^i\partial_i\bar{\phi}\!+\!\frac{1}{2}\left(3v_a^2+x_a^ix_a^j\partial_i\partial_j\right)\bar{\phi}\right]-\sum_a\frac{m_a}{\mpl^2}\int\!\!d t\!\!\intvec{k}e^{i\bm{k}\cdot\bm{x}_a}\varphi_{\bm{k}}\bar{\phi}\, \notag \\
&+\sum_a\frac{m_a}{\mpl}\int\!\!d t\left[\bar{A}_i\!+\!x^j_a\partial_j\bar{A}_i\right]v_a^i+\sum_a\frac{m_a}{2\mpl}\int\!\!d t\bar{\sigma}_{ij}v_a^iv_a^j
\label{eq:Spp.diss.NRG2}
\end{align}
where every radiative field is evaluated in $(x^0,0)$. From this action we find the Feynman rules listed in the next page, tab. \ref{table:ppradmultiNRG}.
\begin{table}[t]
\begin{center}
\begin{tabular}{|c|c|c|}
\hline
Diagrammatic expression & Scaling & Explicit expression \\
\hline
\begin{tikzpicture}[baseline]
\begin{feynman}
\vertex (z);
\vertex [left=0.2cm of z, label=180:$t$] (c');
\vertex [above=0.5cm of c'] (a);
\vertex [above=0.2cm of a] (a');
\vertex [below=0.5cm of c'] (b);
\vertex [below=0.2cm of b] (b');
\vertex [right=0.7cm of z] (c); 
\diagram* {
(a') -- [draw=none] (a) -- [plain] (c') -- [plain] (b) -- [draw=none] (b'),
(c) -- [photon] (c')
}; 
\end{feynman} 
\end{tikzpicture} & $\sim L^{1/2}v^{1/2}$ & \(\displaystyle
-i\frac{m_a}{\mpl}\int\!\!d t\,\bar{\phi} \) \\
\hline
\begin{tikzpicture}[baseline]
\begin{feynman}
\vertex (z);
\vertex [empty dot, minimum size=0.4cm, left=0.2cm of z, label=180:$t$] (c') {1};
\vertex [above=0.5cm of c'] (a);
\vertex [above=0.2cm of a] (a');
\vertex [below=0.5cm of c'] (b);
\vertex [below=0.2cm of b] (b');
\vertex [right=0.7cm of z] (c);
\vertex [right=0.5cm of z] (d); 
\diagram* {
(a') -- [draw=none] (a) -- [plain] (c') -- [plain] (b) -- [draw=none] (b'),
(c) -- [draw=none](d) -- [photon] (c')
}; 
\end{feynman} 
\end{tikzpicture} & $\sim L^{1/2}v^{3/2}$ & \(\displaystyle
-i\frac{m_a}{\mpl}\int\!\!d t\,x_a^i\partial_i\bar{\phi} \) \\
\hline 
\begin{tikzpicture}[baseline]
\begin{feynman}
\vertex (z);
\vertex [empty dot, minimum size=0.4cm, left=0.2cm of z, label=180:$t$] (c') {2};
\vertex [above=0.5cm of c'] (a);
\vertex [above=0.2cm of a] (a');
\vertex [below=0.5cm of c'] (b);
\vertex [below=0.2cm of b] (b');
\vertex [right=0.7cm of z] (c);
\vertex [right=0.5cm of z] (d); 
\diagram* {
(a') -- [draw=none] (a) -- [plain] (c') -- [plain] (b) -- [draw=none] (b'),
(c) -- [draw=none](d) -- [photon] (c')
}; 
\end{feynman} 
\end{tikzpicture} & $\sim L^{1/2}v^{5/2}$ & \(\displaystyle
-i\frac{m_a}{2\mpl}\int\!\!d t\,\left(3v^2_a+x_a^ix_a^j\partial_i\partial_j\right)\bar{\phi} \) \\
\hline 
\begin{tikzpicture}[baseline]
\begin{feynman}
\vertex (z);
\vertex (c');
\vertex [above=0.04cm of c', label=180:$t$] (d');
\vertex [above=0.6cm of c'] (a);
\vertex [above=0.2cm of a] (a');
\vertex [below=0.6cm of c'] (b);
\vertex [below=0.2cm of b] (b');
\vertex [right=0.9cm of z] (c);
\vertex [above=0.04cm of c] (d); 
\diagram* {
(a') -- [draw=none] (a) -- [plain] (c') -- [plain] (b) -- [draw=none] (b'),
(c) -- [photon] (c'),
(d) -- [photon] (d')
}; 
\end{feynman} 
\end{tikzpicture} & $\sim L^{1/2}v^{3/2}$ & 
\(\displaystyle
i\frac{m_a}{\mpl}\int\!\!d t\,\bar{A}_iv_a^i \) \\
\hline 
\begin{tikzpicture}[baseline]
\begin{feynman}
\vertex (z);
\vertex [empty dot, minimum size=0.4cm, left=0.2cm of z, label=180:$t$] (c') {1};
\vertex [right=0.2cm of c'] (d'');
\vertex [above=0.04cm of d''] (d');  
\vertex [above=0.6cm of c'] (a);
\vertex [above=0.2cm of a] (a');
\vertex [below=0.6cm of c'] (b);
\vertex [below=0.2cm of b] (b');
\vertex [right=0.5cm of z] (c);
\vertex [right=0.7cm of z] (f);
\vertex [above=0.04cm of c] (d);  
\diagram* {
(a') -- [draw=none] (a) -- [plain] (c') -- [plain] (b) -- [draw=none] (b'),
(f)--[draw=none] (c) -- [photon] (c'),
(d) -- [photon] (d')
}; 
\end{feynman} 
\end{tikzpicture} & $\sim L^{1/2}v^{5/2}$ & 
\(\displaystyle
i\frac{m_a}{\mpl}\int\!\!d t\,\left(x_a^j\partial_j\bar{A}_i\right)v_a^i \) \\
\hline 
\begin{tikzpicture}[baseline]
\begin{feynman}
\vertex (z);
\vertex [left=0.2cm of z, label=180:$t$] (c');
\vertex [above=0.6cm of c'] (a);
\vertex [above=0.2cm of a] (a');
\vertex [below=0.6cm of c'] (b);
\vertex [below=0.2cm of b] (b');
\vertex [right=0.7cm of z] (c); 
\diagram* {
(a') -- [draw=none] (a) -- [plain] (c') -- [plain] (b) -- [draw=none] (b'),
(c) -- [gluon] (c')
}; 
\end{feynman} 
\end{tikzpicture} & $\sim L^{1/2}v^{5/2}$ & 
\(\displaystyle
i\frac{m_a}{2\mpl}\int\!\!d t\bar{\sigma}_{ij}v_a^iv_a^j \) \\
\hline 
\begin{tikzpicture}[baseline]
\begin{feynman}
\vertex (z);
\vertex [left=0.2cm of z, label=180:$t$, label=75:$\bm{k}$] (c');
\vertex [above=0.7cm of c'] (a);
\vertex [above=0.2cm of a] (a');
\vertex [below=0.7cm of c'] (b);
\vertex [below=0.2cm of b] (b');
\vertex [right=0.7cm of z] (z');
\vertex [above=0.5cm of z'] (c); 
\vertex [below=0.5cm of z'] (d); 
\diagram* {
(a') -- [draw=none] (a) -- [plain] (c') -- [plain] (b) -- [draw=none] (b'),
(c') -- [dotted] (c),
(c') -- [photon] (d)
}; 
\end{feynman} 
\end{tikzpicture} & $\sim v^{5/3}$ & 
\(\displaystyle
-i\frac{m_a}{\mpl^2}\int\!\!d t\!\!\intvec{k}e^{i\bm{k}\cdot\bm{x}_a}\bar{\phi} \) \\
\hline
\end{tabular}
\caption{Point-particles  radiative NRG fields rules with multipole expansion}\label{table:ppradmultiNRG}
\end{center}
\end{table}
\noindent
\subsubsection{The problem of the three-vertex interaction}\label{subsubsec:threevertex}

Similarly to what we said in Sec. \ref{sec:adv}, also in this case we can have many different types of cubic interactions, that could contribute to the dissipative sector of NRGR. For instance, we could have
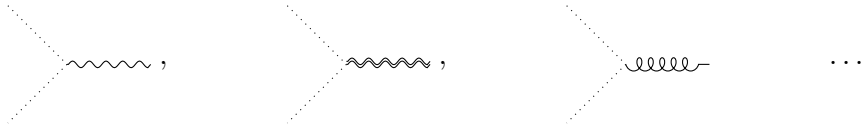
\begin{align*}
\begin{tikzpicture}[baseline]
\begin{feynman}
\vertex (z);
\vertex [above left=1.1cm of z] (a);
\vertex [below left=1.1cm of z] (b);
\vertex [right=1.1cm of z] (c);
\diagram* {
(a) --[dotted] (z) -- [photon] (c),
(z) -- [dotted] (b)
}; 
\end{feynman} 
\end{tikzpicture} \; , & &  \begin{tikzpicture}[baseline]
\begin{feynman}
\vertex (z);
\vertex [above=0.04cm of z] (z');
\vertex [above left=1.1cm of z] (a);
\vertex [below left=1.1cm of z] (b);
\vertex [right=1.1cm of z] (c);
\vertex [above=0.04cm of c] (c');
\diagram* {
(a) --[dotted] (z) -- [photon] (c),
(z') --[photon] (c'),
(z) -- [dotted] (b)
}; 
\end{feynman} 
\end{tikzpicture} \; , & & \begin{tikzpicture}[baseline]
\begin{feynman}
\vertex (z);
\vertex [above left=1.1cm of z] (a);
\vertex [below left=1.1cm of z] (b);
\vertex [right=1.1cm of z] (c);
\diagram* {
(a) --[dotted] (z) -- [gluon] (c),
(z) -- [dotted] (b)
}; 
\end{feynman} 
\end{tikzpicture} & & \dots
\end{align*}
Let's look for the ones that have the lowest scaling in $v$. Since these kind of vertices come from (\ref{eq:SEH.kk}), we expect them to be proportional to the structure $\bar{\Xi}(\partial\xi)(\partial\xi)$, where, since we are looking for the lowest order one in $v$, we decided not to have any derivative acting on the radiation field\footnote{See the scaling in tab. \ref{table:scaling.NRG}, p.~\pageref{table:scaling.NRG}.}. For the same reason, we expect $\bar{\Xi}$ to be the first term of the multipole expansion, hence it is the field evaluated in $(x^0,0)$. Finally, we expect that the lowest order vertex will not contain any time derivative, which implies that the interactions must have the following structure
\begin{equation}
\frac{1}{\mpl}\int\!\!d^4\!x\,\bar{\Xi}\partial_i\xi\partial^i\xi=\frac{1}{\mpl}\int\!\!d t\,\bar{\Xi}\!\!\intvec{k}\,|\bm{k}|^2\xi_{-\bm{k}}\xi_{\bm{k}}\sim\frac{v^{5/2}}{L^{1/2}} \; .
\label{eq:threevertexNRGrad}
\end{equation}
We then understand that the two potential fields have to be two scalars $\varphi_{\bm{k}}$, because, as we can see in eqs. (\ref{eqn:scheme46}), any other potential field would eventually end on a vertex with another power of $v$, so that the diagram would contribute to a higher PN order. 
\begin{table}[t]
\begin{center}
\begin{tabular}{|c|c|c|}
\hline
Diagrammatic expression & Scaling & Explicit expression \\
\hline
\begin{tikzpicture}[baseline]
\begin{feynman}
\vertex (z);
\vertex [right=1cm of z] (c);
\vertex [above left=1cm of z, label=135:$x^0_1$] (a);
\vertex [below left=1cm of z, label=225:$x^0_2$] (b);
\diagram* {
(a) -- [dotted, edge label=$\bm{k}$] (z),
(c) -- [gluon] (z),
(b) -- [dotted, edge label'=$\bm{q}$] (z)
}; 
\end{feynman} 
\end{tikzpicture} & \(\displaystyle \sim \frac{v^{5/2}}{L^{1/2}}\) & 
\(\displaystyle
-\frac{i(2\pi)^3}{2\mpl}\delta(x^0_1\!-\!x_2^0)\frac{\delta^{(3)}\!(\bm{k}\!+\!\bm{q})}{|\bm{k}|^2|\bm{q}|^2}\left[\bar{\sigma}^{ij}k_ik_j\!-\!\frac{|\bm{k}|^2}{2}\bar{\sigma}\right]
\)\\
\hline
\end{tabular}
\caption{Cubic vertex with NRG fields}
\label{table:gravv.NRG}
\end{center}
\end{table}
Looking again at the action (\ref{eq:SEH.kk}), we immediately understand that the only cubic vertex that can contribute at 2.5PN order come from
\begin{align}
\frac{\mpl^2}{2}\int\!\!d^4\!x\sqrt{\gamma}\left\{-2\gamma^{ij}D_i\phi D_j\phi\right\} & &\longrightarrow & & \frac{1}{\mpl}\int\!\!d t\intvec{k}\left(\bar{\sigma}^{ij}-\delta^{ij}\frac{\bar{\sigma}}{2}\right)k_ik_j\varphi_{\bm{k}}\varphi_{-\bm{k}} \; .
\end{align}
From this one can find the Feynman rule written in tab. \ref{table:gravv.NRG}. Therefore, up to 2.5PN order, we can compute
\begin{align}
iS_{\text{diss}} & = \sum_a\begin{tikzpicture}[baseline]
\begin{feynman}
\vertex (z);
\vertex [below=0.1cm of z] (f);
\vertex [above=0.1cm of z] (f');
\vertex [right=0.5cm of f'] (c);
\vertex [right=0.7cm of c] (d);
\vertex [above=0.5cm of d] (d');
\diagram* {
(f') -- [plain] (c) -- [plain] (d),
(f') -- [draw=none] (z) -- [draw=none] (f),
(d') -- [photon] (c)
}; 
\end{feynman} 
\end{tikzpicture} +
\sum_a\begin{tikzpicture}[baseline]
\begin{feynman}
\vertex (z);
\vertex [below=0.1cm of z] (f);
\vertex [above=0.1cm of z] (f');
\vertex [right=0.5cm of f', empty dot, minimum size=0.4cm] (c) {1};
\vertex [right=0.7cm of c] (d);
\vertex [above=0.5cm of d] (d');
\diagram* {
(f') -- [plain] (c) -- [plain] (d),
(f') -- [draw=none] (z) -- [draw=none] (f),
(d') -- [photon] (c)
}; 
\end{feynman} 
\end{tikzpicture} +
\sum_a\begin{tikzpicture}[baseline]
\begin{feynman}
\vertex (z);
\vertex [below=0.1cm of z] (f);
\vertex [above=0.1cm of z] (f');
\vertex [right=0.5cm of f'] (c);
\vertex [above left=0.04cm of c] (c'');
\vertex [right=0.7cm of c] (d);
\vertex [above=0.5cm of d] (d');
\vertex [above left=0.04cm of d'] (d'');
\vertex [left=0.03cm of c] (h);
\diagram* {
(f') -- [plain] (c) -- [plain] (d),
(f') -- [draw=none] (z) -- [draw=none] (f),
(d') -- [photon] (c),
(d'')-- [photon] (c''),
(h) -- [plain] (c'')
}; 
\end{feynman} 
\end{tikzpicture} \notag \\
& +\sum_a\begin{tikzpicture}[baseline]
\begin{feynman}
\vertex (z);
\vertex [below=0.1cm of z] (f);
\vertex [above=0.1cm of z] (f');
\vertex [right=0.5cm of f', empty dot, minimum size=0.4cm] (c) {2};
\vertex [right=0.7cm of c] (d);
\vertex [above=0.5cm of d] (d');
\diagram* {
(f') -- [plain] (c) -- [plain] (d),
(f') -- [draw=none] (z) -- [draw=none] (f),
(d') -- [photon] (c)
}; 
\end{feynman} 
\end{tikzpicture} +
\sum_a\begin{tikzpicture}[baseline]
\begin{feynman}
\vertex (z);
\vertex [below=0.1cm of z] (f);
\vertex [above=0.1cm of z] (f');
\vertex [right=0.5cm of f', empty dot, minimum size=0.4cm] (c) {1};
\vertex [right=0.7cm of c] (d);
\vertex [above=0.5cm of d] (d');
\vertex [above left=0.045cm of d'] (g');
\vertex [above right=0.2cm of c] (h);
\vertex [above left=0.045cm of h] (c'');
\diagram* {
(f') -- [plain] (c) -- [plain] (d),
(f') -- [draw=none] (z) -- [draw=none] (f),
(d') -- [photon] (h),
(g') -- [photon] (c'')
}; 
\end{feynman} 
\end{tikzpicture} +\sum_a\begin{tikzpicture}[baseline]
\begin{feynman}
\vertex (z);
\vertex [below=0.1cm of z] (f);
\vertex [above=0.1cm of z] (f');
\vertex [right=0.5cm of f'] (c);
\vertex [right=0.7cm of c] (d);
\vertex [above=0.5cm of d] (d');
\diagram* {
(f') -- [plain] (c) -- [plain] (d),
(f') -- [draw=none] (z) -- [draw=none] (f),
(d') -- [gluon] (c)
}; 
\end{feynman} 
\end{tikzpicture} \notag \\
& + \begin{tikzpicture}[baseline]
\begin{feynman}
\vertex (z);
\vertex [above=0.7cm of z] (c');
\vertex [below=0.6cm of z] (c);
\vertex [right=1cm of c] (a);
\vertex [left=1cm of c] (b);
\vertex [right=1cm of c'] (a');
\vertex [below=0.52cm of a'] (d);
\vertex [left=1cm of c'] (b');
\diagram* {
(a) -- [plain] (c) -- [plain] (b),
(a') -- [plain] (c') -- [plain] (b'),
(c) -- [dotted] (z) -- [dotted] (c'),
(d) -- [photon] (c')
}; 
\end{feynman} 
\end{tikzpicture}+
\begin{tikzpicture}[baseline]
\begin{feynman}
\vertex (z);
\vertex [right=1cm of z] (d);
\vertex [above=0.7cm of z] (c');
\vertex [below=0.6cm of z] (c);
\vertex [right=1cm of c] (a);
\vertex [left=1cm of c] (b);
\vertex [right=1cm of c'] (a');
\vertex [left=1cm of c'] (b');
\diagram* {
(a) -- [plain] (c) -- [plain] (b),
(a') -- [plain] (c') -- [plain] (b'),
(c) -- [dotted] (z) -- [dotted] (c'),
(d) -- [gluon] (z)
}; 
\end{feynman} 
\end{tikzpicture} +\Ord{L^{1/2}v^{7/2}} \; .
\label{eq:Sdiss.NRG1}
\end{align}
From here we see that the NRG fields parametrisation does not bring any advantages nor disadvantages for the computation in the radiative sector of NRGR. Anyhow the heavily simplifications that happen in the conservative sector make this parametrisation preferable to the one used in the first part of this work.

\subsubsection{Diagrams computation}

Let's now compute for completeness the diagrams in (\ref{eq:Sdiss.NRG1}). The first two lines  follows immediately from the Feynman rules, hence we get
\begin{subequations}
\label{eqn:scheme48}
\begin{align}
\sum_a\begin{tikzpicture}[baseline]
\begin{feynman}
\vertex (z);
\vertex [below=0.1cm of z] (f);
\vertex [above=0.1cm of z] (f');
\vertex [right=0.5cm of f'] (c);
\vertex [right=0.7cm of c] (d);
\vertex [above=0.5cm of d] (d');
\diagram* {
(f') -- [plain] (c) -- [plain] (d),
(f') -- [draw=none] (z) -- [draw=none] (f),
(d') -- [photon] (c)
}; 
\end{feynman} 
\end{tikzpicture} & = -i\sum_{a}\frac{m_a}{\mpl}\int\!\!d t\bar{\phi}  \; ,\\
\sum_a\begin{tikzpicture}[baseline]
\begin{feynman}
\vertex (z);
\vertex [below=0.1cm of z] (f);
\vertex [above=0.1cm of z] (f');
\vertex [right=0.5cm of f', empty dot, minimum size=0.4cm] (c) {1};
\vertex [right=0.7cm of c] (d);
\vertex [above=0.5cm of d] (d');
\diagram* {
(f') -- [plain] (c) -- [plain] (d),
(f') -- [draw=none] (z) -- [draw=none] (f),
(d') -- [photon] (c)
}; 
\end{feynman} 
\end{tikzpicture} & = 0 \; ,\qquad\qquad
\sum_a\begin{tikzpicture}[baseline]
\begin{feynman}
\vertex (z);
\vertex [below=0.1cm of z] (f);
\vertex [above=0.1cm of z] (f');
\vertex [right=0.5cm of f'] (c);
\vertex [above left=0.04cm of c] (c'');
\vertex [right=0.7cm of c] (d);
\vertex [above=0.5cm of d] (d');
\vertex [above left=0.04cm of d'] (d'');
\vertex [left=0.03cm of c] (h);
\diagram* {
(f') -- [plain] (c) -- [plain] (d),
(f') -- [draw=none] (z) -- [draw=none] (f),
(d') -- [photon] (c),
(d'')-- [photon] (c''),
(h) -- [plain] (c'')
}; 
\end{feynman} 
\end{tikzpicture} = 0 \; ,\label{eq:diag=0} \\
\sum_a\begin{tikzpicture}[baseline]
\begin{feynman}
\vertex (z);
\vertex [below=0.1cm of z] (f);
\vertex [above=0.1cm of z] (f');
\vertex [right=0.5cm of f', empty dot, minimum size=0.4cm] (c) {2};
\vertex [right=0.7cm of c] (d);
\vertex [above=0.5cm of d] (d');
\diagram* {
(f') -- [plain] (c) -- [plain] (d),
(f') -- [draw=none] (z) -- [draw=none] (f),
(d') -- [photon] (c)
}; 
\end{feynman} 
\end{tikzpicture} & = -i\sum_a\frac{m_a}{2\mpl}\int\!\!d t\left(3v_a^2\bar{\phi}+x_a^ix_a^j\partial_i\partial_j\bar{\phi}\right)  \; , \\
\sum_a\begin{tikzpicture}[baseline]
\begin{feynman}
\vertex (z);
\vertex [below=0.1cm of z] (f);
\vertex [above=0.1cm of z] (f');
\vertex [right=0.5cm of f', empty dot, minimum size=0.4cm] (c) {1};
\vertex [right=0.7cm of c] (d);
\vertex [above=0.5cm of d] (d');
\vertex [above left=0.045cm of d'] (g');
\vertex [above right=0.2cm of c] (h);
\vertex [above left=0.045cm of h] (c'');
\diagram* {
(f') -- [plain] (c) -- [plain] (d),
(f') -- [draw=none] (z) -- [draw=none] (f),
(d') -- [photon] (h),
(g') -- [photon] (c'')
}; 
\end{feynman} 
\end{tikzpicture} & = i\sum_{a}\frac{m_a}{\mpl}\int\!\!d t\,x_a^j\partial_j \bar{A}_i v_a^i  \; , \\
\sum_a\begin{tikzpicture}[baseline]
\begin{feynman}
\vertex (z);
\vertex [below=0.1cm of z] (f);
\vertex [above=0.1cm of z] (f');
\vertex [right=0.5cm of f'] (c);
\vertex [right=0.7cm of c] (d);
\vertex [above=0.5cm of d] (d');
\diagram* {
(f') -- [plain] (c) -- [plain] (d),
(f') -- [draw=none] (z) -- [draw=none] (f),
(d') -- [gluon] (c)
}; 
\end{feynman} 
\end{tikzpicture} & = i\sum_{a}\frac{m_a}{2\mpl}\int\!\!d t\,\bar{\sigma}_{ij} v_a^iv_a^j \; ,
\end{align}
\end{subequations}
where diagrams in (\ref{eq:diag=0}) are equal to zero because respectively proportional to the total momentum in the center of mass, and the position of the center of mass, which is where we put the center of our frame. 

Then we have to compute the last two diagrams, which, following the rules given in tabs. \ref{table:pot.NRG}, \ref{table:ppradmultiNRG} and \ref{table:gravv.NRG}, are given by
\begin{subequations}
\label{eqn:scheme50}
\begin{align}
\begin{tikzpicture}[baseline]
\begin{feynman}
\vertex (z);
\vertex [above=0.7cm of z] (c');
\vertex [below=0.6cm of z] (c);
\vertex [right=1cm of c] (a);
\vertex [left=1cm of c] (b) {$x_b$};
\vertex [right=1cm of c'] (a');
\vertex [below=0.52cm of a'] (d);
\vertex [left=1cm of c'] (b') {$x_a$};
\diagram* {
(a) -- [plain] (c) -- [plain] (b),
(a') -- [plain] (c') -- [plain] (b'),
(c) -- [dotted] (z) -- [dotted] (c'),
(d) -- [photon] (c')
}; 
\end{feynman} 
\end{tikzpicture} & = \frac{i}{\mpl}\int\!\!d t\frac{\GN m_1 m_2}{r}(2\bar{\phi}) \; , \\
\begin{tikzpicture}[baseline]
\begin{feynman}
\vertex (z);
\vertex [right=1cm of z] (d);
\vertex [above=0.7cm of z] (c');
\vertex [below=0.6cm of z] (c);
\vertex [right=1cm of c] (a);
\vertex [left=1cm of c] (b) {$x_b$};
\vertex [right=1cm of c'] (a');
\vertex [left=1cm of c'] (b') {$x_a$};
\diagram* {
(a) -- [plain] (c) -- [plain] (b),
(a') -- [plain] (c') -- [plain] (b'),
(c) -- [dotted] (z) -- [dotted] (c'),
(d) -- [gluon] (z)
}; 
\end{feynman} 
\end{tikzpicture} & = -\frac{i}{2\mpl}\int\!\!d t\frac{\GN m_1 m_2}{|\bm{x}_{12}|^3}x^i_{12}x^j_{12}\bar{\sigma}_{ij} \notag \\
& =\frac{i}{2\mpl}\int\!\!d t\left[\sum_a m_a x^i_{a}\ddot{x}^j_{a}\right]\bar{\sigma}_{ij} \; ,
\end{align}
\end{subequations}
where in the last step we defined $\bm{x}_{ab} \equiv \abs{\bm{x}_a-\bm{x}_b}$ and we used again the EOM (\ref{eq:EOM_1PN}). Inserting now eqs. (\ref{eqn:scheme48}) and (\ref{eqn:scheme50}) into (\ref{eq:Sdiss.NRG1}) we eventually obtain
\begin{align}
iS_{\text{diss}} & = -i\frac{m}{\mpl}\int\!\!d t\bar{\phi}\!-\frac{i}{\mpl}\int\!\!d t\left\{\left(\sum_a\frac{3}{2}m_av_a^2-2\frac{\GN m_1m_2}{r}\right)\bar{\phi}\!+\!\sum_a\frac{m_a}{2}x_a^ix_a^j\partial_i\partial_j\bar{\phi}\right.\! \notag \\ 
&\;\left.-\!\sum_a m_ax_a^j\partial_j\bar{A}_iv_a^i -\!\sum_a\frac{m_a}{2}v_a^iv_a^j\bar{\sigma}_{ij}-\!\sum_a\frac{m_a}{2}x_a^i\ddot{x}_a^j\bar{\sigma}_{ij}\right\} \!+\!\Ord{L^{1/2}v^{7/2}} \, ,
\label{eq:Sdiss.NRG.first}
\end{align}
where we defined $m\equiv\sum_a m_a$. Even if this does not really resemble the result we found in Sec. \ref{sec:SNR}, we are now going to show that eq. (\ref{eq:Sdiss.NRG.first}) is actually equivalent to eq. (\ref{eq:SNRfin}).

First of all, expanding at linear level the metric in eqs. (\ref{eqn:36}), we see that the relation between NRG fields and the original parametrisation in the radiation sector is given by
\begin{align}
\bar{g}_{\mu\nu} = \eta_{\mu\nu}+\frac{1}{\mpl}\begin{pmatrix}
2\bar{\phi} & -\bar{A}_j \\
-\bar{A}_i & -\bar{\sigma}_{ij}+2\delta_{ij}\bar{\phi}
\end{pmatrix}\equiv \eta_{\mu\nu}+\frac{1}{\mpl}\begin{pmatrix}
\bar{h}_{00} & \bar{h}_{0j} \\
\bar{h}_{i0}  & \bar{h}_{ij} 
\end{pmatrix} \; .
\end{align}
Recalling the definition of $\tilde{\Rc}_{\mu\nu\rho\sigma}$ given in (\ref{eq:R.Mpl.fact}), it is easy to verify that
\begin{align}
\tilde{\Rc}_{0i0j} = -\frac{1}{2}\left(\partial_0\partial_j\bar{A}_i+\partial_0\partial_i\bar{A}_j-\partial^2_0\bar{\sigma}_{ij}+2\partial_i\partial_j\bar{\phi}+2\delta_{ij}\partial_0^2\bar{\phi}\right) \; .
\end{align}
Now let's look separately at each term in the curly brackets of (\ref{eq:Sdiss.NRG.first}):
\begin{itemize}
\item Integrating by parts and using again the EOM given in (\ref{eq:EOM_1PN}) we can see
\begin{align}
\sum_a m_av_a^2\bar{\phi} & \simeq \frac{\GN m_1m_2}{r}+\sum_a\frac{m_a}{2}x_a^ix^j_{a}\delta_{ij}\partial_0^2\bar{\phi} \; ,
\end{align}
where $\simeq$ means equal up to a surface term. So we can write 
\begin{align}
\left(\sum_a\frac{3}{2}m_av_a^2-2\frac{\GN m_1m_2}{r}\right)\bar{\phi} &\simeq E_{\text{N}}\bar{\phi}+\sum_a\frac{m_a}{2}x_a^ix^j_{a}\delta_{ij}\partial_0^2\bar{\phi} \; ,
\label{eq:item1}
\end{align}
where we used again $E_{\text{N}}\equiv(1/2)\sum_am_av_a^2-(\GN m_1m_2)/r$.
\item Integrating by parts the first term in the second line one gets
\begin{align}
\sum_a m_ax_a^iv_a^j\partial_i\bar{A}_j & \simeq \frac{1}{2}\varepsilon^{ijk}L_k\partial_j\bar{A}_i-\frac{1}{2}\sum_a m_a x_a^ix_a^j\partial_0\partial_i\bar{A}_j \; ,
\label{eq:item2}
\end{align}
where we recall that $L_k = \sum_a m_a\varepsilon_{kij}x^i v^j$ is the $k$-th component of the orbital angular momentum of the binary.
\item finally we can easily see that up to a surface term
\begin{align}
\sum_a\frac{m_a}{2}v_a^iv_a^j\bar{\sigma}_{ij}+\sum_a\frac{m_a}{2}x_a^i\ddot{x}_a^j\bar{\sigma}_{ij}\simeq -\frac{1}{2}\sum_a m_ax_a^ix_a^j\partial_0^2\bar{\sigma}_{ij} \; .
\label{eq:item3}
\end{align}
\end{itemize}
Inserting (\ref{eq:item1}), (\ref{eq:item2}) and (\ref{eq:item3}) in (\ref{eq:Sdiss.NRG.first}) we finally obtain the expression (\ref{eq:SNRfin}) for $S_\text{diss}$. From here, one can proceed as we did in the previous part and compute physical quantity like the power loss and the gravitational wave form.

\chapter{NRGR in scalar-tensor theory}\label{ch:conf.coupl}

In the last part of this work we extend NRGR to alternative theories of gravity. In particular, we consider GR conformally coupled to a single massless scalar DOF $\rchi$. This extension has recently been done by A. Kuntz, F. Piazza and F. Vernizzi \cite{NRGR+scalar}. In their paper, the authors considered the theory in the so-called \textit{Einstein frame}, while here we re-derive some of their results using the equivalent description of the theory in the \textit{Jordan frame}. In this way we can not only re-check the results of \cite{NRGR+scalar}, but this is also a good starting point to extend such an approach to higher-order scalar-tensor theories, where finding an Einstein frame is not always possible, see Refs. \cite{EFT_HOST,BeyHord,Galileon}.

We briefly recall\footnote{For a review and the main relevant references on the subject see Refs. \cite{FareseDamour,EFreview}.} that a generic conformally coupled scalar-tensor theory of gravity can be described by the following action
\begin{align}
^{(\text{J})}S \!=\! \frac{\mpl^2}{2}\int\!\!d^4\!x\sqrt{-g}\left\{-F(\rchi)\Rc+Z(\rchi)g^{\mu\nu}\partial_\mu\rchi\partial_\nu\rchi-U(\rchi)\right\} +S_{\text{m}}[\psi_{\text{m}}, g_{\mu\nu}] \; ,
\label{eq:Jordan}
\end{align}
where $U(\rchi)$ is a generic potential for the scalar field $\rchi$, and $S_{\text{m}}$ is the action for the matter fields, here denoted with $\psi_{\text{m}}$. In the above action the scalar DOF is considered as dimensionless. The quantity $^{(\text{J})}S$ denotes what is usually called \textit{Jordan frame action}, i.e. the action in the frame in which matter is minimally coupled to the gravitational metric $g_{\mu\nu}$, which is then called \textit{Jordan metric}. An equivalent description of the theory can be given in the so-called \textit{Einstein frame}, which is the frame in which the kinetic term for the metric is given by the standard Einstein-Hilbert action. The action in the Einstein frame is then
\begin{align}
^{(\text{E})}S \!=\! \frac{\mpl^2}{2}\int\!\!d^4\!x\sqrt{-g_*}\left\{-\Rc_*+g_*^{\mu\nu}\partial_\mu\rchi_*\partial_\nu\rchi_*-V(\rchi_*)\right\} \!+\!S_{\text{m}}[\psi_{\text{m}}, A^2(\rchi_*)g^*_{\mu\nu}] \; ,
\label{eq:EH.einstein}
\end{align}
where we use a star to denote fields in the Einstein frame. Also in eq. (\ref{eq:EH.einstein}), the scalar field $\rchi_*$ is dimensionless. Quantities in the two frame are linked by a conformal transformation, i.e. 
\begin{subequations}
\label{eqn:scheme75}
\begin{align}
g^*_{\mu\nu} = \Omega^2(\rchi)g_{\mu\nu}\; , & &  g_{\mu\nu}= A^2(\rchi_*)g^*_{\mu\nu}\; ,
\label{eq:conf.trans}
\end{align}
where
\begin{equation}
\Omega^2=A^{-2}=F(\rchi) \; .
\end{equation}
\end{subequations}
Performing this transformation on eq. (\ref{eq:EH.einstein}) and imposing the final result to be equal to (\ref{eq:Jordan}), eventually lead to 
\begin{subequations}
\label{eqn:scheme51}
\begin{align}
\left(\frac{d\rchi_*}{d\rchi}\right)^2 & = \frac{3}{2F^2(\rchi)}\left(\frac{dF(\rchi)}{d\rchi}\right)^2+\frac{Z(\rchi)}{F(\rchi)} \; ,
\label{eq:chiE.chiJ} \\
V(\rchi_*) & = \frac{U(\rchi)}{2F^2(\rchi)} \; .
\end{align}
\end{subequations}

In the following we shall consider a scalar-tensor theory in the Jordan frame where $U(\rchi)=0$, and as matter field we consider two point particles, hence we neglect again finite size effects\footnote{See again Ch. \ref{ch:pp}.}. Because of the advantages discussed in the previous chapter, we will use NRG fields to parametrise the Jordan metric, hence $g_{\mu\nu}$ and its inverse are given by eqs. (\ref{eqn:36}).

\section{Gravity action with NRG fields in the Jordan frame}\label{sec:grav_NRG_Jordan}

Let's start with the gravity part of NRGR. In principle one can find the action in the Einstein frame by adding the kinetic term for the scalar field to (\ref{eq:SEH.kk}), and then simply impose the conformal transformation (\ref{eq:conf.trans}). We instead proceed in a slightly different way, eventually using the conformally transformation to check if our result is correct.

Let's consider the derivation of  the action (\ref{eq:SEH.kk}) that we made explicitly in Sec. \ref{sec:EH.NRG}. We can proceed in the same way, and use a non-coordinate frame to find the explicit expression of $\Rc$ in term of NRG fields. However, in order to arrive to our final result, we performed some integrations by parts, which now produce additional terms proportional to the derivatives of $F(\rchi)$. Hence, in order to find the action in the Jordan frame is sufficient to add these extra terms, and then the kinetic term for $\rchi$. 

First of all, we recall that the non-coordinate frame is defined by the vielbein fields $\vartheta{^a}_{\mu}$ and $e{_a}^{\mu}$ introduced in eq. (\ref{eq:theta}). The metric in the new frame is written explicitly in (\ref{eq:non.trivial.metric}). Reviewing the computations of Secs. \ref{sec:explic_comp} and \ref{sec:NRG-fields_expl}, it is not hard to see that, in order to arrive to our first result (\ref{eq:almost_done_SEH_NRG}), we performed three integrations by parts in eqs. (\ref{eq:A}), (\ref{eq:B}) and (\ref{eq:final.C1}). Hence we need to add to (\ref{eq:almost_done_SEH_NRG}) three terms, which are
\begin{subequations}
\label{eqn:scheme53}
\begin{align}
A_F & = -\int\!\!d^4\!x\sqrt{\tilde{\gamma}}\,e^{\phi}\omega^c e{_c}{^\mu}\partial_\mu F(\rchi) \; ,\\
B_F & = \int\!\!d^4\!x\sqrt{\tilde{\gamma}}\,e^{\phi}\hat{\omega}_ag^{ab} e{_b}{^\mu}\partial_\mu F(\rchi) \; ,\\
C_F & = \int\!\!d^4\!x\sqrt{\tilde{\gamma}}\left(\bar{\Gamma}{^\ell}_{j\ell}\tilde{\gamma}^{ji}-\bar{\Gamma}{^i}_{jk}\tilde{\gamma}^{jk}\right)\delta{^{\hat{i}}}_ie{_{\hat{i}}}{^\mu}\partial_\mu F(\rchi) \; ,
\end{align}
\end{subequations}
where we used again the quantities $\omega^c$ and $\hat{\omega}_a$ defined in eqs. (\ref{eqn:scheme4b}). Recalling that we defined $D_i\equiv \partial_i+A_i\partial_t$, one eventually finds
\begin{align}
A_F\!+\!B_F\!+\!C_F \!=\! \int\!\!d^4\!x\sqrt{\tilde{\gamma}}\,e^{\phi}\left\{e^{-2\phi}\tilde{\gamma}^{ij}\dot{\tilde{\gamma}}_{ij}\partial_t F(\rchi)-2\tilde{\gamma}^{ij}\left(D_i\phi+\dot{A}_i\right)D_j F(\rchi)\right\} \; .
\end{align}
Adding this to (\ref{eq:almost_done_SEH_NRG}), we find our first result for the gravity action in the Jordan frame
\begin{align}
^{(\text{J})}S_{\text{EH}} = \frac{\mpl^2}{2}& \int \!\!  d^4\!x\sqrt{\tilde{\gamma}}e^{\phi}F(\rchi)  \bigg\{\frac{e^{2\phi}}{4}\bar{F}^{ij}\bar{F}_{ij} \!+\!\frac{e^{-2\phi}}{4}\left(\tilde{\gamma}^{ik}\tilde{\gamma}^{j\ell}\dot{\tilde{\gamma}}_{ij}\dot{\tilde{\gamma}}_{k\ell}\!-\!\left(\tilde{\gamma}^{ij}\dot{\tilde{\gamma}}_{ij}\right)^2\right)\notag \\
&\;  + \bar{\Rc}[\tilde{\gamma}]-e^{-2\phi}\tilde{\gamma}^{ij}\dot{\tilde{\gamma}}_{ij}\partial_t \log\!F(\rchi) +2\tilde{\gamma}^{ij}\left(D_i\phi+\dot{A}_i\right)D_j\log\!F(\rchi)\bigg\} \; .
\label{eq:first_res_jordan}
\end{align}

Finally we have to impose again $\tilde{\gamma}_{ij} = e^{-2\phi}\gamma_{ij}$. This conformal transformation on the last two new terms of (\ref{eq:first_res_jordan}) is straightforward. Then, we have to redo the same steps that lead us from (\ref{eq:almost_done_SEH_NRG}) to (\ref{eq:SEH.kk}). Also in this case, we performed an integration by parts in eq. (\ref{eq:final.int.parts}), hence we need to add another extra term given by
\begin{align}
D_F = -4\int\!\!d^4\!x\sqrt{\gamma}\gamma^{ij}D_i\phi D_j F(\rchi) \; .
\end{align}
Summing everything up and adding the needed kinetic term for the scalar field, we eventually get the \textit{gravity part of the NRGR action in the Jordan frame written in terms of NRG fields}
\begin{align}
^{(\text{J})}S_{\text{g}} & = \frac{\mpl^2}{2}\int\!\!d^4\!x\sqrt{\gamma}F(\rchi)\bigg\{  -2\gamma^{ij}D_i\phi D_j\phi+\frac{e^{4\phi}}{4}\bar{F}_{ij}\bar{F}^{ij}+\bar{\Rc}_{\gamma}\notag \\
&\qquad\qquad\qquad +\frac{e^{-4\phi}}{4}\left[\dot{\gamma}_{ij}\dot{\gamma}_{k\ell}\gamma^{ik}\gamma^{j\ell}-\left(\gamma^{ij}\dot{\gamma}_{ij}\right)^2\right]-4\gamma^{ij}\dot{A}_iD_j\phi \notag \\
&\qquad\qquad\qquad +e^{-4\phi}\left(2\dot{\phi}\gamma^{ij}\dot{\gamma}_{ij}-6\dot{\phi}^2\right)  - e^{-4\phi}\left(\gamma^{ij}\dot{\gamma}_{ij}-6\dot{\phi}\right)\partial_t\log\!F(\rchi) \notag \\
&\qquad\qquad\qquad -2\gamma^{ij}\left(D_i\phi-\dot{A}_i\right)D_j\log\!F(\rchi)\bigg\} \notag \\
&+\frac{\mpl^2}{2}\int\!\!d^4\!x\sqrt{\gamma}Z(\rchi)\left\{e^{-4\phi}\dot{\rchi}^2-\gamma^{ij}D_i\rchi D_j\rchi\right\} \; .
\label{eq:Sg.NRGR.Jordan}
\end{align}
This is the starting point action for the computations made in the next sections. Since the pure NRG fields part of the action is remained untouched, we expect to have the same advantages discussed in Sec. \ref{sec:adv}. However, from the above equation, we immediately see that the scalar field $\rchi$ and the NRG fields are mixed already at the quadratic order of the action, hence we need to be very careful in defining the various propagators of our EFT theory.

\subsection{Checking the final action}

As we said at the beginning of this section, we now use the conformal transformation to go back to the Einstein frame, and check if the action we have just written is correct. Indeed in the Einstein frame we expect the above action to be simply\footnote{This is only the gravity part of the action. Clearly in the Einstein frame the coupling with matter fields changes, as can be seen from eq. (\ref{eq:EH.einstein}). Moreover we also recall that the link between the Einstein and the Jordan fields are written in eqs. (\ref{eqn:scheme75}a) and (\ref{eqn:scheme51}a). }
\begin{align}
^{(\text{E})}S_{\text{g}} = \frac{\mpl^2}{2}\int\!\!d^4\!x\sqrt{-g_*}\left\{-\Rc_*+g_*^{\mu\nu}\partial_\mu\rchi_*\partial_\nu\rchi_*\right\} \; .
\label{eq:Gravity_action_Eisntein}
\end{align}
According to eqs. (\ref{eqn:scheme75}), we have an explicit relation between the Einstein and the Jordan frame metric. In term of NRG fields we obtain
\begin{align}
\!\!\!\!\begin{pmatrix}
e^{2\phi^*} & -e^{2\phi^*}A^*_j \\
-e^{2\phi^*}A^*_i & -e^{-2\phi^*}\gamma^*_{ij}\!+\!e^{2\phi^*}A^*_iA^*_j
\end{pmatrix} \!=\! F(\rchi)
\begin{pmatrix}
e^{2\phi} & -e^{2\phi}A_j \\
-e^{2\phi}A_i & -e^{-2\phi}\gamma_{ij}\!+\!e^{2\phi}A_iA_j
\end{pmatrix} \; .
\end{align}
From here, we easily find an explicit relation between Einstein and Jordan frame quantities, i.e.
\begin{subequations}
\label{eqn:scheme54}
\begin{align}
\phi^* & = \phi +\frac{1}{2}\log\!F(\rchi)  \; , \\
A^*_i & = A_i \; , & & A_*^i = \frac{1}{F^2(\rchi)}A^i \; ,\\
\gamma^*_{ij} &= F^2(\rchi)\gamma_{ij} \; , & & \gamma_*^{ij} = \frac{1}{F^2(\rchi)}\gamma^{ij} \; .
\end{align}
\end{subequations}
To these identities one has to add also eq. (\ref{eq:chiE.chiJ}), which relates $\rchi^*$ and $\rchi$. Let's then apply this set of transformations to the action (\ref{eq:Sg.NRGR.Jordan}). For most  of the terms the transformation is straightforward, the only term which requires more attention is the one proportional to $\bar{\Rc}$. 

First of all, we need to find a relation between the three dimensional connection $\bar{\Gamma}{^k}{_{ij}}$ defined in eq. (\ref{eq:three_dim_chris_D}), and its counterpart $\bar{\Gamma}{^k_*}{_{ij}}$ in the Einstein frame. It is not hard to see that
\begin{align}
\bar{\Gamma}{^k}{_{ij}} = {\bar{\Gamma}^k_*}{_{ij}}-\left(\delta{^k}{_j}D^*_i\log\!F(\rchi)+\delta{^k}{_i}D^*_j\log\!F(\rchi)-\gamma_*^{k\ell}\gamma^*_{ij}D^*_\ell\log\!F(\rchi)\right) \; ,
\end{align}
where $D^*_i = \partial_i+A^*_i\partial_t$, and ${\bar{\Gamma}^k_*}{_{ij}}$ are given by eq. (\ref{eq:three_dim_chris_D}), replacing all Jordan frame fields with star quantities. From here one eventually finds
\begin{align}
\bar{\Rc} = F^2(\rchi) & \left\{\bar{\Rc}_*+4\gamma_*^{ij}D^*_iD^*_j\log\!F(\rchi)-4\gamma_*^{ij}{\bar{\Gamma}^k_*}{_{ij}}D^*_k\log\!F(\rchi)\right.\notag\\
& \left.\quad-2\gamma_*^{ij}D^*_i\log\!F(\rchi) D^*_j\log\!F(\rchi)\right\} \; ,
\end{align}
where, again, $\bar{\Rc}_*$ is the Ricci scalar for the three dimensional metric $\gamma^*_{ij}$ where every ordinary derivative is replaced with a $D^*_i$. Therefore, from eq. (\ref{eq:Sg.NRGR.Jordan}) we can write
\begin{align}
^{(\text{J})}S_{\text{g}} & = \frac{\mpl^2}{2}\int\!\!d^4\!x\frac{\sqrt{\gamma_*}}{\cancel{F^3(\rchi)}}\cancel{F^3(\rchi)}\bigg\{  -2\gamma_*^{ij}D^*_i\phi^* D^*_j\phi^*+\frac{e^{4\phi^*}}{4}\bar{F}^*_{ij}\bar{F}_*^{ij}+\bar{\Rc}^*_{\gamma}\notag \\
&\qquad\qquad\qquad +\frac{e^{-4\phi^*}}{4}\left[\dot{\gamma}^*_{ij}\dot{\gamma}^*_{k\ell}\gamma_*^{ik}\gamma_*^{j\ell}-\left(\gamma_*^{ij}\dot{\gamma}^*_{ij}\right)^2\right]-4\gamma_*^{ij}\dot{A}^*_iD^*_j\phi^* \notag \\
&\qquad\qquad\qquad +e^{-4\phi^*}\left(2\dot{\phi}^*\gamma_*^{ij}\dot{\gamma}^*_{ij}-6\dot{\phi}_*^2\right) \notag \\
&\qquad\qquad\qquad +4\gamma_*^{ij}\left(D^*_iD^*_j\log\!F(\rchi)-{\bar{\Gamma}^k_*}{_{ij}}D^*_k\log\!F(\rchi)\right) \notag \\
&\qquad\qquad\qquad +4\gamma_*^{ij}\dot{A}_i^*D^*_j\log\!F(\rchi)\bigg\} \notag \\
&+ \frac{\mpl^2}{2}\int\!\!d^4\!x\sqrt{\gamma_*}\frac{3}{2}\left\{e^{-4\phi^*}\left(\partial_t\log\!F(\rchi)\right)^2-\gamma_*^{ij}D^*_i\log\!F(\rchi) D^*_j\log\!F(\rchi)\right\} \notag \\
&+\frac{\mpl^2}{2}\int\!\!d^4\!x\sqrt{\gamma^*}\frac{Z(\rchi)}{F(\rchi)}\left\{e^{-4\phi^*}\dot{\rchi}^2-\gamma_*^{ij}D^*_i\rchi D^*_j\rchi\right\} \; .
\label{eq:step.conf}
\end{align}
Now, integrating by parts as in eq. (\ref{eq:final.int.parts}), we can rewrite the terms in round brackets in the fourth line as 
\begin{align}
\!\!\!\!\!\!\int\!\!d^4\!x\sqrt{\gamma_*} & \left(D^*_iD_*^i\log\!F(\rchi)\!-\!\bar{\Gamma}^k_*D^*_k\log\!F(\rchi)\right)\! \simeq\! -\int\!\!d^4\!x\sqrt{\gamma_*}\gamma_*^{ij}\dot{A}_i^*D^*_j\log\!F(\rchi) \; ,
\end{align}
where $\bar{\Gamma}^k_*\equiv {\bar{\Gamma}^k_*}{_{ij}}\gamma_*^{ij}$ and as usual $\simeq$ means equal up to a surface term. This means that the fourth and the fifth line of eq. (\ref{eq:step.conf}) cancel, leaving us with only the first three lines of the same equation which are nothing but the Einstein-Hilbert action written in term of Einstein-frame NRG fields (see eq. (\ref{eq:SEH.kk})). Therefore, explicitly computing the derivatives of $\log\!F(\rchi)$, we can rewrite eq. (\ref{eq:step.conf}) as
\begin{align}
^{(\text{J})}S_{\text{g}} & = ^{(\text{E})}\!\!S_{\text{EH}} + \frac{\mpl^2}{2}\int\!\!d^4\!x\sqrt{-g_*}\left(\frac{3}{2 F^2(\rchi)}\left(\frac{d F(\rchi)}{d\rchi}\right)^2+\frac{Z(\rchi)}{F(\rchi)}\right)g_*^{\mu\nu}\partial_\mu\rchi\partial_\nu\rchi \; ,
\end{align}
where, for the scalar field part, we also used the following relation
\begin{align}
\sqrt{\gamma^*}\left\{e^{-4\phi^*}\dot{\rchi}^2-\gamma_*^{ij}D^*_i\rchi D^*_j\rchi\right\} = \sqrt{-g_*}g_*^{\mu\nu}\partial_\mu\rchi\partial_\nu\rchi \; .
\end{align}
Recalling eq. (\ref{eq:chiE.chiJ}), we can eventually write
\begin{align}
^{(\text{J})}S_{\text{g}} & = {}^{(\text{E})}S_{\text{EH}} + \frac{\mpl^2}{2}\int\!\!d^4\!x\sqrt{-g_*}\left(\frac{d\rchi_*}{d\rchi}\right)^2g_*^{\mu\nu}\partial_\mu\rchi\partial_\nu\rchi = {^{(\text{E})}S_{\text{g}}} \; .
\end{align}
After the conformal transformation we correctly obtain the gravity part of the action written in the Einstein frame; this is a strong consistency check for the Jordan frame action we found in eq. (\ref{eq:Sg.NRGR.Jordan}).

\subsection{The harmonic gauge fixing}

We now have to find the correct expression for the gauge-fixing action in the Jordan frame. Looking at the action (\ref{eq:Sg.NRGR.Jordan}), we immediately realise that the gravitational and the scalar fields are mixed also at the kinetic level. However, only the pure spin 2 field need a gauge-fixing term in order to have a well defined propagator. Hence, we would need to diagonalise eq. (\ref{eq:Sg.NRGR.Jordan}) to find the real spin 2 field, and then write the gauge-fixing action for the diagonalised field. In practice, we already know the diagonalised version of eq. (\ref{eq:Sg.NRGR.Jordan}): this is nothing but the Einstein-frame action (\ref{eq:Gravity_action_Eisntein}). Therefore, it is sufficient to write the desired gauge-fixing action in the Einstein frame, and then impose the conformal transformation given by eqs. (\ref{eqn:scheme75}) and (\ref{eqn:scheme51}).
 
Let's consider the usual harmonic gauge-fixing in the Einstein frame given in eq. (\ref{eq:GF.for.NRG})
\begin{equation}
^{(\text{E})}S_{\text{GF}} = \frac{\mpl^2}{4}\int\!\!d^4\!x\sqrt{-g^*}g^*_{\mu\nu}\Gamma_*^{\mu}\Gamma_*^{\nu} \; .
\label{eq:GF_Einstein}
\end{equation}
First of all, to rewrite this action in the Jordan frame, we can use that
\begin{align}
\Gamma_*^{\mu} \equiv g_*^{\rho\sigma}{\Gamma_*}{^\mu}_{\rho\sigma} = \frac{1}{F(\rchi)}\Gamma^\mu -\frac{F'(\rchi)}{F^2(\rchi)}g^{\mu\nu}\partial_\nu\rchi \; ,
\label{eq:gamma_harm_einstein}
\end{align}
where $F'(\rchi)\equiv dF(\rchi)/d\rchi$, and $\Gamma^\mu$ is defined analogously to $\Gamma_*^\mu$ as $\Gamma^\mu \equiv g^{\rho\sigma}{\Gamma}{^\mu}_{\rho\sigma}$.
Inserting this into (\ref{eq:GF_Einstein}), we get the general form of the ``harmonic'' gauge-fixing term in the conformally coupled scalar-tensor theory
\begin{align}
^{(\text{J})}S_{\text{GF}} & = \frac{\mpl^2}{4}\int\!\!d^4\!x\sqrt{-g}F(\rchi)g_{\mu\nu}\Gamma^{\mu}\Gamma^{\nu}-\frac{\mpl^2}{2}\int\!\!d^4\!x\sqrt{-g}F'(\rchi)\Gamma^\mu\partial_\mu\rchi\notag \\
&\qquad+\frac{\mpl^2}{4}\int\!\!d^4\!x\sqrt{-g}\frac{\big(F'(\rchi)\big)^2}{F(\rchi)}g^{\mu\nu}\partial_\mu\rchi\partial_\nu\rchi \; .
\label{eq:GF_Jordan_gen}
\end{align}
We note from eq. (\ref{eq:gamma_harm_einstein}) that setting $\Gamma_*^\mu=0$ implies that $\Gamma^\mu \neq 0$. Thus, a harmonic gauge-fixing term in the Einstein frame does not lead to harmonic coordinates, i.e. coordinates  that satisfy $\partial_\mu(\sqrt{-g} g^{\mu \nu} \partial_\nu ) x^\rho=0$, in the Jordan frame. We leave for the future the study of how other gauge choices affect the calculation of the expanded action in the Jordan frame\footnote{The action (\ref{eq:GF_Jordan_gen}) coincides with the one written in Ref. \cite{1loopJordan}, where this gauge-fixing term has been found using another approach in the context of a multi-scalar-tensor theory.}. 

Now we have to write this action in terms of NRG fields. This is easy for the first term of (\ref{eq:GF_Jordan_gen}), because it is completely equivalent to eq. (\ref{eq:SGF.NRG}). For the second term instead we need $\Gamma^\mu$. This can be found starting from $\hat{\Gamma}^a$, given in eqs. (\ref{eqn:scheme71}), and multiplying by the vielbein $e{_a}^{\mu}$. Eventually, one arrives to the final result
\begin{align}
\!\!\!\!^{(\text{J})}S_{\text{GF}} & \!=\! \frac{\mpl^2}{4}\int\!\!d^4\!x\sqrt{\gamma}F(\rchi)\left\{\left(e^{2\phi}\gamma^{ij}D_iA_j\!+\!4e^{-2\phi}\dot{\phi}\!-\frac{e^{-2\phi}}{2}\left(\gamma^{ij}\dot{\gamma}_{ij}\right)\right)^2\!\!\!-\left|\bar{\Gamma}_{i}\!-\!\dot{A}_i\right|^2\right\} \notag \\
& +\frac{\mpl^2}{2}\int\!\!d^4\!x\sqrt{\gamma}F'(\rchi)\left\{\frac{e^{-4\phi}}{2}\left(\gamma^{ij}\dot{\gamma}_{ij}-8\dot{\phi}\right)\dot{\rchi}-\gamma^{ij}D_iA_j\dot{\rchi}+\bar{\Gamma}^iD_i\rchi-\gamma^{ij}\dot{A}_iD_j\rchi\right\} \notag \\
& + \frac{\mpl^2}{4}\int\!\!d^4\!x\sqrt{\gamma}\frac{\big(F'(\rchi)\big)^2}{F(\rchi)}\left\{e^{-4\phi}\dot{\rchi}^2-\gamma^{ij}D_i\rchi D_j\rchi\right\}\; .
\label{eq:GF_harm_Jordan}
\end{align}

Now we have almost everything we need to do computations in the Jordan frame. What we need to discuss now is how the matter action changes in the Jordan frame of this scalar-tensor theory.

\section{Matter action in the Jordan frame}

We need to introduce the source of GWs. As explained in the previous part, we consider a binary of non-spinning BHs treating them as point particles. However, once we integrate out the short modes as  in Ch. \ref{ch:pp}, we realise that, due to the presence of the scalar field, we can now add to the bottom-up constructed action a new set of couplings with $\rchi$ that still respect the symmetries of the theory. These operators take into account the internal gravity of each body.

In practice, we can parametrise this by considering the total mass of each body to depend on the value of the scalar field at its location\footnote{See Ref. \cite{Old-article-Bernard} and references therein.}, i.e. the point-particle action takes the following form
\begin{equation}
S_\text{pp} = -\sum_a \int\!\!d \tau_a\, m_a\!\left(\rchi\right)
\label{eq:firt.pp.J_ST}
\end{equation}
We introduced in this way non-trivial interactions between the matter field $x^\mu_a$ and the scalar $\rchi$. Even if this seems to contradict what we said about the Jordan frame at the beginning of Ch. \ref{ch:conf.coupl}, one has to keep in mind that the Jordan metric minimally couples only to elementary matter fields. When extended self-gravitating object are considered, as in our case, we have to keep track of their gravitational self-energy. That's the reason why also $\rchi$ interacts non trivially with the two point-like BHs.

Let's now be more concrete. We consider variation of the scalar field w.r.t. a constant background $\rchi_0$, i.e.
\begin{equation}
\rchi(x) = \rchi_0 + \delta \rchi(x) \; .
\end{equation}
In this way, expanding $m_a\!\left(\rchi\right)$, the action (\ref{eq:firt.pp.J_ST}) can be rewritten as
\begin{align}
S_\text{pp} = -\sum_a \int\!\!d \tau_a\, m_a\left(1-\alpha_a\delta \rchi-\beta_a\delta \rchi^2 +\dots \right) \; ,
\label{eq:pp_Jordan_first}
\end{align}
where $d \tau_a^2 = g_{\mu\nu}dx_a^\mu dx_a^\nu$, and we defined
\begin{align}
m_a \equiv m_a(\rchi_0) \; , & &  \alpha_a \equiv -\frac{1}{m_a\!(\rchi)}\frac{d m_a\!(\rchi)}{d\rchi}\bigg|_{\rchi = \rchi_0} \; , & & \beta_a \equiv -\frac{1}{2m_a\!(\rchi)}\frac{d^2 m_a\!(\rchi)}{d\rchi^2}\bigg|_{\rchi = \rchi_0} \; .
\end{align}
We stop our expansion at the second order in $\delta \rchi$ because, as we shall see, these are the relevant terms that enter in our computations. 

The last thing we need to do is to explicitly write the proper time in terms of NRG fields. Looking at the action written in eq. (\ref{eq:firt.pp_NRG}), it is not hard to obtain the following result
\begin{align}
S_{\text{pp}} \!=\! & \sum_a m_a\!\!\int\!\!d t \left(-1\!+\frac{1}{2}v_a^2\!+\frac{1}{8}v_a^4\right) \notag \\
&  -\sum_a m_a\!\!\int\!\!d t\bigg(\phi+\frac{\phi^2}{2} -\bm{A}\cdot\bm{v}_a+\frac{3}{2}\phi v_a^2 -\frac{1}{2}\sigma_{ij}v_a^iv_a^j \bigg) \notag \\
& + \sum_a m_a\!\!\int\!\!d t\!\left(\alpha_a\delta \rchi\!-\frac{\alpha_a}{2}v_a^2\delta \rchi +\beta_a \delta \rchi^2 +\alpha_a\phi\delta \rchi\right)+\dots \; .
\label{eq:final_pp_Jordan}
\end{align}
In the above action we split as usual $\gamma_{ij} = \delta_{ij}+\sigma_{ij}$, hence spatial indices are raised/lowered using the three-dimensional Kronecker symbol. We realise that compared to (\ref{eq:PP.kk}), in scalar-tensor theory we have a much richer structure and more possible interaction vertices.

\section{NRGR computations in scalar-tensor theory}

We are now ready to start the NRGR machinery. Since we have already integrated out the short modes associated to the size of the objects\footnote{Recall that we are still considering the two compact objects as point particles.}, we should split each field into potential plus radiation mode. Once we have done this, we can integrate out first the potential modes and find in this way an effective action $\Seff{NR}$ which has essentially the same structure of eq. (\ref{eq:SeffNR}), i.e. 
\begin{equation}
\Seff{NR}[\bar{\Xi},x^\mu_a] = S_\text{cons}[x_a^\mu]+S_{\bar{\Xi}}[\bar{\Xi}]+S_\text{diss}[\bar{\Xi},x^\mu_a] \; ,
\label{eq:SNR_STtheory}
\end{equation}
where we denote with $\bar{\Xi}$ the radiative components of the gravitational and scalar fields. The action $S_{\bar{\Xi}}$ contains the kinetic and self interaction terms of such fields.On the other hand, by integrating out the potential components of the gravitational and scalar fields, we find as final result a pure conservative action $S_\text{cons}$ depending only on the two point particles, plus the dissipative radiative term $S_\text{diss}$. These two actions can be compute as usual using Feynman diagrams having as general structure the same sketched in eqs (\ref{eqn:scheme14}).

Before doing so, however, we first expand the actions (\ref{eq:Sg.NRGR.Jordan}) and (\ref{eq:GF_harm_Jordan}) at quadratic order in the fields and, then, introduce at the end a mode decomposition similar to the one written in eqs. (\ref{eqn:scheme42b}). We proceed in this way in order to highlight the differences with respect to the case discussed in the first part. In the next section we perform the above-mentioned expansion explicitly. Moreover, we shall now fix the conformal factors to be
\begin{align}
F(\rchi) = Z(\rchi) =  e^{\lambda(\rchi-\rchi_0)} \; .
\label{eq:conformal_factor}
\end{align}
Notice also that, being $\rchi_0 = \text{const.}$, we can always simplify $\partial_\mu\delta \rchi = \partial_\mu\rchi$.

\subsection{The quadratic part of the gravity action and the mixing}\label{subsec:triplet}

As we said, we want to find the quadratic part of the gauged-fixed gravity action. So let's start with eq. (\ref{eq:Sg.NRGR.Jordan}). For the moment we keep the conformal factor $F(\rchi)$ and $Z(\rchi)$ still completely generic, and expand to the quadratic order $\sqrt{\gamma}$ and the quantities inside the curly brackets. 

For most terms this expansion is straightforward, we just have to be careful in dealing with $\bar{\Rc}_\gamma$ which, starting from the definition of $\bar{\Gamma}_{ijk}$ given in eq. (\ref{eq:three_dim_chris_D}), has the following form
\begin{equation}
\bar{\Rc}_\gamma \equiv 2 \gamma^{ij} D_{[k|}\bar{\Gamma}{^k}_{i|j]} + 2\gamma^{ij}\bar{\Gamma}{^k}_{\ell [k|}\bar{\Gamma}{^\ell}_{i|j]} \; ,
\label{eq:Ricci_bar_explicit}
\end{equation}
where $D_i \equiv \partial_i + A_i\partial_t$, and we used $\gamma_{ij}$ to raise/lower spatial indices. We have to expand this term at the second order in the fields $\sigma_{ij}$ and $A_i$.  The second term in eq. (\ref{eq:Ricci_bar_explicit}) is not hard, because it is already of second order. On the other hand the first term gives us also a contribution of the following form 
\begin{align}
\!\!2\!\int\!\!d^4\!x F(\rchi) \sqrt{\gamma}\gamma^{ij}D_{[k|}\bar{\Gamma}{^k}_{i|j]} \rightarrow \!\!\int\!\!d^4\!x F(\rchi)\delta^{ij}\partial_{[k|}\left(\partial_i\sigma{^k}_{|j]} + \partial_{|j]}\sigma{^k}_i - \partial^k\sigma_{i|j]}\right) \; ,
\end{align}
where we have used the Kronecker delta $\delta_{ij}$ to raise/lower spatial indices.  We can then integrate by parts so that, up to a surface term,
\begin{align}
\int\!\!d^4\!x F(\rchi)\delta^{ij}\partial_{[k|}\left(\partial_i\sigma{^k}_{|j]} + \partial_{|j]}\sigma{^k}_i - \partial^k\sigma_{i|j]}\right) \simeq \int\!\!d^4\!x \left( \partial^{i}\sigma-\partial_k\sigma^{ik}\right)\partial_i F(\rchi) \; .
\end{align}
where we defined $\sigma \equiv \sigma_{ij}\delta^{ij}$. Now we write $\partial_i F(\rchi) = F'(\rchi)\partial_i\rchi$, where, again, $F'(\rchi)$ is the derivative of the conformal factor w.r.t. the field $\rchi$. We understand that in a scalar-tensor theory also the linear part of the Ricci scalar contributes non trivially to the quadratic part of the action.

Performing a similar procedure for all terms of the gravity action, and for the ones of the gauge-fixing action (\ref{eq:GF_harm_Jordan}), one eventually writes the full gauge-fixed quadratic action for the gravitational sector of the theory
\begin{align}
S\ord{2}_{\text{grav}} & = \frac{\mpl^2}{2}\int\!\!d^4x F(\rchi)\bigg\{2\partial_\mu\phi\partial^\mu\phi\!-\!\frac{1}{2}\partial_\mu A_i\partial^\mu A^i\!+\!\frac{1}{4}\partial_\mu\sigma_{ij}\left(\delta^{ik}\delta^{j\ell}\!-\!\frac{1}{2}\delta^{ij}\delta^{k\ell}\right)\partial^\mu\sigma_{k\ell}\bigg\} \notag \\
& \ +\frac{\mpl^2}{2}\int\!\!d^4x F'(\rchi)\partial_\mu\rchi\partial^{\mu}\left(2\phi\!-\!\frac{\sigma}{2}\right) \notag \\
&\ + \frac{\mpl^2}{2}\int\!\!d^4x \left(Z(\rchi)\!+\!\frac{\left(F'(\rchi)\right)^2}{2F(\rchi)}\right)\partial_\mu\rchi\partial^\mu\rchi \, ,
\label{eq:quad_GF_Jordan_NRG}
\end{align}
where we used the flat Minkowski metric $\eta_{\mu\nu}$ to raise/lower Greek indices. 

As expected, we get a non trivial mixing of NRG fields and the scalar $\rchi$ already at quadratic order. In particular, in this gauge the scalar field has a mixing only with the other two scalars of the theory, i.e. $\phi$ and the trace $\sigma$. To be even more explicit, we split the field $\sigma_{ij}$ into the trace and a trace-free part $\tilde{\sigma}_{ij}$, hence
\begin{equation}
\sigma_{ij} = \tilde{\sigma}_{ij} +\frac{1}{3}\delta_{ij}\sigma \; .
\label{eq:def_tracefree_sigma}
\end{equation}
Clearly $\tilde{\sigma}_{ij}\delta^{ij} = 0 $. At this point, we also expand the conformal factors introduced in eq. (\ref{eq:conformal_factor}), namely
\begin{subequations}
\label{eqn:scheme77}
\begin{align}
F(\rchi) & = F(\rchi_0) + \dots = 1 \; ,\\
F'(\rchi) & = F'(\rchi_0) + \dots = \lambda \; ,\\
Z(\rchi) & = Z(\rchi_ 0) + \dots = 1 \; ,
\end{align}
\end{subequations}
where we stopped at the zero order in the expansion since these terms multiply quantities that are already quadratic in the fields. Finally, similarly to eq. (\ref{eq:norm.field.NRG}), we normalize all fields so that they have mass dimension one, e.g. $\phi\longrightarrow \phi/\mpl$. Inserting eqs. (\ref{eq:def_tracefree_sigma}) and (\ref{eqn:scheme77}) into (\ref{eq:quad_GF_Jordan_NRG}) we eventually obtain
\begin{align}
S\ord{2}_{\text{grav}} & = \frac{1}{2}\int\!\!d^4x\bigg\{2\partial_\mu\phi\partial^\mu\phi\!-\!\frac{1}{2}\partial_\mu A_i\partial^\mu A^i\!+\!\frac{1}{4}\partial_\mu\tilde{\sigma}_{ij}\partial^\mu\tilde{\sigma}^{ij}-\frac{1}{24}\partial_\mu\sigma\partial^\mu\sigma \notag \\
&\qquad\qquad\qquad+\lambda\partial_\mu\rchi\partial^{\mu}\left(2\phi\!-\!\frac{\sigma}{2}\right) + \left(1\!+\!\frac{\lambda^2}{2}\right)\partial_\mu\rchi\partial^\mu\rchi\bigg\} \, .
\label{eq:Final_mixed_quad_action}
\end{align}

Now, in order to get only canonical non-mixed kinetic terms, one can look for a transformation of the fields in order to de-mix the various terms of this action. In fact, we already have such transformation. In fact, the origin of this mixing is in the conformal factors $F(\rchi)$ and $Z(\rchi)$, and we perfectly know that this terms are absorbed in the fields redefinition that brings in the Einstein frame. Then, to remove the mixing at the quadratic order, it is sufficient to take the relations written respectively in eqs. (\ref{eq:chiE.chiJ}) and (\ref{eqn:scheme54}) and write them at the linear order in the fields.  Using the conformal factors of eq. (\ref{eq:conformal_factor}), the de-mixing transformation are explicitly
\begin{subequations}
\label{eqn:scheme78}
\begin{align}
\delta \rchi & = \left(\frac{3}{2}\lambda^2+1\right)^{-\frac{1}{2}}\delta \rchi^* \; , \\ 
\phi & = \phi^*-\frac{\lambda}{2}\delta \rchi \; , & &\sigma = \sigma^* -6\lambda\delta \rchi \; , \\
A_i & = A^*_i \; , &  & A^i = A_*^i \; , \\
\tilde{\sigma}_{ij} & = \tilde{\sigma}^*_{ij} \; , &  & \tilde{\sigma}^{ij} = \tilde{\sigma}_*^{ij} \; .
\end{align}
\end{subequations}
It is indeed not hard to see that these star fields diagonalise the action (\ref{eq:Final_mixed_quad_action}).

However, we can also avoid going back in the Einstein frame, and find a more compact way to deal with this problem. Indeed, if we introduce a triplet of scalar field $\Phi^A$ with $A=1, 2, 3$ such that 
\begin{align}
\Phi^A \equiv \begin{pmatrix}
\phi \\ \sigma \\ \delta \rchi 
\end{pmatrix} \; ,
\end{align} 
then, defining the mixing matrix $\mathcal{M}_{AB}$ as follows
\begin{equation}
\mathcal{M}_{AB} = \begin{pmatrix}
2 & 0 & \lambda \\
0 & -\frac{1}{24} & -\frac{\lambda}{4} \\
\lambda & -\frac{\lambda}{4} & 1+\frac{\lambda^2}{2}
\end{pmatrix} \; ,
\end{equation}
we can rewrite the action (\ref{eq:Final_mixed_quad_action}) as
\begin{equation}
S\ord{2}_{\text{grav}} = \frac{1}{2}\int\!\!d^4x\bigg\{\mathcal{M}_{AB}\partial_\mu\Phi^A\partial^\mu\Phi^B-\frac{1}{2}\partial_\mu A_i\partial^\mu A^i +\frac{1}{4}\partial_\mu\tilde{\sigma}_{ij}\partial^\mu\tilde{\sigma}^{ij} \bigg\} \; .
\label{eq:action_quad_mix_matr}
\end{equation}
In the above equation we assumed that the Einstein summation on repeated indices holds also for the upper case Latin indices. We also stress that these indices can be raised and lowered arbitrarily. With this compact notation, finding the propagators for the three scalars field means simply finding the inverse of the symmetric mixing matrix $\mathcal{M}_{AB}$. 

\subsection{Propagators and scaling}

Now that we have the quadratic part of the action, we can find the propagators of the various fields of the theory. Using the triplet field method, hence the action (\ref{eq:action_quad_mix_matr}), first, we perform the decomposition of fields into potential and radiation modes, i.e.
\begin{subequations}
\label{eqn:scheme79}
\begin{align}
\Phi^A & = \varphi^A +\bar{\Phi}^A = \begin{pmatrix}
\varphi \\ \varsigma \\ \psi 
\end{pmatrix} + 
\begin{pmatrix}
\bar{\phi} \\ \bar{\sigma} \\ \bar{\rchi}  
\end{pmatrix} \; ,\\
A_i & = \mathcal{A}_i + \bar{A}_i \; ,\\
\tilde{\sigma}_{ij} & = \varsigma_{ij} + \bar{\sigma}_{ij} \; .
\end{align}
\end{subequations}

In order to find the propagators for the potential fields, let's first consider the action (\ref{eq:action_quad_mix_matr}) where each field is replaced with its potential mode. Then, to remove again the large fluctuations coming from the spatial derivative acting on a potential field\footnote{See the discussion in Sec. \ref{sec:mod.dec}.}, we perform a partial Fourier transform as in eq. (\ref{eq:frvec_NRG}) for every field. In this way we obtain
\begin{align}
S\ord{2}_{\text{grav}} & = \frac{1}{2}\int\!\!d t\intvec{k}\bigg\{-\modul{k}^2\mathcal{M}_{AB}\varphi_{\bm{k}}^A\varphi_{-\bm{k}}^B+\frac{\modul{k}^2}{2}\mathcal{A}_{\bm{k}\,i}\mathcal{A}_{-\bm{k}}^i -\frac{\modul{k}^2}{4}\varsigma_{\bm{k}\,ij}\varsigma^{ij}_{-\bm{k}}\bigg\} \notag \\
& +\frac{1}{2}\int\!\!d t\intvec{k}\bigg\{\mathcal{M}_{AB}\dot{\varphi}_{\bm{k}}^A\dot{\varphi}_{-\bm{k}}^B-\frac{1}{2}\dot{\mathcal{A}}_{\bm{k}\,i}\dot{\mathcal{A}}_{-\bm{k}}^i +\frac{1}{4}\dot{\varsigma}_{\bm{k}\,ij}\dot{\varsigma}^{ij}_{-\bm{k}}\bigg\}\; .
\label{eq:action_prop_Jordan}
\end{align}
Introducing then the following diagrammatic convention
\begin{align}
\varphi_{\bm{k}}^A \rightarrow \begin{tikzpicture}[baseline]
\begin{feynman}
\vertex (f);
\vertex [above=0.1cm of f] (z);
\vertex [right=1.2cm of z] (w);  
\diagram* {
(z) -- [dotted] (w)
}; 
\end{feynman} 
\end{tikzpicture} \; ,  & & 
\mathcal{A}_{\bm{k}\, i} \rightarrow \begin{tikzpicture}[baseline]
\begin{feynman}
\vertex (f);
\vertex [above=0.1cm of f] (z);
\vertex [right=1.2cm of z] (w);  
\diagram* {
(z) -- [scalar] (w)
}; 
\end{feynman} 
\end{tikzpicture} \; ,  & & 
\varsigma_{\bm{k}\, ij} \rightarrow \begin{tikzpicture}[baseline]
\begin{feynman}
\vertex (f);
\vertex [above=0.1cm of f] (z);
\vertex [right=1.2cm of z] (w);  
\diagram* {
(z) -- [double] (w)
}; 
\end{feynman} 
\end{tikzpicture} \; .
\end{align}
From the first line of eq. (\ref{eq:action_prop_Jordan}) we eventually find the following propagators 
\begin{subequations}
\label{eqn:scheme80}
\begin{align}
\begin{tikzpicture}[baseline]
\begin{feynman}
\vertex (f);
\vertex [above=0.1cm of f, label=295:$B$, label=75:$\bm{q}$, label=180:$t'$] (z);
\vertex [right=1.8cm of z, label=255:$A$, label=105:$\bm{k}$, label=0:$t$] (w);  
\diagram* {
(z) -- [dotted] (w)
}; 
\end{feynman} 
\end{tikzpicture} & = -\frac{i}{\modul{k}^2}(2\pi)^3\delta^{(3)}(\bm{k}\!+\!\bm{q})\delta(t\!-\!t')D_{AB} \; ,\\
\begin{tikzpicture}[baseline]
\begin{feynman}
\vertex (f);
\vertex [above=0.1cm of f, label=75:$\bm{q}$, label=285:$i$, label=180:$t'$] (z);
\vertex [right=1.8cm of z, label=105:$\bm{k}$, label=255:$j$, label=0:$t$] (w); 
\diagram* {
(z) -- [scalar] (w)
}; 
\end{feynman} 
\end{tikzpicture} & = \frac{2i}{\modul{k}^2}(2\pi)^3\delta^{(3)}(\bm{k}\!+\!\bm{q})\delta(t\!-\!t')\delta_{ij} \; ,\\
\begin{tikzpicture}[baseline]
\begin{feynman}
\vertex (f);
\vertex [above=0.1cm of f, label=75:$\bm{q}$, label=285:$k\ell$, label=180:$t'$] (z);
\vertex [right=1.8cm of z, label=105:$\bm{k}$, label=255:$ij$, label=0:$t$] (w); 
\diagram* {
(z) -- [double] (w)
}; 
\end{feynman} 
\end{tikzpicture}  & = -\frac{4i}{\modul{k}^2}(2\pi)^3\delta^{(3)}(\bm{k}\!+\!\bm{q})\delta(t\!-\!t')\delta_{i(k|}\delta_{j|\ell)} \; ,
\end{align}
\end{subequations}
where we defined the inverse of the mixing matrix as
\begin{equation}
D_{AB}\equiv \mathcal{M}^{-1}_{AB} = \frac{1}{2+3\lambda^2}\begin{pmatrix}
1+2\lambda^2 & 6\lambda^2 & -\lambda \\
6\lambda^2 & -48 & -12\lambda \\
-\lambda & -12\lambda & 2
\end{pmatrix} \; .
\end{equation}
From eqs. (\ref{eqn:scheme80}) we can easily understand that all potential fields have the same scaling in $v$ of $\potH{k}{\mu\nu}$ written in tab. \ref{table:scaling} (p.~\pageref{table:scaling}). As we said already in the previous part\footnote{See for instance Sec. \ref{sec:power.count}.}, in the conservative sector of NRGR time derivatives are treated as correction to the instantaneous propagator, i.e. 
\begin{equation}
\begin{tikzpicture}[baseline]
\begin{feynman}
\vertex (f);
\vertex [above=0.1cm of f, label=295:$B$, label=75:$\bm{q}$, label=180:$t'$] (z);
\vertex [right=0.8cm of z, crossed dot, minimum size=0.25cm] (c) {};
\vertex [right=2cm of z, label=255:$A$, label=105:$\bm{k}$, label=0:$t$] (w);  
\diagram* {
(z) -- [dotted] (c) -- [dotted] (w)
}; 
\end{feynman} 
\end{tikzpicture} = -\frac{i}{\modul{k}^4}(2\pi)^3\delta^{(3)}(\bm{k}\!+\!\bm{q})\frac{\partial^2}{\partial t \partial t'}\delta(t\!-\!t')D_{AB} \; ,
\label{eq:corrected_prop_Jordan}
\end{equation}
and analogously for $\mathcal{A}_{\bm{k}\, i}$ and $\varsigma_{\bm{k}\, ij}$. We finally recall that these corrections scale as $v^2$ in the PN expansion.

For the radiative fields, things are a little easier. Indeed the quadratic action is precisely (\ref{eq:action_quad_mix_matr}), where each fields is replaced by its radiative counterpart. Introducing then the following diagrammatic convention
\begin{align}
\bar{\Phi}^A \rightarrow \begin{tikzpicture}[baseline]
\begin{feynman}
\vertex (f);
\vertex [above=0.1cm of f] (z);
\vertex [right=1.2cm of z] (w);  
\diagram* {
(z) -- [photon] (w)
}; 
\end{feynman} 
\end{tikzpicture} \; ,  & & 
\bar{A}_{i} \rightarrow \begin{tikzpicture}[baseline]
\begin{feynman}
\vertex (f);
\vertex [above=0.1cm of f] (z);
\vertex [right=1.2cm of z] (w); 
\vertex [above=0.04cm of z] (z');
\vertex [right=1.2cm of z'] (w');  
\diagram* {
(z) -- [photon] (w),
(z') -- [photon] (w')
}; 
\end{feynman} 
\end{tikzpicture} \; ,  & & 
\bar{\sigma}_{ij} \rightarrow \begin{tikzpicture}[baseline]
\begin{feynman}
\vertex (f);
\vertex [above=0.1cm of f] (z);
\vertex [right=1.2cm of z] (w);  
\diagram* {
(z) -- [gluon] (w)
}; 
\end{feynman} 
\end{tikzpicture} \; ,
\end{align}
one eventually finds
\begin{subequations}
\label{eqn:scheme45double}
\begin{align}
 \begin{tikzpicture}[baseline]
\begin{feynman}
\vertex (f);
\vertex [above=0.1cm of f, label=90:$y$, label=270:$A$] (z);
\vertex [right=1.8cm of z, label=90:$x$, label=270:$B$] (w);  
\diagram* {
(z) -- [photon] (w)
}; 
\end{feynman} 
\end{tikzpicture} & = D(x-y) D_{AB} \; ,\\
\begin{tikzpicture}[baseline]
\begin{feynman}
\vertex (f);
\vertex [above=0.1cm of f, label=90:$y$, label=270:$i$] (z);
\vertex [right=1.8cm of z, label=90:$x$, label=270:$j$] (w); 
\vertex [above=0.04cm of z] (z');
\vertex [above=0.04cm of w] (w');
\diagram* {
(z) -- [photon] (w),
(z') -- [photon] (w')
}; 
\end{feynman} 
\end{tikzpicture} & = -2D(x-y)\delta_{ij} \; ,\\
 \begin{tikzpicture}[baseline]
\begin{feynman}
\vertex (f);
\vertex [above=0.1cm of f, label=90:$y$, label=270:$k\ell$] (z);
\vertex [right=1.8cm of z, label=90:$x$, label=270:$ij$] (w); 
\diagram* {
(z) -- [gluon] (w)
}; 
\end{feynman} 
\end{tikzpicture} & = 4D(x-y)\delta_{i(k|}\delta_{(j|\ell)} \; .
\end{align}
\end{subequations} 
One immediately see that also for all radiative fields the scaling in $v$ is the same as $\radh{\mu\nu}$ written in tab. \ref{table:scaling}.

At this point, we are almost ready to do concrete diagrammatic computations in a conformally coupled scalar-tensor theory in the Jordan frame. The last thing we need are the Feynman rules describing interactions with matter and self interactions of the gravitational field. 

\section{The conservative dynamics up to 1PN order}

Let's start by performing the computations in the conservative sector, postponing the discussion of the radiative part to Sec. \ref{Sec:Diss_Jordan}. Therefore, for the rest of this section we can consistently set to zero  the radiative fields $\bar{\Phi}^A$, $\bar{A}_i$  and $\bar{\sigma}_{ij}$.

\subsection{1 PN order Feynman rules}

Following the procedure described in the first part of this work, we need to find the Feynman rules coming from the matter action (\ref{eq:final_pp_Jordan}). Considering the potential mode only, and performing the needed partial Fourier transform, we find that the relevant part of the point particle action for the conservative sector is given by
\begin{align}
S^{\text{pp}}_{\text{cons}} & = \sum_a m_a\int\!\!d t\left(-1+\frac{v_a^2}{2}+\frac{v^4_a}{8}\right) - \sum_a\frac{m_a}{\mpl}\int\!\!d t\!\!\intvec{k}e^{i\bm{k}\cdot\bm{x}_a}\left(\varphi_{\bm{k}}-\alpha_a\psi_{\bm{k}}\right) \notag \\
& - \sum_a\frac{m_a}{2\mpl}\int\!\!d t\!\!\intvec{k}e^{i\bm{k}\cdot\bm{x}_a}v_a^2\left(3\varphi_{\bm{k}}-\frac{1}{3}\varsigma_{\bm{k}} +\alpha_a\psi_{\bm{k}}\right) + \sum_a\frac{m_a}{\mpl}\int\!\!d t\!\!\intvec{k}e^{i\bm{k}\cdot\bm{x}_a} v^i_a\mathcal{A}_{\bm{k}\, i} \notag\\
& -\sum_a\frac{m_a}{\mpl^2}\int\!\!d t\intvec{k,q}\!\!e^{i(\bm{k}+\bm{q})\cdot\bm{x}_a}\left(\frac{1}{2}\varphi_{\bm{k}}\varphi_{\bm{q}} +\alpha_a \varphi_{\bm{k}}\psi_{\bm{q}} + \beta_a\psi_{\bm{k}}\psi_{\bm{q}}\right) + \dots
\label{eq:final_pp_pot_Jordan}
\end{align}
The first time integral of this action describes just the kinetic contribution of the two point-particles; from the other terms of eq. (\ref{eq:final_pp_pot_Jordan}) we can find the Feynman rules listed in tab. \ref{tab:Rules_Jordan_1PN}. As in Sec. \ref{sec:mod.dec} (p.~\pageref{eq:SGFpot2}), we use again straight line  to represent the point-like BHs. We introduce a Kronecker delta for the triplet of scalar fields recalling that $\varphi^1=\varphi$, $\varphi^2 = \varsigma$ and $\varphi^3 = \psi$.
\begin{table}[t]
\begin{center}
\begin{tabular}{|c|c|c|}
\hline
Diagrammatic expression & Scaling & Explicit expression \\
\hline
\begin{tikzpicture}[baseline]
\begin{feynman}
\vertex (z);
\vertex [left=0.2cm of z, label=180:$t$, label=80:$A$] (c');
\vertex [above=0.6cm of c'] (a);
\vertex [above=0.2cm of a] (a');
\vertex [below=0.6cm of c'] (b);
\vertex [below=0.2cm of b] (b');
\vertex [right=0.7cm of z, label=90:$\bm{k}$] (c); 
\diagram* {
(a') -- [draw=none] (a) -- [plain] (c') -- [plain] (b) -- [draw=none] (b'),
(c') -- [dotted] (c)
}; 
\end{feynman} 
\end{tikzpicture} & $\sim L^{1/2}$ & \(\displaystyle
-\frac{im_a}{\mpl}\int\!\!d t\!\!\intvec{k}e^{i\bm{k}\cdot\bm{x}_a} \left(\delta^{A1}-\alpha_a\delta^{A3}\right)\) \\
\hline 
\begin{tikzpicture}[baseline]
\begin{feynman}
\vertex (z);
\vertex [empty dot, minimum size=0.4cm, left=0.2cm of z, label=180:$t$, label=80:$A$] (c') {2};
\vertex [above=0.7cm of c'] (a);
\vertex [above=0.2cm of a] (a');
\vertex [below=0.7cm of c'] (b);
\vertex [below=0.2cm of b] (b');
\vertex [right=0.7cm of z, label=90:$\bm{k}$] (c); 
\diagram* {
(a') -- [draw=none] (a) -- [plain] (c') -- [plain] (b) -- [draw=none] (b'),
(c') -- [dotted] (c)
}; 
\end{feynman} 
\end{tikzpicture} & $\sim L^{1/2}v^2$ & 
\(\displaystyle
-\frac{im_a}{2\mpl}\int\!\!d t\!\!\intvec{k}e^{i\bm{k}\cdot\bm{x}_a}v^2_a\left(3\delta^{A1}\!-\!\frac{1}{3}\delta^{A2}\!+\!\alpha_a\delta^{A3}\right)\) \\
\hline 
\begin{tikzpicture}[baseline]
\begin{feynman}
\vertex (z);
\vertex [left=0.2cm of z, label=180:$t$, label=80:$\!\!A$, label=285:$\!\!B$] (c');
\vertex [above=0.7cm of c'] (a);
\vertex [above=0.3cm of a] (a');
\vertex [below=0.7cm of c'] (b);
\vertex [below=0.2cm of b] (b');
\vertex [right=0.7cm of z] (z');
\vertex [above=0.5cm of z', label=90:$\bm{k}$] (c); 
\vertex [below=0.5cm of z', label=270:$\bm{q}$] (d); 
\diagram* {
(a') -- [draw=none] (a) -- [plain] (c') -- [plain] (b) -- [draw=none] (b'),
(c') -- [dotted] (c),
(c') -- [dotted] (d)
}; 
\end{feynman} 
\end{tikzpicture} & $\sim v^2$ & 
\(\displaystyle
\begin{aligned}[t]
-\frac{im_a}{\mpl^2} \int\!\!d t\!\!&\intvec{k,q}\!\!e^{i(\bm{k}+\bm{q})\cdot\bm{x}_a}\bigg(\delta^{A1}\delta^{B1} \\
& -2\alpha\delta^{(A|1}\delta^{|B)3} -2\beta_a\delta^{A3}\delta^{B3}\bigg) \\
& \ \end{aligned}\) \\
\hline
\begin{tikzpicture}[baseline]
\begin{feynman}
\vertex (z);
\vertex [left=0.2cm of z, label=15:$i$, label=180:$t$] (c');
\vertex [above=0.7cm of c'] (a);
\vertex [above=0.2cm of a] (a');
\vertex [below=0.7cm of c'] (b);
\vertex [below=0.2cm of b] (b');
\vertex [right=0.7cm of z, label=90:$\bm{k}$] (c); 
\diagram* {
(a') -- [draw=none] (a) -- [plain] (c') -- [plain] (b) -- [draw=none] (b'),
(c') -- [scalar] (c)
}; 
\end{feynman} 
\end{tikzpicture} & $\sim L^{1/2}v$ & 
\(\displaystyle \frac{im_a}{\mpl}\!\int\!\!d t\!\!\intvec{k}e^{i\bm{k}\cdot\bm{x}_a}v_a^i
\)\\
\hline 
\end{tabular}
\caption{Point-particle potential fields rules in Jordan frame}
\label{tab:Rules_Jordan_1PN}
\end{center}
\end{table}

At this point, we have all the necessary ingredients to compute diagrams in the conservative sector of this EFT theory. Indeed, we shall see in the next few sections that the use of NRG fields to parametrise the metric avoids the computation of any self interaction (i.e. cubic) vertex at 1PN order.

\subsection{The Newtonian dynamics}

We split again the conservative part of action (\ref{eq:SNR_STtheory}) action as $S_\text{cons} = S_\text{N} + S_\text{1PN} + \Ord{Lv^4}$, and compute separately these two contributions.

We recall that the conservative effective action at the Newtonian level is given  by the sum of all diagrams that are of order $Lv^0$. Therefore, from tab. \ref{tab:Rules_Jordan_1PN} we understand that we have formally just one diagram to compute which is
\begin{align}
\begin{tikzpicture}[baseline]
\begin{feynman}
\vertex [label=270:$t'$, label=160:$B$, label=20:$\bm{q}$] (c);
\vertex [above=0.6cm of c] (f);
\vertex [above=0.1cm of f] (z);
\vertex [above=0.9cm of z, label=90:$t$, label=200:$A$, label=340:$\bm{k}$] (c');
\vertex [right=1.2cm of c] (a);
\vertex [left=1.2cm of c] (b) {$x_b$};
\vertex [right=1.2cm of c'] (a');
\vertex [left=1.2cm of c'] (b') {$x_a$};
\diagram* {
(a) -- [plain] (c) -- [plain] (b),
(a') -- [plain] (c') -- [plain] (b'),
(c) -- [dotted] (c')
}; 
\end{feynman} 
\end{tikzpicture} & = \frac{i}{2}\sum_{a\neq b}\frac{m_a m_b}{4\pi\mpl^2}\int\!\!d t\frac{1}{\abs*{\bm{x}_{ab}}}\left(D_{11}-(\alpha_a\!+\!\alpha_b)D_{13}+\alpha_a\alpha_b D_{33}\right)  \notag \\
& =  i\int\!\!d t\frac{\tilde{G}_{12}m_1 m_2}{r} \; ,
\label{eq:Newton_Jordan}
\end{align}
where $\bm{x}_{ab} \equiv \bm{x}_a-\bm{x}_b$, and $r$ is its modulus. In the last step we defined the effective gravitational Newton constant 
\begin{equation}
\tilde{G}_{ab} \equiv \frac{2\GN}{2+3\lambda^2}\left[1\!+\!2\lambda^2\!+\!\lambda(\alpha_a+\alpha_2)\!+\!2\alpha_a\alpha_b\right] \; .
\label{eq:eff_Newton}
\end{equation}
Therefore, the Newtonian physics is described by the action
\begin{equation}
S_{\text{N}} = \int\!\!d t\left\{\frac{1}{2}\sum_{a=1}^2m_av_a^2+\frac{\tilde{G}_{12} m_1m_2}{r}\right\} \; .
\end{equation}

The modification of the gravitational constant due to the presence of a fifth-force exchange is a well known result\footnote{See for instance the extensive references \cite{FareseDamour,EFreview} for an alternative derivation of this result.}. However, written in this way, it is not easy to compare it with the one found in Ref. \cite{NRGR+scalar}. Let's then try to find a relation between the charges $\alpha_a$ and $\beta_a$ and the correspondent quantities in the Einstein frame. To this end, let's take the point-particle action written in eq. (\ref{eq:pp_Jordan_first}). We know that the Jordan and the Einstein metric are related by eq. (\ref{eq:conf.trans}), hence
\begin{align}
S_{\text{pp}} & = -\sum_am_a\int\!\!\left(F^{-1}(\rchi)g^*_{\mu\nu} dx_a^\mu dx_a^\nu\right)^{\frac{1}{2}}\left(1-\alpha_a\delta \rchi-\beta_a\delta \rchi^2+\dots\right) \notag \\
& = -\sum_a m_a\int\!\!d \tau^*_a e^{-\frac{\lambda_*\delta \rchi_*}{2}}\left(1-\alpha_a\frac{\lambda_*}{\lambda}\delta \rchi_*-\beta_a\frac{\lambda_*^2}{\lambda^2}\delta \rchi_*^2+\dots\right) \; ,
\label{eq:pp_step_Jordan}
\end{align}
where in the last step we used the conformal factor given in eq. (\ref{eq:conformal_factor}) and defined $\tau^*_a =\sqrt{g^*_\mu\nu dx_a^\mu dx_a^\nu}$. Moreover, recalling the relation between the Einstein and the Jordan scalar field written in eq. (\ref{eqn:scheme78}a), we defined also
\begin{equation}
\lambda_* \equiv \left(\frac{2}{3\lambda^2+2}\right)^{\frac{1}{2}} \lambda \; ,
\label{eq:lambdastar}
\end{equation}
so that $\lambda\delta \rchi = \lambda_*\delta \rchi_*$. Expanding the exponential, we can eventually rewrite eq. (\ref{eq:pp_step_Jordan}) in the Einstein frame as 
\begin{align}
S_\text{pp} = -\sum_a m_a \int\!\!d \tau^*_a\, \left(1-\alpha^*_a\delta \rchi_*-\beta^*_a\delta \rchi_*^2 +\dots \right) \; ,
\label{eq:PP_Einstein_Jordan}
\end{align}
where the Einstein frame scalar charges are
\begin{subequations}
\label{eqn:scheme81}
\begin{align}
\alpha_a^* & \equiv \frac{\lambda_*}{\lambda}\left(\alpha_a+\frac{\lambda}{2}\right) \; ,\\
\beta_a^* & \equiv \frac{\lambda^2_*}{\lambda^2}\left(\beta_a-\frac{\alpha_a\lambda}{2}-\frac{\lambda^2}{8}\right) \; .
\end{align}
\end{subequations}
These are the charges that appear in Ref. \cite{NRGR+scalar}. It is then not hard to see that the effective Newton constant defined in eq. (\ref{eq:eff_Newton}) can be rewritten as
\begin{equation}
\tilde{G}_{ab} = \GN\left(1+2\alpha^*_a\alpha_b^*\right) \; ,
\label{eq:kuntz_eff_Newton}
\end{equation}
which is precisely the result obtained in Ref. \cite{NRGR+scalar}.

\subsection{1PN order action}

We can find also the explicit form of the conservative 1PN order action, by computing all diagrams of order $Lv^2$. From the rules listed in tab. \ref{tab:Rules_Jordan_1PN}, and recalling the propagator correction (\ref{eq:corrected_prop_Jordan}), we immediately see that the diagrams contributing to the 1PN order are 
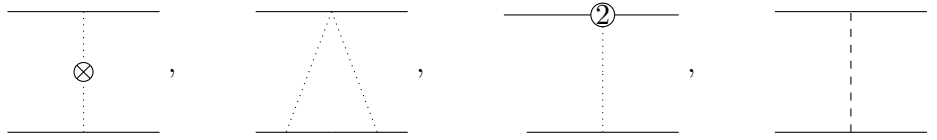
\begin{align*}
\begin{tikzpicture}[baseline]
\begin{feynman}
\vertex [crossed dot, minimum size=0.25cm] (z) {};
\vertex [above=0.8cm of z] (c');
\vertex [below=0.8cm of z] (c);
\vertex [right=1cm of c] (a);
\vertex [left=1cm of c] (b);
\vertex [right=1cm of c'] (a');
\vertex [left=1cm of c'] (b');
\diagram* {
(a) -- [plain] (c) -- [plain] (b),
(a') -- [plain] (c') -- [plain] (b'),
(c) -- [dotted] (z),
(c') -- [dotted] (z)
}; 
\end{feynman} 
\end{tikzpicture} \; , & & 
\begin{tikzpicture}[baseline]
\begin{feynman}
\vertex (z);
\vertex [above=0.8cm of z] (c');
\vertex [below=0.8cm of z] (c);
\vertex [right=1cm of c] (a);
\vertex [left=1cm of c] (b);
\vertex [right=0.6cm of c] (a'');
\vertex [left=0.6cm of c] (b'');
\vertex [right=1cm of c'] (a');
\vertex [left=1cm of c'] (b');
\diagram* {
(a) -- [plain] (c) -- [plain] (b),
(a') -- [plain] (c') -- [plain] (b'),
(c') -- [dotted] (a''),
(c') -- [dotted] (b'')
}; 
\end{feynman} 
\end{tikzpicture} \; , & & 
\begin{tikzpicture}[baseline]
\begin{feynman}
\vertex (z);
\vertex [above=0.6cm of z, empty dot, minimum size=0.3cm] (c') {2};
\vertex [below=0.8cm of z] (c);
\vertex [right=1cm of c] (a);
\vertex [left=1cm of c] (b);
\vertex [right=1cm of c'] (a');
\vertex [left=1.3cm of c'] (b');
\diagram* {
(a) -- [plain] (c) -- [plain] (b),
(a') -- [plain] (c') -- [plain] (b'),
(c) -- [dotted] (z),
(c') -- [dotted] (z)
}; 
\end{feynman} 
\end{tikzpicture}  \; , & & 
\begin{tikzpicture}[baseline]
\begin{feynman}
\vertex (z);
\vertex [above=0.8cm of z] (c');
\vertex [below=0.8cm of z] (c);
\vertex [right=1cm of c] (a);
\vertex [left=1cm of c] (b);
\vertex [right=1cm of c'] (a');
\vertex [left=1cm of c'] (b');
\diagram* {
(a) -- [plain] (c) -- [plain] (b),
(a') -- [plain] (c') -- [plain] (b'),
(c) -- [scalar] (c')
}; 
\end{feynman} 
\end{tikzpicture} \; .
\label{eq:1PN_Jordan}
\end{align*}
We can see that, using the triplet introduced in Sec. \ref{subsec:triplet}, we have to compute only four diagrams, versus the eleven diagrams sketched in Ref. \cite{NRGR+scalar}.  However, in principle nothing prevent us from having also a diagram of the following form
\begin{equation}
\begin{tikzpicture}[baseline]
\begin{feynman}
\vertex (z);
\vertex [below= 0.9cm of z] (c);
\vertex [above=0.9cm of z] (c');
\vertex [right=1.2cm of c] (a);
\vertex [left=1.2cm of c] (b) ;
\vertex [right=0.7cm of c] (a'');
\vertex [left=0.7cm of c] (b'');
\vertex [right=1.2cm of c'] (a');
\vertex [left=1.2cm of c'] (b');
\diagram* {
(a) -- [plain] (c) -- [plain] (b),
(a') -- [plain] (c') -- [plain] (b'),
(c') -- [dotted] (z),
(a'') -- [dotted] (z),
(b'') -- [dotted] (z)
}; 
\end{feynman} 
\end{tikzpicture} \sim Lv^2
\label{eq:three-vertex-Jordan}
\end{equation}
We shall see in Sec. \ref{subsub:three_vertex} that this kind of diagrams can not contributes to this order.

Putting this aside for the moment, let's compute the diagrams sketched above. The first one is explicitly
\begin{align}
\begin{tikzpicture}[baseline]
\begin{feynman}
\vertex [label=270:$t$, label=115:$B$, label=75:$\bm{q}$](c);
\vertex [crossed dot, minimum size=0.25cm, above=0.8cm of c] (z) {};
\vertex [above=0.8cm of z, label=90:$t$, label=255:$A$, label=295:$\bm{k}$] (c');
\vertex [right=1cm of c] (a);
\vertex [left=1cm of c] (b) {$x_b$};
\vertex [right=1cm of c'] (a');
\vertex [left=1cm of c'] (b') {$x_a$};
\diagram* {
(a) -- [plain] (c) -- [plain] (b),
(a') -- [plain] (c') -- [plain] (b'),
(c) -- [dotted] (z),
(c') -- [dotted] (z)
}; 
\end{feynman} 
\end{tikzpicture}  & = \frac{i}{2}\sum_{a\neq b}\frac{m_a m_b}{\mpl^2}\int\!\!d tv_a^i v_a^j\intvec{k}\frac{k_i k_j}{\modul{k}^2}e^{-i\bm{k}\cdot\bm{x}_{ab}}\notag \\
&\qquad\qquad\qquad\times\left(D_{11}-(\alpha_a\!+\!\alpha_b)D_{13}+\alpha_a\alpha_b D_{33}\right) \notag \\
& = \frac{\tilde{G}_{12} m_1 m_2}{2r}\left[(\bm{v}_1\cdot\bm{v}_2)-\frac{(\bm{v}_1\cdot\bm{r})(\bm{v}_2\cdot\bm{r})}{r^2}\right] \; , 
\label{eq:first_final_Jordan}
\end{align}
where in the last step we used the integral written in eq. (\ref{eq:wkresult}). Moreover, we realised that the terms inside round brackets of the second line are precisely the same as in eq. (\ref{eq:Newton_Jordan}). Therefore, we reconstructed the effective gravitation Newton constant given in eq. (\ref{eq:eff_Newton}). 

Then, we can compute
\begin{align}
\begin{tikzpicture}[baseline]
\begin{feynman}
\vertex (c);
\vertex [above=0.8cm of c, label=200:$\bm{k}$, label=340:$\bm{q}$](z);
\vertex [above=0.8cm of z, label=90:$t$, label=225:$A$, label=295:$B$] (c');
\vertex [right=1cm of c] (a);
\vertex [left=1cm of c] (b) {$x_b$};
\vertex [right=0.6cm of c, label=270:$t''$, label=75:$B'$] (a'');
\vertex [left=0.6cm of c, label=270:$t'$, label=125:$A'$] (b'');
\vertex [right=1cm of c'] (a');
\vertex [left=1cm of c'] (b') {$x_a$};
\diagram* {
(a) -- [plain] (c) -- [plain] (b),
(a') -- [plain] (c') -- [plain] (b'),
(c') -- [dotted] (a''),
(c') -- [dotted] (b'')
}; 
\end{feynman} 
\end{tikzpicture} & \!=\! -\frac{i}{2}\sum_{a\neq b}\frac{m_a m_b^2}{\mpl^4}\int\!\!d t\!\!\left[\intvec{k}\frac{e^{-i\bm{k}\cdot\bm{x}_{ab}}}{\modul{k}^2}\right]^2\!\bigg\{(D_{11})^2\!-\!2\alpha_b D_{11}D_{13}\!+\!\alpha_b(D_{13})^2\notag \\
&\qquad\qquad -\beta_a\left((D_{13})^2-2\alpha_b D_{13}D_{33}+\alpha_b^2(D_{33})^2\right) \notag \\
& \qquad\qquad -2\alpha_a\left[D_{11}D_{13}-\alpha_b\left(D_{11}D_{33}\!+\!(D_{13})^2\right)+2\alpha^2_b D_{13}D_{33}\right]\bigg\} \notag \\
& = -\frac{i}{2}\sum_{a\neq b}\int\!\!d t\,\frac{4\GN^2 m_a m_b^2}{\abs*{\bm{x}_{ab}}^2}\big\{ \mathcal{A}(\lambda, \alpha_a) - \mathcal{B}(\lambda, \alpha_a, \beta_a)\big\} \; ,  
\label{eq:first_diag_Jordan_1PN}
\end{align}
where we defined for simplicity
\begin{subequations}
\label{eqn:scheme82}
\begin{align}
\mathcal{A}(\lambda, \alpha_a)  & \equiv \frac{\left(1\!+\!2\lambda^2\!+\!\lambda(\alpha_a+\alpha_2)\!+\!2\alpha_a\alpha_b\right)^2}{(2\!+\!3\lambda^2)^2} = \left(\frac{1+2\alpha^*_a\alpha^*_b}{2}\right)^2\; ,\\
\mathcal{B}(\lambda, \alpha_a, \beta_a) & \equiv \frac{(\lambda^2+2\alpha_b)^2}{(2+3\lambda^2)^2}(\alpha_a^2+2\beta_a) = (\alpha_a^*\alpha_b^*)^2 + 2\beta_a^*\alpha_b^*{}^2\; .
\end{align}
\end{subequations} 
Here we used the relation with the Einstein charges defined in eqs. (\ref{eqn:scheme81}) in order to write these coefficients in a more compact way, and to make the comparison with known result easier. At this point, we introduce the symmetric matrix
\begin{equation}
f_{ab} \equiv \beta_a^*\alpha_b^*{}^2 + \beta_b^*\alpha_a^*{}^2 + \left(\frac{m_a-m_b}{m_a+m_b}\right)\left(\beta_a^*\alpha_b^*{}^2 - \beta_b^*\alpha_a^*{}^2\right) \; ,
\end{equation}
so that we can rewrite the last term of $ \mathcal{B}(\lambda, \alpha_a, \beta_a)$ as
\begin{equation}
\sum_{a\neq b} 2m_a m_b^2\beta_a^*\alpha_b^*{}^2 = \sum_{a\neq b} m_a m_b^2 f_{ab} \; .
\label{eq:step_f_Jordan}
\end{equation}
Inserting eqs. (\ref{eqn:scheme82}) in (\ref{eq:first_diag_Jordan_1PN}) and using eq. (\ref{eq:step_f_Jordan}), we obtain, eventually,
\begin{equation}
\begin{tikzpicture}[baseline]
\begin{feynman}
\vertex (z);
\vertex [below=0.8cm of z](c);
\vertex [above=0.8cm of z] (c');
\vertex [right=1cm of c] (a);
\vertex [left=1cm of c] (b) {$x_b$};
\vertex [right=0.6cm of c] (a'');
\vertex [left=0.6cm of c] (b'');
\vertex [right=1cm of c'] (a');
\vertex [left=1cm of c'] (b') {$x_a$};
\diagram* {
(a) -- [plain] (c) -- [plain] (b),
(a') -- [plain] (c') -- [plain] (b'),
(c') -- [dotted] (a''),
(c') -- [dotted] (b'')
}; 
\end{feynman} 
\end{tikzpicture} = \frac{i}{2}\int\!\!d t\frac{\tilde{G}_{12} m_1 m_2(m_1+m_2)}{2}\left(1-2\beta_{12}\right) \; ,
\label{eq:seagull_Jordan_final}
\end{equation}
where in the last step we defined the PPN parameter\footnote{See Ref. \cite{FareseDamour} for a generalization of this parameter in multi-tensor-scalar theories.} introduced in Ref. \cite{NRGR+scalar}
\begin{equation}
\beta_{ab} \equiv 1-2\frac{\alpha_a^*{}^2\alpha_b^*{}^2+f_{ab}}{\left(1+2\alpha^*_a\alpha^*_b\right)^2} \; .
\end{equation}

The third diagram is given by
\begin{align}
\begin{tikzpicture}[baseline]
\begin{feynman}
\vertex [ label=270:$t'$, label=125:$B$] (c);
\vertex [above=0.8cm of c, label=0:$\bm{k}$] (z);
\vertex [above=0.6cm of z, empty dot, minimum size=0.3cm, label=90:$t$, label=245:$A$] (c') {2};
\vertex [right=1cm of c] (a);
\vertex [left=1cm of c] (b) {$x_b$};
\vertex [right=1cm of c'] (a');
\vertex [left=1.3cm of c'] (b') {$x_a$};
\diagram* {
(a) -- [plain] (c) -- [plain] (b),
(a') -- [plain] (c') -- [plain] (b'),
(c) -- [dotted] (z),
(c') -- [dotted] (z)
}; 
\end{feynman} 
\end{tikzpicture} &  = \frac{i}{2}\sum_{a\neq b}\frac{m_a m_b}{4\pi\mpl^2}\int\!\!d t\frac{v_a^2}{\abs*{\bm{x}_{ab}}}\bigg\{3D_{11}-\frac{D_{21}}{3}\!+\!(\alpha_a\!-\!3\alpha_b)D_{13}\notag \\ 
&\qquad\qquad\qquad\qquad\qquad\qquad -\!\alpha_a\alpha_b D_{33}\!+\!\frac{\alpha_b}{3}D_{23}\bigg\}  \notag \\
& =\frac{i}{2}\sum_{a\neq b}\GN m_a m_b\int\!\!d t\frac{v_a^2}{\abs*{\bm{x}_{ab}}}\mathcal{C}(\lambda, \alpha_a) \; ,
\label{eq:step_2_Jordan}
\end{align}
where we defined
\begin{equation}
\mathcal{C}(\lambda, \alpha_a) \equiv \frac{1}{2+3\lambda^2}\left\{6\!+\!8\lambda^2\!-\!4\alpha_a\alpha_b\!-\!2\lambda(\alpha_a\!+\!\alpha_b)\right\} = \left(1\!+\!2\alpha^*_a\alpha^*_b\right)\left(1\!+\!2\gamma_{ab}\right)\; .
\end{equation}
Again, this term looks more compact using the charges defined in (\ref{eqn:scheme81}).  Here we have also introduced the second PPN parameter
\begin{equation}
\gamma_{ab} \equiv 1-4\frac{\alpha_a^*\alpha_b^*}{1+2\alpha_a^*\alpha_b^*} \; .
\label{eq:PPNgamma}
\end{equation}
Therefore, recalling the definition of the effective Newton constant (\ref{eq:eff_Newton}), we can rewrite eq. (\ref{eq:step_2_Jordan}) as
\begin{equation}
\begin{tikzpicture}[baseline]
\begin{feynman}
\vertex (z);
\vertex [below=0.8cm of z] (c);
\vertex [above=0.6cm of z, empty dot, minimum size=0.3cm] (c') {2};
\vertex [right=1cm of c] (a);
\vertex [left=1cm of c] (b) {$x_b$};
\vertex [right=1cm of c'] (a');
\vertex [left=1.3cm of c'] (b') {$x_a$};
\diagram* {
(a) -- [plain] (c) -- [plain] (b),
(a') -- [plain] (c') -- [plain] (b'),
(c) -- [dotted] (z),
(c') -- [dotted] (z)
}; 
\end{feynman} 
\end{tikzpicture}  = \frac{i}{2}\int\!\!d t\frac{\tilde{G}_{12}m_1m_2}{r}(v_1^2+v_2^2)\left(1+2\gamma_{12}\right) \; .
\label{eq:2_Jordan_final}
\end{equation}

Finally, we have the diagram with th exchange of a vector potential $\mathcal{A}_i$. Since this field does mix with the scalar $\rchi$, this diagram gives the same result written in eq. (\ref{eqn:scheme73}d). However, in order to rewrite it in a more useful way, from eq.(\ref{eq:eff_Newton}) we can easily see that
\begin{equation}
2\GN = \tilde{G}_{ab}\left(1+\frac{1-2\alpha_a^*\alpha_b^*}{1+2\alpha_a^*\alpha_b^*}\right)  = \tilde{G}_{ab}\left(1+\gamma_{ab}\right) \; .
\end{equation} 
Therefore, the last diagrams gives us 
\begin{align}
\begin{tikzpicture}[baseline]
\begin{feynman}
\vertex (z);
\vertex [above=0.8cm of z] (c');
\vertex [below=0.8cm of z] (c);
\vertex [right=1cm of c] (a);
\vertex [left=1cm of c] (b) {$x_b$};
\vertex [right=1cm of c'] (a');
\vertex [left=1cm of c'] (b') {$x_a$};
\diagram* {
(a) -- [plain] (c) -- [plain] (b),
(a') -- [plain] (c') -- [plain] (b'),
(c) -- [scalar] (c')
}; 
\end{feynman} 
\end{tikzpicture} & =-2i\int\!\!d t\frac{\tilde{G}_{12}m_1m_2}{r}\bm{v}_1\!\cdot\!\bm{v}_2 \left(1+\gamma_{12}\right) \; .
\label{eq:last_diag_Jordan}
\end{align}

Summing eqs. (\ref{eq:first_final_Jordan}), (\ref{eq:seagull_Jordan_final}), (\ref{eq:2_Jordan_final}), (\ref{eq:last_diag_Jordan}) and adding the needed kinetic term, we eventually obtain
\begin{align}
S_{1\text{PN}} & = \int\!\!d t\bigg\{\sum_a\frac{m_a}{8}v_a^4 +\notag \\
&\qquad +\frac{\tilde{G}_{12}m_1 m_2}{2r}\bigg[(v_1^2\!+\!v_2^2)-3\bm{v}_1\!\cdot\!\bm{v}_2 -\frac{(\bm{v}_1\!\cdot\!\bm{r})(\bm{v}_2\!\cdot\!\bm{r})}{r^2}+2\gamma_{12}\left(\bm{v}_1\!-\!\bm{v}_2\right)^2\bigg]\notag \\
&\qquad -\frac{\tilde{G}^2_{12}m_1 m_2(m_1\!+\!m_2)}{2r^2}(2\beta_{12}-1)\bigg\} 
\label{eq:SEIH_Jordan}
\end{align}
This result coincides with the one derived in Ref. \cite{NRGR+scalar}, and it gives us again the EIH action (\ref{eq:SEIH}) once we set to zero the charges $\alpha_a$, $\beta_a$ and to one the conformal factors, i.e. $\lambda = 0$.

\subsection{The problem of the three-vertex interaction}\label{subsub:three_vertex}

At this point we need to explain why we did not include in our computations diagrams like the one sketched in eq. (\ref{eq:three-vertex-Jordan}). In fact, since we have already obtained the correct result in eq. (\ref{eq:SEIH_Jordan}), we expect any other diagrams that might contribute to this order to give a vanishing contribution. To see why this is the case, let's not use the triplet methods, but go in the Einstein Frame, where we have the three potential scalars $\psi^*$, $\varphi^*$ and $\varsigma^*$, analogous to the one defined in eq. (\ref{eqn:scheme79}a) in the Jordan frame.

The point-particle action is given in eq. (\ref{eq:PP_Einstein_Jordan}). It has the same form of the one written in the Jordan frame, thus, the Feynman rules are exactly equal to the one written in tab. \ref{tab:Rules_Jordan_1PN}, but with the three scalar fields de-mixed.

On the other hand, the gravity part of the total action has the same form of eq. (\ref{eq:Gravity_action_Eisntein}), hence, as expected, the NRG fields and the scalar $\psi^*$ does not get mixed at quadratic order. The part of the action proportional to the Ricci scalar has been written explicitly in eq. (\ref{eq:SEH.kk}). From here, we understand that we can have cubic vertices involving only $\varphi^*$ and $\varsigma^*$, i.e. 
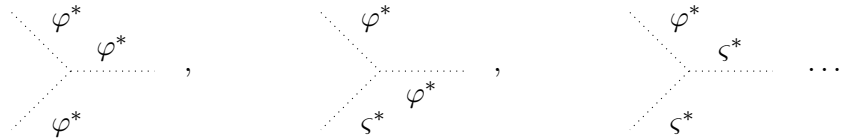
\begin{align*}
\begin{tikzpicture}[baseline]
\begin{feynman}
\vertex (z);
\vertex [above left=1.1cm of z] (a);
\vertex [below left=1.1cm of z] (b);
\vertex [right=1.1cm of z] (c);
\diagram* {
(a) --[dotted, edge label=$\varphi^*$] (z) -- [dotted, edge label=$\varphi^*$] (c),
(z) -- [dotted, edge label=$\varphi^*$] (b)
}; 
\end{feynman} 
\end{tikzpicture} \quad , & &  \begin{tikzpicture}[baseline]
\begin{feynman}
\vertex (z);
\vertex [above left=1.1cm of z] (a);
\vertex [below left=1.1cm of z] (b);
\vertex [right=1.1cm of z] (c);
\diagram* {
(a) --[dotted, edge label=$\varphi^*$] (z),
(c) -- [dotted, edge label=$\varphi^*$] (z),
(z) -- [dotted, edge label=$\varsigma^*$] (b)
}; 
\end{feynman} 
\end{tikzpicture} \quad , & & \begin{tikzpicture}[baseline]
\begin{feynman}
\vertex (z);
\vertex [above left=1.1cm of z] (a);
\vertex [below left=1.1cm of z] (b);
\vertex [right=1.1cm of z] (c);
\diagram* {
(a) --[dotted, edge label=$\varphi^*$] (z) -- [dotted, edge label=$\varsigma^*$] (c),
(z) -- [dotted, edge label=$\varsigma^*$] (b)
}; 
\end{feynman} 
\end{tikzpicture} \quad \dots \; 
\end{align*}
However, in Sec. \ref{sec:adv} we saw that with the NRG fields parametrisation of the metric, this kind of vertices enter in higher PN order diagrams.

Therefore, since from (\ref{eq:Gravity_action_Eisntein}) we can not have a vertex involving three $\psi$, the only possible diagrams that could contribute to the 1PN order action are
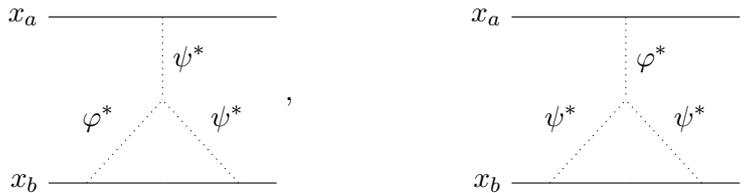
\begin{align*}
\begin{tikzpicture}[baseline]
\begin{feynman}
\vertex (z);
\vertex [below= 1.1cm of z] (c);
\vertex [above=1.1cm of z] (c');
\vertex [right=1.5cm of c] (a);
\vertex [left=1.5cm of c] (b) {$x_b$};
\vertex [right=1cm of c] (a'');
\vertex [left=1cm of c] (b'');
\vertex [right=1.5cm of c'] (a');
\vertex [left=1.5cm of c'] (b') {$x_a$};
\diagram* {
(a) -- [plain] (c) -- [plain] (b),
(a') -- [plain] (c') -- [plain] (b'),
(c') -- [dotted, edge label=$\psi^*$] (z),
(a'') -- [dotted, edge label'=$\psi^*$] (z),
(b'') -- [dotted, edge label=$\varphi^*$] (z)
}; 
\end{feynman} 
\end{tikzpicture} \; , & & 
\begin{tikzpicture}[baseline]
\begin{feynman}
\vertex (z);
\vertex [below= 1.1cm of z] (c);
\vertex [above=1.1cm of z] (c');
\vertex [right=1.5cm of c] (a);
\vertex [left=1.5cm of c] (b) {$x_b$};
\vertex [right=1cm of c] (a'');
\vertex [left=1cm of c] (b'');
\vertex [right=1.5cm of c'] (a');
\vertex [left=1.5cm of c'] (b') {$x_a$};
\diagram* {
(a) -- [plain] (c) -- [plain] (b),
(a') -- [plain] (c') -- [plain] (b'),
(c') -- [dotted, edge label=$\varphi^*$] (z),
(a'') -- [dotted, edge label'=$\psi^*$] (z),
(b'') -- [dotted, edge label=$\psi^*$] (z)
}; 
\end{feynman} 
\end{tikzpicture} \; .
\end{align*}
Diagrams involving a scalar $\varsigma$ eventually end on a vertex of order $v^2$, making the whole diagrams higher order in the PN expansion. The two diagrams sketched above come from the kinetic term of the scalar field, hence
\begin{equation}
\frac{\mpl^2}{2}\int\!\!d^4 x\sqrt{\gamma^*}\left\{e^{-4\phi^*}\dot{\rchi_*^2}+\gamma^{ij}D^*_i\rchi_*D^*_j\rchi_*\right\}\longrightarrow -\frac{2}{\mpl}\int\!\!d^4x\, \varphi\dot{\psi}^2 \; ,
\end{equation}
where in the last step we normalized every field as in eq. (\ref{eq:norm.field.NRG}). Similarly to what happened in pure NRGR (see eq. (\ref{eq:three_vertex_phi})) the above term contributes to a higher PN order, because it contains two times derivatives. Hence, we understand that also in this scalar-tensor theory we do not have any cubic vertex topology diagram at 1PN order. In a sense, the advantages of NRG fields described in Sec. \ref{sec:adv} seems to hold also in scalar-tensor theories.

For what we have just said, it could be convenient to introduce from the beginning the de-mixed fields written in eqs. (\ref{eqn:scheme78}), and then define again the triplet $\varphi^A$ with the three scalar $\varphi^*, \varsigma^*$ and $\psi^*$. In this case we have a mixing matrix which is diagonal, and reduce the number of diagrams to display at every PN order.

\section{The dissipative sector up to 2.5PN order}\label{Sec:Diss_Jordan}

Let's now turn on again the radiative fields $\bar{\Xi} =\{\bar{\Phi}^A, \bar{A}_i, \bar{\sigma}_{ij}\}$, and compute also the dissipative part of the action (\ref{eq:SNR_STtheory}). As explain in Sec. \ref{sec:multipole}, we need to perform a multipole expansion similar to eq. (\ref{eq:multipole_generic}) for each radiation field. Once we have done this, we can isolate the relevant part for the dissipative contribution from the point-particle action (\ref{eq:final_pp_Jordan}), i.e.
\begin{align}
S^{\text{pp}}_{\text{diss}} = & -\sum_a\frac{m_a}{\mpl}\int\!\!d t\!\left[\bar{\phi}\!+\!x_a^i\partial_i\bar{\phi}\!+\!\frac{1}{2}\left(3v_a^2+x_a^ix_a^j\partial_i\partial_j\right)\bar{\phi}\right]-\sum_a\frac{m_a}{\mpl^2}\int\!\!d t\!\!\intvec{k}e^{i\bm{k}\cdot\bm{x}_a}\varphi_{\bm{k}}\bar{\phi}\, \notag \\
&+\sum_a\frac{m_a}{\mpl}\int\!\!d t\left[\bar{A}_i\!+\!x^j_a\partial_j\bar{A}_i\right]v_a^i+\sum_a\frac{m_a}{2\mpl}\int\!\!d t\left(\bar{\sigma}_{ij}+\frac{\bar{\sigma}}{3}\delta_{ij}\right)v_a^iv_a^j  \notag \\
& + \sum_a\frac{m_a}{\mpl}\alpha_a\int\!\!d t\left[\bar{\rchi}\!+\!x_a^i\partial_i\bar{\rchi}\!+\!\frac{1}{2}\left(-v_a^2+x_a^ix_a^j\partial_i\partial_j\right)\bar{\rchi}\right] 
\notag \\
& + \sum_a\frac{m_a}{\mpl}\int\!\!d t\!\!\intvec{k}\left[2\beta_a\psi_{\bm{k}}\bar{\rchi}+\alpha_a\left(\phi_{\bm{k}}\bar{\rchi}+\psi_{\bm{k}}\bar{\phi}\right)\right]e^{i\bm{k}\cdot\bm{x}_a} \; ,
\label{eq:Spp.diss.NRG2-2}
\end{align}
where every radiative field is evaluated in $(x^0,0)$. From this action we find the Feynman rules listed in tab. \ref{table:ppradmultiNRGST}.
\begin{table}[t]
\begin{center}
\begin{tabular}{|c|c|c|}
\hline
Diagrammatic expression & Scaling & Explicit expression \\
\hline
\begin{tikzpicture}[baseline]
\begin{feynman}
\vertex (z);
\vertex [left=0.2cm of z, label=180:$t$] (c');
\vertex [above=0.5cm of c'] (a);
\vertex [above=0.2cm of a] (a');
\vertex [below=0.5cm of c'] (b);
\vertex [below=0.2cm of b] (b');
\vertex [right=0.7cm of z] (c); 
\diagram* {
(a') -- [draw=none] (a) -- [plain] (c') -- [plain] (b) -- [draw=none] (b'),
(c) -- [photon] (c')
}; 
\end{feynman} 
\end{tikzpicture} & $\sim L^{1/2}v^{1/2}$ & \(\displaystyle
i\frac{m_a}{\mpl}\int\!\!d t\,\left(\alpha_a\bar{\rchi}-\bar{\phi}\right) \) \\
\hline
\begin{tikzpicture}[baseline]
\begin{feynman}
\vertex (z);
\vertex [empty dot, minimum size=0.4cm, left=0.2cm of z, label=180:$t$] (c') {1};
\vertex [above=0.5cm of c'] (a);
\vertex [above=0.2cm of a] (a');
\vertex [below=0.5cm of c'] (b);
\vertex [below=0.2cm of b] (b');
\vertex [right=0.7cm of z] (c);
\vertex [right=0.5cm of z] (d); 
\diagram* {
(a') -- [draw=none] (a) -- [plain] (c') -- [plain] (b) -- [draw=none] (b'),
(c) -- [draw=none](d) -- [photon] (c')
}; 
\end{feynman} 
\end{tikzpicture} & $\sim L^{1/2}v^{3/2}$ & \(\displaystyle
i\frac{m_a}{\mpl}\int\!\!d t\,\left(\alpha_ax_a^i\partial_i\bar{\rchi}-x_a^i\partial_i\bar{\phi}\right) \) \\
\hline 
\begin{tikzpicture}[baseline]
\begin{feynman}
\vertex (z);
\vertex [empty dot, minimum size=0.4cm, left=0.2cm of z, label=180:$t$] (c') {2};
\vertex [above=0.5cm of c'] (a);
\vertex [above=0.2cm of a] (a');
\vertex [below=0.5cm of c'] (b);
\vertex [below=0.2cm of b] (b');
\vertex [right=0.7cm of z] (c);
\vertex [right=0.5cm of z] (d); 
\diagram* {
(a') -- [draw=none] (a) -- [plain] (c') -- [plain] (b) -- [draw=none] (b'),
(c) -- [draw=none](d) -- [photon] (c')
}; 
\end{feynman} 
\end{tikzpicture} & $\sim L^{1/2}v^{5/2}$ & \(\displaystyle \begin{aligned}[t]
i\frac{m_a}{2\mpl}&\int\!\!d t \bigg(\alpha_ax_a^ix_a^j\partial_i\partial_j\bar{\rchi}-\alpha_av_a^2\bar{\rchi}+\frac{v_a^2}{3}\bar{\sigma}\\
& \qquad\quad  -3v_a^2\bar{\phi}-x_a^ix_a^j\partial_i\partial_j\bar{\phi}\bigg)  \end{aligned}\) \\
\hline 
\begin{tikzpicture}[baseline]
\begin{feynman}
\vertex (z);
\vertex (c');
\vertex [above=0.04cm of c', label=180:$t$] (d');
\vertex [above=0.6cm of c'] (a);
\vertex [above=0.2cm of a] (a');
\vertex [below=0.6cm of c'] (b);
\vertex [below=0.2cm of b] (b');
\vertex [right=0.9cm of z] (c);
\vertex [above=0.04cm of c] (d); 
\diagram* {
(a') -- [draw=none] (a) -- [plain] (c') -- [plain] (b) -- [draw=none] (b'),
(c) -- [photon] (c'),
(d) -- [photon] (d')
}; 
\end{feynman} 
\end{tikzpicture} & $\sim L^{1/2}v^{3/2}$ & 
\(\displaystyle
i\frac{m_a}{\mpl}\int\!\!d t\,\bar{A}_iv_a^i \) \\
\hline 
\begin{tikzpicture}[baseline]
\begin{feynman}
\vertex (z);
\vertex [empty dot, minimum size=0.4cm, left=0.2cm of z, label=180:$t$] (c') {1};
\vertex [right=0.2cm of c'] (d'');
\vertex [above=0.04cm of d''] (d');  
\vertex [above=0.6cm of c'] (a);
\vertex [above=0.2cm of a] (a');
\vertex [below=0.6cm of c'] (b);
\vertex [below=0.2cm of b] (b');
\vertex [right=0.5cm of z] (c);
\vertex [right=0.7cm of z] (f);
\vertex [above=0.04cm of c] (d);  
\diagram* {
(a') -- [draw=none] (a) -- [plain] (c') -- [plain] (b) -- [draw=none] (b'),
(f)--[draw=none] (c) -- [photon] (c'),
(d) -- [photon] (d')
}; 
\end{feynman} 
\end{tikzpicture} & $\sim L^{1/2}v^{5/2}$ & 
\(\displaystyle
i\frac{m_a}{\mpl}\int\!\!d t\,\left(x_a^j\partial_j\bar{A}_i\right)v_a^i \) \\
\hline 
\begin{tikzpicture}[baseline]
\begin{feynman}
\vertex (z);
\vertex [left=0.2cm of z, label=180:$t$] (c');
\vertex [above=0.6cm of c'] (a);
\vertex [above=0.2cm of a] (a');
\vertex [below=0.6cm of c'] (b);
\vertex [below=0.2cm of b] (b');
\vertex [right=0.7cm of z] (c); 
\diagram* {
(a') -- [draw=none] (a) -- [plain] (c') -- [plain] (b) -- [draw=none] (b'),
(c) -- [gluon] (c')
}; 
\end{feynman} 
\end{tikzpicture} & $\sim L^{1/2}v^{5/2}$ & 
\(\displaystyle
i\frac{m_a}{2\mpl}\int\!\!d t\bar{\sigma}_{ij}v_a^iv_a^j \) \\
\hline 
\begin{tikzpicture}[baseline]
\begin{feynman}
\vertex (z);
\vertex [left=0.2cm of z, label=180:$t$, label=75:$\!\!A\quad \bm{k}$] (c');
\vertex [above=0.7cm of c'] (a);
\vertex [above=0.2cm of a] (a');
\vertex [below=0.7cm of c'] (b);
\vertex [below=0.2cm of b] (b');
\vertex [right=0.7cm of z] (z');
\vertex [above=0.5cm of z'] (c); 
\vertex [below=0.5cm of z'] (d); 
\diagram* {
(a') -- [draw=none] (a) -- [plain] (c') -- [plain] (b) -- [draw=none] (b'),
(c') -- [dotted] (c),
(c') -- [photon] (d)
}; 
\end{feynman} 
\end{tikzpicture} & $\sim v^{5/3}$ & 
\(\displaystyle \begin{aligned}[t]
i\frac{m_a}{\mpl^2}\int\!\!d t\!\!\intvec{k}e^{i\bm{k}\cdot\bm{x}_a}\bigg[&\left(\alpha_a\delta^{A3}-\delta^{A1}\right)\bar{\phi} \\
& +\left(\alpha_a\delta^{A1}+2\beta_a\delta^{A3}\right)\bar{\rchi}\bigg] \\
&\  \end{aligned}\) \\
\hline
\end{tabular}
\caption{Point-particles radiative fields rules}\label{table:ppradmultiNRGST}
\end{center}
\end{table}

\subsection{The cubic interaction vertex}

Similarly to what we said in Sec. \ref{sec:Rad_sec_NRG}, in particular at p.~\pageref{subsubsec:threevertex}, in order to get the action in the radiative sector at 2.5PN order we also need to compute a diagram involving a cubic interaction between gravity and scalar fields. Indeed, similarly to eq. (\ref{eq:threevertexNRGrad}), we can easily see that the lowest order cubic interaction vertex has the following structure and scaling
\begin{equation}
\begin{tikzpicture}[baseline]
\begin{feynman}
\vertex (z);
\vertex [above left=1.1cm of z, label=270:$\!\!\varphi^A$] (a);
\vertex [below left=1.1cm of z, label=90:$\!\!\varphi^B$] (b);
\vertex [right=1.1cm of z, label=90:$\bar{\Xi}$] (c);
\diagram* {
(a) --[dotted] (z) -- [gluon] (c),
(z) -- [dotted] (b)
}; 
\end{feynman} 
\end{tikzpicture} \rightarrow \frac{1}{\mpl}\int\!\!d^4\!x\,\bar{\Xi}\partial_i\varphi^A\partial^i\varphi^B \sim \frac{v^{5/2}}{L^{1/2}} \; ,
\label{eq:threevertexNRGRJordan}
\end{equation}
 where we recall that $\bar{\Xi}$ represents a generic radiative field. Looking at the gravity action (\ref{eq:Sg.NRGR.Jordan}) and its gauge-fixing (\ref{eq:GF_Jordan_gen}), we immediately see that a contribution to such a vertex can come from
\begin{subequations}
\label{eqn:scheme90}
\begin{align}
\mathcal{C}_1 & = \frac{\mpl^2}{2}\int\!\!d^4\!x\sqrt{\gamma}F(\rchi)\left(-2\gamma^{ij}D_i\phi D_j\phi\right) \; , \\
\mathcal{C}_2 & = \frac{\mpl^2}{2}\int\!\!d^4\!x\sqrt{\gamma}F'(\rchi)\left(-2\gamma^{ij}D_i\phi D_j\rchi\right) \; , \\
\mathcal{C}_3 & = \frac{\mpl^2}{2}\int\!\!d^4\!x\sqrt{\gamma}\left[Z(\rchi)+\frac{\big(F'(\rchi)\big)^2}{2F(\rchi)}\right]\left(-\gamma^{ij}D_i\rchi D_j\rchi\right) \; , \\
\mathcal{C}_4 & = \frac{\mpl^2}{2}\int\!\!d^4\!x\sqrt{\gamma}\left\{F(\rchi)\left[\bar{\Rc}_\gamma-\frac{1}{2}\gamma_{ij}\bar{\Gamma}^i\bar{\Gamma}^j\right]+F'(\rchi)\bar{\Gamma}^i D_i\rchi\right\} \; , 
\end{align}
\end{subequations}
where the conformal factors $F(\rchi)$ and $Z(\rchi)$ are given by eq. (\ref{eq:conformal_factor}). Expanding these terms, we eventually find that the Feynman rule for the vertex sketched in eq. (\ref{eq:threevertexNRGRJordan}) can be found from
\begin{align}
\sum_{\kappa=1}^{4}\mathcal{C}_\kappa\rightarrow & \frac{1}{\mpl}\int\!\!d t\!\!\intvec{k} k_ik_j\bar{\sigma}^{ij}\bigg[\varphi_{\bm{k}}\varphi_{-\bm{k}}\!+\!\lambda\varphi_{\bm{k}}\psi_{-\bm{k}}\!+\!\frac{1}{2}\left(1\!+\!\frac{\lambda^2}{2}\right)\psi_{\bm{k}}\psi_{-\bm{k}}\notag \\
&\qquad\qquad\qquad\qquad\quad -\frac{1}{48}\varsigma_{\bm{k}}\varsigma_{-\bm{k}}-\frac{\lambda}{4}\varsigma_{\bm{k}}\psi_{-\bm{k}}\bigg]\notag \\
-& \frac{1}{2\mpl}\int\!\!d t\!\!\intvec{k} \modul{k}^2\bigg\{\left(\frac{\bar{\sigma}}{3}+2\lambda\bar{\rchi}\right)\left[\varphi_{\bm{k}}\varphi_{-\bm{k}}\!+\!\lambda\varphi_{\bm{k}}\psi_{-\bm{k}}\!+\!\frac{1}{2}\left(1\!+\!\frac{\lambda^2}{2}\right)\psi_{\bm{k}}\psi_{-\bm{k}}\right]\notag \\
&\qquad\qquad\qquad\quad+\left(\frac{\bar{\sigma}}{2}-\lambda\bar{\rchi}\right)\frac{1}{24}\varsigma_{\bm{k}}\varsigma_{-\bm{k}}+\left(\frac{\bar{\sigma}}{3}-2\lambda\bar{\rchi}\right)\frac{\lambda}{4}\varsigma_{\bm{k}}\psi_{-\bm{k}}\bigg\} \; .
\end{align}
We write the two final Feynman rules in tab. \ref{table:gravv.NRGJordan}.

\begin{table}[t]
\begin{center}
\begin{tabular}{|c|c|c|}
\hline
Diagrammatic expression & Scaling & Explicit expression \\
\hline
\begin{tikzpicture}[baseline]
\begin{feynman}
\vertex (z);
\vertex [right=1cm of z, label=90:$\left\{\bar{\sigma}\text{,}\,\bar{\rchi}\right\}$] (c);
\vertex [above left=1cm of z, label=135:$t$, label=225:$A$] (a);
\vertex [below left=1cm of z, label=225:$t'$, label=135:$B$] (b);
\diagram* {
(a) -- [dotted, edge label=$\bm{k}$] (z),
(c) -- [photon] (z),
(b) -- [dotted, edge label'=$\bm{q}$] (z)
}; 
\end{feynman} 
\end{tikzpicture} & \(\displaystyle \sim \frac{v^{5/2}}{L^{1/2}}\) & 
\(\displaystyle  \begin{aligned}[t]
&\frac{i(2\pi)^3}{2\mpl}\frac{\delta^{(3)}\!(\bm{k}\!+\!\bm{q})}{|\bm{k}|^2}\bigg\{\left(\frac{\bar{\sigma}}{3}\!+\!2\lambda\bar{\rchi}\right)\bigg[2D_{1A}D_{B1} \\
& \; +2\lambda D_{1(A}D_{B)3}\!+\!\left(1\!+\!\frac{\lambda^2}{2}\right)D_{3A}D_{B3}\bigg] \\
&\; +\left(\frac{\bar{\sigma}}{2}\!-\!\lambda\bar{\rchi}\right)\frac{D_{2A}D_{B2}}{12}\\
&\; +\left(\frac{\bar{\sigma}}{3}\!-\!2\lambda\bar{\rchi}\right)\frac{\lambda}{2}D_{2(A}D_{B)3}\bigg\}\delta(t\!-\!t') \\
& \ \end{aligned}\) \\
\hline
\begin{tikzpicture}[baseline]
\begin{feynman}
\vertex (z);
\vertex [right=1cm of z] (c);
\vertex [above left=1cm of z, label=135:$t$, label=225:$A$] (a);
\vertex [below left=1cm of z, label=225:$t'$, label=135:$B$] (b);
\diagram* {
(a) -- [dotted, edge label=$\bm{k}$] (z),
(c) -- [gluon] (z),
(b) -- [dotted, edge label'=$\bm{q}$] (z)
}; 
\end{feynman} 
\end{tikzpicture} & \(\displaystyle \sim \frac{v^{5/2}}{L^{1/2}}\) & 
\(\displaystyle \begin{aligned}[t]
&-\frac{i(2\pi)^3}{\mpl}\frac{\delta^{(3)}\!(\bm{k}\!+\!\bm{q})}{|\bm{k}|^2|\bm{q}|^2}k_ik_j\bar{\sigma}^{ij}\bigg\{2D_{1A}D_{B1} \\
& \; +2\lambda D_{1(A}D_{B)3}\!+\!\left(1\!+\!\frac{\lambda^2}{2}\right)D_{3A}D_{B3}\\
&\; -\frac{1}{24}D_{2A}D_{B2}-\frac{\lambda}{4}D_{2(A}D_{B)3}\bigg\}\delta(t\!-\!t') \\
& \ \end{aligned}\) \\
\hline
\end{tabular}
\caption{Cubic vertices}
\label{table:gravv.NRGJordan}
\end{center}
\end{table}

Having these, it is straightforward to see that what we call $S_{\text{diss}}$ in eq. (\ref{eq:SNR_STtheory}) is given diagrammatically by
\begin{align}
iS_{\text{diss}} & = \sum_a\begin{tikzpicture}[baseline]
\begin{feynman}
\vertex (z);
\vertex [below=0.1cm of z] (f);
\vertex [above=0.1cm of z] (f');
\vertex [right=0.5cm of f'] (c);
\vertex [right=0.7cm of c] (d);
\vertex [above=0.5cm of d] (d');
\diagram* {
(f') -- [plain] (c) -- [plain] (d),
(f') -- [draw=none] (z) -- [draw=none] (f),
(d') -- [photon] (c)
}; 
\end{feynman} 
\end{tikzpicture} +
\sum_a\begin{tikzpicture}[baseline]
\begin{feynman}
\vertex (z);
\vertex [below=0.1cm of z] (f);
\vertex [above=0.1cm of z] (f');
\vertex [right=0.5cm of f', empty dot, minimum size=0.4cm] (c) {1};
\vertex [right=0.7cm of c] (d);
\vertex [above=0.5cm of d] (d');
\diagram* {
(f') -- [plain] (c) -- [plain] (d),
(f') -- [draw=none] (z) -- [draw=none] (f),
(d') -- [photon] (c)
}; 
\end{feynman} 
\end{tikzpicture} +
\sum_a\begin{tikzpicture}[baseline]
\begin{feynman}
\vertex (z);
\vertex [below=0.1cm of z] (f);
\vertex [above=0.1cm of z] (f');
\vertex [right=0.5cm of f'] (c);
\vertex [above left=0.04cm of c] (c'');
\vertex [right=0.7cm of c] (d);
\vertex [above=0.5cm of d] (d');
\vertex [above left=0.04cm of d'] (d'');
\vertex [left=0.03cm of c] (h);
\diagram* {
(f') -- [plain] (c) -- [plain] (d),
(f') -- [draw=none] (z) -- [draw=none] (f),
(d') -- [photon] (c),
(d'')-- [photon] (c''),
(h) -- [plain] (c'')
}; 
\end{feynman} 
\end{tikzpicture} \notag \\
& +\sum_a\begin{tikzpicture}[baseline]
\begin{feynman}
\vertex (z);
\vertex [below=0.1cm of z] (f);
\vertex [above=0.1cm of z] (f');
\vertex [right=0.5cm of f', empty dot, minimum size=0.4cm] (c) {2};
\vertex [right=0.7cm of c] (d);
\vertex [above=0.5cm of d] (d');
\diagram* {
(f') -- [plain] (c) -- [plain] (d),
(f') -- [draw=none] (z) -- [draw=none] (f),
(d') -- [photon] (c)
}; 
\end{feynman} 
\end{tikzpicture} +
\sum_a\begin{tikzpicture}[baseline]
\begin{feynman}
\vertex (z);
\vertex [below=0.1cm of z] (f);
\vertex [above=0.1cm of z] (f');
\vertex [right=0.5cm of f', empty dot, minimum size=0.4cm] (c) {1};
\vertex [right=0.7cm of c] (d);
\vertex [above=0.5cm of d] (d');
\vertex [above left=0.045cm of d'] (g');
\vertex [above right=0.2cm of c] (h);
\vertex [above left=0.045cm of h] (c'');
\diagram* {
(f') -- [plain] (c) -- [plain] (d),
(f') -- [draw=none] (z) -- [draw=none] (f),
(d') -- [photon] (h),
(g') -- [photon] (c'')
}; 
\end{feynman} 
\end{tikzpicture} +\sum_a\begin{tikzpicture}[baseline]
\begin{feynman}
\vertex (z);
\vertex [below=0.1cm of z] (f);
\vertex [above=0.1cm of z] (f');
\vertex [right=0.5cm of f'] (c);
\vertex [right=0.7cm of c] (d);
\vertex [above=0.5cm of d] (d');
\diagram* {
(f') -- [plain] (c) -- [plain] (d),
(f') -- [draw=none] (z) -- [draw=none] (f),
(d') -- [gluon] (c)
}; 
\end{feynman} 
\end{tikzpicture} \notag \\
& + \begin{tikzpicture}[baseline]
\begin{feynman}
\vertex (z);
\vertex [above=0.7cm of z] (c');
\vertex [below=0.6cm of z] (c);
\vertex [right=1cm of c] (a);
\vertex [left=1cm of c] (b);
\vertex [right=1cm of c'] (a');
\vertex [below=0.52cm of a'] (d);
\vertex [left=1cm of c'] (b');
\diagram* {
(a) -- [plain] (c) -- [plain] (b),
(a') -- [plain] (c') -- [plain] (b'),
(c) -- [dotted] (z) -- [dotted] (c'),
(d) -- [photon] (c')
}; 
\end{feynman} 
\end{tikzpicture}+
\begin{tikzpicture}[baseline]
\begin{feynman}
\vertex (z);
\vertex [right=1cm of z] (d);
\vertex [above=0.7cm of z] (c');
\vertex [below=0.6cm of z] (c);
\vertex [right=1cm of c] (a);
\vertex [left=1cm of c] (b);
\vertex [right=1cm of c'] (a');
\vertex [left=1cm of c'] (b');
\diagram* {
(a) -- [plain] (c) -- [plain] (b),
(a') -- [plain] (c') -- [plain] (b'),
(c) -- [dotted] (z) -- [dotted] (c'),
(d) -- [photon] (z)
}; 
\end{feynman} 
\end{tikzpicture} +
\begin{tikzpicture}[baseline]
\begin{feynman}
\vertex (z);
\vertex [right=1cm of z] (d);
\vertex [above=0.7cm of z] (c');
\vertex [below=0.6cm of z] (c);
\vertex [right=1cm of c] (a);
\vertex [left=1cm of c] (b);
\vertex [right=1cm of c'] (a');
\vertex [left=1cm of c'] (b');
\diagram* {
(a) -- [plain] (c) -- [plain] (b),
(a') -- [plain] (c') -- [plain] (b'),
(c) -- [dotted] (z) -- [dotted] (c'),
(d) -- [gluon] (z)
}; 
\end{feynman} 
\end{tikzpicture} +\Ord{L^{1/2}v^{7/2}} \; .
\label{eq:Sdiss.NRG}
\end{align}

\subsection{Diagrams computation}

The first six diagrams of eq. (\ref{eq:Sdiss.NRG}) follow directly from their Feynman rules (see tab. \ref{table:ppradmultiNRGST}), i.e. 
\begin{subequations}
\label{eqn:scheme92}
\begin{align}
\sum_a\begin{tikzpicture}[baseline]
\begin{feynman}
\vertex (z);
\vertex [below=0.1cm of z] (f);
\vertex [above=0.1cm of z] (f');
\vertex [right=0.5cm of f'] (c);
\vertex [right=0.7cm of c] (d);
\vertex [above=0.5cm of d] (d');
\diagram* {
(f') -- [plain] (c) -- [plain] (d),
(f') -- [draw=none] (z) -- [draw=none] (f),
(d') -- [photon] (c)
}; 
\end{feynman} 
\end{tikzpicture} & = i\sum_{a}\frac{m_a}{\mpl}\int\!\!d t\left(\alpha_a\bar{\rchi}-\bar{\phi}\right)  \; ,\\
\sum_a\begin{tikzpicture}[baseline]
\begin{feynman}
\vertex (z);
\vertex [below=0.1cm of z] (f);
\vertex [above=0.1cm of z] (f');
\vertex [right=0.5cm of f', empty dot, minimum size=0.4cm] (c) {1};
\vertex [right=0.7cm of c] (d);
\vertex [above=0.5cm of d] (d');
\diagram* {
(f') -- [plain] (c) -- [plain] (d),
(f') -- [draw=none] (z) -- [draw=none] (f),
(d') -- [photon] (c)
}; 
\end{feynman} 
\end{tikzpicture} & = i\sum_{a}\frac{m_a}{\mpl}\int\!\!d t\,x_a^i\left(\alpha_a\partial_i\bar{\rchi}-\partial_i\bar{\phi}\right) =  i\sum_{a}\frac{\alpha_a m_a}{\mpl}\int\!\!d t\,x_a^i\partial_i\bar{\rchi}  \; , \\
\sum_a\begin{tikzpicture}[baseline]
\begin{feynman}
\vertex (z);
\vertex [below=0.1cm of z] (f);
\vertex [above=0.1cm of z] (f');
\vertex [right=0.5cm of f'] (c);
\vertex [above left=0.04cm of c] (c'');
\vertex [right=0.7cm of c] (d);
\vertex [above=0.5cm of d] (d');
\vertex [above left=0.04cm of d'] (d'');
\vertex [left=0.03cm of c] (h);
\diagram* {
(f') -- [plain] (c) -- [plain] (d),
(f') -- [draw=none] (z) -- [draw=none] (f),
(d') -- [photon] (c),
(d'')-- [photon] (c''),
(h) -- [plain] (c'')
}; 
\end{feynman} 
\end{tikzpicture} &= -i\sum_a\frac{m_a}{\mpl}\int\!\!d t\,v_a^i \bar{A}_i=0 \; , \\
\sum_a\begin{tikzpicture}[baseline]
\begin{feynman}
\vertex (z);
\vertex [below=0.1cm of z] (f);
\vertex [above=0.1cm of z] (f');
\vertex [right=0.5cm of f', empty dot, minimum size=0.4cm] (c) {2};
\vertex [right=0.7cm of c] (d);
\vertex [above=0.5cm of d] (d');
\diagram* {
(f') -- [plain] (c) -- [plain] (d),
(f') -- [draw=none] (z) -- [draw=none] (f),
(d') -- [photon] (c)
}; 
\end{feynman} 
\end{tikzpicture} & = -i\sum_a\frac{m_a}{2\mpl}\int\!\!d t\!\left[v_a^2\left(3\bar{\phi}\!+\!\alpha_a\bar{\rchi}\!-\!\frac{\bar{\sigma}}{3}\right)\!+\!x_a^ix_a^j\partial_i\partial_j\left(\bar{\phi}\!-\!\alpha_a\bar{\rchi}\right)\right]  \; , \\
\sum_a\begin{tikzpicture}[baseline]
\begin{feynman}
\vertex (z);
\vertex [below=0.1cm of z] (f);
\vertex [above=0.1cm of z] (f');
\vertex [right=0.5cm of f', empty dot, minimum size=0.4cm] (c) {1};
\vertex [right=0.7cm of c] (d);
\vertex [above=0.5cm of d] (d');
\vertex [above left=0.045cm of d'] (g');
\vertex [above right=0.2cm of c] (h);
\vertex [above left=0.045cm of h] (c'');
\diagram* {
(f') -- [plain] (c) -- [plain] (d),
(f') -- [draw=none] (z) -- [draw=none] (f),
(d') -- [photon] (h),
(g') -- [photon] (c'')
}; 
\end{feynman} 
\end{tikzpicture} & = i\sum_{a}\frac{m_a}{\mpl}\int\!\!d t\,x_a^j\partial_j \bar{A}_i v_a^i  \; , \\
\sum_a\begin{tikzpicture}[baseline]
\begin{feynman}
\vertex (z);
\vertex [below=0.1cm of z] (f);
\vertex [above=0.1cm of z] (f');
\vertex [right=0.5cm of f'] (c);
\vertex [right=0.7cm of c] (d);
\vertex [above=0.5cm of d] (d');
\diagram* {
(f') -- [plain] (c) -- [plain] (d),
(f') -- [draw=none] (z) -- [draw=none] (f),
(d') -- [gluon] (c)
}; 
\end{feynman} 
\end{tikzpicture} & = i\sum_{a}\frac{m_a}{2\mpl}\int\!\!d t\,\bar{\sigma}_{ij} v_a^iv_a^j \; ,
\end{align}
\end{subequations}
where, since we put the origin of our frame in the CoM of the binary, in eqs. (\ref{eqn:scheme92}b) and (\ref{eqn:scheme92}c) we set to zero the terms proportional to respectively the position of the CoM and the total momentum in the CoM. 

Comparing these with eqs. (\ref{eqn:scheme48}), we immediately see an important difference w.r.t the pure GR computations: in the case $\alpha_1\neq\alpha_2$, i.e. when the two bodies carry a different scalar charges, from eq. (\ref{eqn:scheme92}b) one expects a non-zero dipole scalar emission in this scalar-tensor theory. We will briefly discuss the consequences of such a term in Sec. \ref{Sec:Obs-scalar}. 

The last three diagrams of $S_\text{diss}$ involve non linearities, hence their computations are a bit more involved. 
\begin{align}
\begin{tikzpicture}[baseline]
\begin{feynman}
\vertex [label=135:$B$, label=45:$\bm{q}$, label=270:$t'$] (c);
\vertex [above=0.9cm of c](z);
\vertex [above=1cm of z, label=90:$A\text{, } t$, label=225:$\bm{k}$] (c');
\vertex [right=1.3cm of c] (a);
\vertex [left=1.3cm of c, label=180:$x_b$] (b);
\vertex [right=1.3cm of c'] (a');
\vertex [below=0.82cm of a'] (d);
\vertex [left=1.3cm of c', label=180:$x_a$] (b');
\diagram* {
(a) -- [plain] (c) -- [plain] (b),
(a') -- [plain] (c') -- [plain] (b'),
(c) -- [dotted] (z) -- [dotted] (c'),
(d) -- [photon] (c')
}; 
\end{feynman} 
\end{tikzpicture} & = i\sum_{a\neq b}\frac{m_am_b}{\mpl^3}\int\!\!d t\!\!\intvec{k}\frac{e^{-i\bm{k}\cdot\bm{x}_{ab}}}{\modul{k}^2}\bigg\{\bar{\phi}\left[D_{11}\!-\!\left(\alpha_a\!+\!\alpha_b\right)D_{13}\!+\!\alpha_a\alpha_bD_{33}\right]\notag \\
&\qquad\qquad -\bar{\rchi}\left[\alpha_aD_{11}\!+\!\left(2\beta_a\!+\!\alpha_a\alpha_b\right)D_{13}\!-\!2\beta_a\alpha_bD_{33}\right]\bigg\} \notag \\
& =\frac{i}{\mpl}\sum_{a\neq b}\int\!\!d t\frac{\tilde{G}_{ab} m_am_b}{|\bm{x}_{ab}|}\left\{\bar{\phi}+\frac{\lambda}{2}\bar{\rchi}\!+\frac{4\beta_a^*\alpha_b^*-\alpha_a^*}{1\!+\!2\alpha_a^*\alpha_b^*}\frac{\lambda}{\lambda^*}\bar{\rchi}\right\} \; ,
\label{eq:nonlinNRGJ}
\end{align}
where we defined as usual $\bm{x}_{ab}\equiv \bm{x}_{a}-\bm{x}_{b}$, and used relations (\ref{eq:lambdastar}), (\ref{eqn:scheme81}) and (\ref{eq:kuntz_eff_Newton}) to write the result in a more compact way using the Einstein-frame charges.

Then we have to compute the two diagrams involving the cubic gravitational and scalar interactions. Let's first analyse the one with an emission of a scalar field, i.e.
\begin{align}
\begin{tikzpicture}[baseline]
\begin{feynman}
\vertex [label=135:$B$, label=45:$\bm{q}$, label=270:$t'$] (c);
\vertex [above=0.9cm of c] (z);
\vertex [right=1.2cm of z] (d);
\vertex [above=0.9cm of z, label=90:$t$, label=225:$A$, label=315:$\bm{k}$] (c');
\vertex [right=1.2cm of c] (a);
\vertex [left=1.2cm of c, label=180:$x_b$] (b);
\vertex [right=1.2cm of c'] (a');
\vertex [left=1.2cm of c', label=180:$x_a$] (b');
\diagram* {
(a) -- [plain] (c) -- [plain] (b),
(a') -- [plain] (c') -- [plain] (b'),
(c) -- [dotted] (z) -- [dotted] (c'),
(d) -- [photon] (z)
}; 
\end{feynman} 
\end{tikzpicture} & = -\frac{i}{2\mpl}\sum_{a\neq b}\int\!\!d t\frac{\GN m_a m_b}{|\bm{x}_{ab}|}\bigg\{\left(\frac{\bar{\sigma}}{3}\!+\!2\lambda\bar{\rchi}\right)\mathcal{D}_1(\lambda, \alpha_a)\notag \\
& \qquad\qquad +\!\left(\frac{\bar{\sigma}}{2}-\lambda\bar{\rchi}\right)\mathcal{D}_2(\lambda, \alpha_a)+\!\left(\frac{\bar{\sigma}}{3}-2\lambda\bar{\rchi}\right)\mathcal{D}_3(\lambda, \alpha_a)\bigg\} \; .
\label{eq:nonlinNRGHordan1step}
\end{align}
After some lengthy calculations, one can see that the three coefficients $\mathcal{D}_1(\lambda, \alpha_a)$, $\mathcal{D}_2(\lambda, \alpha_a)$ and $\mathcal{D}_3(\lambda, \alpha_a)$ are given by
\begin{subequations}
\label{eqn:scheme93}
\begin{align}
\mathcal{D}_1(\lambda, \alpha_a) & = \frac{1+2\alpha_a^*\alpha_b^*}{2+3\lambda^2}\left(\frac{2+3\lambda^2}{2}-\frac{3\lambda^2}{4}\left(1-\gamma_{ab}\right)\right) \; ,\\
\mathcal{D}_2(\lambda, \alpha_a)& = \frac{1+2\alpha_a^*\alpha_b^*}{2+3\lambda^2}\left(\frac{3\lambda^2}{2}\left(1-\gamma_{ab}\right)\right) = - \mathcal{D}_3(\lambda, \alpha_a) \; ,
\end{align}
\end{subequations}
where $\gamma_{ab}$ is the PPN parameter defined in eq. (\ref{eq:PPNgamma}). Inserting these two equations in (\ref{eq:nonlinNRGHordan1step}) we eventually arrive to 
\begin{align}
\begin{tikzpicture}[baseline]
\begin{feynman}
\vertex (z);
\vertex [below=0.9cm of z] (c);
\vertex [right=1.2cm of z] (d);
\vertex [above=0.9cm of z] (c');
\vertex [right=1.2cm of c] (a);
\vertex [left=1.2cm of c, label=180:$x_b$] (b);
\vertex [right=1.2cm of c'] (a');
\vertex [left=1.2cm of c', label=180:$x_a$] (b');
\diagram* {
(a) -- [plain] (c) -- [plain] (b),
(a') -- [plain] (c') -- [plain] (b'),
(c) -- [dotted] (z) -- [dotted] (c'),
(d) -- [photon] (z)
}; 
\end{feynman} 
\end{tikzpicture} = -\frac{i}{2\mpl}\int\!\!d t\frac{\tilde{G}_{12}m_1 m_2}{r}\left(\frac{\bar{\sigma}}{3}\!+\!2\lambda\bar{\rchi}\right) \; ,
\label{eq:nonlinNRGJ1}
\end{align}
where $r$ is the modulus of $\bm{r} = \bm{x}_1-\bm{x}_2$. 

Finally, we have to evaluate the last diagram, i.e. the one that involves the emission of a graviton through a cubic interaction vertex.  The computation are very similar to the ones we have just done for the above diagram, hence one gets eventually
\begin{align}
\begin{tikzpicture}[baseline]
\begin{feynman}
\vertex [label=135:$B$, label=45:$\bm{q}$, label=270:$t'$] (c);
\vertex [above=0.9cm of c] (z);
\vertex [right=1.2cm of z] (d);
\vertex [above=0.9cm of z, label=90:$t$, label=225:$A$, label=315:$\bm{k}$] (c');
\vertex [right=1.2cm of c] (a);
\vertex [left=1.2cm of c, label=180:$x_b$] (b);
\vertex [right=1.2cm of c'] (a');
\vertex [left=1.2cm of c', label=180:$x_a$] (b');
\diagram* {
(a) -- [plain] (c) -- [plain] (b),
(a') -- [plain] (c') -- [plain] (b'),
(c) -- [dotted] (z) -- [dotted] (c'),
(d) -- [gluon] (z)
}; 
\end{feynman} 
\end{tikzpicture} & = -\frac{i}{2\mpl}\sum_{a\neq b}\int\!\!d t\frac{\GN m_a m_b}{|\bm{x}_{ab}|}\bar{\sigma}_{ij}\frac{x^i_{ab}x^j_{ab}}{|\bm{x}_{ab}|^2}\bigg\{\mathcal{D}_1(\lambda, \alpha_a)-\frac{\mathcal{D}_2(\lambda, \alpha_a)}{2}\bigg\}\notag \\
& = -\frac{i}{2\mpl}\int\!\!d t\sum_{a\neq b}\frac{\tilde{G}_{ab} m_a m_b}{|\bm{x}_{ab}|^3}x^i_{ab}x^j_{ab}\bar{\sigma}_{ij} \; ,
\label{eq:nonlinNRGJ2}
\end{align}
where the coefficients $\mathcal{D}_1$ and $\mathcal{D}_2$ are the same written in eqs. (\ref{eqn:scheme93}a) and (\ref{eqn:scheme93}b).

Summing eqs. (\ref{eqn:scheme92}), (\ref{eq:nonlinNRGJ}), (\ref{eq:nonlinNRGJ1}) and (\ref{eq:nonlinNRGJ2}) we eventually obtain the dissipative part of action $\Seff{NR}$, namely
\begin{align}
S_{\text{diss}} & = \sum_a\frac{m_a}{\mpl}\int\!\!d t\bigg\{\left(\alpha_a\bar{\rchi}-\bar{\phi}\right)\!+\!\alpha_ax_a^i\partial_i\bar{\rchi}\bigg\}+\sum_a\frac{m_a}{2\mpl}\int\!\!d t\bigg\{-v_a^2\left(3\bar{\phi}+\alpha_a\bar{\rchi}\right)\notag \\
&\qquad\qquad\qquad+\!x_a^ix_a^j\partial_i\partial_j\left(\alpha_a\bar{\rchi}-\bar{\phi}\right)+2x_a^j\partial_j\bar{A}_iv_a^i +v_a^iv_a^j\left(\bar{\sigma}_{ij}+\frac{\bar{\sigma}}{3}\delta_{ij}\right)\bigg\}\! \notag \\ 
&+\frac{1}{\mpl}\!\sum_{a\neq b}\int\!\!d t\frac{\tilde{G}_{ab} m_am_b}{|\bm{x}_{ab}|}\left\{\bar{\phi}\!+\!\frac{\lambda}{2}\bar{\rchi}\!+\frac{4\beta_a^*\alpha_b^*-\alpha_a^*}{1\!+\!2\alpha_a^*\alpha_b^*}\frac{\lambda}{\lambda^*}\bar{\rchi}\right\} \notag \\
&-\frac{1}{4\mpl}\!\sum_{a\neq b}\int\!\!d t\frac{\tilde{G}_{ab} m_am_b}{|\bm{x}_{ab}|^3}x^i_{ab}x^j_{ab}\left(\bar{\sigma}_{ij}+\frac{\bar{\sigma}}{3}\delta_{ij}\right)\!+\!\!\Ord{L^{1/2}v^{7/2}} \, .
\label{eq:Sdiss.NRGJ}
\end{align}

Let's now rewrite it using Einstein-frame fields defined in eqs. (\ref{eqn:scheme78}). In this way we will be able to compare directly our result with Ref. \cite{NRGR+scalar}, and, being the gravitational and scalar field decoupled, it will be easier to derive our final result: the computation of the power loss of the system. It is not hard to see that, in terms of $\bar{\rchi}^*$, $\bar{\phi}^*$, $\bar{\sigma}^*$ $\bar{A}^*_i$ and $\bar{\sigma}^*_{ij}$ the action (\ref{eq:Sdiss.NRGJ}) becomes
\begin{equation}
S_{\text{diss}} = S^{\rchi}_{\text{diss}}+S^{\text{g}}_{\text{diss}} \; ,
\end{equation}
where explicitly
\begin{subequations}
\label{eqn.scheme94}
\begin{align}
S^{\rchi}_{\text{diss}} & =\frac{1}{\mpl}\int\!\!d t\Bigg\{\sum_a m_a\alpha^*_a\left(1-\frac{v_a^2}{2}\right)\bar{\rchi}^*+\sum_{a\neq b}\frac{\GN m_am_b}{|\bm{x}_{ab}|}\left(4\beta_a^*\alpha_b^*-\alpha_a^*\right)\bar{\rchi}^*\notag \\
&\qquad\qquad\qquad +\sum_am_a\alpha_a^*x_a^i\partial_i\bar{\rchi}^*+\sum_a\frac{m_a}{2}\alpha_a^*x_a^ix_a^j\partial_i\partial_j\bar{\rchi}^*\Bigg\} \; ,\\
S^{\text{g}}_{\text{diss}} & = -\frac{1}{\mpl}\int\!\!d t\left\{\left(\sum_a m+\sum_a\frac{3}{2}m_av_a^2\!-\!2\frac{\tilde{G}_{12} m_1m_2}{r}\right)\bar{\phi}^*\!+\!\sum_a\frac{m_a}{2}x_a^ix_a^j\partial_i\partial_j\bar{\phi}^*\right.\! \notag \\ 
&\qquad\left.-\!\sum_a m_ax_a^j\partial_j\bar{A}^*_iv_a^i -\!\sum_a\frac{m_a}{2}v_a^iv_a^j\tilde{\sigma}^*_{ij}-\!\sum_a\frac{m_a}{2}x_a^i\ddot{x}_a^j\tilde{\sigma}^*_{ij}\right\}  \, ,
\end{align}
\end{subequations}
where in $S^{\text{g}}_{\text{diss}}$ we introduced $\tilde{\sigma}^*_{ij} = \bar{\sigma}^*_{ij}+\delta_{ij}\bar{\sigma}^*/3$, and we used the following EOM at lowest order
\begin{align}
\ddot{x}_1^i=\frac{\tilde{G}_{12} m_2}{|\bm{x}_{12}|^3}x_{21}^i +\Ord{v} \; ,& & \ddot{x}_2^i=\frac{\tilde{G}_{12} m_1}{|\bm{x}_{12}|^3}x_{12}^i+\Ord{v} \; ,
\label{eq:EOM_1PNJ}
\end{align}
to simplify the last term. We immediately realise that $S^{\text{g}}_{\text{diss}}$ coincides with eq. (\ref{eq:Sdiss.NRG.first}), therefore it is completely equivalent\footnote{See the brief discussion after eq. (\ref{eq:Sdiss.NRG.first}).}  to the last part of eq. (\ref{eq:SNRfin}). The only difference is that here the Newtonian energy is given by $E_\text{N} = \sum_am_av_a^2/2-\tilde{G}_{12} m_1m_2/r$.

In order to be able to compare our result with the one of Ref. \cite{NRGR+scalar}, we have to rewrite eq. (\ref{eqn.scheme94}a) as a multipole expansion with STF tensor. The monopole and dipole terms are already written in term of STF tensor. In order to have a STF tensor also in front of the quadrupole, we need to add and subtract to the action the following term
\begin{equation}
\sum_a\frac{m_a\alpha_a^*}{6\mpl}\int\!\!d tx_a^2\delta^{ij}\partial_i\partial_j\bar{\rchi}^* \; ,
\label{eq:sumsubNRGJ}
\end{equation}
where $x_a^2 = \delta_{ij}x_a^ix_a^j$. Using the leading order EOM of the scalar field
\begin{equation}
\ddot{\bar{\rchi}}^*-\partial_j\partial^j\bar{\rchi}^* = \frac{J}{\mpl} \; ,
\end{equation}
where $J$ is the source of the scalar field, i.e. the two-point particles, we can rewrite eq. (\ref{eq:sumsubNRGJ}) as follow
\begin{equation}
\sum_a\frac{m_a\alpha_a^*}{6\mpl}\int\!\!d tx_a^2\delta^{ij}\partial_i\partial_j\bar{\rchi}^* = \sum_a\frac{m_a\alpha_a^*}{3\mpl}\int\!\!d t\left(v_a^2+\delta_{ij}x^i_a\ddot{x}^j_a\right)\bar{\rchi}^* +\dots \; ,
\end{equation}
where we integrated by parts twice, and we omitted contact terms that would contribute to the renormalization of point-particle masses. Using again the EOM (\ref{eq:EOM_1PNJ}), we can rewrite the last term of the above equation as
\begin{equation}
\sum_a m_a\alpha_a^*\delta_{ij}x^i_a\ddot{x}^j_a = -\frac{\tilde{G}_{12}m_1m_2}{2r}\left(\alpha_1^*+\alpha_2^*+\frac{m_1-m_2}{m_1+m_2}\left(\alpha^*_2-\alpha_1^*\right)\right) \; .
\end{equation}
Putting all together we can finally rewrite $S^{\rchi}_{\text{diss}}$ as follows
\begin{equation}
S^{\rchi}_{\text{diss}} =\frac{1}{\mpl}\int\!\!d t\left\{I_\rchi\bar{\rchi}^*+I^i_\rchi\partial_i\bar{\rchi}^*+\frac{1}{2}I^{ij}_\rchi\partial_i\partial_j\bar{\rchi}^*\right\} \; ,
\label{eq:finalSdissScalar}
\end{equation} 
where $I_\rchi$, $I^i_\rchi$ and $I^{ij}_\rchi$ are STF tensor whose explicit expressions are
\begin{subequations}
\label{eqn:scheme95}
\begin{align}
I_\rchi & = \sum_a m_a\alpha_a^*-\frac{1}{6}\sum_a m_a\alpha_a^*v_a^2+\text{\textit{g}}_{\!12}\frac{\GN m_1m_2}{r}\; ,\\
I^i_\rchi & = \sum_a m_a\alpha_a^*x_a^i\; ,\\
I^{ij}_\rchi & =  \sum_a m_a\alpha_a^*\left(x_a^ix_a^j-\frac{x_a^2}{3}\delta^{ij}\right)\; ,
\end{align}
\end{subequations}
where the symmetric coefficient $\text{\textit{g}}_{ab}$ is given by
\begin{equation}
\text{\textit{g}}_{ab} \!\equiv\! \alpha^*_a(4\beta_b^*\!-\!1)+\alpha^*_b(4\beta_a^*\!-\!1)-\frac{1+2\alpha_a^*\alpha_b^*}{6}\left(\alpha_a^*+\alpha_b^*+\frac{m_a-m_b}{m_a+m_b}\left(\alpha^*_b-\alpha_a^*\right)\right) \; .
\end{equation}
The expressions of these multipole moments are precisely the ones obtained in Ref. \cite{NRGR+scalar}; hence, also for the scalar part of the dissipative dynamics we find consistent results.

\section{Observables and Leading-Order power loss}\label{Sec:Obs-scalar}

At this point one can start computing observables using the actions we have just written. For instance, imposing retarded boundary conditions\footnote{Which basically means using the in-in formalism, see Sec. \ref{sec:GWandGR}.}, one can compute the gravitational and scalar waveform, and compare it to experimental results. See \cite{NRGR+scalar} for a derivation of this observable at leading order, and \cite{2PNGwave} for the waveform in scalar-tensor theory at 2PN order.

Then, one can integrate out also the radiation fields and find the final effective action that we called $\Seff{NRGR}$. Ignoring non-linear interaction between radiation fields and using again the diagrammatic convention we introduced in Sec. \ref{sec:power_leading} we expect the final action to contain
\begin{align}
i\Seff{NRGR}\supset\begin{tikzpicture}[baseline]
\begin{feynman}
\vertex (f);
\vertex [above=0.1cm of f] (z);
\vertex [right=0.6cm of z, dot, label=270:$I_g^L(t)$] (c) {};
\vertex [right=1.5cm of c, dot, label=270:$I_g^K(t')$] (c') {};
\vertex [right=0.6cm of c'] (w);  
\diagram* {
(z) -- [double] (c) -- [double] (c') -- [double] (w),
(c') -- [gluon, half right, edge label'=$\bar{\phi}^*\text{, } \bar{A}^*_i\text{, } \bar{\sigma}^*_{ij}$] (c)
}; 
\end{feynman} 
\end{tikzpicture}+\begin{tikzpicture}[baseline]
\begin{feynman}
\vertex (f);
\vertex [above=0.1cm of f] (z);
\vertex [right=0.6cm of z, dot, label=270:$I_\rchi^L(t)$] (c) {};
\vertex [right=1.5cm of c, dot, label=270:$I_\rchi^K(t')$] (c') {};
\vertex [right=0.6cm of c'] (w);  
\diagram* {
(z) -- [double] (c) -- [double] (c') -- [double] (w),
(c') -- [photon, half right, edge label'=$\bar{\rchi}^*$] (c)
}; 
\end{feynman} 
\end{tikzpicture} \; ,
\end{align}
where we used again the compact notation for spatial indices, and denoted with $I_g^L(t)$ the STF multipole moments for the gravitational fields. For the gravity part computations are completely equivalent to the ones we did in Secs. \ref{sec:power_leading} and \ref{sec:bott.up}, just rewritten in terms of NRG fields.

For the scalar part contribution, it is not hard to see that
\begin{align}
\!\!\begin{tikzpicture}[baseline]
\begin{feynman}
\vertex (f);
\vertex [above=0.1cm of f] (z);
\vertex [right=0.6cm of z, dot, label=270:$I_\rchi^L(t)$] (c) {};
\vertex [right=1.5cm of c, dot, label=270:$I_\rchi^K(t')$] (c') {};
\vertex [right=0.6cm of c'] (w);  
\diagram* {
(z) -- [double] (c) -- [double] (c') -- [double] (w),
(c') -- [photon, half right, edge label'=$\bar{\rchi}^*$] (c)
}; 
\end{feynman} 
\end{tikzpicture} \propto \int\!\!d td t'I_\rchi^{L}(t)I_\rchi^{K}(t')\bangle*{\Tprod{\partial_L\bar{\rchi}^*(t, \bm{x})\partial_K\bar{\rchi}^*(t', \bm{x'})}}\bigg|_{\bm{x} = \bm{x}'= 0}  \; .
\label{eq:Scalar_emission_diag}
\end{align}
Since the final result should be rotational invariant, then we expect the multipole indices to be contracted with invariant tensors, i.e. with either $\delta_{ij}$ or $\epsilon_{ijk}$. Because we wrote everything in terms of STF tensors, we understand that eq. (\ref{eq:Scalar_emission_diag}) is non-vanishing only when the two multipole moments are equal, hence when $L = K$.

From this, using again the optical theorem given in eqs. (\ref{eq:Opt.th}) and (\ref{eq:Ploss}), one can compute the power loss. In this scalar-tensor theory we expect it to be given by 
\begin{equation}
\mathcal{P} = \mathcal{P}_g+\mathcal{P}_\rchi \; ,
\end{equation}
where $\mathcal{P}_g$ and $\mathcal{P}_\rchi$ are respectively the graviton and the scalar contributions to the emitted power. Again for the gravitational part, things are completely equivalent to what we have done in the previous part. Given what we said above, it is not hard to compute also the scalar contribution to the total emitted power which is\footnote{In part II of Ref. \cite{Ross-multi} one can find the expression of this power loss at all multipole order, modulo an overall factor of $\mpl^{-2}$.}
\begin{equation}
\mathcal{P}_\rchi = 2\GN\left(\braket{\dot{I}^2_\rchi}+\frac{1}{3}\braket{\ddot{I}^i_\rchi\ddot{I}^i_\rchi}+\frac{1}{30}\braket{\dddot{I}^{\,ij}_\rchi\dddot{I}^{\,ij}_\rchi}\right) \; .
\end{equation}
Therefore, beside the quadrupole, also the monopole and the dipole contribute to the total scalar power emission. For a complete discussion and calculations on this quantity we redirect the reader to Refs. \cite{NRGR+scalar} and \cite{FareseDamour}. Finally we stress that these kind of terms would enter before the usual quadrupole emission previously computed in the PN expansion, making these kind of theories easily distinguishable from pure GR. In fact, no significant deviation from the quadrupole emission has been observed at present, which suggests the need of a ``screening mechanism'' for these theories\footnote{See for instance Refs. \cite{Van-original,Van-Rev}. See also Ref. \cite{Constr-scalar} for some constraints precisely coming from GWs observations.}. We leave for future studies the implementation of such mechanism in this context.

\chapter*{Conclusions of the second part}
\addcontentsline{toc}{chapter}{Conclusions of the second part} 

In the second part of this work, we extended the EFT approach to the binary inspiral problem described in the previous part to alternative theories of gravity. First, In Ch. \ref{ch:NRGR_NRG}, we introduced a useful tool used to perform more compact computations: the so-called NRG fields parametrisation of the metric. Based on a Kaluza-Klein temporal reduction, this metric was first applied to the binary inspiral problem in Refs. \cite{KK1, KK2}. 

We wrote the Einstein-Hilbert action, and the correspondent harmonic gauge-fixing action, in terms of NRG fields, and then, in Sec. \ref{sec:adv}, we saw the advantages of such a parametrisation in computing the Feynman diagrams in NRGR at 1PN order. More specifically, we saw that NRG fields allows to eliminate the cubic interaction topology diagrams at this order in the PN expansion.

This cancellation has been proved to hold at all PN orders in the conservative sector, and NRG fields have been used to compute the conservative dynamic of the binary up to 4PN order,  see Refs. \cite{2PN,3PN,4PN-1,4PN-2}, and to build the public Mathematica package \textit{EFTofPNG} \cite{EFTofPNG}.

With these advantages in mind, we decided to use NRG fields also in Ch. \ref{ch:conf.coupl}, where we extended NRGR to a conformally coupled scalar-tensor theory of gravity. This extension has been recently done in Ref. \cite{NRGR+scalar} in the Einstein frame, therefore, we decided to work in the equivalent Jordan frame. 

Thus, in Sec. \ref{sec:grav_NRG_Jordan} we derived for the first time the gravity action for a scalar-tensor theory in the Jordan frame, written in terms of the  NRG fields. We checked our result performing the conformal transformation to go back to the Einstein frame, correctly obtaining the action written in Ref. \cite{KK2}, plus the kinetic term of the scalar field.

Then, we started the EFT procedure for a PN study of the binary inspiral problem. We expanded the obtained action to the quadratic order in the fields. As expected, we found a mixing between the gravity NRG fields and the scalar field already at this level. Using mixed state propagators, we wrote the Feynman rules for the conservative sector at 1PN order, and for the dissipative sector at 2.5PN order. Our result coincides with the one found in Refs. \cite{NRGR+scalar, EFreview}.

Once again the EFT approach has been proved to be a powerful and useful way of studying physical systems. Of course, this work can now be extended to higher-order scalar-tensor theory, like Horndeski and beyond Horndeski theories \cite{BeyHord,EFT_HOST,Galileon}. We actually attempted to realise such an extension, but this discussion would deserve a more thorough discussion which is beyond the scope of this first work.

\chapter*{Acknowledgements}
\addcontentsline{toc}{chapter}{Acknowledgements} 

I would like to express my gratitude to my supervisor, Dr. Filippo Vernizzi of IPhT, for making this work possible and for all the support, mentorship and training he provided throughout the project. Sincere thanks also to my internal supervisor, Prof. Claudio Destri of university of Milano - Bicocca, for attentively reading this work and for detailed and accurate suggestions.

I gratefully acknowledge University of Milano-Bicocca, the Erasmus Programme and the CEA-IPhT for financial support and access to all the necessary resources for the fulfilment of this project.

I would like also to deeply thanks my parents Giuseppe and Annunciata, my brothers and sisters, Francesco, Erica, Maddalena, Giovanni and Stefano and all my friends for the patience, support, encouragement and happiness they gave me during all my life. Finally, my warmest thanks go to Umberto, for all his unique help, support and kindness provided during these last two years.

\appendix

\part*{Appendix}
\addcontentsline{toc}{part}{Appendix}  

\chapter{The PN Feynman rules}\label{App:action}

We would like to have an explicit expression for the diagrammatic Feynman rules of NRGR. In order to do so, we expand in an (almost entirely) explicit manner the action used to describe NRGR.  See also Ref. \cite{Ddim-Rules} for general Feynman rules in $D$ dimensional NRGR.

In \ref{sec:EHe} we expand the Einstein-Hilbert action for a metric $g_{\mu\nu}$ proceeding as follows:
\begin{itemize}
\item first we expand at order $\Ord{h^3}$ where $h_{\mu\nu}$ is such that $g_{\mu\nu} = \eta_{\mu\nu}+h_{\mu\nu}/\Mpl$
\item then we split into potential and radiation modes $h_{\mu\nu} =H_{\mu\nu}+\radh{\mu\nu}$, and we add also the needed gauge-fixing term. 
\end{itemize} 
Since we will not consider any quantum effect, we do not need any ghost fields, then in \ref{sec:PPe} we work out the expression for the point-particle action:
\begin{itemize}
\item at order $\Ord{v^4}$ in the PN expansion parameter, and at $\Ord{h^3}$
\item then we explicit again the potential and radiation modes, and we also perform the multipole expansion of $\radh{\mu\nu}$. 
\end{itemize} 

\section{Einstein-Hilbert action expansion}\label{sec:EHe}

\subsection{First step: flat background expansion}\label{sec:EHflatb}

Let's start by
\begin{equation}
S_{\text{EH}} = -2\Mpl^2\!\int\!\!d^4\!x\sqrt{-g}\Rc \; ,
\label{eq:ASeh}
\end{equation}
see the conventions at p.~\pageref{Ch:convent} for the definition of the Ricci scalar $\Rc$. We first perform the usual expansion of the Einstein-Hilbert action on a flat background metric. This kind of computations had already been performed largely in the literature, see for instance \cite{Grav.vert, Grav.vert2, rev.QG}.

We start by redefining the metric $g_{\mu\nu}$ as
\begin{equation}
g_{\mu\nu} = \eta_{\mu\nu} +\frac{h_{\mu\nu}}{\Mpl} \; .
\end{equation}
The first thing we need to find is the inverse metric $g^{\mu\nu}$. At order $\Ord{h^3}$ this is given by
\begin{equation}
g^{\mu\nu} = \eta^{\mu\nu} -\frac{h^{\mu\nu}}{\Mpl}+\frac{h^{\mu\alpha}h_\alpha{^\nu}}{\Mpl^2}-\frac{h^{\mu\alpha}h_{\alpha\beta}h^{\beta\nu}}{\Mpl^3} \; .
\end{equation}
On the right hand side of this expression we used the flat background metric to raise/lower indices, e.g. $h^{\mu\nu} =\eta^{\mu\alpha}h_{\alpha\beta}\eta^{\beta\nu}$. The expression for the inverse metric allows us to write the Levi-Cvita connection as
\begin{align}
\Gamma{^\rho}_{\mu\nu} = {\ord{1}}\Gamma{^\rho}_{\mu\nu}+{\ord{2}}\Gamma{^\rho}_{\mu\nu}+{\ord{3}}\Gamma{^\rho}_{\mu\nu} \; ,
\end{align}
where explicitly
\begin{subequations}
\label{eqn:scheme4}
\begin{align}
{\ord{1}}\Gamma{^\rho}_{\mu\nu} & = \frac{1}{2\Mpl}(\partial_\mu h{^\rho}_{\nu}+\partial_{\nu}h_\mu{^\rho}-\partial^{\rho}h_{\mu\nu})  \; ,\\
{\ord{2}}\Gamma{^\rho}_{\mu\nu} & = -\frac{1}{2\Mpl^2}h^{\rho\sigma}(\partial_\mu h_{\sigma\nu}+\partial_{\nu}h_{\mu\sigma}-\partial{_\sigma}h_{\mu\nu}) =-\frac{1}{\Mpl}h{^\rho}_{\alpha}{\ord{1}}\Gamma{^\alpha}_{\mu\nu} \; , \\
{\ord{3}}\Gamma{^\rho}_{\mu\nu} & = -\frac{1}{2\Mpl^2}h^{\rho\alpha}h_{\alpha}{^\sigma}(\partial_\mu h_{\sigma\nu}\!+\!\partial_{\nu}h_{\mu\sigma}\!-\!\partial{_\sigma}h_{\mu\nu}) =-\frac{1}{\Mpl}h{^\rho}_{\alpha}{\ord{2}}\Gamma{^\alpha}_{\mu\nu} \; .
\end{align}
\end{subequations}
From here one can compute the Riemann tensor. We will not do this computation explicitly, we just say that one eventually arrives to 
\begin{equation}
\Rc{^\rho}_{\mu\sigma\nu} = {\ord{1}}\Rc{^\rho}_{\mu\sigma\nu}+{\ord{2}}\Rc{^\rho}_{\mu\sigma\nu}+{\ord{3}}\Rc{^\rho}_{\mu\sigma\nu} \; .
\end{equation}

Then we need to expand the determinant of the metric present in (\ref{eq:ASeh}). One should proceed as follows
\begin{align}
\sqrt{-g} & = \exp\left\{\frac{1}{2}\Tr{\log[-g_{\mu\nu}]}\right\} = \exp\left\{\frac{1}{2}\Tr{\log\left[-\eta_{\mu\sigma}\left(\delta{^\sigma}_{\nu}+\frac{h{^\sigma}_{\nu}}{\Mpl}\right)\right]}\right\} \notag \\
& = \left\{1+\frac{1}{2}\Tr{\frac{h{^\sigma}_{\nu}}{\Mpl}-\frac{1}{2}\frac{h{^\sigma}_{\alpha}h{^\alpha}_{\nu}}{\Mpl^2}}+\frac{1}{8}\left[\Tr{\frac{h{^\sigma}_{\nu}}{\Mpl}-\frac{1}{2}\frac{h{^\sigma}_{\alpha}h{^\alpha}_{\nu}}{\Mpl^2}}\right]^2\dots\right\} \notag \\
& = 1+\frac{1}{2\Mpl}h-\frac{1}{4\Mpl^2}h^{\mu\nu}h_{\mu\nu}+\frac{1}{8\Mpl^2}h^2+\Ord{h^3} \; .
\end{align}
In the last step we defined $h\equiv h_{\mu\nu}\eta^{\mu\nu}$. Here we stopped at second order because the Ricci scalar is at least of order $\Ord{h}$. One eventually gets
\begin{align}
S_{\text{EH}} & = -2\Mpl^2\int\!\!d^4\!x\left(1+\frac{h}{2\Mpl}-\frac{h^{\mu\nu}h_{\mu\nu}}{4\Mpl^2}+\frac{h^2}{8\Mpl^2}\right)\left({\ord{1}}\Rc+{\ord{2}}\Rc+{\ord{3}}\Rc\right) \notag \\
& = \int\!\!d^4\!x\left(\DLagr_1+\DLagr_2+\DLagr_3\right) \; ,
\end{align}
where
\begin{subequations}
\label{eqn:scheme5}
\begin{align}
\DLagr_1 & = -2\Mpl^2{\ord{1}}\Rc =-2\Mpl\,\partial_\mu\left(\partial_\nu h^{\mu\nu}-\partial^{\mu}h\right) \; ,\\
\DLagr_2 & = -2\Mpl^2\left(\frac{h}{2\Mpl}{\ord{1}}\Rc+{\ord{2}}\Rc\right) \notag \\
& = \frac{1}{2}\partial_{\mu}h_{\rho\sigma}\partial^{\mu}h^{\rho\sigma}-\frac{1}{2}\partial_\mu h\partial^\mu h +\partial_\mu h\partial_\nu h^{\mu\nu}-\partial_\mu h_{\rho\sigma}\partial^{\rho} h^{\mu\sigma} \; ,
\label{eq:quad.grav} 
\end{align}
\begin{align}
\DLagr_3 & = -2\Mpl^2{\ord{3}}\Rc-\Mpl\frac{h}{2}\left(\frac{h}{2\Mpl}{\ord{1}}\Rc+{\ord{2}}\Rc\right)+\frac{h^{\mu\nu}h_{\mu\nu}}{2}\,{\ord{1}}\Rc  \notag \\
& = \frac{1}{\Mpl}\bigg(2h_{\mu\nu}\partial_\rho h^{\mu\sigma}\partial^{\nu}h{^\rho}_{\sigma}-h_{\mu\nu}\partial_{\sigma}h^{\mu\rho}\partial^{\sigma}h{^\nu}_{\rho}+h_{\mu\nu}\partial_{\sigma}h^{\mu\rho}\partial_{\rho}h^{\sigma\nu}\notag \\ 
&\qquad\qquad +h_{\mu\nu}\partial^\rho h^{\mu\nu}\partial_\rho h+h_{\mu\rho}h{^\rho}_\nu\partial^\mu\partial^\nu h+\frac{1}{2}h_{\mu\nu}\partial^\mu h_{\rho\sigma}\partial^{\nu}h^{\rho\sigma} \notag \\
&\qquad\qquad -\frac{1}{2}h\partial_\mu h_{\nu\rho}\partial^{\nu}h^{\mu\rho}+\frac{1}{4}h\partial_{\mu}h_{\rho\sigma}\partial^{\mu}h^{\rho\sigma}-\frac{1}{4}h\partial_\mu h\partial^\mu h\notag \\
&\qquad\qquad -\frac{1}{2}h h_{\mu\nu}\partial^{\mu}\partial^{\nu}h+h_{\mu\nu}h_{\rho\sigma}\partial^\rho\partial^\sigma h^{\mu\nu}\bigg) \; .
\label{eq:lagrh3}
\end{align}
Since $\DLagr_1$ is a total derivative, we can neglect it from now on.
\end{subequations}

\subsection{Second step: potential + radiation modes decomposition}\label{subsec:po+radEH}

Now we have to split the graviton field in (UV) potential modes $\potH{}{\mu\nu}$ and (IR) radiation modes $\radh{\mu\nu}$. We recall that
\begin{itemize}
\item The potential gravitons are always off-shell, and they are such that
\begin{align}
k_H^0\sim\frac{v}{r}\; , & & |\bm{k}_H|\sim\frac{1}{r} \quad \longrightarrow \quad \partial_0 \potH{}{\mu\nu}\sim\frac{v}{r}\potH{}{\mu\nu} \; , & &   \partial_i \potH{}{\mu\nu}\sim \frac{1}{r}\potH{}{\mu\nu} 
\end{align}
\item The radiation gravitons are on-shell, and they are such that
\begin{align}
k_{\radh{}}^0\sim\frac{v}{r} \; , & & |\bm{k}_{\radh{}}|\sim\frac{v}{r} \quad \longrightarrow \quad \partial_\rho \radh{\mu\nu}\sim\frac{v}{r}\radh{\mu\nu} \; .
\end{align}
\end{itemize}
As usual we then perform a partial Fourier transform of $\potH{}{\mu\nu}$, hence we define $\potH{k}{\mu\nu}$ as in eq. (\ref{eq:Pot.Fourier}). Since in these work we stop at 1PN order, from now on we can neglect time derivatives of $\potH{k}{}$ and derivatives of $\radh{}$.
After this decomposition, we expect
\begin{equation}
S_{\text{EH}}[h]\longrightarrow S[\radh{}]+S[\potH{}{}]+S[\radh{},\potH{}{}]
\end{equation} 

\subsubsection{Potential part}

Since in our EFT the potential modes are integrated out first in the background of the radiation ones, let's focus on $\potH{}{}$. We define for convenience the following action
\begin{equation}
S_{\text{pot}}[\potH{}{\mu\nu},\radh{\mu\nu}] \equiv S_{\potH{}{}}[\potH{}{\mu\nu}]+S_{\potH{}{}\radh{}}[\potH{}{\mu\nu},\radh{\mu\nu}]+S^{\text{GF}}_{H}[\potH{}{\mu\nu},\radh{\mu\nu}] \; ,
\label{eq:Spot1}
\end{equation}
where $S_{\potH{}{}}$ and $S_{H\radh{}}$ can be read from eqs.\footnote{Imposing the EOM of $\bar{h}_{\mu\nu}$, terms linear in $\potH{}{\mu\nu}$ cancel.} (\ref{eqn:scheme5})
\begin{subequations}
\label{eqn:scheme7}
\begin{align}
S_{\potH{}{}} & = \int\!\!d^4\!x\left[\frac{1}{2}\partial_{\mu}H_{\rho\sigma}\partial^{\mu}H^{\rho\sigma}\!-\!\frac{1}{2}\partial_\mu H\partial^\mu H \!+\!\partial_\mu H\partial_\nu H^{\mu\nu}\!-\!\partial_\mu H_{\rho\sigma}\partial^{\rho} H^{\mu\sigma}\right] \notag \\
& +\frac{1}{\Mpl}\int\!\!d^4\!x\bigg[2H_{\mu\nu}\partial_\rho H^{\mu\sigma}\partial^{\nu}H{^\rho}_{\sigma}-H_{\mu\nu}\partial_{\sigma}H^{\mu\rho}\partial^{\sigma}H{^\nu}_{\rho} \notag \\ 
&\qquad\qquad +H_{\mu\nu}\partial_{\sigma}H^{\mu\rho}\partial_{\rho}H^{\sigma\nu}+H_{\mu\nu}\partial^\rho H^{\mu\nu}\partial_\rho H+H_{\mu\rho}H{^\rho}_\nu\partial^\mu\partial^\nu H \notag \\
&\qquad\qquad +\frac{1}{2}H_{\mu\nu}\partial^\mu H_{\rho\sigma}\partial^{\nu}H^{\rho\sigma}-\frac{1}{2}H\partial_\mu H_{\nu\rho}\partial^{\nu}H^{\mu\rho}+\frac{1}{4}H\partial_{\mu}H_{\rho\sigma}\partial^{\mu}H^{\rho\sigma} \notag \\
&\qquad\qquad -\frac{1}{4}H\partial_\mu H\partial^\mu H-\frac{1}{2}H H_{\mu\nu}\partial^{\mu}\partial^{\nu}H+H_{\mu\nu}H_{\rho\sigma}\partial^\rho\partial^\sigma H^{\mu\nu}\bigg] \; , \label{eq:steppot1}
\end{align}
\begin{align}
S_{H\radh{}} & = \int\!\!d^4\!x\frac{\radhin{\alpha\beta}}{\Mpl}\bigg[2\partial_\mu H_{\alpha\sigma}\partial_{\beta}H^{\mu\sigma}-\partial_{\sigma}H_{\alpha\rho}\partial^{\sigma}H{^\rho}_{\beta}+ \partial^{\sigma}H_{\alpha\rho}\partial^{\rho}H_{\sigma\beta} \notag \\ 
&\qquad\qquad\qquad +\partial^\rho H_{\alpha\beta}\partial_\rho H+2H_{\beta\mu}\partial_\mu\partial^\nu H+\frac{1}{2}\partial_\alpha H_{\mu\nu}\partial_{\beta}H^{\mu\nu} \notag \\
&\qquad\qquad\qquad -\frac{1}{2}H\partial_\alpha\partial_\beta H +H_{\rho\sigma}\partial^{\rho}\partial^\sigma H_{\alpha\beta}+H_{\mu\nu}\partial_{\alpha}\partial_\beta H^{\mu\nu} \notag \\
&\qquad\qquad\qquad -\frac{\eta_{\alpha\beta}}{2}\bigg(\partial_\mu H_{\sigma\nu}\partial^\nu H^{\mu\sigma}+H_{\mu\nu}\partial^\mu\partial^\nu H \notag \\ 
&\qquad\qquad\qquad\qquad\qquad\qquad+\frac{1}{2}\partial_\mu H\partial^\mu H-\frac{1}{2}\partial_\mu H_{\rho\sigma}\partial^\mu H^{\rho\sigma}\bigg)\bigg] \; .
\label{eq:steppot2}
\end{align}
\end{subequations} 
To write (\ref{eq:steppot2}) we substituted $h_{\mu\nu} =\potH{}{\mu\nu}+\radh{\mu\nu}$ in (\ref{eq:lagrh3}), and we discarded all terms proportional to a derivative of $\radh{\mu\nu}$. Indeed these terms would contribute at a higher PN perturbative order.

$S^{\text{GF}}_{H}$ is the gauge-fixing term for the field $\potH{}{\mu\nu}$. As usual we are free to work in the harmonic gauge. At this stage we are still working in the background of the radiation modes; therefore, to preserve gauge invariance, we will chose $S^{\text{GF}}_{H}$ to be invariant under general coordinate transformations of the background metric $\bar{g}_{\mu\nu} =\eta_{\mu\nu}+\radh{\mu\nu}/\Mpl$. The explicit expression of the gauge-fixing action is given in eqs. (\ref{eq:SGFpot}) and (\ref{eq:SGFpot2}). We recall that we are expanding up to order $\Ord{h^3}$. Hence, since in (\ref{eq:SGFpot}) $\Gamma_{\mu}\Gamma_{\nu}$ is for sure at least of second order, it is enough to have
\begin{subequations}
\label{eqn:scheme6}
\begin{align}
\bar{g}^{\mu\nu} & = \eta^{\mu\nu}-\frac{1}{\Mpl}\radhin{\mu\nu}+\dots \\
\sqrt{-\bar{g}} & = 1+\frac{\bar{h}}{2\Mpl} + \dots \\
\bar{\Gamma}{^\rho}_{\mu\nu} & =\frac{\eta^{\rho\sigma}}{2\Mpl}\left(\partial_\mu\radh{\sigma\nu}+\partial_\nu\radh{\mu\sigma}-\partial_\sigma\radh{\mu\nu}\right)+\dots \label{eq:levcivrad}
\end{align}
\end{subequations}
Again we can neglect terms proportional to the derivatives of $\radh{\mu\nu}$, hence we can neglect all terms coming from (\ref{eq:levcivrad}). This is equivalent to replace covariant derivatives with ordinary one in the gauge-fixing term (\ref{eq:SGFpot2}). Therefore the relevant part of $\Gamma_\mu$ is
\begin{equation}
\Gamma_\mu = \frac{1}{\Mpl}\left[\left(\partial_\sigma\potHmi{\ }{\sigma}{\mu}-\frac{1}{2}\partial_\mu\potH{}{}\right)-\frac{\radhin{\rho\sigma}}{\Mpl}\left(\partial_\sigma\potH{}{\rho\mu}-\frac{1}{2}\partial_\mu\potH{}{\rho\sigma}\right)\right]+\dots
\end{equation}
where indices are raised/lowered using the flat metric $\eta_{\mu\nu}$, and $H\equiv H_{\mu\nu}\eta^{\mu\nu}$. From here one can finds
\begin{align}
S^{\text{GF}}_{H} & =\int\!\!d^4\! x\left(1+\frac{\bar{h}}{2\Mpl}\right)\left(\eta^{\mu\nu}-\frac{\radhin{\mu\nu}}{\Mpl}\right) \notag \\
&\qquad\times\left[\left(\partial_\sigma\potHmi{\ }{\sigma}{\mu}\!-\!\frac{1}{2}\partial_\mu\potH{}{}\right)\!-\!\frac{\radhin{\rho\mu}}{\Mpl}\left(\partial_\sigma\potH{}{\rho\sigma}\!-\!\frac{1}{2}\partial_\mu\potH{}{\rho\sigma}\right)\right] \notag \\
&\qquad\times\left[\left(\partial_\beta\potHmi{\ }{\beta}{\nu}-\frac{1}{2}\partial_\nu\potH{}{}\right)-\frac{\radhin{\alpha\beta}}{\Mpl}\left(\partial_\beta\potH{}{\alpha\nu}-\frac{1}{2}\partial_\nu\potH{}{\alpha\beta}\right)\right]+\dots
\end{align}
Discarding all terms higher than $\Ord{h^3}$ one obtains
\begin{align}
S^{\text{GF}}_{H} & \simeq \int\!\!d^4\!x\left(\partial^\sigma\potH{}{\sigma\mu}\partial_\rho\potHin{}{\rho\mu}-\partial^\sigma\potH{}{\sigma\mu}\partial_\mu\potHin{}{}+\frac{1}{4}\partial_\mu H\partial^\mu H\right)\notag \\
& +\int\!\!d^4\!x\,\frac{\radhin{\alpha\beta}}{\Mpl}\bigg[\!-2\partial^\sigma\potH{}{\sigma\mu}\partial_{\alpha}\potHmi{\ }{\mu}{\beta}\!+\!\partial^\sigma\potH{}{\sigma\mu}\partial^\mu\potH{}{\alpha\beta}\!+\!\partial_\mu H\partial_\alpha\potHmi{\ }{\mu}{\beta}\notag \\
&\qquad\qquad\quad -\!\frac{1}{2}\partial_\mu H\partial^\mu\potH{}{\alpha\beta}\!-\!\partial^\sigma\potH{}{\sigma\alpha}\partial^{\rho}\potH{}{\rho\beta}\!+\!\partial^\sigma\potH{}{\sigma\alpha}\partial_\beta H \!-\!\frac{1}{4}\partial_\alpha H\partial_\beta H\notag \\
&\qquad\qquad\quad+\!\frac{\eta_{\alpha\beta}}{2}\left(\partial^\sigma\potH{}{\sigma\mu}\partial_\rho\potHin{}{\rho\mu}-\partial^\sigma\potH{}{\sigma\mu}\partial_\mu\potHin{}{}+\frac{1}{4}\partial_\mu H\partial^\mu H\right)\bigg] \; .
\label{eq:steppot3}
\end{align}
From this expression we understand that the gauge-fixing part of the action corrects the quadratic part in $\potH{}{}$ of the action (\ref{eq:steppot1}), as well as the interaction terms of the form $\radh{}\potH{}{}\potH{}{}$ in (\ref{eq:steppot2}). 

Rewriting then eq. (\ref{eq:Spot1}) as
\begin{equation}
S_{\text{pot}}[\potH{}{\mu\nu},\radh{\mu\nu}] = S^{(2)}_{H}[\potH{}{\mu\nu}]+S^{(3)}_{H}[\potH{}{\mu\nu}]+S^{(3)}_{H\radh{}}[\potH{}{\mu\nu},\radh{\mu\nu}] \; ,
\end{equation}
we can find  $S^{(2)}_{H}$ by summing the quadratic part in $H_{\mu\nu}$ of (\ref{eq:steppot1}) and (\ref{eq:steppot3})
\begin{align}
S^{(2)}_{H} & = \frac{1}{2}\int\!\!d^4\!x\,\partial_{\alpha}H_{\rho\sigma}\left(\eta^{\rho\mu}\eta^{\sigma\nu}-\frac{1}{2}\eta^{\rho\sigma}\eta^{\mu\nu}\right)\partial^{\alpha}H_{\mu\nu} \; .
\end{align}
The action $S^{(3)}_{H}$ can easily be constructed by taking the cubic term in (\ref{eq:steppot1}). Then we compute $S^{(3)}_{H\bar{h}}$ by summing (\ref{eq:steppot2}) and the ``cubic part'' (meaning terms of the form $HH\radh{}$) of (\ref{eq:steppot3})
\begin{align}
S^{(3)}_{H\bar{h}} & = \int\!\!d^4\!x\frac{\radhin{\alpha\beta}}{\Mpl}\bigg[-\partial_\sigma H_{\alpha\rho}\partial^{\sigma}H_{\beta}{^\rho}-\frac{1}{2}\partial_\alpha H_{\mu\nu}\partial_\beta H^{\mu\nu}+\frac{1}{2}\partial_\mu H\partial^\mu H_{\alpha\beta}+\notag \\
&\qquad\qquad\qquad+\!\frac{1}{4}\partial_\alpha H\partial_\beta H\!+\!\frac{\eta_{\alpha\beta}}{4}\bigg(\partial_\mu H_{\rho\sigma}\partial^\mu H^{\rho\sigma}\!-\frac{1}{2}\partial_\mu H\partial^\mu H\bigg)\bigg] \; .
\end{align}
Finally, we perform a partial Fourier transform in order to write everything in term of $\potH{k}{\mu\nu}$. Discarding terms proportional to the time derivatives of $\potH{k}{}$ in $S^{(3)}_{H}$ (they contribute to a higher PN order), one obtains
\begin{align}
S_{\text{pot}}[\potH{k}{\mu\nu},\radh{\mu\nu}] & = S^{(2)}_{\potH{k}{}}[\potH{k}{\mu\nu}]+S^{(3)}_{\potH{k}{}}[\potH{k}{\mu\nu}]+S^{(3)}_{\potH{k}{}\radh{}}[\potH{k}{\mu\nu},\radh{\mu\nu}] = \notag \\
& = \int\!\!d t\left(\Lagr^{(2)}_{\potH{k}{}}[\potH{k}{\mu\nu}]+\Lagr^{(3)}_{\potH{k}{}}[\potH{k}{\mu\nu}]+\Lagr^{(3)}_{\potH{k}{}\radh{}}[\potH{k}{\mu\nu},\radh{\mu\nu}]\right) \, ,
\end{align}
where
\begin{subequations}
\label{eqn:scheme8}
\begin{align}
\!\!\Lagr^{(2)}_{\potH{k}{}} & \!=\! \intvec{k}\left(-\frac{\modul{k}^2}{2}\right)\potH{k}{\rho\sigma}\left(\eta^{\rho\mu}\eta^{\sigma\nu}-\frac{1}{2}\eta^{\rho\sigma}\eta^{\mu\nu}\right)\potH{-k}{\mu\nu} \notag \\
&\qquad+\frac{1}{2}\intvec{k}\partial_0\potH{k}{\rho\sigma}\left(\eta^{\rho\mu}\eta^{\sigma\nu}-\frac{1}{2}\eta^{\rho\sigma}\eta^{\mu\nu}\right)\partial_0\potH{-k}{\mu\nu}\; ,\label{eq:quadH}
\end{align}
\begin{align}
\Lagr^{(3)}_{\potH{k}{}} & = \frac{1}{\Mpl}\intvec{k,q,p}(2\pi)^3\delta^{(3)}\!\left(\bm{k}\!+\!\bm{q}\!+\!\bm{p}\right)\bigg\{\bm{q}\cdot\bm{p}\bigg[-\!\potH{k}{\mu\nu}\potHin{q}{\mu\rho}\potHmi{p}{\nu}{\rho}\!+\!\potH{k}{\mu\nu}\potHin{q}{\mu\nu}\potH{p}{}\notag \\
&\qquad\qquad\qquad\qquad\qquad\qquad\qquad\qquad\qquad\quad+\frac{1}{4}\potH{k}{}\potH{q}{\rho\sigma}\potHin{p}{\rho\sigma}-\frac{1}{4}\potH{k}{}\potH{q}{}\potH{p}{}\bigg]\notag \\
&\qquad\qquad\qquad\quad-\left(p^ip^j\right)\bigg[\potH{k}{i\rho}\potHmi{q}{\rho}{j}\potH{p}{}+\potH{k}{\mu\nu}\potH{q}{ij}\potHin{p}{\mu\nu}-\frac{1}{2}\potH{k}{}\potH{q}{ij}\potH{p}{}\bigg] \notag \\
&\qquad\qquad\qquad\quad-\left(q^ip^j\right)\bigg[2\potH{k}{\mu j}\potHin{q}{\mu\sigma}\potH{p}{\sigma i}+\potH{k}{\mu\nu}\potHmi{q}{\mu}{j}\potHmi{p}{\nu}{i}\notag \\
&\qquad\qquad\qquad\qquad\qquad\qquad\quad+\frac{1}{2}\potH{k}{ij}\potH{q}{\rho\sigma}\potHin{p}{\rho\sigma}-\frac{1}{2}\potH{k}{}\potHmi{q}{j}{\rho}\potHin{p}{i\rho}\bigg]\bigg\} \; ,\label{eq:3vertH}\\
\Lagr^{(3)}_{\potH{k}{}\radh{}} & = \frac{1}{\Mpl}\intvec{k}\bigg\{\modul{k}^2\radh{00}\bigg[-\frac{1}{4}\potH{k}{\mu\nu}\potHin{-k}{\mu\nu}+\frac{1}{8}\potH{k}{}\potH{-k}{}+\potH{k}{0\mu}\potHin{-k}{0\mu}-\frac{1}{2}\potH{k}{00}\potH{-k}{}\bigg]\notag \\
&\qquad\qquad\qquad\quad -\modul{k}^2\radh{0i}\left[2\potH{k}{\mu 0}\potHin{-k}{\mu i}-\potHin{k}{0i}\potH{-k}{}\right] \notag \\
&\qquad\qquad\qquad\quad +\radh{ij}\bigg[\modul{k}^2\left(\potHmi{k}{i}{\mu}\potHin{-k}{j\mu}-\frac{1}{2}\potHin{k}{ij}\potH{-k}{}\right)\notag \\
&\qquad\qquad\qquad\qquad\qquad-\frac{k^i k^j}{2}\left(\potH{k}{\mu\nu}\potHin{-k}{\mu\nu}-\frac{1}{2}\potH{k}{}\potH{-k}{}\right)\notag \\
&\qquad\qquad\qquad\qquad\qquad-\modul{k}^2\frac{\eta^{ij}}{4}\left(\potH{k}{\mu\nu}\potHin{-k}{\mu\nu}-\frac{1}{2}\potH{k}{}\potH{-k}{}\right)\bigg]\bigg\} \; .
\label{eq:3vertHhbar}
\end{align}
\end{subequations}
From these Lagrangian one finds the corresponding Feynman rules needed in NRGR. We shall list the complete set of rules in Sec. \ref{App:FRscale}.

\subsubsection{Radiation part} 

For the radiation field things are much easier. In fact, for the computations we made in this work we do not need anything else but the propagator of $\radh{\mu\nu}$, hence we need to find only the explicit expression of the (gauge-fixed) quadratic part of the action, i.e.
\begin{equation}
S_{\text{rad}}[\radh{\mu\nu}] = S_{\radh{}}[\radh{\mu\nu}]+S^{\text{GF}}_{\radh{}}[\radh{\mu\nu}] \; .
\end{equation} 
Again we work in the harmonic gauge; the explicit expression of $S^{\text{GF}}$ is given in eq. (\ref{eq:SGFrad}). From eq. (\ref{eq:quad.grav}) it is then trivial to find
\begin{equation}
S_{\text{rad}}[\radh{\mu\nu}] =\frac{1}{2}\int\!\!d^4x\,\partial_\alpha\radh{\rho\sigma}\left(\eta^{\mu\rho}\eta^{\nu\sigma}-\frac{1}{2}\eta^{\mu\nu}\eta^{\rho\sigma}\right)\partial^\alpha\radh{\mu\nu} \; .
\label{eq:quad.hbar}
\end{equation}
From the above equation we can find the propagator of the radiation graviton.

\section{Point-particle action expansion}\label{sec:PPe}

Now we expand at order $\Ord{v^4}$ the point-particle action
\begin{equation}
S_{\text{pp}}[g_{\mu\nu},x_a^\mu]=-\sum_a\,m_a\int\!\!\sqrt{g_{\mu\nu}dx_a^\mu dx_a^\nu} \; ,
\end{equation}
where $a=1,2$. In order to simplify the notation, we will do the expansion for just one particle, hence dropping the index $a$.

\subsection{First step: flat background expansion}\label{sec:PPflatb}

As in the previous section, we start by splitting the metric as $g_{\mu\nu} = \eta_{\mu\nu}+h_{\mu\nu}/\Mpl$, whence we immediately obtain
\begin{align}
S_{\text{pp}} & = -\,m\int\!\!\left(\eta_{\mu\nu}dx^\mu dx^\nu+\frac{h_{\mu\nu}}{\Mpl}dx^\mu dx^\nu\right)^{1/2} \notag \\
& = \!-m\!\int\!\!d \bar{\tau}\!-\!\frac{m}{2\Mpl}\int\!\!d \bar{\tau}h_{\mu\nu}\frac{dx^\mu}{d \bar{\tau}} \frac{dx^\nu}{d \bar{\tau}}\!+\!\frac{m}{8\Mpl^2}\int\!\!d \bar{\tau}\left(h_{\mu\nu}\frac{dx^\mu}{d \bar{\tau}} \frac{dx^\nu}{d \bar{\tau}}\right)^2\!+\dots
\label{eq:Spp1}
\end{align}
where we defined $d \bar{\tau}^2\equiv\eta_{\mu\nu}dx^\mu dx^\nu$. Now we need to explicit the dependence on the PN parameter $v$. It is straightforward to see that up to $\Ord{v^4}$
\begin{subequations}
\label{eqn:scheme9}
\begin{align}
  d \bar{\tau} & = d t(1-v^2)^{1/2}\simeq d t\left(1-\frac{1}{2}v^2-\frac{1}{8}v^4\right) \; , \\
  \frac{dx^\mu}{d \bar{\tau}} & = \frac{1}{(1-v^2)^{1/2}}\frac{dx^\mu}{d t} \simeq \left(1+\frac{1}{2}v^2+\frac{3}{8}v^4\right)(1,\bm{v}) \; .
\end{align}
\end{subequations}
Therefore, eq. (\ref{eq:Spp1}) immediately becomes
\begin{align}
S_{\text{pp}} & = - m\int\!\!d t\left(1-\frac{1}{2}v^2-\frac{1}{8}v^4\right)\notag \\
 & \ -\!\frac{m}{2\Mpl}\int\!\!d t\left(h_{00}\!+\!2h_{0i}v^i\!+\!h_{ij}v^iv^j\!+\!\frac{h_{00}}{2}v^2\!+\!h_{0i}v^iv^2\!+\!\frac{h_{ij}}{2}v^iv^jv^2\!+\!\frac{3}{8}h_{00}v^4\right)  \notag \\
 &\ +\frac{m}{8\Mpl^2}\int\!\!d t\bigg[h_{00}^2+4h_{00}h_{0i}v^i+4\left(h_{0i}v^i\right)^2+2h_{00}h_{ij}v^iv^j+\frac{3}{2}h^2_{00}v^2\notag \\
 &\qquad\qquad\qquad +4h_{0i}h_{jk}v^iv^jv^k+6h_{00}h_{0i}v^iv^2+\left(h_{ij}v^iv^j\right)^2+6\left(h_{0i}v^i\right)^2v^2\notag \\
  &\qquad\qquad\qquad +3h_{00}h_{ij}v^iv^jv^2+\frac{15}{8}h_{00}v^4\bigg] \; .
\end{align}

\subsection{Second step: potential + radiation modes decomposition}

The next step is to split the $h_{\mu\nu}$ field into potential and radiation mode. Recalling the scaling rules we wrote in tab. \ref{table:scaling}, we shall write all terms that are at most of order $v^{4}$. It is not hard to rewrite
\begin{equation}
S_{\text{pp}}[\potH{k}{\mu\nu},\radh{\mu\nu},x^\mu] \!=\! S^{0}_{\text{pp}}[x^\mu]\!+\!S^{H}_{\text{pp}}[\potH{k}{\mu\nu},x^\mu]\!+\!S^{\radh{}}_{\text{pp}}[\radh{\mu\nu},x^\mu]\!+\!S^{H\radh{}}_{\text{pp}}[\potH{k}{\mu\nu},\radh{\mu\nu},x^\mu] \; ,
\end{equation}
where
\begin{subequations}
\label{eqn:scheme10}
\begin{align}
S^{0}_{\text{pp}} & = - m\int\!\!d t\left(1-\frac{1}{2}v^2-\frac{1}{8}v^4\right) \; ,\label{eq:Spp0}
\end{align}
\begin{align}
S^{H}_{\text{pp}} & = -\frac{m}{2\Mpl}\int\!\!d t\intvec{k}e^{i\bm{k}\cdot\bm{x}}\bigg(\potH{k}{00} + 2\potH{k}{0i}v^i + \potH{k}{ij}v^iv^j +	\frac{\potH{k}{00}}{2}v^2 \notag \\
&\qquad\qquad\qquad\qquad\qquad\qquad+ \potH{k}{0i}v^iv^2 + \frac{\potH{k}{ij}}{2}v^iv^jv^2 + \frac{3}{8}\potH{k}{00}v^4\bigg) \notag \\
&\ +\frac{m}{8\Mpl^2}\int\!\!d t\intvec{k,q}e^{i(\bm{k}+\bm{q})\cdot\bm{x}}\bigg(\potH{k}{00}\potH{q}{00}\!+\!4\potH{k}{00}\potH{q}{0i}v^i\!+\!4\potH{k}{0i}\potH{q}{0j}v^iv^j\! \notag \\
&\qquad\qquad\qquad\qquad\qquad\qquad +2\potH{k}{00}\potH{q}{ij}v^iv^j\!+\frac{3}{2}\potH{k}{00}\potH{q}{00}v^2\!\notag \\
&\qquad\qquad\qquad\qquad\qquad\qquad+ \!4\potH{k}{0i}\potH{q}{j\ell }v^iv^jv^\ell \!+\!6\potH{k}{00}\potH{q}{0i}v^iv^2\bigg) \; , \label{eq:SppH}\\
S^{\radh{}}_{\text{pp}} & = -\frac{m}{2\Mpl}\int\!\!d t\left(\radh{00}+2\radh{0i}v^i+\radh{ij}v^iv^j+\frac{\radh{00}}{2}v^2+\radh{0i}v^iv^2\right) \notag \\
 &\qquad +\frac{m}{8\Mpl^2}\int\!\!d t\bigg(\radh{00}^2+4\radh{00}\radh{0i}v^i\bigg) \; , \label{eq:Spphrad}\\
 S^{H\radh{}}_{\text{pp}} & = \frac{m}{8\Mpl^2}\int\!\!d t\intvec{k}e^{i\bm{k}\cdot\bm{x}}\bigg[2\potH{k}{00}\radh{00}+4\left(\potH{k}{00}\radh{0i}+\radh{00}\potH{k}{0i}\right)v^i\bigg] \; .
 \label{eq:SppHhrad}
\end{align}
\end{subequations}

Actually, as explained in Sec. \ref{sec:multipole}, for the radiation field we need to perform also a multipole expansion, see eq. (\ref{eq:multipole_generic}). Therefore eqs. (\ref{eq:Spphrad}) and (\ref{eq:SppHhrad}) now becomes
\begin{subequations}
\label{eqn:scheme11}
\begin{align}
S^{\radh{}}_{\text{pp}} & = -\frac{m}{2\Mpl}\int\!\!d t\bigg\{\radh{00}+\left[\delta x^i\partial_i\radh{00}+2\radh{0i}v^i\right]+\bigg[\frac{1}{2}\delta x^i\delta x^j\partial_i\partial_j\radh{00}+2\delta x^\ell \partial_\ell \radh{0i}v^i\notag \\
&\qquad\qquad\qquad\qquad+\radh{ij}v^iv^j+\frac{\radh{00}}{2}v^2\bigg]+\bigg[\frac{1}{6}\delta x^i\delta x^j\delta x^k\partial_i\partial_j\partial_k\radh{00} \notag \\
&\qquad\qquad\qquad\qquad+\delta x^j\delta x^k\partial_j\partial_k\radh{0i}v^i+\delta x^\ell \partial_\ell \radh{ij}v^iv^j \notag \\
&\qquad\qquad\qquad\qquad +\frac{1}{2}\delta^i\partial_i\radh{00}v^2+\radh{0i}v^iv^2\bigg]\bigg\}\notag \\
 &\qquad +\frac{m}{8\Mpl^2}\int\!\!d t\bigg\{\radh{00}^2+\left[2\delta x^i\radh{00}\partial_i\radh{00}+4\radh{00}\radh{0i}v^i\right]\bigg\} \label{eq:Spphradmult} \; ,\\
 S^{H\radh{}}_{\text{pp}} & = \frac{m}{8\Mpl^2}\int\!\!d t\intvec{k}e^{i\bm{k}\cdot\bm{x}}\bigg\{2\potH{k}{00}\radh{00}\notag \\
 &\qquad\qquad\qquad\qquad\quad + \big[2\delta x^i\potH{k}{00}\partial_i\radh{00}+4\left(\potH{k}{00}\radh{0i}+\radh{00}\potH{k}{0i}\right)v^i\big]\bigg\} \; ,\label{eq:SppHhradmult}
\end{align}
\end{subequations}
where we decided to do the multipole expansion in the CoM $\bm{x}_{\text{cm}}$, hence every $\radh{\mu\nu}$ is evaluated in $(x^0,\bm{x}{_\text{cm}})$, and we defined $\delta x^i \equiv x^i-x^i_{\text{cm}}$.

\section{Feynman Rules and their scaling}\label{App:FRscale}

Using the actions written in the previous section and tab. \ref{table:scaling}, we now derive the Feynman rules\footnote{See appendix A of Ref. \cite{Ddim-Rules} for a more complete list of Feynman rules for NRGR in general dimension.} and the corresponding scaling in the PN parameter $v$. For the rest of this appendix, we will adopt the drawing conventions of Sec. \ref{sec:mod.dec}.

\subsection{Rules from the Einstein-Hilbert action}

Let's start by the results we got in section \ref{sec:EHe}. The first thing we can find are the propagators of the potential and radiation modes. These has been explicitly derived in Sec. \ref{sec:power.count}, see eqs. (\ref{eq:Hprop}) and (\ref{eq:hbarprop}). Recall that for the potential propagator we also have an infinite series of corrections, the first of which is written in eq. (\ref{eq:corr_prop}). To be more complete, we rewrite this first correction in the first line of tab. \ref{table:gravv}.

In the same table we write also the interaction vertices that we need for the computations made in this work. For the potential part, we only need the interaction vertex between three potential gravitons with polarization $00$, that can be found form eq. (\ref{eq:3vertH}). Then, we also need the non linear interaction vertex between two potential gravitons with polarization $00$ and one radiation graviton considered as an external particle and already multipole expanded. This rule is obtained from the Lagrangian (\ref{eq:3vertHhbar}).
\begin{table}[t]
\begin{center}
\begin{tabular}{|c|c|c|}
\hline
Diagrammatic expression & Scaling & Explicit expression \\
\hline
\begin{tikzpicture}[baseline]
\begin{feynman}
\vertex [label=75:$\bm{q}$, label=285:$\rho\sigma$, label=180:$x^0_2$] (z);
\vertex [crossed dot, right=1.2cm of z] (z') {};
\vertex [right=1.4cm of z', label=105:$\bm{k}$, label=255:$\mu\nu$, label=0:$x^0_1$] (w); 
\diagram* {
(z) -- [scalar] (z') -- [scalar] (w)
}; 
\end{feynman} 
\end{tikzpicture} & $\sim v^2$ & $
-\dfrac{i}{|\bm{k}|^4}(2\pi)^3\delta^{(3)}\left(\bm{k}\!+\!\bm{q}\right)\dfrac{\partial^2}{\partial x^0_2 \partial x^0_1}\delta(x^0_1\!-\!x^0_2)P_{\mu\nu\rho\sigma}
$ \\
\hline  
\begin{tikzpicture}[baseline]
\begin{feynman}
\vertex (z);
\vertex [right=1cm of z, label=0:$x^0_3$] (c);
\vertex [above left=1cm of z, label=135:$x^0_1$] (a);
\vertex [below left=1cm of z, label=225:$x^0_2$] (b);
\diagram* {
(a) -- [scalar, edge label=$\bm{k}_1$] (z) -- [scalar, edge label'=$\bm{k}_3$] (c),
(b) -- [scalar, edge label=$\bm{k}_2$] (z)
}; 
\end{feynman} 
\end{tikzpicture} & 
$\begin{aligned}[t] 
\sim \dfrac{v^2}{L^{1/2}}
\end{aligned}$  & 
\(\displaystyle\begin{aligned}[t]
&-\dfrac{1}{4\Mpl}\delta(x^0_1-x^0_2)\delta(x^0_1-x^0_3) \\
&\times(2\pi)^3\delta^{(3)}\left(\sum_{r=1}^3\bm{k}_r\right)\frac{|\bm{k}_1|^2\!+\!|\bm{k}_2|^2\!+\!|\bm{k}_3|^2}{|\bm{k}_1|^2|\bm{k}_2|^2|\bm{k}_3|^2}\\
&\ 
\end{aligned}\)\\
\hline
\begin{tikzpicture}[baseline]
\begin{feynman}
\vertex (z);
\vertex [right=1cm of z] (c);
\vertex [above left=1cm of z, label=135:$x^0_1$] (a);
\vertex [below left=1cm of z, label=225:$x^0_2$] (b);
\diagram* {
(a) -- [scalar, edge label=$\bm{k}_1$] (z),
(c) -- [gluon] (z),
(b) -- [scalar, edge label=$\bm{k}_2$] (z)
}; 
\end{feynman} 
\end{tikzpicture} & \(\displaystyle \sim \frac{v^{5/2}}{L^{1/2}}\) & 
\(\displaystyle
\begin{aligned}[t]
 &-\frac{i}{2\Mpl}(2\pi)^3\delta(x^0_1\!-\!x_2^0)\frac{\delta^{(3)}\!(\bm{k}_1\!+\!\bm{k}_2)}{|\bm{k}_1|^2|\bm{k}_2|^2} \\
 &\times\left[\frac{3}{2}|\bm{k}_1|^2\radh{00}-\radh{ij}\left(\frac{\eta^{ij}}{2}|\bm{k}_1|^2+k_1^ik_1^j\right)\right] \\
&\ 
\end{aligned}
\)\\
\hline
\end{tabular}
\caption{Pure graviton Feynman rules}\label{table:gravv}
\end{center}
\end{table}

\subsection{Rules from the point-particle action}
Compact objects in NRGR are external sources of gravitational waves, hence we do not have any propagator associated to them. Dropping the constant term from (\ref{eq:Spp0}), needed only when we renormalize the theory, from eq. (\ref{eq:Spp0}) we can immediately find the Feynman rules written in tab. \ref{table:freepp}

\begin{table}[t]
\begin{center}
\begin{tabular}{|c|c|c|}
\hline
Diagrammatic expression & Scaling & Explicit expression \\
\hline
\begin{tikzpicture}[baseline]
\begin{feynman}
\vertex (z);
\vertex [left=0.2cm of z] (c');
\vertex [above=0.6cm of c'] (a);
\vertex [above=0.2cm of a] (a');
\vertex [below=0.6cm of c'] (b);
\vertex [below=0.2cm of b] (b');
\vertex [right=0.7cm of z] (c); 
\diagram* {
(a') -- [draw=none] (a) -- [plain] (c') -- [plain] (b) -- [draw=none] (b'),
(c') -- [draw=none] (c)
}; 
\end{feynman} 
\end{tikzpicture} & $\sim L$ & \(\displaystyle
i\int\!\!d t\,\frac{m_a}{2}v^2_a \) \\
\hline 
\begin{tikzpicture}[baseline]
\begin{feynman}
\vertex (z);
\vertex [empty dot, minimum size=0.4cm, left=0.2cm of z] (c') {2};
\vertex [above=0.7cm of c'] (a);
\vertex [above=0.2cm of a] (a');
\vertex [below=0.7cm of c'] (b);
\vertex [below=0.2cm of b] (b');
\vertex [right=0.7cm of z] (c); 
\diagram* {
(a') -- [draw=none] (a) -- [plain] (c') -- [plain] (b) -- [draw=none] (b'),
(c') -- [draw=none] (c)
}; 
\end{feynman} 
\end{tikzpicture} & $\sim L v^2$ & 
\(\displaystyle
i\int\!\!d t\,\frac{m_a}{8}v^4_a \) \\
\hline 
\end{tabular}
\caption{Free point particle Feynman rules}\label{table:freepp}
\end{center}
\end{table}

\begin{table}[t]
\begin{center}
\begin{tabular}{|c|c|c|}
\hline
Diagrammatic expression & Scaling & Explicit expression \\
\hline
\begin{tikzpicture}[baseline]
\begin{feynman}
\vertex (z);
\vertex [left=0.2cm of z, label=180:$t$, label=75:$\bm{k}$, label=290:$\mu\nu$] (c');
\vertex [above=0.6cm of c'] (a);
\vertex [above=0.2cm of a] (a');
\vertex [below=0.6cm of c'] (b);
\vertex [below=0.2cm of b] (b');
\vertex [right=0.7cm of z] (c); 
\diagram* {
(a') -- [draw=none] (a) -- [plain] (c') -- [plain] (b) -- [draw=none] (b'),
(c') -- [scalar] (c)
}; 
\end{feynman} 
\end{tikzpicture} & $\sim L^{1/2}$ & \(\displaystyle
-i\frac{m_a}{2\Mpl}\int\!\!d t\!\!\intvec{k}e^{i\bm{k}\cdot\bm{x}_a}\eta^{0\mu}\eta^{0\nu} \) \\
\hline 
\begin{tikzpicture}[baseline]
\begin{feynman}
\vertex (z);
\vertex [empty dot, minimum size=0.4cm, left=0.2cm of z, label=180:$t$, label=75:$\bm{k}$, label=290:$\mu\nu$] (c') {1};
\vertex [above=0.7cm of c'] (a);
\vertex [above=0.2cm of a] (a');
\vertex [below=0.7cm of c'] (b);
\vertex [below=0.2cm of b] (b');
\vertex [right=0.7cm of z] (c); 
\diagram* {
(a') -- [draw=none] (a) -- [plain] (c') -- [plain] (b) -- [draw=none] (b'),
(c') -- [scalar] (c)
}; 
\end{feynman} 
\end{tikzpicture} & $\sim L^{1/2}v$ & 
\(\displaystyle
-i\frac{m_a}{\Mpl}\int\!\!d t\!\!\intvec{k}e^{i\bm{k}\cdot\bm{x}_a}v_a^i\eta^{0(\mu}\eta{^{\nu)}}{_i} \) \\
\hline 
\begin{tikzpicture}[baseline]
\begin{feynman}
\vertex (z);
\vertex [empty dot, minimum size=0.4cm, left=0.2cm of z, label=180:$t$, label=75:$\bm{k}$, label=290:$\mu\nu$] (c') {2};
\vertex [above=0.7cm of c'] (a);
\vertex [above=0.2cm of a] (a');
\vertex [below=0.7cm of c'] (b);
\vertex [below=0.2cm of b] (b');
\vertex [right=0.7cm of z] (c); 
\diagram* {
(a') -- [draw=none] (a) -- [plain] (c') -- [plain] (b) -- [draw=none] (b'),
(c') -- [scalar] (c)
}; 
\end{feynman} 
\end{tikzpicture} & $\sim L^{1/2}v^2$ & 
\(\displaystyle
\begin{aligned}[t]
-i\frac{m_a}{2\Mpl}\!\int\!\!d t\!\!\intvec{k}e^{i\bm{k}\cdot\bm{x}_a}\!\bigg(&\eta{_i}{^\mu}\eta{_j}{^\nu}v_a^iv_a^j \\
& \quad+\!\frac{1}{2}\eta^{0\mu}\eta^{0\nu}v^2_a\bigg) \\
& \ 
\end{aligned}
\)\\
\hline 
\begin{tikzpicture}[baseline]
\begin{feynman}
\vertex (z);
\vertex [left=0.2cm of z, label=180:$t$, label=75:$\bm{\!\!k}$, label=290:$\bm{\!\!q}$] (c');
\vertex [above=0.7cm of c'] (a);
\vertex [above=0.2cm of a] (a');
\vertex [below=0.7cm of c'] (b);
\vertex [below=0.2cm of b] (b');
\vertex [above right=0.8cm of z, label=90:$\mu\nu$] (c); 
\vertex [below right=0.8cm of z, label=270:$\rho\sigma$] (d); 
\diagram* {
(a') -- [draw=none] (a) -- [plain] (c') -- [plain] (b) -- [draw=none] (b'),
(c') -- [scalar] (c),
(c') -- [scalar] (d)
}; 
\end{feynman} 
\end{tikzpicture} & $\sim v^2$ & 
\(\displaystyle
i\frac{m_a}{4\Mpl^2}\int\!\!d t\!\!\intvec{k,q}e^{i(\bm{k}+\bm{q})\cdot\bm{x}_a}\eta^{0\mu}\eta^{0\nu}\eta^{0\rho}\eta^{0\sigma} \) \\
\hline
\end{tabular}
\caption{Point particle - potential graviton interactions}\label{table:pppot}
\end{center}
\end{table}

Then we consider the interaction between the compact objects and the potential gravitons; from eq. (\ref{eq:SppH}) we get the rules listed in the tab \ref{table:pppot}.

Next, we consider the interactions between the compact objects and the radiation gravitons, which, we recall, are on-shell particles. From eq. (\ref{eq:Spphrad}) one gets the rules listed in tab. \ref{table:pprad}.
\begin{table}[H]
\begin{center}
\begin{tabular}{|c|c|c|}
\hline
Diagrammatic expression & Scaling & Explicit expression \\
\hline
\begin{tikzpicture}[baseline]
\begin{feynman}
\vertex (z);
\vertex [left=0.2cm of z, label=180:$t$] (c');
\vertex [above=0.5cm of c'] (a);
\vertex [above=0.2cm of a] (a');
\vertex [below=0.5cm of c'] (b);
\vertex [below=0.2cm of b] (b');
\vertex [right=0.7cm of z] (c); 
\diagram* {
(a') -- [draw=none] (a) -- [plain] (c') -- [plain] (b) -- [draw=none] (b'),
(c) -- [gluon] (c')
}; 
\end{feynman} 
\end{tikzpicture} & $\sim L^{1/2}v^{1/2}$ & \(\displaystyle
-i\frac{m_a}{2\Mpl}\int\!\!d t\,\radh{00} \) \\
\hline 
\begin{tikzpicture}[baseline]
\begin{feynman}
\vertex (z);
\vertex [empty dot, minimum size=0.4cm, left=0.2cm of z, label=180:$t$] (c') {1};
\vertex [above=0.6cm of c'] (a);
\vertex [above=0.2cm of a] (a');
\vertex [below=0.6cm of c'] (b);
\vertex [below=0.2cm of b] (b');
\vertex [right=0.7cm of z] (c); 
\diagram* {
(a') -- [draw=none] (a) -- [plain] (c') -- [plain] (b) -- [draw=none] (b'),
(c) -- [gluon] (c')
}; 
\end{feynman} 
\end{tikzpicture} & $\sim L^{1/2}v^{3/2}$ & 
\(\displaystyle
-i\frac{m_a}{\Mpl}\int\!\!d t\,\radh{0i}v_a^i \) \\
\hline 
\begin{tikzpicture}[baseline]
\begin{feynman}
\vertex (z);
\vertex [empty dot, minimum size=0.4cm, left=0.2cm of z, label=180:$t$] (c') {2};
\vertex [above=0.6cm of c'] (a);
\vertex [above=0.2cm of a] (a');
\vertex [below=0.6cm of c'] (b);
\vertex [below=0.2cm of b] (b');
\vertex [right=0.7cm of z] (c); 
\diagram* {
(a') -- [draw=none] (a) -- [plain] (c') -- [plain] (b) -- [draw=none] (b'),
(c) -- [gluon] (c')
}; 
\end{feynman} 
\end{tikzpicture} & $\sim L^{1/2}v^{5/2}$ & 
\(\displaystyle
-i\frac{m_a}{2\Mpl}\int\!\!d t\left(\radh{ij}v_a^iv_a^j+\frac{\radh{00}}{2}v_a^2\right)\) \\
\hline
\end{tabular}
\caption{Point particle - radiation interactions}\label{table:pprad}
\end{center}
\end{table}
\noindent At this stage, the graviton $\radh{\mu\nu}$ is still not multipole expanded. These rules are the one used in the bottom up approach implemented in Sec. \ref{sec:bott.up}.

Then, performing the needed multipole expansion, we derive from eq. (\ref{eqn:scheme11}a) the Feynman rules listed in tab \ref{table:ppradmulti}. These are used for the computations in Ch. \ref{ch:first} and Ch. \ref{ch:rad}, Sec. \ref{sec:SNR}. Because of the multipole expansion $\bar{h}_{\mu\nu}$ is evaluated at $(x^0, \bm{x}_{\text{cm}})$.
\begin{table}[H]
\begin{center}
\begin{tabular}{|c|c|c|}
\hline
Diagrammatic expression & Scaling & Explicit expression \\
\hline
\begin{tikzpicture}[baseline]
\begin{feynman}
\vertex (z);
\vertex [left=0.2cm of z, label=180:$t$] (c');
\vertex [above=0.5cm of c'] (a);
\vertex [above=0.2cm of a] (a');
\vertex [below=0.5cm of c'] (b);
\vertex [below=0.2cm of b] (b');
\vertex [right=0.7cm of z] (c); 
\diagram* {
(a') -- [draw=none] (a) -- [plain] (c') -- [plain] (b) -- [draw=none] (b'),
(c) -- [gluon] (c')
}; 
\end{feynman} 
\end{tikzpicture} & $\sim L^{1/2}v^{1/2}$ & \(\displaystyle
-i\frac{m_a}{2\Mpl}\int\!\!d t\,\bar{h}_{00} \) \\
\hline 
\begin{tikzpicture}[baseline]
\begin{feynman}
\vertex (z);
\vertex [empty dot, minimum size=0.4cm, left=0.2cm of z, label=180:$t$] (c') {1};
\vertex [above=0.6cm of c'] (a);
\vertex [above=0.2cm of a] (a');
\vertex [below=0.6cm of c'] (b);
\vertex [below=0.2cm of b] (b');
\vertex [right=0.7cm of z] (c); 
\diagram* {
(a') -- [draw=none] (a) -- [plain] (c') -- [plain] (b) -- [draw=none] (b'),
(c) -- [gluon] (c')
}; 
\end{feynman} 
\end{tikzpicture} & $\sim L^{1/2}v^{3/2}$ & 
\(\displaystyle
-i\frac{m_a}{2\Mpl}\int\!\!d t\,\left(\delta x_a^i\partial_i\bar{h}_{00}+2\bar{h}_{0i}v_a^i\right) \) \\
\hline 
\begin{tikzpicture}[baseline]
\begin{feynman}
\vertex (z);
\vertex [empty dot, minimum size=0.4cm, left=0.2cm of z, label=180:$t$] (c') {2};
\vertex [above=0.6cm of c'] (a);
\vertex [above=0.2cm of a] (a');
\vertex [below=0.6cm of c'] (b);
\vertex [below=0.2cm of b] (b');
\vertex [right=0.7cm of z] (c); 
\diagram* {
(a') -- [draw=none] (a) -- [plain] (c') -- [plain] (b) -- [draw=none] (b'),
(c) -- [gluon] (c')
}; 
\end{feynman} 
\end{tikzpicture} & $\sim L^{1/2}v^{5/2}$ & 
\(\displaystyle \begin{aligned}[t]
-i\frac{m_a}{2\Mpl}&\int\!\!d t \bigg(\frac{1}{2}\bar{h}_{00}v_a^2+\bar{h}_{ij}v_a^i v_a^j\\
& +2\delta x_a^\ell \partial_\ell  \bar{h}_{0i}v^i_a + \frac{1}{2}\delta x_a^i\delta x_a^j\partial_i\partial_j\bar{h}_{00}\bigg) \\
& \ \end{aligned}\) \\
\hline 
\end{tabular}
\caption{Point particle - radiation interaction with multipole expansion}\label{table:ppradmulti}
\end{center}
\end{table}

Finally, from eq. (\ref{eq:SppHhrad}) one obtains the needed Feynman rule for the interaction between the point-particle and one potential and one radiation graviton. 
\begin{table}[H]
\begin{center}
\begin{tabular}{|c|c|c|}
\hline
Diagrammatic expression & Scaling & Explicit expression \\
\hline 
\begin{tikzpicture}[baseline]
\begin{feynman}
\vertex (z);
\vertex [left=0.2cm of z, label=180:$t$, label=75:$\!\!\bm{k}$] (c');
\vertex [above=0.6cm of c'] (a);
\vertex [above=0.2cm of a] (a');
\vertex [below=0.6cm of c'] (b);
\vertex [below=0.2cm of b] (b');
\vertex [above right=0.8cm of z, label=90:$\mu\nu$] (c); 
\vertex [below right=0.8cm of z] (d); 
\diagram* {
(a') -- [draw=none] (a) -- [plain] (c') -- [plain] (b) -- [draw=none] (b'),
(c') -- [scalar] (c),
(d) -- [gluon] (c')
}; 
\end{feynman} 
\end{tikzpicture} & $\sim v^{5/2}$ & 
\(\displaystyle
i\frac{m_a}{4\Mpl^2}\int\!\!d t\!\!\intvec{k}e^{i\bm{k}\cdot\bm{x}_a}\bar{h}_{00}\eta^{0\mu}\eta^{0\nu} \) \\
\hline
\end{tabular}
\caption{Point particle - potential - radiation interaction vertex}\label{table:pppotrad}
\end{center}
\end{table}
In this case the rule has the same form before and after the multipole expansion. The only difference is that, once we have multipole expanded the action, $\radh{00}$ is evaluated at $(x^0, \bm{x}_{\text{cm}})$.

\chapter{Coupling to conserved quantities}\label{App:cons.quant}

Let's consider the following diagram
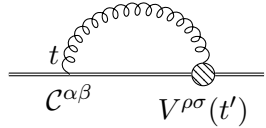
\begin{center}
\begin{tikzpicture}[baseline]
\begin{feynman}
\vertex (f);
\vertex [above=0.1cm of f] (z);
\vertex [right=0.8cm of z, label=270:$\mathcal{C}^{\alpha\beta}$, label=100:$t$] (c);
\vertex [right=1.6cm of c, blob, minimum size=0.3cm, label=270:$V^{\rho\sigma}(t')$] (c') {};
\vertex [right=0.8cm of c'] (w);  
\diagram* {
(z) -- [double] (c) -- [double] (c') -- [double] (w),
(c') -- [gluon, half right] (c)
}; 
\end{feynman} 
\end{tikzpicture} 
\end{center}
where $\mathcal{C}^{\alpha\beta}$ is a conserved quantity (to a certain PN order) and $V^{\rho\sigma}(t)$ is a completely generic vertex. We here report the computation made in \cite{rad.rec} to prove that these kind of diagrams gives a vanishing contribution to the effective action. 

\section{Master integral}\label{sec:mast.int}

First we need to understand what is the behaviour of the following integral
\begin{equation}
I^{(D)}_{\text{M}}(n,p,q) = \int_{-\infty}^{+\infty}\!\!d s\int\!\!\frtr{D-1}{\bm{k}}P\!\!\int\!\!\frac{d k^0}{2\pi}\frac{e^{ik^0s}}{(k^0)^2-\modul{k}^2}s^n(k^0)^pk_{i_1}\dots k_{i_q} \; ,
\label{eq:IM.1}
\end{equation} 
where $P$ stands for the Cauchy principal value. If $q$ is odd, then the above integral is zero. When $q$ is even we can consider
\begin{equation}
\int\!\!d s\,s^ne^{ik^0s} = \frac{1}{(i)^n}\frac{d^n}{d (k^0)^n}\int\!\!d s\,e^{ik^0s} = \frac{2\pi}{(i)^n}\frac{d^n}{d (k^0)^n}\delta(k^0) \; ,
\end{equation}
then, eq. (\ref{eq:IM.1}) becomes
\begin{align}
I^{(D)}_{\text{M}}(n,p,q) & = (-i)^n\int\!\!\frtr{D-1}{\bm{k}}P\int\!\!d k^0\frac{(k^0)^pk_{i_1}\dots k_{i_q}}{(k^0)^2-\modul{k}^2}\frac{d^n}{d (k^0)^n}\delta(k^0) = \notag \\
& = (-)^n(-i)^n\left.\left\{\frac{d^n}{d (k^0)^n}\left[(k^0)^p\int\!\!\frtr{D-1}{\bm{k}}\frac{k_{i_1}\dots k_{i_q}}{(k^0)^2-\modul{k}^2}\right]\right\}\right|_{k^0=0} \; .
\end{align}
Under the integral we can rewrite
\begin{equation}
k_{i_1}\dots k_{i_q} =\frac{\modul{k}^q}{(q+1)!!}\delta_{i_1\dots i_q} \; ,
\end{equation}
where we used the shorthand notation
\begin{equation}
\delta_{i_1\dots i_q} \equiv \Big(\delta_{i_1i_2}\dots\delta_{i_{q-1}i_q}+\left\{\text{all other possible parings}\right\}\,\Big)  \; .
\end{equation}

Therefore going in spherical coordinates we get
\begin{align}
I^{(D)}_{\text{M}}(n,p,q) & =\frac{(i)^n}{(q+1)!!}\left\{\frac{d^n}{d (k^0)^n}\left[\frac{(k^0)^p}{(2\pi)^{D-1}}\int\!d \Omega_{D-1} \right.\right.\notag \\
&\qquad\qquad\qquad\times\left.\left.\left.\!\int_0^{\infty}\!\!d \modul{k}\frac{\modul{k}^{D-2+q}}{(k^0)^2-\modul{k}^2}\delta_{i_1\dots i_q}\right]\right\}\right|_{k^0=0} \notag \\
& = \frac{(i)^{n-D+3}\sec\left(\frac{\pi D}{2}\right)}{2^{D-1}\pi^{(D-3)/2}\Gamma\left(\frac{D-1}{2}\right)}\frac{\delta_{i_1\dots i_q}}{(q+1)!!}\notag \\
&\qquad\qquad\qquad\left.\times\frac{(p+q+D-3)!}{(p+q+D-3-n)!}(k^0)^{p+q+D-3-n}\right|_{k^0=0} \; .
\end{align}
So eventually we arrive to our final result
\begin{equation}
I^{(D)}_{\text{M}}(n,p,q) = \begin{cases}
0 & \text{if } n\neq p+q+D-3 \\
\ & \ \\
\dfrac{(i)^{n-D+3}\sec\left(\frac{\pi D}{2}\right)}{2^{D-1}\pi^{(D-3)/2}\Gamma\left(\frac{D-1}{2}\right)}\dfrac{\delta_{i_1\dots i_q}}{(q+1)!!}\,n! & \text{if } n = p+q+D-3
\end{cases}
\end{equation}
In $D=4$ dimension in particular
\begin{equation}
I^{(4)}_{\text{M}}(n,p,q) = \begin{cases}
0 & \text{if } n\neq p+q+1 \\
\ & \ \\
\dfrac{i^{n-1}n!}{4\pi(q+1)!!}\delta_{i_1\dots i_q} & \text{if } n = p+q+1
\end{cases} \; .
\label{eq:mast.int}
\end{equation}

\section{Computation of the diagram}

Let's now go back to the computation of 
\begin{align}
\begin{tikzpicture}[baseline]
\begin{feynman}
\vertex (f);
\vertex [above=0.1cm of f] (z);
\vertex [right=0.8cm of z, label=270:$\mathcal{C}^{\alpha\beta}$, label=100:$t$] (c);
\vertex [right=1.6cm of c, blob, minimum size=0.3cm, label=270:$V^{\rho\sigma}(t')$] (c') {};
\vertex [right=0.8cm of c'] (w);  
\diagram* {
(z) -- [double] (c) -- [double] (c') -- [double] (w),
(c') -- [gluon, half right] (c)
}; 
\end{feynman} 
\end{tikzpicture} & =\int\!\!d td t'i\mathcal{C}^{\alpha\beta}P_{\alpha\beta\rho\sigma}D(t-t',0)iV^{\rho\sigma}(t')  \notag \\
& = -i\mathcal{C}^{\alpha\beta}P_{\alpha\beta\rho\sigma}\int\!\!d td t'\intvec{k}\!\int\!\!\frac{d k^0}{(2\pi)}\frac{e^{-ik^0(t-t')}}{(k^0)^2-\modul{k}^2+i\varepsilon}V^{\rho\sigma}(t')  \notag \\
& = -i\mathcal{C}^{\alpha\beta}P_{\alpha\beta\rho\sigma}\int\!\!d td t'\Bigg\{\intvec{k}P\!\!\int\!\!\frac{d k^0}{(2\pi)}\frac{e^{-ik^0(t-t')}}{(k^0)^2-\modul{k}}V^{\rho\sigma}(t') \notag \\
&\qquad -i\pi\intvec{k}\!\int\!\!\frac{d k^0}{(2\pi)}\,e^{-ik^0(t-t')}\delta\Big((k^0)^2-\modul{k}^2\Big)V^{\rho\sigma}(t') \Bigg\} \notag \\
& = -i\mathcal{C}^{\alpha\beta}P_{\alpha\beta\rho\sigma}\left\{\mathcal{I}^{\rho\sigma}_1+\mathcal{I}^{\rho\sigma}_2\right\} \; ,
\label{eq:cons.quant.comp}
\end{align}
where the radiation propagator is evaluated in $\bm{x} = 0$ because of the multipole expansion, we used the usual relation (\ref{eq:Pl.Sok}) to go from the second to the third line, and defined
\begin{align}
\mathcal{I}^{\rho\sigma}_1 & \equiv \int\!\!d td t'\intvec{k}P\!\!\int\!\!\frac{d k^0}{(2\pi)}\frac{e^{-ik^0(t-t')}}{(k^0)^2-\modul{k}}V^{\rho\sigma}(t') \; ,\\
\mathcal{I}^{\rho\sigma}_2 & \equiv -i\pi\int\!\!d td t'\!\intvec{k}\!\int\!\!\frac{d k^0}{(2\pi)}\,e^{-ik^0(t-t')}\delta\Big((k^0)^2-\modul{k}^2\Big)V^{\rho\sigma}(t') \; .
\end{align}
Now we solve the two integrals separately:
\begin{itemize}
\item replacing $t'=t+s$ in $\mathcal{I}^{\rho\sigma}_1$ we get
\begin{align}
\mathcal{I}^{\rho\sigma}_1 & =\! \int\!\!d td s\intvec{k}P\!\!\int\!\!\frac{d k^0}{(2\pi)}\frac{e^{-ik^0s}}{(k^0)^2-\modul{k}}V^{\rho\sigma}(t+s) \notag \\
& = \int\!\!d t\sum_{n=0}^{\infty}\frac{1}{n!}\frac{d^n}{d t^n}V^{\rho\sigma}(t)\int\!\!d s\intvec{k}P\int\!\!\frac{d k^0}{2\pi}\frac{s^ne^{ik^0s}}{(k^0)^2-\modul{k}^2} \; ,
\end{align}
where in the second step we Taylor expanded $V^{\rho\sigma}(t\!+\!s)$ and we sent $k^0\to -k^0$. In the second line of the above equation we recognise the integral $I^{(4)}_{\text{M}}(n,0,0)$. Thus, according to (\ref{eq:mast.int}), only $n=1$ gives a non-vanishing contribution
\begin{align}
\mathcal{I}^{\rho\sigma}_1 \!=\! \frac{1}{4\pi}\int\!\!d t\frac{d}{d t}V^{\rho\sigma}(t) \; .
\end{align}
Being a surface term, $\mathcal{I}^{\rho\sigma}_1$ does not contribute to (\ref{eq:cons.quant.comp})
\item the integral $\mathcal{I}^{\rho\sigma}_2$ is straightforward to compute
\begin{align}
\mathcal{I}^{\rho\sigma}_2 & = -i\pi\int\!\!d t\!\intvec{k}\int\!\!\frac{d k^0}{(2\pi)}e^{-ik^0t}\delta\Big((k^0)^2\!-\!\modul{k}^2\Big)\int\!\!d t'e^{ik^0t'}V^{\rho\sigma}(t') \notag \\
& = -i\pi\int\!\!\frac{d k^0}{(2\pi)}\!\intvec{k}\!\left(\int\!\!d te^{-ik^0t}\right)\delta\Big((k^0)^2\!-\!\modul{k}^2\Big)\tilde{V}^{\rho\sigma}(k^0)  \notag \\
& = -i\pi\int\!\!\frac{d k^0}{(2\pi)}\!\intvec{k}(2\pi)\delta(k^0)\delta\Big((k^0)^2\!-\!\modul{k}^2\Big)\tilde{V}^{\rho\sigma}(k^0)  \notag \\
& = -i\pi\tilde{V}^{\rho\sigma}(0)\intvec{k}\delta\Big(\modul{k}^2\Big) = 0 \; ,
\end{align}
where we denote with a tilde the time Fourier transform of $V^{\rho\sigma}$.
\end{itemize}
Therefore we get the final result that we used Sec. \ref{sec:power_leading}
\begin{equation}
\begin{tikzpicture}[baseline]
\begin{feynman}
\vertex (f);
\vertex [above=0.1cm of f] (z);
\vertex [right=0.8cm of z, label=270:$\mathcal{C}^{\alpha\beta}$, label=100:$t$] (c);
\vertex [right=1.6cm of c, blob, minimum size=0.3cm, label=270:$V^{\rho\sigma}(t')$] (c') {};
\vertex [right=0.8cm of c'] (w);  
\diagram* {
(z) -- [double] (c) -- [double] (c') -- [double] (w),
(c') -- [gluon, half right] (c)
}; 
\end{feynman} 
\end{tikzpicture} = 0 \; .
\end{equation}

\chapter{Concerning \textit{in-out} and \textit{in-in} formalism}\label{App:inininout}
 
In this appendix we briefly discuss in-out and in-in path integral formalisms in order to justify at least the result of our main text. For a more complete discussion on this subject we suggest Ref. \cite{rad.rec} and references therein.

Before proceeding, we recall the different Green functions we can have in a field theory. Given
\begin{align}
\Delta^+(x) =\int\!\!\frtr{3}{\bm{k}}\frac{e^{-i\omega_{\bm{k}}t+i\bm{k}\cdot\bm{x}}}{2\omega_{\bm{k}}} \; , & & \Delta^-(x) = \Delta^+(-t,\bm{x}) = \Delta^+(-x) \; ,
\end{align}
where as usual $\omega_{\bm{k}} = \sqrt{\modul{k}^2+m^2}$, we can define:
\begin{subequations}
\label{eqn:scheme23}
\begin{itemize}
\item the Feynman  and anti-Feynman (or Dyson) two-point functions
\begin{align}
D(x,y) & = \vartheta\Big(x^0\!-\!y^0\Big)\Delta^+(x-y)+\vartheta\Big(-(x^0\!-\!y^0)\Big)\Delta^-(x-y) \; ,\\
D_{\text{D}}(x,y) & = \vartheta\Big(-(x^0\!-\!y^0)\Big)\Delta^+(x-y)+\vartheta\Big(x^0\!-\!y^0\Big)\Delta^-(x-y) \; .
\end{align}
\item The Hadamard and anti-Hadamard two point-functions
\begin{align}
D_{\text{H}}(x,y) & = \Delta^+(x-y)+\Delta^-(x-y) \; ,\label{eq:had.2point}\\
D_{\text{C}}(x,y) & = \Delta^+(x-y)-\Delta^-(x-y) \; .
\end{align}
\item The retarded and advanced two point-functions
\begin{align}
D_{\text{Rt}}(x,y) & = i\vartheta\Big(x^0-y^0\Big)D_{\text{C}}(x-y) \; ,\\
D_{\text{Ad}}(x,y) & = -i\vartheta\Big(-(x^0-y^0)\Big)D_{\text{C}}(x-y) \; .
\end{align}
\end{itemize}
\end{subequations}
From the above definitions, we can find the following useful identities
\begin{subequations}
\label{eqn:scheme24}
\begin{align}
-iD_{\text{Rt}}(x,y) & = D(x,y)-\Delta^-(x,y) = \Delta^+(x,y)-D_{\text{D}}(x,y) \; , \\
iD_{\text{Ad}}(x,y) & = D_{\text{D}}(x,y)-\Delta^-(x,y) = \Delta^+(x,y) - D(x,y) \; .
\end{align}
\end{subequations}

\section{One-point Green function in the in-out formalism}

In what follows we will consider a much simpler theory, namely scalar field theory in flat space coupled linearly to a source $Q(x)$ 
\begin{equation}
S = \int\!\!d^4\!x\left[\frac{1}{2}\left(\partial_\mu\varphi\partial^\mu\varphi-m^2\varphi^2\right)+Q\varphi\right] \; .
\end{equation}
Then the generating functional in this case is define as
\begin{equation}
Z[J] = e^{iW[J]} = \Path\varphi\exp\left\{i\int\!\!d^4\!x\left[\frac{1}{2}\left(\partial_\mu\varphi\partial^\mu\varphi-m^2\varphi^2\right)+Q\varphi+J\varphi\right]\right\} \; .
\end{equation}
In the interaction picture, we know that the evolution of a state is realised by the operator
\begin{equation}
\mathcal{U}_{J+Q}(t,t')\equiv  \Tprod{\exp\left\{i\int_{t'}^t\!\!d t\!\int\!\!d^3\!\bm{x}(Q+J)\varphi\right\}\,} \; .
\end{equation}
Therefore we can immediately rewrite 
\begin{align}
Z[J] = e^{iW[J]} & = \Path\varphi\,\mathcal{U}_{J+Q}(\infty,-\infty)\exp\left\{i\int\!\!d^4\!x\frac{1}{2}\left(\partial_\mu\varphi\partial^\mu\varphi-m^2\varphi^2\right)\right\}  \notag \\
& = {_\text{out}}\bangle{0\left|\mathcal{U}_{J+Q}(\infty,-\infty)\right|0}_{\text{in}} \; .
\label{eq:gen.func.scal}
\end{align}
$\ket{0}_{\text{in}}$ and $\ket{0}_{\text{out}}$ are the vacuum state respectively in the far past and in the far future. In principle they do not coincide; however when $Q$ and $J$ are adiabatically turned on/off, as usual in QFT, we finds that they do coincide and that the vacuum state can be chosen to be the preferred one selected by the Poincaré symmetry $\ket{0}$. 

From eq. (\ref{eq:gen.func.scal}) we can easily find the one point Green function as
\begin{equation}
\bangle{\varphi(x)}_{\text{in-out}} = \left.\frac{\delta W[J]}{\delta J(x)}\right|_{J=0} \; .
\end{equation}
We shall now see that this is not the vacuum expectation value of the field in the Heisenberg picture. Indeed we know that the relation between the Heisenberg and the interaction picture is the following
\begin{equation}
\varphi(x) = \mathcal{U}_{J+Q}(t, \infty)\varphi_\text{H}(x)\mathcal{U}_{J+Q}(-\infty,t) \; .
\end{equation}
Therefore, we obtain 
\begin{align}
\bangle{\varphi(x)}_{\text{in-out}} & = \left.\frac{\delta W[J]}{\delta J(x)}\right|_{J=0}   \notag \\
& = \left.\bangle{0\left|\mathcal{U}_{Q+J}(\infty,t)\varphi(x)\mathcal{U}_{Q+J}(t',-\infty)\right|0}\right|_{J=0}  \notag \\
& = \bangle{0\left|\mathcal{U}_{Q}(\infty,-\infty)\varphi_{\text{H}}(x)\right|0}\neq \bangle{0\left|\varphi_{\text{H}}(x)\right|0} \; .
\end{align}
As expected, $\bangle{\varphi(x)}_{\text{in-out}}$ is not the vacuum expectation value of the field. Moreover in the in-out formalism we do not get a causal propagation of the field. In fact it is easy to show that
\begin{equation}
\bangle{\varphi(x)}_{\text{in-out}} = i\int\!\!d^4\!x\,D(x-x')Q(x') \; .
\end{equation}
This is not a retarded (hence causal) propagation.

\section{The in-in formalism: expectation value and causality}\label{sec:inin.app}

We want now to construct a formalism such that
\begin{equation}
\bangle{\varphi(x)}_{\text{in-in}} = \bangle{0\left|\varphi_{\text{H}}(x)\right|0} \; .
\end{equation}
Unless otherwise noted, from now on the vacuum state is the in-vacuum $\ket{0}_{\text{in}}$. This is accomplished by using the so-called in-in path integral formalism, also known as the \textit{closed-time-path} formalism (CTP). 

The idea of this new formalism is to introduce two different currents $J_1$ and $J_2$ to construct the generator of connected diagrams; these currents are such that:
\begin{itemize}
\item $J_1$ couples to the field $\varphi_1$ in forward evolution
\item $J_2$ couples to the field $\varphi_2$  for backward evolution.
\end{itemize}
Then we have to add the condition $\varphi_1=\varphi_2$ at $t=\infty$, hence the name CTP formalism. Using the notation introduced in the previous section, we can say that
\begin{align}
e^{iW[J_1,J_2]} & = \bangle{0\left|\mathcal{U}_{J_2+Q}(-\infty,\infty)\mathcal{U}_{J_1+Q}(\infty,-\infty)\right|0} \notag \\
& = \int\!\!\Diff\varphi_1\Diff\varphi_2\exp\left\{iS[\varphi_1]-iS[\varphi_2]+i\int\!\!d^4\!xJ_1\varphi_1-i\int\!\!d^4\!xJ_2\varphi_2\right\} \; .
\label{eq:gen.inin}
\end{align}
From here one can immediately see that
\begin{subequations}
\label{eqn:scheme25}
\begin{align}
\bangle{\varphi_1(x)}_{\text{in-in}} & = \left.\frac{\delta W[J_1,J_2]}{\delta J_1(x)}\right|_{J_1=J_2=0} \notag \\
& = \bangle{0\left|\mathcal{U}_{Q}(-\infty,\infty)\mathcal{U}(\infty,t)\varphi(x)\mathcal{U}_{Q}(t, -\infty)\right|0} \notag \\
& =  \bangle{0\left|\mathcal{U}_{Q}(-\infty,t)\varphi(x)\mathcal{U}_{Q}(t, -\infty)\right|0}  \; ,\\
\bangle{\varphi_2(x)}_{\text{in-in}} & = \left.\frac{\delta W[J_1,J_2]}{\delta J_2(x)}\right|_{J_1=J_2=0} \notag \\
& = \bangle{0\left|\mathcal{U}_{Q}(-\infty,t)\varphi(x)\mathcal{U}_{Q}(t,\infty)\mathcal{U}_{Q}(\infty,-\infty)\right|0} = \notag \\
& =  \bangle{0\left|\mathcal{U}_{Q}(-\infty,t)\varphi(x)\mathcal{U}_{Q}(t, -\infty)\right|0} \; .
\end{align}
\end{subequations}
Hence, in the in-in formalism we get the ``right'' vacuum expectation value
\begin{equation}
\bangle{\varphi_1(x)}_{\text{in-in}} = \bangle{\varphi_2(x)}_{\text{in-in}} =\bangle{\varphi(x)}_{\text{in-in}} = \bangle{0\left|\varphi_{\text{H}}(x)\right|0} \; .
\end{equation}

Following the usual procedure that one can find in any QFT book \cite{Srednicki,Itsy}, from eq. (\ref{eq:gen.inin}) it is not hard to see that 
\begin{equation}
W[J_1,J_2] = \frac{i}{2}\int\!\!d^4\!x d^4\!y\big(J_A(x)+Q_A(x)\big)D^{AB}(x,y)\big(J_B(y)+Q_B(y)\big) \; ,
\label{eq:W.inin}
\end{equation}
where $A,B=1,2$, and
\begin{equation}
D^{AB} = \left(\begin{matrix}
D & -\Delta^- \\
-\Delta^+ & D_{\text{D}}
\end{matrix}\right) \; .
\end{equation}
The Einstein convention on repeated indices holds also for the upper case labels; in this case one uses the identity $\delta_{AB}$ to raise/lower indices. From here we can immediately find that
\begin{align}
\bangle{\varphi(x)}_{\text{in-in}} & = \frac{i}{2}\int\!\!d^4\!y\left(D^{1B}(x-y)Q_B(y)+Q_A(y)D^{A1}(y-x)\right) \; .
\end{align}
In this case $Q$ does not depend on dynamical quantities, hence 
\begin{equation}
Q_1=Q_2=Q \; ,
\label{eq:Q_Q_inin}
\end{equation}
therefore, eventually,
\begin{align}
\bangle{\varphi(x)}_{\text{in-in}} & = \frac{i}{2}\int\!\!d^4\!y\left(2D(x-y)-(\Delta^-(x\!-\!y)+\Delta^+(y\!-\!x)\right)Q(y) \notag \\
& = i\int\!\!d^4\!y\big(D(x-y)-\Delta^-(x\!-\!y)\big)Q(y) \notag \\
& = \int\!\!d^4\!y\,D_{\text{Rt}}(x-y)Q(y) \; .
\label{eq:one-poit_inin}
\end{align}
The in-in formalism then solve also the problem of non causal propagation.

\subsection{The Keldysh representation}\label{subsec:Keldysh}

A very useful representation for the in-in formalism is the so-called \textit{Keldysh representation}. This consists in reperametrising $J_1$ and $J_2$ by defining 
\begin{align}
J_- \equiv J_1-J_2 \; , & & J_+\equiv\frac{1}{2}(J_1+J_2) \; ,
\end{align}
and analogously for all the other variables. it is easy to see then that $W[J_+,J_-]$ has the same form as (\ref{eq:W.inin}), where now $A,B=+,-$ and
\begin{equation}
D^{AB} = \left(\begin{matrix}
D^{++} & D^{+-} \\
D^{-+} & D^{--}
\end{matrix}\right) = \left(\begin{matrix}
0 & -iD_{\text{Ad}} \\
-iD_{\text{Rt}} & \frac{1}{2}D_{\text{H}}
\end{matrix}\right) \; .
\label{eq:matrix.prop}
\end{equation}
For the Keldysh indices $A, B$, however, we cannot use the identity to raise/lower indices, but we need the symmetric matrix
\begin{align}
C_{AB} = \left(\begin{matrix}
C_{++} & C_{+-} \\
C_{-+} & C_{--}
\end{matrix}\right)=\left(\begin{matrix}
0 & 1 \\
1 & 0
\end{matrix}\right)\; ,  & & C^{AB} = \left(\begin{matrix}
C^{++} & C^{+-} \\
C^{-+} & C^{--}
\end{matrix}\right)=\left(\begin{matrix}
0 & 1 \\
1 & 0
\end{matrix}\right)  \; .
\end{align}
This implies, in particular, that $J_{\pm} = J^{\mp}$. In this representation, starting from (\ref{eq:W.inin}), we can easily find $W[J_+, J_-]$  to be
\begin{equation}
W[J_+,J_-] = \frac{i}{2}\int\!\!d^4\!x d^4\!y\big(J_A(x)+Q_A(x)\big)D^{AB}(x,y)\big(J_B(y)+Q_B(y)\big) \; .
\end{equation}
Recalling eq. (\ref{eq:Q_Q_inin}), we can find that the expectation value of the one point function in the Keldysh representation is given by
\begin{equation}
\bangle{\varphi(x)}_{\text{in-in}} = \left.\frac{\delta W[J_1,J_2]}{\delta J_-(x)}\right|_{J_-=J_+=0} = \left.\frac{\delta W[J_1,J_2]}{\delta J^+(x)}\right|_{J_-=J_+=0} \; .
\label{eq:one_point_inin_K}
\end{equation}
It is not hard to show that this value is indeed equal to  (\ref{eq:one-poit_inin}). 

\chapter{The Cartan equations and non-zero torsion}\label{ch.forms.torsion}

In the first part of this appendix, starting by the definition of torsion and curvature tensor (\ref{eqn:schem58}), we derive the \textit{two Cartan structure equation}, showing that the methods we used in the text to compute these objects in a non-coordinate basis is completely equivalent to the one employed in \cite{KK2}. Since, as we saw in Sec. \ref{susec:non-coord-basis} we end up with an apparent non-zero torsion, we devote Sec. \ref{sec:torsion} to a brief discussion on GR with torsion. 

\section{The two Cartan structure equations}\label{sec:Cart-eq}

Let's work in a non coordinate basis $\overline{e}_a$ and $\underline{\vartheta}^a$, defined through vielbein fields as in eq. (\ref{eq:new_basis}). Denoting with $\omega{^c}_{ab}$ the affine connection, with $T{_{ab}}^{c}$ the torsion and with $\Rc{^d}_{cab}$ the curvature, we can introduce the connection one form $\underline{\omega}{^c}_a$, the torsion two form ${}\ord{2}\underline{T}^c$ and the curvature two from ${}\ord{2}\underline{\Rc}{^d}_a$ as follows
\begin{align}
\underline{\omega}{^c}_a \equiv \omega{^c}_{ab} \underline{\vartheta}^b \; , &&
{}\ord{2}\underline{T}^c \equiv \frac{1}{2}T{_{ab}}{^c} \underline{\vartheta}^a\wedge\underline{\vartheta}^b \; ,& &
{}\ord{2}\underline{\Rc}{^d}_a \equiv \frac{1}{2}\Rc{^d}_{cab} \underline{\vartheta}^a\wedge\underline{\vartheta}^b  \; ,
\label{eq:bunch_of_eq}
\end{align}
These quantities are connected by the so-called \textit{Cartan structure equations}
\begin{subequations}
\label{eqn:scheme43}
\begin{align}
{}\ord{2}\underline{T}^c & = d\underline{\vartheta}^{c}+\underline{\omega}{^c}_{b}\wedge\underline{\vartheta}^{b} \; , \\
{}\ord{2}\underline{\Rc}{^d}_a & = d\underline{\omega}{^d}_{a} +\underline{\omega}{^d}_e\wedge\underline{\omega}{^e}_a \; .
\end{align}
\end{subequations}
We will now show that these two equations used in \cite{KK2} are completely equivalent to eqs. (\ref{eqn:scheme62}), that we used to find the affine connection and the curvature in the non-coordinate basis. Since eqs. (\ref{eqn:scheme62}) are just the components version of eqs. (\ref{eqn:schem58}), we understand that the Cartan equation are nothing but another way of defining the curvature and the torsion. 

\subsubsection{The first Cartan Structure equation} 

Let's consider eq. (\ref{eqn:scheme62}a) and let's multiply both sides of this equation by $(1/2)\underline{\vartheta}^a\wedge \underline{\vartheta}^b$. On the right-hand-side we simply get the torsion two form ${}\ord{2}\underline{T}^c$ defined in (\ref{eq:bunch_of_eq}). On the left-hand-side instead we get two contributions: the first is given by
\begin{equation}
\omega{^c}_{[ba]}\underline{\vartheta}^a\wedge \underline{\vartheta}^b = \underline{\omega}{^c}_{b}\wedge \underline{\vartheta}^b \; ,
\label{eq:first_LHS}
\end{equation}
where we used the antisymmetric property of the wedge product of forms, and we recognised the connection one form defined again in (\ref{eq:bunch_of_eq}). The second term of the left-hand-side instead becomes
\begin{align}
\frac{1}{2}\Omega{_{ab}}^c\underline{\vartheta}^a\wedge \underline{\vartheta}^b & = -e{_a}{^\mu}\partial_\mu e{_b}{^\nu}\vartheta{^c}_\nu \underline{\vartheta}^a\wedge \underline{\vartheta}^b \notag \\
& = e{_a}{^\mu}e{_b}{^\nu}\left(\partial_\mu\vartheta{^c}_\nu\right)\vartheta{^a}_\rho\vartheta{^b}_\sigma dx^\rho\wedge dx^\sigma \notag \\
& = \partial_\rho\vartheta{^c}_\sigma dx^\rho\wedge dx^\sigma = d\underline{\vartheta}^c \; ,
\label{eq:second_LHS}
\end{align}
where to go from the first to the second line we integrated by parts and used one of the identity written in (\ref{eq:delta_vielbein}). Putting together eqs. (\ref{eq:first_LHS}) and (\ref{eq:second_LHS}) we eventually arrive to our final result
\begin{equation}
{}\ord{2}\underline{T}^c = \underline{\omega}{^c}_{b}\wedge \underline{\vartheta}^b + d\underline{\vartheta}^c \; .
\end{equation}

This is precisely the first Cartan structure equation that we were looking for. As we said at the beginning, this is nothing but a redefinition of the torsion tensor given in (\ref{eqn:schem58}a).

\subsubsection{The second Cartan Structure equation} 

We now follow a similar procedure to derive the second Cartan structure eq. (\ref{eqn:scheme43}b). We multiply again both sides of eq. (\ref{eqn:scheme62}b) by $(1/2)\underline{\vartheta}^a\wedge \underline{\vartheta}^b$. Again on the right-hand-side we have the definition of the curvature two form given in (\ref{eq:bunch_of_eq}). On the left-hand-side instead we have again two contributions: the first given by
\begin{align}
\left(\partial_{[a|}\omega{^d}{_{c|b]}} + \omega{^d}_{cf}\Omega{_{ab}}{^e}\right)\underline{\vartheta}^a\wedge\underline{\vartheta}^b & = \left(\vartheta{^b}_\nu\partial_\mu\omega{^d}_{cb} + \omega{^d}_{cb}\partial_\mu\vartheta{^b}_\nu\right)dx^\mu\wedge dx^\nu  \notag \\
& = \partial_\mu\left(\omega{^d}_{cb}\vartheta{^b}_\nu\right)dx^\mu\wedge dx^\nu  = d\underline{\omega}{^d}_c \; ,
\label{eq:first_LHS_c}
\end{align}
where to go from the first to the second step we used the definition (\ref{eq:def_der_frame}) and the result we got in eq. (\ref{eq:second_LHS}). The second contribution come from 
\begin{equation}
\omega{^d}{_{e[a|}}\omega{^e}{_{c|b]}}\,\underline{\vartheta}^a\wedge\underline{\vartheta}^b = \omega{^d}{_{ea}}\underline{\vartheta}^a\wedge\omega{^e}{_{cb}}\underline{\vartheta}^b = \underline{\omega}{^d}_e\wedge \underline{\omega}{^e}_c \; .
\label{eq:second_LHS_c}
\end{equation}
Putting together eqs. (\ref{eq:first_LHS_c}) and (\ref{eq:second_LHS_c}), we rewrite (\ref{eqn:scheme43}b) as
\begin{equation}
{}\ord{2}\underline{\Rc}{^d}_a = d\underline{\omega}{^d}_c + \underline{\omega}{^d}_e\wedge \underline{\omega}{^e}_c
\end{equation}

We correctly re-derived also the second Cartan structure equation, showing that it is indeed equivalent to the definition of curvature (\ref{eqn:schem58}b).

\section{Generic system with non-zero torsion}\label{sec:torsion}

Now let's use again the canonical coordinate basis $\partial_\mu$ and $dx^\mu$. Let $g_{\mu\nu}$ be the metric of a generic manifold, and let's call its affine connection $\omega{^\rho}_{\mu\nu}$. We know that if the manifold has a non-zero torsion, then, starting from (\ref{eqn:schem58}a), we can find the components $T{_{\mu\nu}}{^\rho}$ as
\begin{equation}
T{_{\uwidehat{\mu\nu}}}{^\rho} = 2\omega{^\rho}_{[\nu\mu]} \; .
\label{eq:torsion.def}
\end{equation}
In this section we use $\uwidehat{\mu\nu\dots}$ to stress that a group of indices in a tensor is completely antisymmetric; similarly we use $\underline{\mu\nu\dots}$ to denote a group of completely symmetric indices.

We impose then the metricity condition $\nabla_\rho g_{\mu\nu} = 0$. From here we can write 
\begin{subequations}
\label{eqn:scheme38}
\begin{align}
\nabla_{\rho}g_{\mu\nu} & = \partial_\rho g_{\mu\nu}-\omega{^\sigma}_{\mu\rho}g_{\sigma\nu}-\omega{^\sigma}_{\nu\rho}g_{\mu\sigma}=0 \; ,\\
\nabla_{\nu}g_{\mu\rho} & = \partial_\nu g_{\mu\rho}-\omega{^\sigma}_{\mu\nu}g_{\sigma\rho}-\omega{^\sigma}_{\rho\nu}g_{\mu\sigma}=0 \; ,\\
-\nabla_{\mu}g_{\rho\nu} & = -\partial_\mu g_{\rho\nu}+\omega{^\sigma}_{\rho\mu}g_{\sigma\nu}+\omega{^\sigma}_{\nu\mu}g_{\rho\sigma}=0 \; .
\end{align}
\end{subequations}
Summing up these three equations one gets eventually
\begin{align}
\left(\partial_\nu g_{\mu\rho} +\partial_\rho g_{\mu\nu}-\partial_\mu g_{\rho\nu}\right) & + g_{\sigma\nu}\left(\omega{^\sigma}_{\rho\mu}-\omega{^\sigma}_{\mu\rho}\right)\notag \\
& + g_{\rho\sigma}\left(\omega{^\sigma}_{\nu\mu}-\omega{^\sigma}_{\mu\nu}\right) \notag \\
& -g_{\mu\sigma}\left(\omega{^\sigma}_{\nu\rho}+\omega{^\sigma}_{\rho\nu}\right) = 0 \; .\label{eq:step.tor}
\end{align}
The first term in round brackets is nothing but two times the Christoffel symbol $2\Gamma_{\mu\underline{\nu\rho}}$, which is of course symmetric in the last two indices; recalling the definition of the torsion tensor we can rewrite (\ref{eq:step.tor}) as
\begin{equation}
2\Gamma_{\mu\nu\rho}+T_{\mu\rho\nu}+T_{\mu\nu\rho}-2\omega_{\mu\nu\rho}-T_{\nu\rho\mu} = 0 \; .
\end{equation}
From here we can easily find the explicit expression of the connection
\begin{equation}
\omega_{\mu\nu\rho} = \Gamma_{\mu\underline{\nu\rho}} +\frac{1}{2}\left(T_{\uwidehat{\mu\nu}\rho}-T_{\uwidehat{\nu\rho}\mu}-T_{\uwidehat{\rho\mu}\nu}\right) \; .
\end{equation}
It is customary to introduce the so call contorsion tensor
\begin{equation}
\Sigma_{\uwidehat{\mu\nu}\rho} \equiv \frac{1}{2}\left(T_{\uwidehat{\mu\nu}\rho}-T_{\uwidehat{\nu\rho}\mu}-T_{\uwidehat{\rho\mu}\nu}\right) \; .
\label{eq:cont.tens}
\end{equation}
it is not hard to see that, by definition, also the contorsion tensor is antisymmetric in the first two indices. Therefore eventually we write the affine connection as
\begin{equation}
\omega_{\mu\nu\rho} = \Gamma_{\mu\underline{\nu\rho}}+\Sigma_{\uwidehat{\mu\nu}\rho} \; .
\label{eq:aff_con_cont}
\end{equation}

We immediately see that eq. (\ref{eq:nedd_app_tor}) is exactly equivalent to (\ref{eq:torsion.def}), with the torsion given by the anholonomy coefficients. We can then redo exactly the steps we wrote above in order to find the components of the affine connection as in eq. (\ref{eq:aff_con_cont}). 

\subsection{Different definitions of the torsion}\label{sec:diff_tors_def}

We finally stress that different authors use different conventions in defining the torsion tensor components or covariant derivative. Using a different convention eventually leads to a different definition of the contorsion tensor which differ from (\ref{eq:cont.tens}), and this may be a source of confusion and errors. In the following table we list some possible conventions we found in the most common general relativity books.

\begin{table}[H]
\begin{center}
\begin{tabular}{|c|c|c|}
\hline
Covariant Derivative & Torsion & Reference \\
\hline
\(\displaystyle \begin{aligned}[c] \ & \\ \nabla_{\mu}v^{\nu} & = \partial_\mu v^{\nu}+\omega{^\nu}_{\rho\mu}v^\rho \\ \nabla_{\mu}w_{\nu} & = \partial_\mu w_{\nu}-\omega{^\rho}_{\nu\mu}w_\rho \\ & \ \end{aligned}\) & \(\displaystyle T{_{\uwidehat{\mu\nu}}}{^\rho} = 2\omega{^\rho}_{[\nu\mu]}\) & Misner Thorne and Wheeler's book \cite{MTW} \\
\hline
 \(\displaystyle \begin{aligned}[c] \ & \\ \nabla_{\mu}v^{\nu} & = \partial_\mu v^{\nu}+\omega{^\nu}_{\mu\rho}v^\rho \\ \nabla_{\mu}w_{\nu} & = \partial_\mu w_{\nu}-\omega{^\rho}_{\mu\nu}w_\rho \\ \ & \end{aligned}\) & \(\displaystyle T{_{\uwidehat{\mu\nu}}}{^\rho} = 2\omega{^\rho}_{[\mu\nu]}\) & Carrol's book \cite{Sean-Carrol} \\
\hline 
\(\displaystyle \begin{aligned}[c] \  & \\ \nabla_{\mu}v^{\nu} & = \partial_\mu v^{\nu}+\omega{^\nu}_{\mu\rho}v^\rho \\ \nabla_{\mu}w_{\nu} & = \partial_\mu w_{\nu}-\omega{^\rho}_{\mu\nu}w_\rho \\ \ & \end{aligned}\) & \(\displaystyle T{^\rho} {_{\uwidehat{\mu\nu}}}= 2\omega{^\rho}_{[\mu\nu]}\) & Wald's book \cite{Wald} \\
\hline 
\end{tabular}
\end{center}
\end{table}

Given this, one can easily find the corresponding definition of the contorsion tensor imposing the metricity condition as in eqs. (\ref{eqn:scheme38}).

\nocite{*}
\newpage
\thispagestyle{empty}
\bibliographystyle{utphys}
\bibliography{biblio_eft}

\end{document}